\documentclass[3p,sort&compress]{elsarticle}
 \usepackage[colorlinks=false]{hyperref}
 \hypersetup{hidelinks=true,}
 \def\url#1{}
 \def\urlprefix{ }
\usepackage{epstopdf}
\usepackage{fleqn}
\usepackage[polutonikogreek,english]{babel}

\usepackage{graphicx}
\usepackage{amssymb,latexsym,amsthm,amsmath,dsfont,mathtools}
\DeclareMathOperator{\sgn}{sgn}
\usepackage{color}
\usepackage[utf8x]{inputenc}
\usepackage[T1]{fontenc}
\usepackage{mathdots}
\usepackage{multirow}

\newcommand{\greek}[1]{{\selectlanguage{polutonikogreek}#1}}

\usepackage{bm} % bold math
\newcommand{\be}{\begin{equation}}
\newcommand{\ee}{\end{equation}}
\newcommand{\ba}{\begin{eqnarray}}
\newcommand{\ea}{\end{eqnarray}}
\def\bea{\begin{eqnarray}}
\def\eea{\end{eqnarray}}

\newcommand{\avk}{\langle k\rangle}

\usepackage{titlesec}
\titlespacing\section{0pt}{14pt}{14pt}
\titlespacing\subsection{0pt}{14pt}{14pt}
\titlespacing\subsection{0pt}{6pt}{6pt}

\journal{Physics Reports}

\begin{document}

\begin{frontmatter}

\title{Explosive transitions in complex networks' structure and dynamics: percolation and synchronization}

\author[Shangai,SB,SB2]{S. Boccaletti}
\address[Shangai]{Department of Physics, East China Normal University, Shanghai 200062, China}
\address[SB]{CNR- Institute of Complex Systems, Via Madonna del Piano, 10, 50019 Sesto Fiorentino, Florence, Italy}
\address[SB2]{The Italian Embassy in Israel, 25 Hamered st., 68125 Tel  Aviv, Israel}
\ead{stefano.boccaletti@gmail.com}

\author[URJC,CTB]{J.A. Almendral}
\address[URJC]{Complex Systems Group, Universidad Rey Juan Carlos,  28933 M\'ostoles, Madrid, Spain}
\address[CTB]{Center for Biomedical Technology, Universidad
  Polit\'ecnica de Madrid,  28223 Pozuelo de Alarc\'on, Madrid, Spain}
\ead{juan.almendral@gmail.com}

\author[Shangai]{S. Guan\corref{corrauthor}}
\cortext[corrauthor]{Corresponding author at the Department of Physics, East China Normal University, Shanghai 200062, China}
\ead{sgguan@phy.ecnu.edu.cn}

\author[URJC,CTB]{I. Leyva\corref{corrauthor2}}
\cortext[corrauthor2]{Corresponding author at the Complex Systems Group, Universidad Rey Juan Carlos,  28933 M\'ostoles, Madrid, Spain}
\ead{inmaculada.leyva@gmail.com}

\author[Shangai]{Z. Liu\corref{corrauthor}}
\ead{zhliu@phy.ecnu.edu.cn}

\author[URJC,CTB]{I. Sendi\~na-Nadal}
\ead{irene.sendina@urjc.es}

\author[Xian]{Z. Wang}
\address[Xian]{School of Automation, Northwestern Polytechnical University, Xi'an 710072, China}
\ead{zhenwang0@gmail.com}
\author[Shangai]{Y. Zou}
\ead{yzou@phy.ecnu.edu.cn}

\begin{abstract}
Percolation and synchronization are two phase transitions that have been extensively studied since already long ago.
A classic result is that, in the vast majority of cases, these
transitions are of the second-order type, i.e. continuous and
reversible. Recently, however, explosive phenomena have been reported in complex networks' structure and dynamics,
which rather remind first-order (discontinuous and irreversible) transitions.
Explosive percolation, which was discovered in 2009, corresponds to an abrupt change in the
network's structure, and explosive synchronization (which is concerned,
instead, with the abrupt emergence of a collective state in the
networks' dynamics) was studied as early as the first models of globally coupled phase oscillators
were taken into consideration.
The two phenomena have stimulated investigations and debates,
attracting attention in many relevant fields. So far, various substantial contributions and progresses
(including experimental verifications) have been made, which have provided insights on what structural and
dynamical properties are needed for inducing such abrupt
transformations, as well as have greatly enhanced our
understanding of phase transitions in networked systems. Our intention
is to offer here a monographic review on the main-stream
literature, with the twofold aim of summarizing the existing results
and pointing out possible directions for future research.
\end{abstract}
% \begin{keyword}
% %\MSC[2010] 00-01\sep  99-00
% \end{keyword}
\end{frontmatter}
\tableofcontents
%\listoffigures
%\linenumbers
%
%%%%%%%%%%%%%%%%%%%%%%%%%%%%%%%%%%%%%%%%%%%%%%%%%%%%%%%%
\section{Introduction}

\subsection{Phase transitions}

With the term {\it phase} of a system physicists actually denote one of its states,
characterized by a set of physical properties that can be there considered
as uniform over a macroscopic length scale.

Now, from the melting of ice into water to ferro-magnetism and from
superconductivity to protein folding, nature is daily exhibiting {\it phase
transitions}, i.e. processes in which systems drastically change one or many
of their physical properties when a minimal variation in one of their control variables occurs.

In simpler cases, such transitions are described by the variation of the so
called {\it order parameter} (a scalar, or a vector, or even a tensor measuring
the degree of order in the system, whose magnitude normally ranges between
zero in one phase and nonzero in the other), as the control variable (or the
{\it control parameter}) is gradually and adiabatically changed.

More precisely, let us exemplify the simplest case of a system having only two
phases (say phase 1 and phase 2), and passing from one to another with a process
involving scalar control and order parameters. To gather a quantitative
description and characterization of the transition, the following procedure is
usually run after: {\it i)} the system is prepared in one of its two phases (say
phase 1), which actually corresponds to set a specific initial value of the control
parameter; {\it ii)} the control parameter is then gradually changed (increased, or
decreased, depending on the particular case) in very small steps, up to the
point at which the other phase is attained; {\it iii)} the corresponding
variations of the order parameter are monitored. This procedure can be followed
by starting either from phase 1 or from phase 2, producing the curves that
describe the {\it forward} and {\it backward} transitions, respectively.

Now, a fundamental concept in statistical mechanics is that of {\it reversibility}
of a given process, which, indeed, can be carried out reversibly or irreversibly.
Reversibility can be attributed to processes which, after having
taken place from an initial system's phase, can be {\it fully reversed}, i.e. they can be brought back to
their original phase via the exactly opposite
transition, and with no changes in either the systems or their surroundings.
In our case (as well as it happens in many other circumstances) reversibility consists in a full equivalence of the two curves obtained
during the forward and backward transitions. When, instead, transitions are
irreversible, the two forward and backward curves differ, and a wealth of
interesting and relevant events, such as hysteretic loops, occur.

Given the ubiquitous presence of phase transitions in the real world, it is
not strange that physicists started, since the very beginning of statistical mechanics,
to delve into discovering universal and critical
behaviors (or scaling properties in the proximity of the transition points), as
well as to pay effort in classifying phase transitions. The first attempt
of providing a classification was given by Ehrenfest \cite{Jaeger1998}, but more modern
classification schemes consider phase transitions as divided into two broad
categories: the first-order and the second-order ones.

In thermodynamics,
first-order phase transitions are those that involve a latent heat, i.e. those
during which the system either absorbs or releases a typically large amount of
energy per volume. Normally, they are characterized (at the transition point)
by a ``mixed-phase regime'', in which some parts of the system have completed the
transition and others have not, this way featuring a sort of {\it coexistence} of the two
system's phases. Signatures of such transitions are the abrupt and discontinuous
behavior of the order parameter in the proximity of the transition point, as
well as the intrinsic irreversibility of the transition, with the presence
(in the majority of the cases) of hysteretic loops.
At variance, second-order
phase transitions (also called continuous phase transitions) are reversible, and
correspond to order parameters displaying a continuous behavior near
criticality. Examples of the former processes are the melting of ice, or the
boiling of water, or the super-cooling and super-heating transitions. Examples
of the latter processes are the ferromagnetic transition, the super-conducting
transition, and the super-fluid transition.

\subsection{Percolation and synchronization}

Our main scope is reviewing recent progresses on two classical
phase transitions taking place in networked units (percolation and
synchronization), as well as on how they can in fact be rendered {\it explosive},
i.e. {\it first-order-like}.

Percolation (from the Latin {\it percolare}, which means ``to filter'' or ``to
trickle through'') is a transition involving a drastic change in the network’s
structure, due to the emergence of a giant connected component whose size is of
the order of the network’s size. By means of percolation, a phase of the network
where fragmented portions, or clusters, are isolated is changed into a fully
connected structure, where nodes can globally interact. The process can be
conducted in two ways: either starting from isolated nodes and adding gradually
links (what is called {\it bond percolation}), or sequentially adding nodes to
the graph (what is called {\it site percolation}).

Synchronization (from the Greek \greek{σύγ}=together and \greek{χρóνος}=time) is
instead a process having to do with network's dynamics. There, dynamical
systems adjust some properties of their trajectories (due to their interactions, or to a
driving force) so that the network eventually operates in a collective and
macroscopically coherent way.

Percolation and synchronization are nowadays classical fields of study in
statistical mechanics, and an interested reader can find orientation within the wealth of currently available
results in various monographic papers and books
\cite{Stauffer1992,Bollobas2006,Blekhman1988,Boccaletti2002,Strogatz2003,Boccaletti2008,Pikovsky2001}.

A common result is that the vast majority of transitions to
percolation and synchronization are of the second-order type,
continuous and reversible.
However, as soon as networked units with complex architectures of interaction are taken into consideration,
two abrupt and explosive phenomena have been recently reported, namely
explosive percolation \cite{Achlioptas2009,Cho2011a,daCosta2010,Grassberger2011,Riordan2011}
and explosive synchronization \cite{Pazo2005,Gomez-Gardenes2011,Leyva2012,Leyva2013a,Leyva2013b,Zhang2014,Zou2014,Zhang2015,Sendina-Nadal2015},
which rather remind first-order like transitions.
Explosive percolation (a sudden materialization of a giant connected component in the structure of the network) was discovered in 2009 \cite{Achlioptas2009}, while explosive synchronization (an abrupt emergence of a collective, synchronized, behavior in network's dynamics) was first described already in 1984 \cite{Kuramoto1984}. In literature, the use of the jargon {\em first-order like} purports that those explosive processes fulfill some of the properties of a first order transition in thermodynamics (as, for instance, discontinuity and irreversibility) and yet they fail featuring {\em all} the attributes, qualities and traits of a true first-order transition.

Unveiling the main mechanisms at the origin of these two abrupt transitions
is of fundamental relevance for a better understanding of networks' structure and dynamics.
For instance, explosive percolation is considered to be at the basis of phenomena such as cascading failures,
which, on their turn, may have devastating effects (and cause huge economic losses) in various circumstances: electrical blackouts in power grids,
financial crises in the network of global financial market, and the spreading of information and rumors through
online social networks, such as Facebook or Twitter.
On the other hand, synchronization is of primary importance in sustaining basic brain functions such as emotions,
complex thoughts, memory, language comprehension, consciousness, etc., while clinical evidence seems to indicate that explosive synchronization is actually the mechanism underlying the transition from normal to pathological brain behavior during epilepsy, one of the world's most prominent brain disorders.

Furthermore, when synchronization occurs through an irreversible transition, one interesting and ingenious application is the construction of
{\it magnetic-like} states of synchronization in networked oscillators: setting the coupling strength inside the hysteresis region of the transition, and entraining (through a unidirectional coupling with an external pacemaker) the phases of the oscillators, the system can be forced to pass from the unsynchronized to the phase-locked configuration, which (once the pacemaker is later switched off) remains actually unalterable (like the case of a permanent magnet).

Given, therefore, their relevance to practical applications, explosive percolation and synchronization have attracted a comprehensive and lasting attention in various fields, and have stimulated a large amount of exciting
works and debates. The result is that many substantial contributions and progresses (including experimental verifications) are today
available, which provide a rather deep insight on the crucial structural and dynamical
mechanisms at the basis of the abruptness of the two transitions.

It is therefore now the time to offer a comprehensive review
on the current state of the art on the subject, which would offer a reflected and thought
out viewpoint on the many achievements and developments (summarizing the existing
models and results), as well as a weighted and meditated outlook to the still open questions and
to the possible directions for future research.

\subsection{Outline of the report}

After the present introduction, the review is organized as follows. In Chapter 2, the main results on explosive percolation are summarized and discussed. Starting from the seminal work by Achlioptas, the Chapter gives a survey on the various processes that have been proposed in order to produce abruptness in the percolation transition, as well as it critically examines the essential physical mechanisms which are at the origin of the explosiveness. The Chapter ends with accounting for the debate stirred within the scientific community, and reports some rigorous conclusions that have been ultimately made on the nature of the transition in the thermodynamic limit (i.e. for systems whose size tends to infinity).

Chapter 3 contains a brief excursus on classical synchronization in both identical and non-identical networked systems. An interested reader
can definitely  find a much more detailed description of the subject in various other books and review articles, like those of Refs. \cite{Blekhman1988,Boccaletti2002,Strogatz2003,Boccaletti2008,Pikovsky2001}, whereas Chapter 3 concentrates on the main ideas and concepts which are then conveniently recalled at the moment of describing those processes where synchronization emerges, instead, abruptly.

In Chapter 4, we provide a detailed overview of explosive synchronization. Starting from the early evidences of first order transitions in the thermodynamic limit of models of globally coupled phase oscillators (as well as from the seminal work of Ref.~\cite{Pazo2005}), the Chapter
describes the cases where the transition to synchronization is first-order-like, and discusses the generality of explosive synchronization in networked systems with complex architectures and topologies.
The Chapter, furthermore, gives an account of some recently discovered and unveiled coherent states (the Bellerophon states), whose existence is intimately related to the possibility for the system to feature explosive transition to synchronization.

Chapter 5 is an exhaustive journey on the applications of explosive percolation and synchronization. In particular, applications of explosive percolation are presented in cascading failures of inter-connected multi-layer networks (of relevance in technology and biology), and applications of explosive synchronization are described in power and smart grids, and in brain dynamics. The Chapter accounts also for experimental realizations of explosive synchronization in controlled laboratory systems.

Finally, Chapter 6 presents our conclusive remarks and perspective ideas on yet open problems to solve, and on possible research lines for the future of the subject.

For the reference of the reader, we summarize here below (in Table~\ref{tab:symabbr}) the list of major symbols and abbreviations which are used all throughout the Manuscript.

\begin{table}\renewcommand{\arraystretch}{1.}
  \caption{Nomenclature and abbreviations used in the
    manuscript.   \label{tab:symabbr}}
  \begin{center}
%\centering
  \small\begin{tabular*}{\textwidth}{|p{1.1cm}@{\extracolsep{\fill}}p{7.5cm} @{\extracolsep{\fill}}p{1.3cm}@{\extracolsep{\fill}}p{5.5cm}|}\hline
%\multicolumn{2}{|l|}{ }  \\[2pt]
\multicolumn{2}{|l}{ \textbf {Nomenclature}}  & \multicolumn{2}{l|}{\textbf {Abbreviations}} \\[5pt]
$A$	&	Adjacency matrix	&	2D	&	Two dimensional\\
$k$	&	Node degree	&	3D	&	Three dimensional\\
$\langle k\rangle$	&	Network average degree	&	AP	&	Achlioptas process\\
$P(k)$	&	Degree distribution	&	BA	&	Barab\'asi-Albert\\
$\cal{A}$	&	Degree-degree correlation 	&	BF	&	Bohman-Frieze \\
$\lambda_i$	&	Eigenvalues of the adjacency matrix &	BFW	&	Bohman-Frieze-Wormald \\
$E$	&	Number of edges	&	CA	&	Coauthorship\\
$\lambda$ 	&	Coupling strength, synchronization control parameter 	&	CM	&	Configuration Model\\
$N$	&	System size	&	CS	&	Complete Synchronization\\
$\theta$	&	Oscillator phase	&	CPT	&	Continuous Phase Transition\\
$\phi$	&	Oscillator phase in the rotating frame	&	DPT	&	Discontinuous Phase Transition\\
$\nu$	&	Frequency	&	dCDGM	&	da Costa-Dorogovtsev-Goltsev-Mendes\\
$\omega$	&	Natural frequency	&	DLCA	&	Diffusion-limited cluster aggregation\\
$g(\omega)$	&	Natural frequency distribution	&	EP	&	Explosive Percolation\\
$\psi$	&	Average phase	&	ER	&	Erd\"os-R\'enyi \\
$\Omega$	&	Average frequency	&	ES	&	Explosive Synchronization\\
$r$	&	Kuramoto Order Parameter	&	GAP	&	Generalized Achlioptas process\\
$S$	&	Percolation order parameter	&	H-PHN	&	Human protein homology network \\
$s_{max}$	&	Size of the largest cluster	&	LCC	&	Largest Connected Component\\
$s$	&	Size of a network cluster	&	MC	&	Minimum Cluster\\
$t$	&	Time, percolation control parameter ($t=E/N$)	&	MSF	&	Master Stability Function\\
    &		&	PR	&	Product Rule\\
	&		&	PT	&	Phase Transition\\
	&		&	SCA	&	Spanning Cluster-Avoiding\\
	&		&	SF	&	Scale-Free\\
	&		&	SP	&	Suppression Principle\\
	&		&	SR	&	Sum Rule\\
	&		&	SW	&	Small-world\\
	&		&	TR	&	Triangle Rule\\
	&		&	TW	&	Traveling Wave\\
\hline  \end{tabular*}
  \end{center}
\end{table}

%%%%%%%%%%%%%%%%%%%%%%%%%%%%%%%%%%%%%%%%%%%%%%%%%%%%%%%%%
\section{Explosive percolation}\label{sec:percol}
\subsection{Introduction and historical overview}

Percolation transition refers to the emergence of a giant connected cluster (or giant connected component)
in a lattice, or a network, when bonds/sites are gradually occupied at random. For simplicity (and unless mentioned otherwise), throughout this review we always consider
bond percolation (i.e. the process through which the nodes of a network are fixed, and links, or bonds, are gradually added).

In bond percolation, one then starts from $N$ unconnected nodes. At
each time step, an edge is added between two nodes selected according
to a given rule. The number of edges ($E$) added to the system at a certain
time step divided by the system size ($N$) is the control parameter ($t=E/N$) that describes the phase transition (PT). As for the order parameter (usually denoted as $S(t)$) one
can take the fraction of nodes belonging to  the giant cluster (the largest cluster) in the network. As time increases, more and more edges are formed in the network, which causes the order parameter to increase from zero. In the thermodynamic limit ($N \to \infty$), $S(t)$ exhibits a PT from zero to $O(1)$ at a critical point $t_c$.

Classical percolation is a typical geometrical PT, which has been extensively
studied since 1940's in many fields, including mathematics,
statistical physics, and engineering.
Numerous percolation models, such as invasion percolation,
first-passage percolation, directed percolation,
bootstrap percolation, $k$-core percolation have been
developed, and it has been shown that PT is closely related
to various applications, such as conducting materials,
fractality of coastlines, turbulence,
magnetic models, colloids, growth models, watersheds of landscapes, the spin quantum
Hall transition, etc. The reader is addressed to recent reviews in Refs.~\cite{Araujo2014,Bastas2014,Saberi2015,DSouza2015} and references therein for an overview of the main results and concepts about classical percolation.

In modern statistical physics, PTs are traditionally classified into two types, namely,
the continuous PT (CPT) and the discontinuous PT (DPT), which are also known as the second-order PT and the first-order PT,
respectively. CPT exhibits universal characteristics, and can thus be categorized into different universality classes. At variance, DPT is non-universal. There are many ingredients that determine whether a phase transition occurs in a system via a CPT or DPT, such as, for instance, dimensionality, the type of the dynamical rules, as well as the underlying network's structures.
%The universality of critical transitions is tightly related to the divergence of a correlation length.

For decades, most transitions in classical (or ordinary) percolation  have been shown to be of CPT  (the second-order) type.
One example is the  prototypical bond percolation in Erd\"{o}s-R{\'e}nyi (ER) networks, for which the size of the giant cluster
smoothly increases as the control parameter exceeds the percolation threshold.
For $t>t_c$, the order parameter $S(t)$ increases with $t$ as $S(t)\propto(t-t_c)^{1/2}$.
%in 1991, bootstrap percolation on hypercubic lattices was shown to exhibit discontinuous transition for certain bounds \cite{Adler1991}.
% review DPT before 2009
It should be pointed out, however, that there are also rare examples of percolation models that exhibit DPT, such as the bootstrap percolation \cite{Adler1991,Holroyd2003},
the $k$-core percolation \cite{Chalupa1979,Adler1991,Dorogovtsev2006},
and the jamming percolation \cite{Echenique2005}.

In 2009, a competitive percolation process was introduced, later known as
the Achlioptas process (AP) \cite{Achlioptas2009}.
AP is essentially different from traditional uncorrelated percolation models. Indeed, at each  time step, two edges are first
randomly selected as potential candidates. Then, a product rule (PR) is applied: the candidate edge that actually minimizes the product  of the sizes of the two clusters containing its end points is eventually established, while the other one is disregarded.
An alternative is the sum rule (SR) in which the established link is the one that minimizes the sum of the two clusters' sizes.
In AP, a giant cluster emerges after a number of steps that is much smaller than the system size. As a consequence,
 the order parameter exhibits an extremely abrupt ``jump'' at the percolation point.
At first glance, such a  behavior resembles a discontinuous
transition and this is the reason why AP was termed ``explosive percolation'' (EP) \cite{Achlioptas2009}.
As we will see later on (in Section~\ref{sec:EPc}), EP can actually be a CPT, or a genuine DPT. %works following AP in different networks

After the initial work by Achlioptas \cite{Achlioptas2009}, extensive studies on EP have been carried out in various configurations, such as 2D lattices \cite{Ziff2009,Radicchi2010,Ziff2010,Bastas2011,
Grassberger2011,Choi2011,Angst2012,Li2012a,Choi2012,Reis2012,
Choi2014}, 3D or high-D lattices \cite{Choi2014}, Bethe lattices \cite{Chae2012},
ER networks \cite{Achlioptas2009,Radicchi2010,Cho2010b,Kim2010,Riordan2011,Grassberger2011,Nagler2011,Cho2011,Li2012a},
scale-free (SF) networks \cite{Cho2009,Radicchi2009,Radicchi2010}, and real-world networks \cite{Rozenfeld2010,Squires2013}.

%{\bf Generalized Achlioptas process (GAP)}
The key factor in AP is the competition (at each time step) between two potential edges before the actual selection of the one which is  added to the network. Such a competition mechanism can be straightforwardly extended from two to more edges,
leading to the so called ``best-of-$m$'' rule, where $m \ge 2$ potential edges are first chosen at each time step. Then, the candidates are evaluated using either PR or SR, and the one minimizing the product or the sum of the two end clusters is eventually
established. For this reason, the algorithm is also known as the ``min-cluster-$m$'' rule.
In the limit case where $m=N$, the ``min-cluster-$m$'' rule becomes the smallest cluster  model \cite{Friedman2009},
i.e., at each time step the two smallest clusters in the network are identified to connect. So far, the ``best-of-$m$'' rule has been studied in 2D lattices \cite{Araujo2011,AndradeJr2011,Giazitzidis2013}, 3D lattices \cite{Giazitzidis2013},
ER networks \cite{Friedman2009,Riordan2011,Nagler2011,
Hooyberghs2011,Qian2012},
and real-world networks \cite{Pan2011,Squires2013}.

Other generalized Achlioptas processes (GAP) have been proposed and investigated, including the da Costa-Dorogovtsev-Goltsev-Mendes (dCDGM) model \cite{daCosta2010} and the
$l$-vertex model \cite{Nagler2012,Riordan2012a,Zhang2013}, which has (for $l=3$) several variants,
such as the adjacent edge  rule \cite{DSouza2010}, the triangle rule (TR) \cite{DSouza2010,Nagler2012},
and the clique-3 competition rule \cite{Nagler2011}.
All  GAP  models have been studied in
2D lattices \cite{Reis2012}, in ER networks \cite{daCosta2010,DSouza2010,Nagler2011,Riordan2011,Cho2011,
Li2012a,Nagler2012,Riordan2012a,Zhang2013,Chen2013,
daCosta2014a,daCosta2014b,daCosta2015a,daCosta2015b},
and in real-world networks \cite{Yi2013}.

The relevant literature regarding AP and GAP models is summarized in Table~\ref{tab:AP}, where the reader can visually (and easily) find information on which specific model has been adopted in which specific network's topology.

%----------------Table I--------------
\begin{table}
\renewcommand{\arraystretch}{1.3}
\caption{Relevant literature concerning AP and GAP models in lattices, random networks, SF networks, and other networks (real-world, directed, and growing networks), including PR, SR, best-of-$m$ rule, dCDGM, and $l$-vertex rules.
Notice that for the best-of-$m$ rule ($m$-min-cluster rule), both PR and SR can be used and they are no longer distinguishable. \label{tab:AP}}
\begin{center}
\footnotesize \begin{tabular*}{\textwidth}{p{5.5em}@{\extracolsep{\fill}}p{8em}@{\extracolsep{\fill}}p{7em}@{\extracolsep{\fill}}p{6.5em}@{\extracolsep{\fill}}p{11em}@{\extracolsep{\fill}}p{5em}@{\extracolsep{\fill}}p{4em}}
\hline
Model & Lattice ($d=2$) & Lattice ($d\ge3$)  & Other Lattices& Random & SF & Others\\
\hline
Product rule &\cite{Ziff2009,Radicchi2010,Ziff2010,Grassberger2011,
Choi2011,Angst2012,Li2012a,Choi2012,Reis2012,Choi2014}
&\cite{Choi2014}
&\cite{Chae2012}
&\cite{Achlioptas2009,Radicchi2010,Cho2010b,Kim2010,Riordan2011,
Grassberger2011,Nagler2011,Cho2011,Li2012a} &\cite{Cho2009,Radicchi2009,Radicchi2010}
&\cite{Rozenfeld2010,Squires2013}\\
%\tabularnewline \hline
Sum rule
&\cite{Bastas2011,Choi2012,Choi2014}
&\cite{Choi2014}
&\cite{Chae2012}
&\cite{Achlioptas2009,Riordan2011,Grassberger2011,Cho2011} &&\\
%\tabularnewline \hline
Best-of-$m$
&\cite{Araujo2011,AndradeJr2011,Giazitzidis2013}
&\cite{Giazitzidis2013}
&
&\cite{Friedman2009,Riordan2011,Nagler2011,Hooyberghs2011,Qian2012}
&
&\cite{Pan2011,Squires2013}\\
%\tabularnewline \hline
dCDGM
&&&
&\cite{daCosta2010,Cho2011,Li2012a,daCosta2014a,daCosta2014b,
daCosta2015a,daCosta2015b}
&&\cite{Yi2013}\\
%\tabularnewline \hline
$l$-vertex
&&&
&\cite{DSouza2010,Nagler2011,Riordan2011,Nagler2012,
Riordan2012a,Zhang2013,Chen2013a}
&&\\ \hline
\end{tabular*}\end{center}
\end{table}

The fundamental difference between classical percolation
and AP is that the latter follows a competitive (or correlated) rule,
which involves non-local (or even global) information on the graph.
It is well known that introducing long-range interactions during percolation can change the universality class of the PT, or even the transition type. Thus, it is not surprising that a percolation model with non-local competition (like AP) can give rise to abrupt transitions.
Following the same motivation, many other models were also investigated, in which various constraints on the occupation of edges were proposed.
These models include the Bohman-Frieze-Wormald (BFW) model
\cite{Riordan2011,Chen2011,Schrenk2012,Zhang2012,Chen2012,
Chen2013a,Chen2014,Waagen2014,Chen2015},
the probability model
\cite{Moreira2010,Araujo2010,Cho2010a,Manna2011,Araujo2011,
Schrenk2011,Hooyberghs2011,Chung2013}),
the hybrid model \cite{Araujo2011,Cao2012,Fan2012,Bastas2014},
the diffusion-limited cluster aggregation (DLCA) model
\cite{Cho2011a,Cho2012},
the spanning cluster-avoiding (SCA) model \cite{Cho2013,Ziff2013},
the two-species cluster aggregation  model \cite{Cho2015},
the bootstrap model \cite{Baxter2010},
the $k$-core model \cite{Cao2012,Liu2012,Zhao2013,Cellai2013},
the cascading failure model in inter-dependent networks
\cite{Parshani2010,Parshani2011,Gao2011,
Son2011,Gao2012,Li2012b,Shekhtman2014,Baxter2012,Li2013},
and others \cite{Panagiotou2011,Boettcher2012,Matsoukas2015,Maksymenko2015}.
The relevant literature regarding all these other models is summarized in Table~\ref{tab:Others}. Like in the case of Table~\ref{tab:AP}, the reader can find orientation on the various adopted models and network structures.

Let us now briefly discuss the mechanisms leading to EP.
In classical percolation, nodes are progressively chosen at random to connect. Thus,
the probability to establish a specific edge is proportional to the product of the sizes of the clusters where the edge's end points reside.
The result is that a giant cluster can quickly form, and this in turn amplifies the probability for its further expansion.
It is therefore expected that CPT always occurs under this scheme.

In  AP and GAP models, instead, competition among potential edges is introduced at the basis of the cluster-merging processes.
The key factor leading to EP is here that the competition mechanism systematically suppresses the formation of large components
\cite{Araujo2010,Moreira2010,Cho2011,Cho2013}, and such a suppression principle generates the necessary ``powder keg'' (i.e., abundant
small-sized clusters) in a specific range  before the onset of the transition \cite{Friedman2009}.
Typically, after a number of steps, such small/medium sized clusters merge to each other, and a giant cluster suddenly emerges.
Similar suppressive mechanisms have also been pointed out to occur in the DLCA model \cite{Cho2011a}, the spanning cluster-avoiding model
\cite {Cho2013}, and the explosive synchronization model \cite{Zhang2014}.
%overtaking

%-----------Table II---------------------
\begin{table}
\renewcommand{\arraystretch}{1.3}
\caption{Summary of the relevant literature concerning other
  algorithms leading to EP in lattices, random networks, and complex
  networks: small-world (SW), SF, and inter-dependent networks,
  including the BFW model, the probability model, the hybrid model,
  the DLCA model, the spanning cluster-avoiding (SCA) model, the
  hierarchical model, the two-species cluster aggregation model (TCA), the bootstrap model, the $k$-core model, and the cascading failure model. \label{tab:Others}
}
\begin{center}
\footnotesize \begin{tabular*}{\textwidth}{p{5.5em}@{\extracolsep{\fill}}p{8em}@{\extracolsep{\fill}}p{7em}@{\extracolsep{\fill}}p{6.5em}@{\extracolsep{\fill}}p{11em}@{\extracolsep{\fill}}p{5em}@{\extracolsep{\fill}}p{4em}@{\extracolsep{\fill}}p{7em}}
\hline
Model   &  Lattice ($d=2$) & Lattice ($d\ge3$) & Other Lattices& Random & SW  & SF& Inter-dependent
\tabularnewline \hline
BFW
&\cite{Schrenk2012}
&\cite{Schrenk2012}
&\cite{Schrenk2012}
&\cite{Riordan2011,Chen2011,Waagen2014,Zhang2012,Chen2012,
Chen2013a,Chen2014}
&&\cite{Chen2015}&
\\
Probability  &\cite{Moreira2010,Araujo2010,Manna2011,Araujo2011} &\cite{Schrenk2011}
&\cite{Schrenk2011}
&\cite{Cho2010a,Moreira2010,Manna2011,Hooyberghs2011,
Chung2013}&&&
\\
Hybrid
&\cite{Araujo2011,Cao2012,Bastas2014}
&&&
\cite{Fan2012,Cao2012,Bastas2014} &&&
\\
DLCA
&\cite{Cho2011a,Cho2012}
&\cite{Cho2012}
&&&&&
\\
SCA
&&&&\cite{Cho2013,Ziff2013}   &&&
\\
TCA
&&&
&\cite{Cho2015}   &&&
\\
Bootstrap
&&&& \cite{Baxter2010} &&&
\\
$k$-core
&\cite{Cao2012,Zhao2013}&\cite{Zhao2013}&
&\cite{Cao2012,Zhao2013,Cellai2013}
&&&\cite{Liu2012}
\\
Cascading failure
&&&&&&
&\cite{Buldyrev2010,Parshani2010,Parshani2011,Gao2011,
Son2011,Gao2012,Li2012b,Shekhtman2014,Baxter2012,Li2013}
\\
Others
&&&
&\cite{Panagiotou2011,Maksymenko2015}
&\cite{Boettcher2012,Matsoukas2015}&&\\
 \hline
\end{tabular*}\end{center}
\end{table}

In BFW models \cite{Chen2012,Chen2013a}, another underlying mechanism for EP (namely, the growth by overtaking) was revealed.
In these models, the growth of the largest cluster is so severely limited by the edge selection rules that it can only
merge  with isolated nodes. As a consequence, all significant changes in the size of the giant cluster result from two smaller components merging together and overtaking the previous largest cluster to become the new largest component.

The central issue is  whether or not EP is really a discontinuous transition, especially in the thermodynamic limit. Because of
the extremely slow convergence to an asymptotic behavior (as the system size increases),
it is difficult to determine the transition nature of EP with the only help of numerical simulations.
Therefore, whether or not EP is a genuinely discontinuous transition became, for some time, the object of a controversial issue \cite{daCosta2010,Lee2011,Grassberger2011,Cho2011,Riordan2011,
Kim2010,Choi2011,Bastas2011}. Later, it was eventually revealed that the EP transition of Ref.~\cite{Achlioptas2009} is in fact continuous in the thermodynamic limit \cite{daCosta2010}.

The continuous character of EP was initially asserted on the basis of numerical results for a specific model \cite{Grassberger2011}, and later on  it was further pointed out that EP under AP is continuous, but with a unusual finite size behavior. Namely EP under AP belongs to a new universality class with an extremely small exponent of the order parameter ~\cite{daCosta2010} .
Numerical results give a power law $S(t)\propto (t-t_c )^\beta$ with $\beta=0.0555$ (close to $1/18$), for $t_c = 0.9232$.
It is therefore such a littleness of the exponent $\beta$ that leads to a large discrete jump in the order parameter, resembling
a seeming discontinuity at the critical point.

Finally, based on numerical investigations regarding the average size of medium-sized cluster \cite{Friedman2009},  Riordan and Warnke
provided a mathematical proof that all AP models on ER networks give rise actually to continuous percolation transitions \cite{Riordan2011}.
The proof reveals that the number of clusters participating in the merging process (needed to generate a macroscopic-sized cluster) is not sub-extensive to the system size, and therefore it cannot bring out a DPT in the thermodynamic limit.
On the other hand, Ref.~\cite{Riordan2011} also showed that, for a random network, if the number of competitive edges $m$ is allowed to increase with the system size $N$ in a way that $m\to \infty$ as $N \to \infty$ [for example, $m \propto \log(\log N)$],
a true DPT can occur also in the thermodynamic limit.

It is today accepted that EP under Achlioptas processes are actually CPT (second-order), but with a universality class different from all the others previously observed \cite{Grassberger2011}. The interested reader is addressed to Ref.~\cite{Bastas2014} for a
critical and exhaustive review on all critical
exponents. While other GAP models (such as minimal cluster model \cite{Friedman2009} and dCDGM model \cite{daCosta2010}) turn out to generate explosive (but actually continuous) percolation transitions in finite size systems,
it has been discovered that many alternative algorithms lead to genuinely DPT. These latter models include
the BFW model \cite{Chen2011},
the Gaussian model \cite{Araujo2010},
the spanning cluster-avoiding model \cite{Cho2013}, the
hierarchical model \cite{Boettcher2012}, and the two-species cluster aggregation model \cite{Cho2015}.

The next Sections will expand the discussion on all the concepts, ideas and methods summarized so far.

\subsection{Explosive percolation in the Achlioptas process}
\label{gingillino}

The percolation process introduced by Achlioptas can be described as follows \cite{Achlioptas2009}.
The starting point is a network with $N$ vertices and no edges, and edges are added one by one.
Precisely, at each time step, two potential edges $e_1$ and $e_2$ are arbitrarily chosen, each of which is supposed to connect two clusters
(i.e. the two end points of each edge are supposed to belong to two distinct clusters). The size of a cluster is defined here as the number of nodes in it. Now, let us denote the sizes of the two end clusters linked by $e_1$ as $s_1$ and $s'_1$, and the sizes of the two end clusters linked by $e_2$ by $s_2$ and $s'_2$. In order to select the edge to be actually added to the network, $s_1$, $s'_1$, $s_2$, and $s'_2$ are compared among them, by means of the use of certain rules.
For instance,  one can compare
the products $p_1=s_1  s'_1$ and $p_2=s_2  s'_2$. If $p_1<p_2$ ($p_1>
p_2$), then edge $e_1$ ($e_2$) is established and $e_2$ ($e_1$) is
discarded. This rule is called ``the product rule'', as one always establishes those links minimizing the product of the sizes of the two end clusters. If one instead replaces the ``product'' by the ``sum'' of the clusters' sizes, the resulting rule is called the ``sum rule''.

\begin{figure}
\begin{center}
\includegraphics[width=0.8\textwidth]{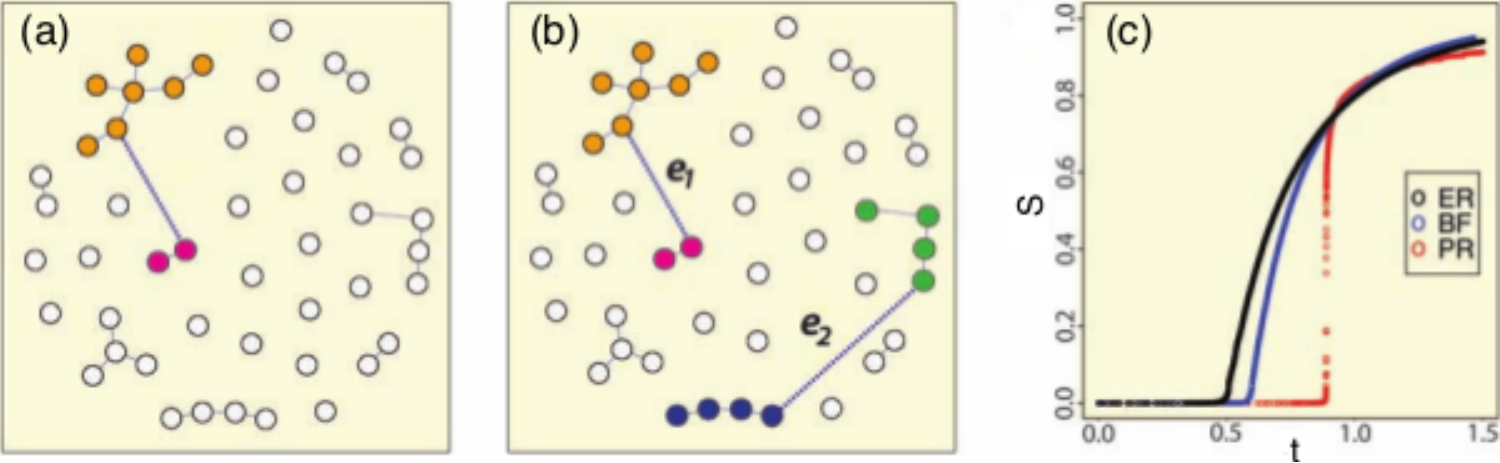}
 \caption{(Color online). Comparison between random percolation and AP
   percolation \cite{daCosta2014b}. (a) Classical ER percolation,
   where edges are randomly added into the network. (b) AP
   percolation, where at each step two potential edges compete to be
   established. (c) Order parameter (the relative size of the giant
   cluster) vs. control parameter (the number of added edges
   normalized by system size). For the ER process, PT is continuous
   (black solid line). For the BF process, it is still continuous with
   a postponed threshold (blue solid line). For the AP process with
   the use of the PR rule (red dotted line), EP is observed. From Ref.~\cite{Achlioptas2009}. Reprinted with permission from the American Association for the Advancement of Science.  \label{fig-AP} }
\end{center}
\end{figure}

Figure~\ref{fig-AP} compares classical percolation and AP, and shows that the latter exhibits two distinguished characteristics. One is that the percolation threshold is significantly postponed (Fig.~\ref{fig-AP}(c)). Another, and most important, is that the order parameter shows an extremely steep ``jump'' at the percolation point which, at first glance, seemingly resembles a DPT (Fig.~\ref{fig-AP}(c)).
In order to characterize the jump, Ref.~\cite{Achlioptas2009} counts the fraction of vertices
added in the network when the giant cluster grows from size at most $\sqrt{N}$ to size at least $N/2$. It is shown that at most $2N^{2/3}$ steps are needed. This implies that, at the phase transition, a constant fraction of the vertices is accumulated into a single giant cluster within a sub-linear number of steps. %In the thermodynamic limit $N\to\infty$, the number of steps for the jump of the order parameter vanishes.

It should be emphasized that not all competitive rules lead to EP.  Traditionally, the selecting rules in percolation models can be classified into two types. One is the ``bounded-size rule'',
in which decisions depend only on the sizes of the clusters, and all  clusters with size greater than a certain
constant $K$ are regarded as the same. For example, $K=1$ is  the Bohman-Frieze (BF) rule,  where $e_1$ is chosen if it joins two isolated vertices, otherwise $e_2$ is chosen.
As shown in Fig.~\ref{fig-AP}(c), the percolation under the BF rule turns out to be still continuous,
but with a postponed threshold. In fact,
it has been conjectured  that the percolation transition is  continuous for all bounded-size rules \cite{Spencer2007}.

At variance, an ``unbounded-size rule'', like the PR or SR in Ref.~\cite{Achlioptas2009}, treats all clusters of distinct sizes
uniquely.  Such unbounded-size rules typically lead to interesting transition behaviors, such as shifting the transition
threshold and even yielding EP. Not all AP models lead to EP.
For example, in the ``largest-cluster rule'' \cite{Achlioptas2009,Friedman2009}, at each step one selects the two largest
clusters in the network to connect.
Apparently, the process seems to be similar to AP (due to the considered competitive rule for adding edges). However, the percolation turns out to be continuous under this rule.

PR and SR can be straightforwardly extended
to the case where at each time step $m > 2$ potential edges
are chosen to compete. This leads to the ``best-of-$m$ rule'' or the ``min-cluster-$m$ rule'',
which typically leads to EP \cite{Friedman2009}.
Note that when $m=1$ the ``best-of-$m$ rule'' recovers the ordinary random percolation, which is continuous. When $m=2$, it
specializes the PR/SR model studied in Ref.~\cite{Achlioptas2009}, which leads to EP.
With increasing $m$, the process becomes more  and more competitive.
If $m=N$, where $N$ is the total number of nodes, we have global competition among all unoccupied links in the network.
This limit situation corresponds to the ``smallest-cluster rule'' \cite{Friedman2009}, where at
each step two smallest clusters in the
network are identified to be connected, and global information is used in order to identify
the two smallest clusters \cite{Nagler2011}.

\subsection{Explosive percolation in generalized Achlioptas processes} %223

Stimulated by the observation that non-local or global competition rules during cluster-merging processes
can lead to EP, a number of generalized Achlioptas processes (GAP) have been studied.
In the following, we briefly describe two generalizations of the original AP.

\subsubsection{The dCDGM model} 
\begin{figure}
\begin{center}
\includegraphics[width=0.45\textwidth]{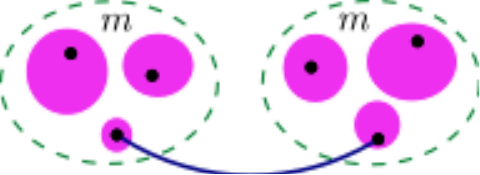}
 \caption{(Color online). Illustration of the dCDGM algorithm \cite{daCosta2014b}. At each time step, two sets of $m$ nodes are chosen at
random. Within each set, the node in the smallest cluster is selected,
and the pair of selected nodes are connected. Reprinted with
permission from Ref.~\cite{daCosta2014b}. $\copyright$ 2014 by the American Physical Society. \label{fig-dCDGM}}
\end{center}
\end{figure}

The  model has been studied in Refs.~\cite{daCosta2010,Cho2011,Li2012a,Yi2013,daCosta2014a,daCosta2014b,
daCosta2015a,daCosta2015b}. At each step, a pair of clusters are initially randomly selected, and the smaller cluster is kept.
Then the same process repeats, and another (smaller) cluster is picked.
In order to establish the link, one actually chooses two random nodes (one from each of the chosen clusters) and connects them.
The above procedure is sequentially repeated. The dCDGM model provides even more stringent selection of small components during cluster-merging
processes than usual AP, since it guarantees that the product of the sizes of the two merging
clusters is the smallest of the four possibilities. This algorithm can be naturally extended to the case of $m > 1$ nodes.
For instance, Fig.~\ref{fig-dCDGM} illustrates this process for $m=3$ \cite{daCosta2014b}.

\subsubsection{The $l$-vertex rule} In this second model,  $l$  vertices  are randomly chosen at each step, and two of them are connected
with an edge using certain rules \cite{Riordan2012a,Riordan2012b,Nagler2012,Zhang2013}.
Notice that the resulting percolation process includes the ER model, which is obtained for $l=2$.
Also, Achlioptas processes can be regarded as  a special case of the 4-vertex rule.
When $l=3$, the rule has several variants, such as the
triangle rule (TR) \cite{DSouza2010}, the adjacent edge  rule \cite{DSouza2010}, and the
clique-3 competition rule \cite{Nagler2011}.
In all these variants, three distinct vertices
are randomly selected at each step. Then, one examines the three possible edges (TR and clique-3 competition) or
two adjacent edges in the triangle, and selects the edge that connects the two smallest components.

\begin{figure}
\begin{center}
\includegraphics[width=0.4\textwidth]{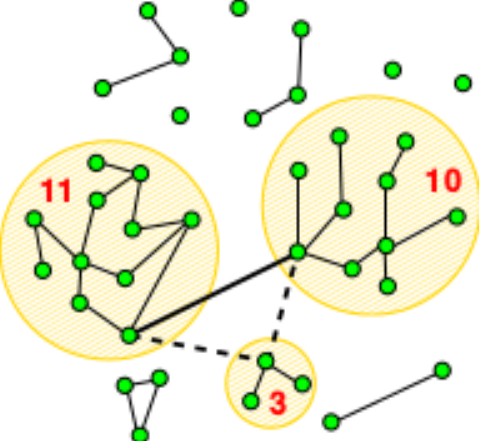}
 \caption{(Color online). Illustration of the algorithm in a variant of the TR (3-vertex) model  \cite{Nagler2012}.
At each step, three vertices are selected and those two vertices that
reside in clusters whose size difference is minimal are
connected. Reprinted from Ref.~\cite{Nagler2012}, published under Creative Commons Attribution 3.0 License. \label{fig-3-vertex}}
\end{center}
\end{figure}

The bounded-size rule variant occurs when all components have size above the bound $K$, and one chooses a random edge
out of the three; if two components have size above the
bound $K$, one chooses a random edge out of the two adjacent
of the smallest component.
Furthermore, Nagler et al. have proposed a variant of the TR \cite{Nagler2012}. There, one first chooses three vertices
at random and connects those two vertices that reside in clusters whose size difference is minimal (see Fig.~\ref{fig-3-vertex}).
Notice that this latter model does not suppress the growth of the largest cluster, in contrast to many other Achlioptas processes.
It is found that the continuous phase transition can have the shape of an incomplete devil's staircase
with discontinuous steps in arbitrary vicinity of the transition point \cite{Nagler2012}.

\subsection{Explosive percolation under other algorithms}

Beyond ordinary percolation, many other models have been developed, where various constraints on the occupation of edges
have been considered in order to produce EP.
In the following, we briefly review some of these models.

\subsubsection{The BFW model}
In Ref.~\cite{Chen2011} a model is introduced (based on the work of Bohman, Frieze, and Wormald (BFW) \cite{Bohman2004}), where
the members of a set of potential edges are considered one at a time, and either added to the network or rejected.
Specifically, the algorithm of BFW is as follows.
Initially, there are $N$ isolated nodes and all bonds are unoccupied (empty). Therefore, there are $N$ clusters of unitary size.
At each time step, one edge is added to the network. Let $u$ be the total number of selected bonds, $t$ the number of occupied bonds,
and $k$ the stage of the process, initially set to $k = 2$. The first bond is occupied at random, such that $t = u = 1$.

Then, at each step $u$,
\begin{enumerate}
\item  One edge is randomly selected out of all edges (or the unoccupied ones);
\item  The maximum cluster size $l$ is measured, if the selected edge is occupied;
\item If $l \le k$, go to 4. Else, if $t/u \ge g(k) = 1/2 +
\sqrt{1/(2k)}$, go to 5. Else, increment $k$ by one and go to (3);
\item Occupy the selected edge;
\item Increment $u$ by one.
\end{enumerate}

The procedure is applied iteratively from $t = 0$ (corresponding
to a bond occupation fraction $p = t/(2N) =
0$) to $t = 2N$ (where all bonds in the system are occupied, i.e., $p = 1$).
Using such a model, Chen et al.  have shown that multiple stable giant clusters can
coexist, and the percolation transition is strongly discontinuous \cite{Chen2011}.
The BFW model has been extensively studied in various networks,
including lattices and tree-like networks \cite{Schrenk2012},
ER networks \cite{Riordan2011,Chen2011,Waagen2014,Zhang2012,Chen2012,
Chen2013a,Chen2014}, and
SF networks \cite{Chen2015}.
Notice that in Refs.~\cite{Bohman2004,Chen2011,Chen2012,Chen2013a,Chen2014,Chen2015}  edges are randomly chosen from all edges (i.e. regardless on whether or not they are occupied), while in Refs.~\cite{Schrenk2012,Zhang2012} edges are only sampled from the set of unoccupied bonds.
In Refs.~\cite{Chen2011,Schrenk2012,Zhang2012,Chen2015}, $g(k) = 1/2 + \sqrt{1/(2k)}$.
Later, generalizations of this function have been considered as
$g(k) = 1/2 +(2k)^{-\beta}$ \cite{Chen2012}, and
$g(k) = \alpha +(2k)^{-\beta}$ \cite{Chen2013a,Chen2014}.

\medskip

\subsubsection{Probability models}
In this type of models, an edge between a pair of clusters is occupied with certain probability at each time step.
This process differs substantially from AP, since here all unoccupied connections can potentially be chosen at each step,
though some of them with a much larger probability \cite{Moreira2010}.
For instance, in cluster aggregation model \cite{Cho2010a,Manna2011}, all currently unoccupied edges are considered.
Let one denote the sizes of the clusters to be connected by the edge  $ij$ as $s_i$ and $s_j$.
Each of these vacant edges is given a weight $(s_is_j)^\alpha$, and the sum over all the weights is calculated as a normalization constant $w$. Then the edge $ij$ is occupied with probability $(s_is_j)^\alpha/w$.

In the Hamiltonian model of Ref.~\cite{Moreira2010}, instead, one starts with a network of $N$ vertices without edges,
so each vertex initially belongs to a different cluster.
First, a simple Hamiltonian $H$ is defined as a function of the cluster sizes and of the number of redundant
edges added to the clusters. Then, a new edge $b$ between any pair of vertices not yet
connected is added, with a probability proportional to $\exp(-\beta\Delta H_b)$,
where $\Delta H_b$ is the energy change after adding the edge.

Another model belonging to this class is the largest cluster model (or the Gaussian model) \cite{Araujo2010}.
There, a link is randomly selected
among the unoccupied ones. If its occupation would not lead to the formation or growth of the largest cluster,
it is always occupied; otherwise, it is occupied with probability
min$\{1,\exp[-\alpha(\frac{s-\bar{s}}{\bar{s}})]\}$.
For $\alpha>0$, this probability suppresses the formation of a cluster significantly larger than the average,
inducing a homogenization of cluster sizes, which is important to generate EP.

Such probability models have found applications in
2D lattices \cite{Moreira2010,Araujo2010,Manna2011,Araujo2011},
3D and high-D lattices \cite{Schrenk2011},
as well as ER networks \cite{Cho2010a,Moreira2010,Manna2011,Hooyberghs2011,
Chung2013}.

\medskip

\subsubsection{Hybrid models}
Hybrid models are those for which the rule for the occupation of edges is a mixture of ordinary
percolation and competitive rules: at each time step, an edge can be chosen randomly with probability $p$,
while it can also be chosen following a given competitive rule with probability $1-p$.
For example, in Refs.~\cite{Fan2012,Bastas2014}, PR is used;
in Ref.~\cite{Araujo2011}, the best-of-ten PR is used; in Ref.~\cite{Cao2012},
a mixture of $k = 2$- and $k = 3$-core on the random network and
a mixture of $k = 3$-core and jamming percolation models in 2D lattice are studied.
Hybrid models have been studied in 2D lattices \cite{Araujo2011,Cao2012,Bastas2014} and
ER networks \cite{Fan2012,Cao2012,Bastas2014}.

\medskip

\subsubsection{The diffusion-limited cluster aggregation (DLCA) model}
In this model, studied in Refs.~\cite{Cho2011a,Cho2012,DSouza2015}, $N$ single particles are initially randomly placed in a $L \times L$ square lattice.
The particles are assumed to be Brownian, so that the velocity of a cluster is inversely proportional
to the square root of its size, i.e. the velocity of an $s$-sized cluster is given in a general form
as $v_s \propto s^{-1/2}$. As a consequence, the larger clusters move considerably more slowly.
When two clusters become nearest neighbors, they merge and form a larger cluster. The Brownian motion suppresses the
mobility of the largest clusters, impeding their growth, and leading to the discontinuous emergence of a giant
cluster. Generalized Brownian motion, for which the velocity is inversely proportional to the size of the
cluster to a power $\eta$ (i.e., $v_s \propto s^{\eta}$)  gives rise to a tri-critical point that
separates discontinuous from
continuous transitions as a function of $\eta$.
Namely, as soon as $\eta_c$ is a tri-critical point, it is shown that when $\eta <\eta_c$  ($\eta>\eta_c$),  the PT is discontinuous (continuous) \cite{Cho2011a}.

\begin{figure}
\begin{center}
\includegraphics[width=0.6\textwidth]{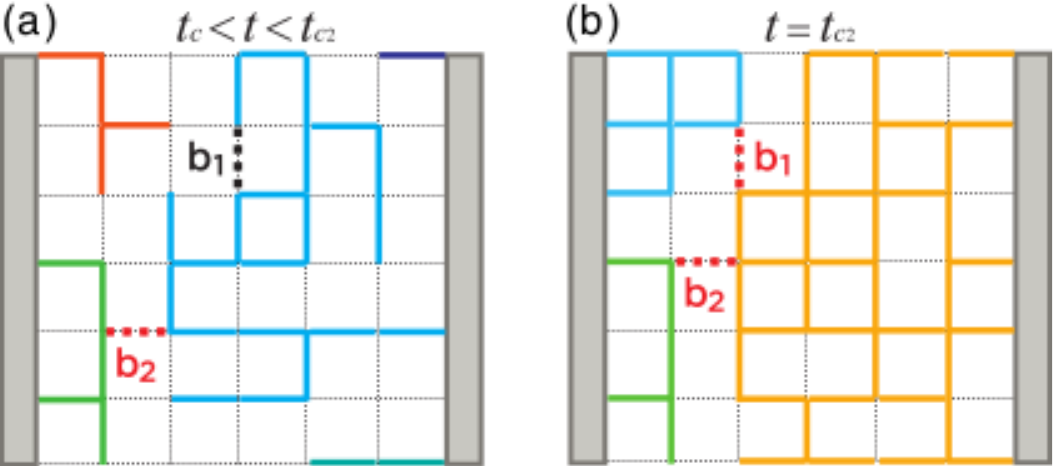}
\caption{(Color online). Illustration of the algorithm in the SCA model with $m = 2$ on a square lattice \cite{Cho2013}.
(a) At each time step, two empty bonds, $b_1$ and
$b_2$ (shown as dashed lines), are randomly selected. If one of them is a bridge bond (like $b_2$ in this case), then the non-bridge bond ($b_1$) is chosen. (b) At $t = t_{c2}$, two bridge bonds
can be selected for the first time. Then, one of them is taken
randomly and a spanning cluster is formed. From
Ref.~\cite{Cho2013}. Reprinted with permission from the American Association for the Advancement of Science. \label{fig-SCA}}
\end{center}
\end{figure}

\medskip

\subsubsection{The spanning cluster-avoiding (SCA) model}
The SCA model, studied in Refs.~\cite{Cho2013,Ziff2013}, considers percolation on an Euclidean lattice, where the emergence of a spanning cluster is concerned. In a spanning cluster, a path of activated
links exists that connect sites from one side of the lattice to the other.
Inspired by the best-of-$m$ rule, Cho et al. introduced a new global constraint for the dynamics.
As shown in Fig.~\ref{fig-SCA}, at each time step $t$,  $m$ unoccupied edges are chosen randomly and classified
into two types: bridge and non-bridge edges. Bridge (non-bridge) edges are actually those that would (would not) form a spanning cluster when occupied.
The SCA model strictly avoids bridge edges to be occupied, and therefore one of the non-bridge edges is randomly selected and occupied. If the $m$ potential edges are all bridge edges, then one of them is randomly chosen and occupied.
Once a spanning cluster is formed, no more
restrictions are imposed on the occupation of
edges.  This procedure continues until all edges are occupied.
Numerical simulations and theoretical results have shown that the EP in SCA model can be either discontinuous or continuous for $d < d_c = 6$, depending on the number of eligible edges $m$.
For $d \ge d_c$, %i.e., in the mean-field limit,
the transition is shown to be continuous for any finite and fixed value of $m$. However if $m$ varies with the system size $N$, a discontinuous transition can also take place for $d \ge d_c$.
%This analytic result leads to the following conclusion: the EP transition can be either continuous or discontinuous,
%depending on the number of multiple options, if the spatial dimension is less than the upper critical dimension, and is always continuous otherwise. Subsequently, it was
%concluded that the transition of the ordinary EP model is continuous as a mean-field solution of the SCA model.

\medskip

\subsubsection{The two-species cluster aggregation model}
In this model, studied in Ref.~\cite{Cho2015}, $N$ initially isolated nodes are randomly separated into two populations:
half of them is colored in black (B) and the other half in white (W), with all nodes in the same cluster keeping the same color.
At each time step, the three possible combinations BB, BW, and WW are given selection probabilities
$1/(1 + 2p)$, $p/(1 + 2p)$, and $p/(1 + 2p)$, respectively (where $p$ is
a model parameter in the range $0 < p \le 1$). Two clusters are then selected
independently of the cluster sizes, and two nodes (one from each cluster) are selected randomly
and connected, which causes the two clusters to merge. If the two selected clusters are the same, then
two distinct nodes from that cluster are connected.
Meanwhile, the  colors of all the nodes in the resulting merged cluster have to be updated according to a certain rule.
See Fig.~3 in Ref.~\cite{Cho2015} for an illustration of the algorithm.
The cluster aggregation model can exhibit both type-I and -II DPT, as the model parameter changes \cite{Cho2015}.

\medskip

\subsubsection{The bootstrap models}
In the context of EP, some bootstrap models have been proposed \cite{Baxter2010}.
The standard bootstrap percolation process on a lattice assumes that sites are either active or inactive,
and the state of a site depends on its neighbors.
Initially, each site is activated with a probability $p$, and inactivated with probability $1-p$. Every activated site remains in its state, while each inactivated site can become active (and remain active forever) if its $k$ nearest neighbors are active (with $k=2,3,\cdots$).
The procedure is continued until the system reaches the stable configuration which does not change anymore.
In the final state of the process, one is concerned on whether or not a spanning giant cluster of active sites exists.
Bootstrap percolation can show either continuous or discontinuous transitions,
depending on the type of networks (lattice or random), and on the dimensionality of the lattice.

\medskip

\subsubsection{The $k$-core model}
$k$-core percolation \cite{Cao2012,Liu2012,Zhao2013,Cellai2013} is a model of correlated percolation beyond the ordinary one.
The $k$-core of a graph is the maximal subgraph for which all vertices have at least $k$ neighbors.
It can be found as an asymptotic structure obtained by
recursively removing all nodes whose neighbors are less than $k$.
For  example, if $k= 2$, all leaves (nodes with only a connection) are removed in a first step. This might turn into some nodes (which originally had 2 or more neighbors)
becoming leaves. Then, these latter nodes are removed in a second round, and if this again turns nodes into leaves, they
are removed again, and so on. In this way, one ends up either with an empty cluster, or with a cluster where all nodes have $k \ge 2$ neighbors, i.e., the 2-core. It has been proved that $k$-core percolation on random networks exhibits a CPT for $k \le 2$, and a DPT for $k \ge 3$ \cite{Pittel1996}. However, unlike a typical DPT, not one but several diverging length scales exist such that the transition has an unusual nature \cite{Schwarz2006}.
%This is the reason that hinders the indepth studies for this model.

\medskip

\subsubsection{Cascading failure model in inter-dependent networks}
This type of models aims at studying processes such as cascading-failure dynamics (and related properties, as resilience and robustness) in  inter-dependent systems \cite{Buldyrev2010,Parshani2010,Parshani2011,Gao2011,Son2011,Gao2012,Li2012b,Shekhtman2014,Baxter2012,Li2013}. For example,
the infrastructure network of a country includes both
power grids and computer networks. The latter are needed for
controlling the power stations, while the former are needed to provide power to the computers. If a node fails in one of the two networks,
this can lead to a huge back and forth cascade of failures, as historically happened in  many countries.
See Fig.~2 of Ref.~\cite{Buldyrev2010} for an illustration of
a cascade of failures in inter-dependent networks which resembles a first-order transition.
Several other types of interdependencies were also studied.
%In all cases with first-order transitions, these are indeed again hybrid.
For example, first-order percolation has been investigated in  a single
network with two sets of links: the dependency links  and the connectivity links \cite{Parshani2011}.
Figure~1 in Ref.~\cite{Parshani2011} illustrates the cascading failures of such a latter case.
It should be pointed out that, in this DPT, the order parameter is taken as the fraction of nodes in the mutually connected giant cluster,
which is not the same as the standard ones.
Inter-dependencies of networks, and robustness of power-grids will be also discussed in Subsection~\ref{sec:powergrid}, from the viewpoint of explosive synchronization.

\medskip

\subsubsection{Other models}
%Half-restricted
Several other models for explosive percolation have been proposed. In the half-restricted ER model, two vertices are connected by an edge at each step, but one of them is arbitrarily chosen and the other is
chosen randomly from a restricted set \cite{Panagiotou2011}. This variant of random percolation can exhibit a DPT.
%Hierarchy
In the hierarchical model \cite{Boettcher2012}, Boettcher et al. studied the effect of an ordinary (uncorrelated) bond-adding rule on a network with a recursive, hierarchical structure. It is shown that the hierarchical adding of SW bonds into a 1D lattice can induce DPT.
%In contrast to the AP-like constraints leading to EP, which are typically correlated,
%experiments
Furthermore, two works are highly relevant for experimental realizations of EP: in  Ref.~\cite{Kim2010}, the formation of single-walled nanotube bundles with uniform diameter is investigated, and by applying the cluster repulsion process to stick percolation, it is shown that the transition becomes explosive. In Ref.~\cite{Oliveira2014}, the electric breakdown of a substrate was studied, on which highly conducting particles
are adsorbed and desorbed with a probability that depends on the local electric field.
It is revealed that the electric breakdown
due to pollution with metallic powder can become
explosive under certain conditions.
This model is thus an example of a possible
experimental realization exhibiting a true DPT.

\subsection{Explosive percolation in 2D lattices}

The phase transition corresponding to EP is established to be continuous in the
mean-field level, or in higher dimensions. An interesting issue is therefore determining the transition
nature for EP in lower dimensions. To answer the question, one needs to understand what differentiates
percolation in the cases of higher and lower dimensions. We take here ER networks, and two
dimensional (2D) lattices, as illustrative examples.
ER networks are, indeed, usually treated as infinite dimensional systems,
while 2D lattices are typical low dimensional ones.

One of the main features of complex networks
is the SW property, which guarantees a long-range correlation among the nodes. Thus,
in the growing process of ER, it is easy to generate a core and then the core gradually grows up.
When an AP is applied, the growing of the core is suppressed, and a sharp jump in the growth dynamics
occurs for finite-size systems. However, when the size $N$ goes to infinity, the AP rule
cannot suppress the growing of the core completely, resulting again in a continuous transition. At variance, for a 2D
lattice, the nodes feature only short-range correlations, and therefore it is not clear whether or not an AP
would produce a discontinuous transition in these circumstances. %In this sense, we would like to know what will happen when the AP rule is applied or
%whether the AP rule still produces a discontinuous transition in the 2D percolation model.

A lot of attention has, indeed, been paid to
the case of 2D lattices \cite{Ziff2009,Kim2010,Radicchi2010,Ziff2010,Bastas2011,Choi2011,Choi2012,
Schrenk2012,Cho2013,Ziff2013,Choi2014}. Another reason is that many interaction networks, whether physical or social, are  spatially
constrained, and bonds can form only between close
neighbors. Therefore, 2D percolation models represent possibly the strongest form of such spatial restrictions.

Both bond percolation and site percolation have been studied in 2D lattices.
In the former process, bonds are formed randomly and independently throughout the system, and, at a critical
concentration $p_c$ of occupied bonds, a finite fraction of the sites (nodes) are connected together,
and percolation takes place through a continuous or second-order transition. The latter process corresponds
to consider randomly occupied sites on a given lattice with a probability $p$, with neighboring occupied sites
that merge to form clusters. A remarkable finding here is that, for both bond and site percolation (and together with
the existence of the sharp jump in the growth dynamics of the 2D lattices), a hysteretic loop can be observed, which is the signature of an irreversible transition. However, this finding does
not mean that all the the transitions observed in 2D lattices are discontinuous, and the controversy is still
standing on whether the 2D lattices percolation is continuous \cite{Bastas2011,Choi2014} or discontinuous
\cite{Kim2010,Radicchi2010,Ziff2010,Choi2011,Choi2012}. %The controversy should come from ambiguity in the details of explosive percolation models on lattices.

\medskip

\subsubsection{Bond percolation} The AP process in bond percolation of 2D lattices was first considered by Ziff in Ref.~\cite{Ziff2009}, which started from a $L \times L$ square lattice
with periodic boundary conditions in both directions, and with $n=L^2$ sites.
Connected sites are considered as one cluster. As initial sites are fully disconnected, the starting point is a set of
 $n$ clusters. At each time step, only one bond is added, which always connects different clusters, this way
 reducing the number of clusters by one.

 Let the time $t$ represent the number of bonds successfully added, then
the number of clusters $N_c$ in the system is reduced to $n-t$. Let $s$
represent the mass in a cluster, i.e. the number of sites. The key point is that each successful new bond is
chosen by the minority product rule (PR) of Ref.~\cite{Achlioptas2009}. That is, one randomly chooses two
bonds at the same time, and only the bond that minimizes the product of the masses of the two joined clusters
is preferentially chosen to become occupied, while the other is discarded. %By this approach, Ziff
%found that in the PR model, the transition is quite sharp, indicating the first-order transition.

\begin{figure}
\centering \includegraphics[width=0.6\linewidth]{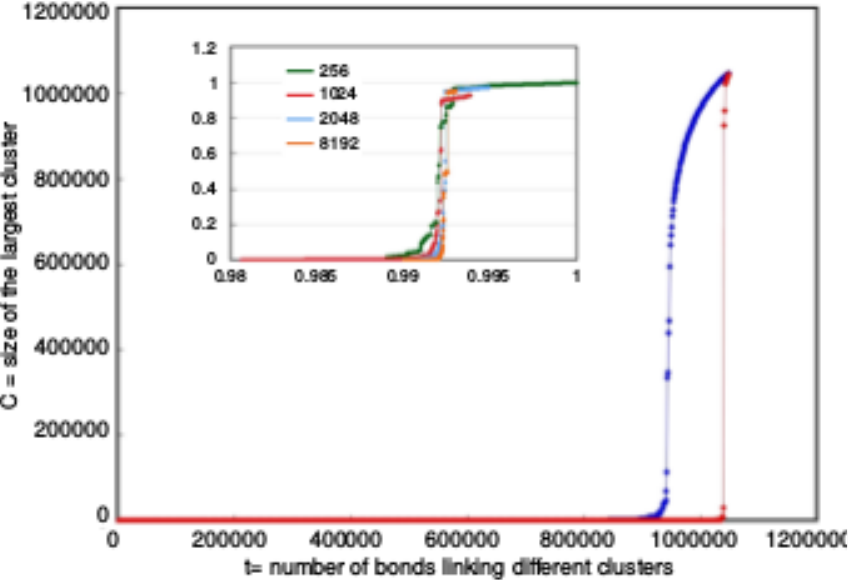}
\caption{(Color online). Regular (blue) and the
PR (red) bond percolation on a lattice of size $1,024 \times 1,024$. Notice the delayed and explosive growth
in the PR case. Points are reported every $1,024$ time steps. The inset zooms on the PR model, for
systems of sizes $L = 256$, $1,024$, $2,048$, and $8,192$ (color codes in the legend),  with scaled axes
$C/n$ vs $t/n$ with $n = L^2$. Reprinted with permission from Ref.~\cite{Ziff2009}. $\copyright$  2009 by the American Physical Society. \label{Fig:Ziff2009}}
\end{figure}

Figure~\ref{Fig:Ziff2009} shows the results obtained in Ref.~\cite{Ziff2009}, and reports the maximum cluster size $C$ as a function of $t$. It is evident that a jump  occurs over a small number of time steps, as shown in the
inset of Fig.~\ref{Fig:Ziff2009} for different system sizes (here the axes are scaled by $n$). For
comparison, regular percolation is also reported in Fig.~\ref{Fig:Ziff2009} with a blue curve, pointing to a much smoother transition.
% figure 1

A way to distinguish the difference in the transition between the regular  and the PR percolation is by accounting
for two  different times: the time $t_0$ at which $C=\sqrt{n}$, and the time at which $t_1=n/2$ \cite{Achlioptas2009}.
By calling $\Delta=t_1-t_0$, Achlioptas, D'Souza, and Spencer found that $\Delta\sim n$ for unbiased
growth of ER networks, while $\Delta\sim n^{2/3}$ for the PR growth, and therefore the two transitions
are qualitatively different \cite{Achlioptas2009}. Ziff measured the same
quantities for regular and PR growths on square lattices with different sizes, and also found that the
relationship between $\Delta$ and $n$ is quite different. For regular percolation,
$\Delta/n\sim L^{-36/91}$ was found, while for the PR model $\Delta/n\sim L^{-0.683}$ \cite{Ziff2009}.
Notice that  $\Delta/n\rightarrow 0$ as $n\rightarrow \infty$ for both the PR and random growth models. However, it
was later discovered that, with a different criterion for the upper end of the gap $\Delta$ (say $C=0.7n$), then one would find
that $\Delta/n\rightarrow const.$ as $n\rightarrow \infty$ for the random growth, while still $\Delta/n\rightarrow 0$ for the
PR model \cite{Ziff2010}. %So with this criterion, the two models are qualitatively different on the 2D lattice just as for the ER network.

Kim et al. considered the formation of single-walled nano-tube bundles with uniform diameter
\cite{Kim2010}. Except for the growing process, they also measured the changes in the order parameter during the stick
removal process, and found that a hysteresis exists. This result clearly shows that the transition
of the stick system with cluster repulsion is discontinuous.
AP in such a nano-stick system is defined as follows \cite{Kim2010}:
\begin{enumerate}
\item[(I)] One first places two sticks
$\alpha$ and $\beta$ at random positions with random orientations.
\item[(II)] Then one lets
$\{ s_{\alpha_1}, s_{\alpha_2}, \cdots, s_{\alpha_n}\}$
($\{ s_{\beta_1}, s_{\beta_2}, \cdots, s_{\beta_m}\}$) be the sizes of the clusters which would merge into a
bigger one of size $\sum_{k=1}^n s_{\alpha_k}+1$ ($\sum_{k=1}^m s_{\beta_k}+1$) by the placement of the
stick $\alpha$ ($\beta$). Here, the size is defined by the number of sticks in the cluster.
A PR is adopted by calculating the products:
\begin{equation}
\label{nanostick-1}
\pi_{\alpha}=\prod_{i=1}^n s_{\alpha_i}, \quad \pi_{\beta}=\prod_{j=1}^m s_{\beta_j}.
\end{equation}
\item[(III)] If $\pi_{\alpha}\le\pi_{\beta}$
($\pi_{\alpha}\ge\pi_{\beta}$) then the stick $\alpha$ ($\beta$) is kept and the stick $\beta$ ($\alpha$)
is removed. Processes (II) and (III) favor the connection between small clusters,
which suppresses the growth of larger ones.
\end{enumerate}

% figure 2
\begin{figure}
\centering \includegraphics[width=0.5\linewidth]{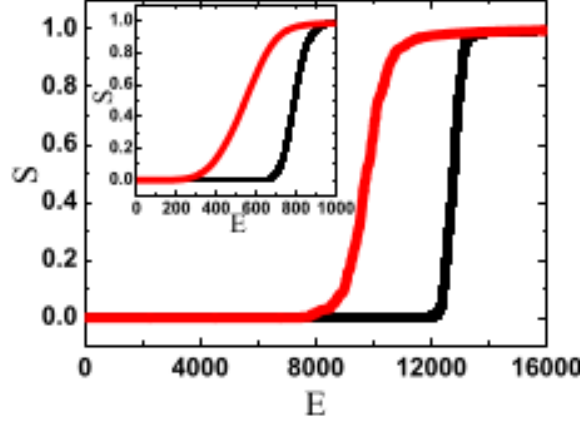}
\caption{(Color online). $S$ vs. $E$, for $L=40$. Black (red) solid line indicates $S$
when sticks are added (removed). The two curves manifest the presence of hysteresis. Inset: the same plot but
for $L=10$. Reprinted with permission from Ref.~\cite{Kim2010}. $\copyright$  2010 by the American Physical Society.\label{Fig:kim2010} }
\end{figure}

An order parameter $S$ is introduced to describe the probability that a stick belongs to a spanning
(largest) cluster \cite{Kim2010},
\begin{equation}
\label{nanostick-2}
S=\frac{E_{\infty}}{E},
\end{equation}
where $E_{\infty}$ is the number of sticks in the spanning cluster, and $E$ is the total number of sticks placed
in a given square. In general, EP is marked by the existence of a jump in the order parameter at
the critical connecting probability $p$. For convenience, this probability $p$ is here replaced by the total
number of sticks $E$ \cite{Kim2010}. Although these two approaches are similar, some differences between the
variables $p$ and $E$ have to be noticed. For the former, $p$ is a state variable and is thus independent of the
evolutionary process. A small increase of $\Delta p$ will result in an increase of approximately $\Delta pL^2(L^2-1)/2$
new sticks, which is enough to induce a significant increase of $E_{\infty}$.
For the latter, instead, $E$ is an evolutionary variable and is equivalent
to the time $t$. That is, $E$ will increase one by one, and this will result in a gradual increase
of $E_{\infty}$. In this sense, we cannot expect a jumping of $S$ in Eq.~(\ref{nanostick-2}) at the
critical point, but only a fast or explosive growing of $S$.

A first-order like phase transition is usually accompanied
with a property of history-dependence, i.e. the existence of a hysteresis loop, because of the presence of meta-stable states.
Therefore, one of the most generally accepted (and simplest) methods to verify whether or not
an observed transition is first-order like is the existence of a hysteresis loop \cite{Landau2000,Chaikin1995}.
To measure the hysteresis, one needs to consider both processes of adding and removing sticks. Results are shown in Fig.~\ref{Fig:kim2010}, which clearly manifests the existence of a hysteresis, indicating a first-order like transition.

\subsubsection{Site percolation}

An alternative mechanism is site percolation \cite{Bastas2011,Choi2011}. It refers to a situation
where, initially, all the sites of a square are empty. During the growing process,
sites are sequentially occupied, and two occupied neighboring sites are considered to belong to the same cluster.
The AP rule is applied as follows. One chooses two candidates at each time step, and investigates which one of them
would lead to the smaller clustering. This one is then kept and the other discarded.

Bastas, Kosmidis, and Argyrakis found that this procedure considerably slows down
the emergence of the giant component \cite{Bastas2011}. Their algorithm for the case of the sum rule
proceeds as follows: initially, one starts from an empty lattice and randomly occupies a single site. Then, at
each time step, one randomly chooses two unoccupied sites, say $A$ and $B$, and calculates the sizes $s_A$ and
$s_B$ of the resulting clusters to which $A$ and $B$ belong, respectively. When  $s_A < s_B$ ($s_B < s_A$), site $A (B)$ is permanently occupied, and site $B (A)$ discarded. When $s_A = s_B$, one randomly selects either $A$ or $B$ and permanently
occupies it. With $S_{max}$ representing the largest cluster at time $t$, the process is repeated until the entire lattice
is covered.

\begin{figure}
\centering \includegraphics[width=0.6\linewidth]{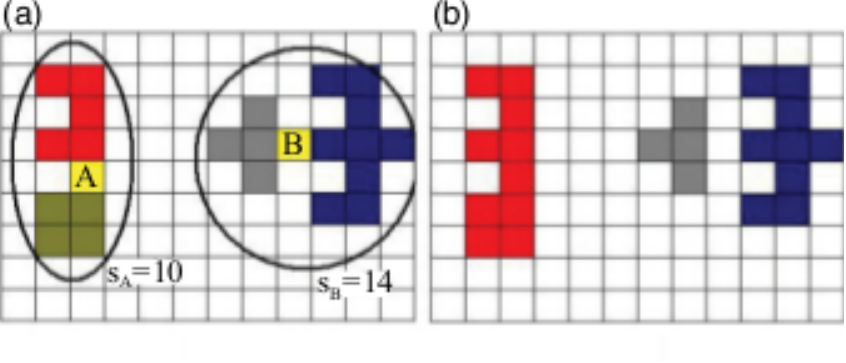}
\caption{(Color online). AP according to the SR for site percolation. White cells correspond to unoccupied
sites, while colored cells correspond to occupied sites. Different
colors (red, green, gray, and blue) indicate different clusters. (a) One
randomly selects two unoccupied sites (yellow), denoted by $A$ and
$B$. The size of the clusters after the addition of $A$ and $B$, $s_A$ and $s_B$, respectively, is evaluated.
In this example $s_A = 10$ and $s_B = 14$. (b) According to AP, $A$ is kept and $B$ is discarded. Reprinted with permission from Ref.~\cite{Bastas2011}. $\copyright$  2011 by the American Physical Society. \label{Fig:Bastas2011}}
\end{figure}

Figure \ref{Fig:Bastas2011} shows an illustrative example of the algorithm applied on an $L \times L$ square lattice
with periodic boundary conditions, where the white (colored) cells correspond to unoccupied (occupied) sites,
and where different colors (red, green, gray, and blue) indicate different clusters. In Fig.~\ref{Fig:Bastas2011}(a)
one randomly selects a pair of unoccupied sites (yellow), say $A$ and
$B$, one at a time, and evaluates the size of the clusters that are formed
when $A$ and $B$ are considered, $s_A$ and $s_B$,
respectively. One finds that $s_A = 10$ and $s_B = 14$. Therefore, $A$ is kept (because it leads to the smaller
cluster)  and $B$ is discarded. The result of this operation is shown in Fig.~\ref{Fig:Bastas2011}(b).

% figure 3

Similarly to the case of bond percolation, one can introduce here an order parameter $S$ to track the explosive nature
of the transition, which is defined as the ratio of the sites that belong to the largest cluster $s_{max}$ to the total
number of sites in the lattice,
\begin{equation}
\label{site-2}
S=\frac{s_{max}}{N},
\end{equation}
where $N=t$. By monitoring this quantity, Bastas et al. \cite{Bastas2011} found an explosive behavior
similar to that of Figs.~\ref{Fig:Ziff2009} and ~\ref{Fig:kim2010}.

Later on, Choi, Yook, and Kim studied the explosive site percolation
under AP with a product rule in a 2D square, and also found
the existence of a hysteresis loop in the order parameter \cite{Choi2011}. The algorithm is similar to
the one in Eq.~(\ref{nanostick-1}) with the word ``sticks'' replaced by ``sites''. The reverse AP process is also
the same as in Ref.~\cite{Kim2010}. The main difference is that $n(m)$ in
Eq.~(\ref{nanostick-1}) is now at most $4$, while $n (m)$ for bond percolation is always $2$.
Choi et al. noticed a qualitative difference in the hysteresis loops between the SR and PR cases.
Let $A(L)$ represent the area enclosed by $P_{max}$ for the increasing and decreasing processes
on $L$. The SR shows that $A(L)\rightarrow 0$ as $L\rightarrow \infty$ \cite{Bastas2011}, while the PR
shows that $A(L)$ increases when $L$ increases and then saturates to a nonzero value \cite{Choi2011}.
In a 2D lattice, the PR induces a  transition whose nature is completely different from that induced by the SR, as the percolation transition caused by the PR is discontinuous.

\medskip

\subsection{Mechanisms for explosive percolation}
\begin{figure}
	\centering
	\includegraphics[width=0.55\columnwidth]{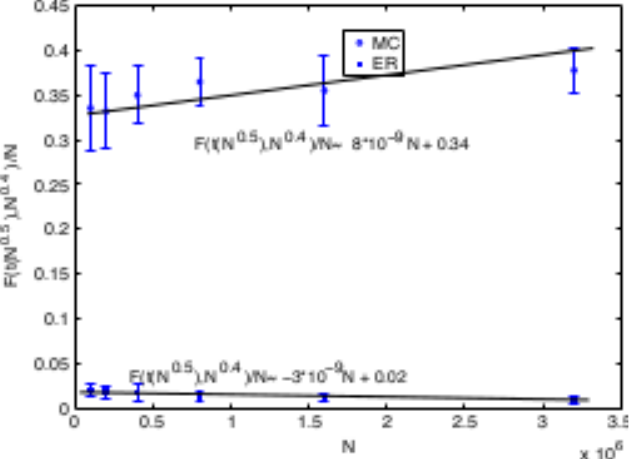}
\caption{Size of the powder keg ($\frac{1}{N}F(t(N^{0.5}), N^{0.4})$) as a function of $N$, for
the ER (squares) and AP (circles) sum rules (cluster minimized).  Reprinted with
permission from Ref.~\cite{Friedman2009}. $\copyright$  2009 by American Physical Society. \label{fig:powderK}}
\end{figure}

The first studies and predictions on EP (with Achlioptas processes on top of ER random networks) were mainly based on numerical
simulations  \cite{Achlioptas2009}. Subsequent studies, however, tried to dwell more deeply upon the generation mechanisms for the observed explosiveness in the transition.

\subsubsection{Powder keg}

First, Friedman and Landsberg tried to relate EP with a sort of ``powder keg'' mechanism
taking place immediately before the explosive transition in ER-type models \cite{Friedman2009}.
More specifically, before the onset of the transition, a fixed non-zero fraction of
nodes in a small-sized cluster constitutes  a ``powder keg''.
Let therefore $F(\tau, a)$ be the number of nodes in clusters of size $\geq a$ after the
addition of the $\tau$-th edge. In AP processes with sum rules, one finds that the
powder keg
\begin{equation}
W = \frac{1}{N} F(t(N^\alpha), N^{1-\beta})
\end{equation}
approaches a non-zero constant in the large $N$ limit for some $0 < \alpha < 1$,
and $\beta < 1$. In other words, at time $t(N^\alpha)$ (which is the beginning
of the phase transition) a fixed non-zero fraction of nodes are contained in
small clusters with sizes ranging from $N^{1-\beta}$ to $N^\alpha$. This set is
termed as the powder keg, and has a size approaching  a non-zero constant
in the large $N$-limit under the AP rule, while the powder keg is empty in the
large $N$ limit for ER rules. Such a conclusion was based on a linear regression
of data, comparing the size of the powder keg (estimated numerically) and the system size $N$, as
shown in Fig.~\ref{fig:powderK} for $\alpha = 1/2$ and $\beta = 0.6$.

Although each individual cluster in this set of small-sized clusters contains
only a vanishing small fraction of nodes, if taken together these nodes are ignitable and collectively
enable an explosive transition. Based on identified powder keg of
appropriate sizes, a heuristic criterion has been proposed to predict
whether or not a given random network will display EP. In
particular, it is conjectured that any ``reasonable'' network model for which
the probability of creating a cluster of size $a$ at time $t$ is
proportional to $(F(t, a/2)/N)^p$ with $p > 1$ will produce a powder keg. A more
simplified criterion based on powder keg has been proposed in
Ref.~\cite{Hooyberghs2011}, which concludes that, under general conditions, a
percolation process will be explosive if the mean number of nodes per cluster
diverges at the onset of the phase transition in the thermodynamic limit.

\subsubsection{Suppression principle}

\begin{figure}
	\centering
	\includegraphics[width=0.35\textwidth]{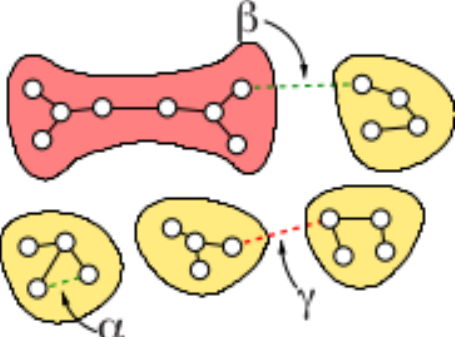}
\caption{Three candidate bonds are considered for possible addition at a given step, namely $\alpha, \beta$ and $\gamma$. Among these bonds, $\alpha$ has the
smallest probability to be added due to condition (ii) of the main text, $\beta$ can not be accepted due to condition (i), and the most probable bond is the $\gamma$ bond.
Reprinted with permission from Ref.~\cite{Moreira2010}. $\copyright$  2010 by American Physical Society. \label{fig:mergingBonds}}
\end{figure}

The original model of AP  essentially identifies the dynamics
that prevents the creation of a giant cluster by choosing one edge from a given
number of randomly selected potential candidates \cite{Achlioptas2009}. The
following two ingredients are sufficient for obtaining an explosive transition
\cite{Moreira2010}: (i) the size of all growing clusters should be kept
approximately the same, and (ii) the merging bonds should dominate with respect
to the redundant bonds. The term ``merging bonds'' indicates those edges that connect
vertices in distinct clusters, while ``redundant bonds'' are edges connecting
vertices in the same cluster, as shown in Fig.~\ref{fig:mergingBonds}.
These two conditions are sufficient for obtaining a first-order like
transition in a growth process where bonds are included one by one. More
precisely, merging bonds must be introduced with much higher probability than
redundant ones.

\begin{figure}
	\centering
	\includegraphics[width=0.45\columnwidth]{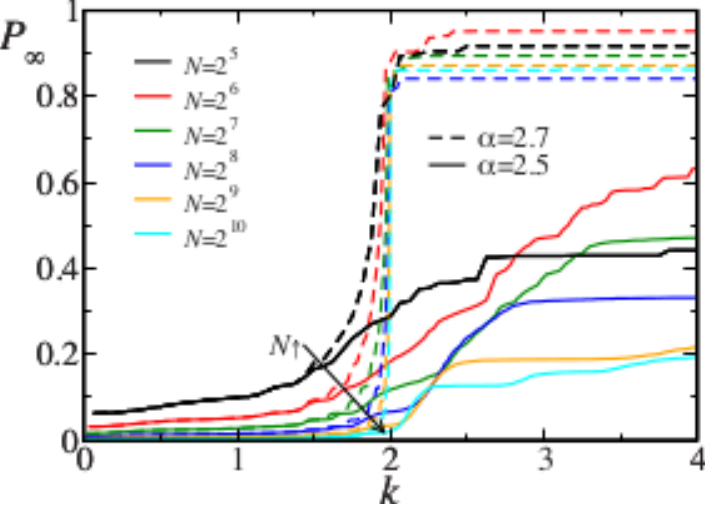}
\caption{(Color online). Transition to EP:  the process shows (i) a slow
continuous growth for the largest cluster when it favors redundant bonds for
$\alpha = 2.5$ (continuous lines), and (ii) an abrupt transition when the
merging bonds are favored for $\alpha = 2.7$ (dashed lines). Legend indicates the color code of the curves.
Reprinted with permission from Ref.~\cite{Moreira2010}. $\copyright$  2010 by American Physical Society.\label{fig:C2DTHpre}}
\end{figure}

By defining a Hamiltonian $H(G)$ of a graph $G$ that includes the two ingredients,
one can explicitly study the distinct roles of merging and redundant bonds.
$H(G)$ reads as
\begin{equation} \label{eq:HG}
H(G) = \sum_{i \in \bf{C}} s_{i}^2 + l_i s_i^\alpha,
\end{equation}
where the sum is over the entire set of clusters $\bf {C}$, $s_i$ is the number
of vertices in cluster $i$, and $l_i$ is the number of redundant bonds
added to the cluster. The parameter $\alpha$ controls the probability of
redundant bonds to be added to the system. For small values of $\alpha$,
redundant bonds are favored over merging bonds, and one might expect that a new
merging bond will be included only after the addition of all possible redundant
bonds. The situation becomes different when $\alpha$ is large, and
redundant bonds are not included. By tuning the parameter $\alpha$ in Eq.~\eqref{eq:HG} from
small to large values (from $\alpha = 2.5$ to $2.7$), a change from a slow continuous growth to a sharp transition is observed, as
shown in Fig.~\ref{fig:C2DTHpre}, which reports  the fraction occupied
by the largest cluster $P_{\infty}$ vs. the average connectivity $k$.
% , which
% shows that in the case of large value of $\alpha=2.7$, the transition is delayed
%  to $k\approx 2$ abruptly.

\begin{figure}
	\centering
	\includegraphics[width=0.7\columnwidth]{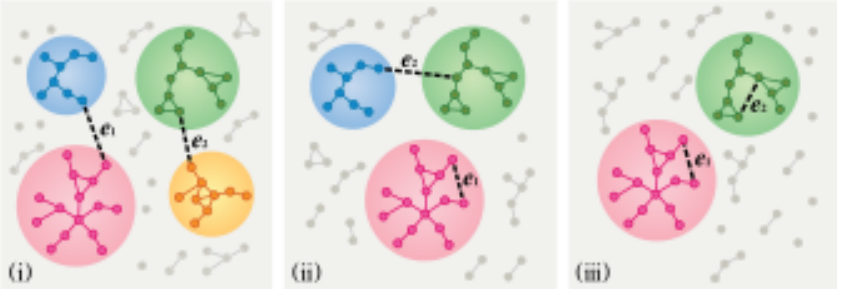}
\caption{Classification of types of edge candidate pairs. (i) Both candidates
$e_1$ and $e_2$ are inter-cluster edges. (ii) $e_1$ is an intra-cluster edge and
 $e_2$ is an inter-cluster edge. (iii) Both $e_1$ and $e_2$
are intra-cluster edges. Reprinted with permission from Ref.~\cite{Cho2011}. $\copyright$  2011 by American Physical Society. \label{fig:SR_RPL20011}}
\end{figure}

The same idea is also referred to as the suppression principle
(SP) \cite{Cho2011}. %  that can systematically supresses the growth of clusters \cite{Cho2011a},
% which shows that not all AP processes explicitly follow the basic idea of suppression
% principle.
As schematically reported in  Fig.~\ref{fig:SR_RPL20011}, one first classifies
the types of edge candidate pairs $e_1$ and $e_2$ as follows:
(i) both $e_1$ and $e_2$ connects clusters of potentially different sizes.
Clusters of sizes $s_{1a}^{(i)}$ and
$s_{1b}^{(i)}$ are connected by the edge $e_1$, and clusters of sizes
$s_{2a}^{(i)}$ and $s_{2b}^{(i)}$ are connected by the edge $e_2$; (ii) one edge
(say, $e_1$) is an intra-cluster edge (in a cluster of size $s_1^{(ii)}$), while the
 other edge ($e_2$) is an inter-cluster edge and connects the two clusters of sizes
$s_{2a}^{(ii)}$ and $s_{2b}^{(ii)}$; and (iii) both $e_1$ and $e_2$ are
intra-cluster edges in either the same cluster or two distinct clusters.

\begin{figure}
	\centering
	\includegraphics[width=0.5\columnwidth]{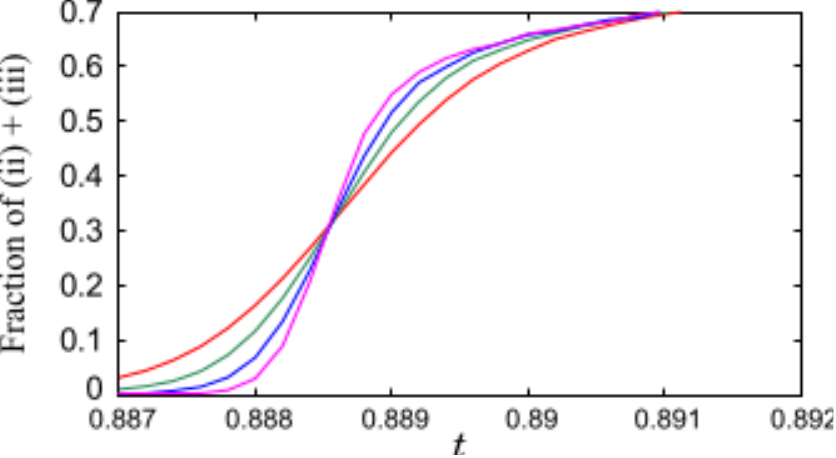}
\caption{The fractions of potential edges falling in case (ii) and case (iii) as  a function
of time $t$ for different system sizes: $N/10^4=32$ (red curve), $64$
(green curve), $128$ (blue curve), and $256$ (magenta curve)  from
the top (bottom) in the small-$t$ (large-$t$) region. As $N$ increases, the fraction
increases dramatically. Adapted with permission from Ref.~\cite{Cho2011}. Courtesy of B. Kahng. \label{fig:SRtypeII} }
\end{figure}

Based on the above classification, several variants of dynamic rules
have been proposed to determine whether the SP is fulfilled. In a class of
models (models of type A) one edge is an intra-cluster edge (cases (ii) or (iii)).
In another class (models of type B) when the case (ii) is considered
and the intra-cluster edge is selected, so that the cluster size does not increase.
When the two potential edges are both intra-cluster edges [case (iii)], one of them is
randomly selected. In a third class of models (models of type C), the dynamics proceeds
via only inter-cluster  connections [case (i)]. Models of type C may be regarded to be
the same as models of type B, because the clusters do not grow when the
intra-cluster edge is selected. The difference between type B and type C models
is that time  advances in type B model but not in type C model.
Type A models can fail to follow the SP. For instance,
when $s_1^{(ii)} = 5$, $s_{2a}^{(ii)}=3$, and
$s_{2b}^{(ii)}=7$, the edge $e_2$ is selected in type A models, and therefore, the size
of the resulted cluster is $10$. On the other hand, if edge $e_1$ is selected,
then none of the clusters would increase in size. As time approaches the
percolation threshold, intra-cluster edges can be selected more frequently, which
leads to a more frequent failure of suppression principles, as shown in Fig.~\ref{fig:SRtypeII}.

In addition, based on the same  classification, a fundamental difference
between the product and sum rules can be identified.
In particular, the PR may not satisfy the suppression principle. A simple example
 of two inter-cluster connections can be considered, in which
$s_{1a}^{(i)} = 2$, $s_{1b}^{(i)} = 7$, $s_{2a}^{(i)} = 4$, and $s_{2b}^{(i)} =
4$. The product of the two inter-clusters is respectively $P_1^{(i)} = 14$
and $P_2^{(i)} = 16$, and therefore, edge $e_1$ is added to the system.
However, the resulting cluster size is $9$ in this case, which is larger than the resulting
size $8$ when $e_2$ is added. Therefore the SP is not fulfilled.

\begin{figure}
	\centering
	\includegraphics[width=0.8\columnwidth]{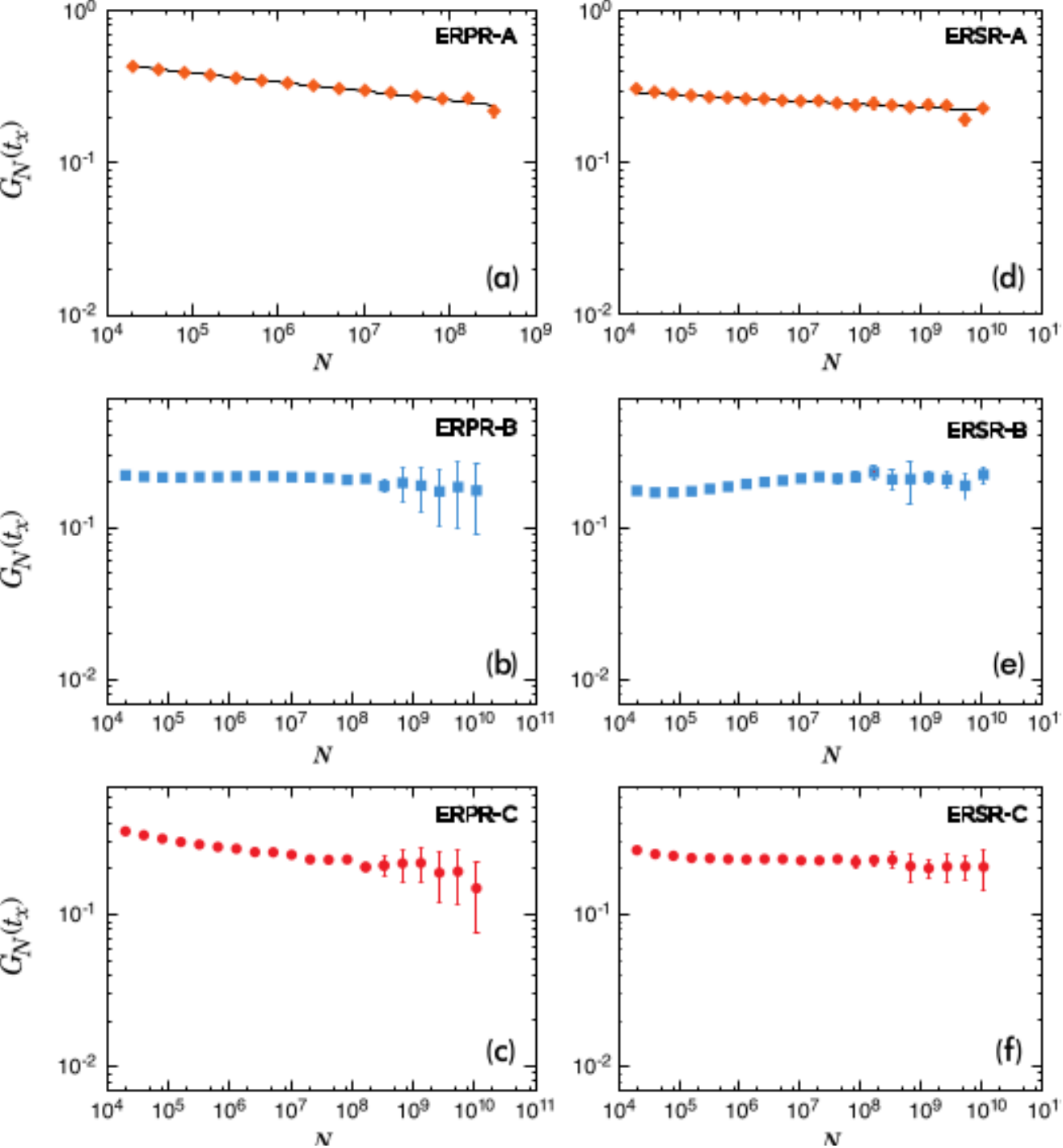}
\caption{$G_N(t_x)$ vs. $N$ for models of product rules (a)-(c), and sum rules
(d)-(f). (a) and (d) refer to type A models, (b) and (e) refer to type B models, and (c) and (f) refer to type C models. Best linear fits of the data occur for slopes of  $-0.06$ (a),  $-0.05$ (b),  $-0.05$ (c),  $-0.02$ (d),  $0$ (e), and  $0$ (f).  In the case of sum rules in
type B and C models, $G_N(t_x)$ converges to finite values in the thermodynamic
limit. Adapted with permission from Ref.~\cite{Cho2011}.  Courtesy of B. Kahng.\label{fig:SRFitting}}
\end{figure}

The SR  may remove the drawback inherent to the PR, yielding
discontinuous percolation transition as observed numerically \cite{Cho2011}. To
evaluate the discontinuity of the transition, the following two criteria have been
proposed: (a) the values $(t_x(N), G_N(t_x))$ remain finite as $N \to
\infty$, where the time $t_x(\infty)$ is regarded as the transition point $t_c$
in the thermodynamic limit, and $G_N(t_x)$ represents the $G$ components at time
$t_x$; (b) the tangent of the curve of $G_N(t)$ with respect to $t$ at $t_x(N)$
diverges as $N$ increases. More specifically, as shown in
Fig.~\ref{fig:SRFitting}(a)-(c), for all models with product rules,
$G_N(t_x)$ decreases with increasing $N$ in the whole considered range,
suggesting that $G_N(t_c) \to 0$ in the limit $N\to \infty$. Furthermore, when
the SR is implemented in type A models, a similar convergence of $G_N(t_c) \to 0$ in
the limit $N\to \infty$ is observed in Fig.~\ref{fig:SRFitting}(d). This
indicates that these cases violate the SP, and lead to a continuous percolation
transition. However, when the SR is implemented in type B and C models,
the asymptotic values of $G_N(t_x)$ look relatively flat, yielding discontinuous percolation transitions within the range of the
numerical data (see Figs.~\ref{fig:SRFitting}(e)-(f)).

The original dCDGM model \cite{daCosta2010} does not follow
the suppression principle, because it does not take into
account the natural choice of intra-cluster edges. The selection of an
intra-cluster edge becomes actually more and more frequent as time approaches the percolation
point, and this yields a continuous transition. However, once
the dCDGM model is properly modified, discontinuities in the transitions can be
observed. The suppression principle plays a fundamental role in achieving
discontinuous percolation on random graphs,  inter-dependent networks
\cite{Gao2011,Baxter2012}, adaptive networks
\cite{Echenique2005,Gross2006}, and fractal networks \cite{Boettcher2012}.

\begin{figure}
	\centering
	\includegraphics[width=0.5\columnwidth]{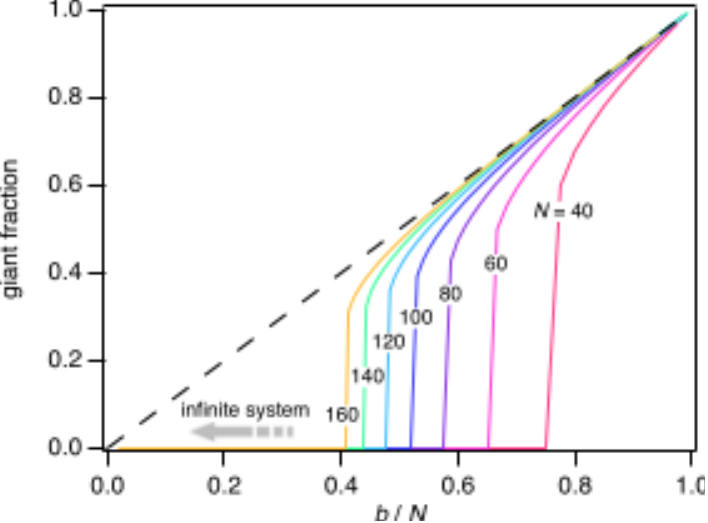}
\caption{Effect of system size on the discontinuity of one dimensional models. In
the limit $N \to \infty$, the discontinuity evaporates since the giant cluster
forms at $b / N \to 0$, where $b$ is the number of edges that have been added to
the system. Reprinted with permission from Ref.~\cite{Matsoukas2015}. $\copyright$  2015 by American Physical Society. \label{fig:finiteSR}}
\end{figure}

For a finite size system, suppressive rules are not a necessary
condition for the sharpness of the transition at the percolation point
\cite{Matsoukas2015}. Rather, a finite size system with aggressive tendency to form a
giant cluster may exhibit an instability that is
relieved through an abrupt and discontinuous transition to the stable branch.
Although the Author of Ref.~\cite{Matsoukas2015} has tracked analytically
such an abrupt transition to
percolation, it is worth noticing that discontinuity is especially pronounced
in small size networks, but it disappears when the size tends to infinite, as
shown in Fig.~\ref{fig:finiteSR}.
This suggests that in the thermodynamic limit the system
actually shows a continuous transition to percolation, a point which will be further
discussed in Section~\ref{sec:EPc}.

Generally speaking, the idea of AP is that of choosing the best among randomly given
multiple options, in order to avoid the formation of a certain target pattern
\cite{Riordan2011}. In Ref.~\cite{Cho2013} the spanning cluster
rather than the giant cluster is chosen as the target pattern. Such a latter choice
enables one to determine analytically the order of the EP
transition in the thermodynamic limit, and in particular
for percolation in a Euclidean space. Spanning cluster
avoiding models feature either discontinuous or continuous percolation
transition, depending on the dimension of the Euclidean space in which the
dynamics is implemented. More specifically, the EP is continuous or
discontinuous if $d < d_c =6$, and it is always continuous if $d \ge d_c$ (where
$d$ is the spatial dimension), this way explaining some of the confusing and contradictory
behavior previously observed \cite{Ziff2013}.

\subsubsection{Growth by overtaking}

The growth by dominant overtaking has been interpreted as the
mechanism responsible for the discontinuous percolation transition of the
Bohman-Frieze-Wormald (BFW) model \cite{Chen2012}. More specifically, the
BFW model has been extended to the case where  various constraints exist
on the occupation of competitive edges. Such a simple stochastic graph evolution
examines only one edge at a time, which leads to a discontinuous transition
provided the control parameter $\beta$ is small enough. The overtaking mechanism
means that the size of the largest component does not change by direct growth,
but it changes when two smaller components merge and become the new largest
component. On the other hand, a continuous transition appears if $\beta$ is
large enough to allow substantial direct growth of the giant component. Notice
that the BFW model has been originally proposed to study discontinuous
percolation with multiple giant clusters in Ref.~\cite{Chen2011}, and was further
studied on finite dimensional square, simple-cubic lattices, and SF
networks obtaining similar results
\cite{Schrenk2012,Zhang2012,Zhang2013b,Chen2015b}. Furthermore, in the
generalized BFW model, it has been shown that the emergence of global connectivity
is announced by microscopic transitions of the largest component \cite{Chen2014}.

\subsubsection{Recursive and hierarchical structures}

Networks with a recursive and hierarchical structure induce explosive
percolation dynamics as well, although the formation of an extensive cluster is
restricted to ordinary (uncorrelated) and random bond addition
\cite{Boettcher2012}. This suggests that a hierarchy of SW bonds
grafted onto a one-dimensional lattice results in an EP
transition, even if bonds are added sequentially in an uncorrelated manner.

\subsection{Debates on explosive percolation}
\label{sec:EPc}

The results of Ref.~\cite{Achlioptas2009} have triggered a big debate in the community,
aiming at definitely clarifying whether the observed transition are discontinuous or not.
In  equilibrium systems, discontinuous
phase transitions are usually identified as first-order transitions, while its
counterpart of continuous transitions are called second-order transitions.
However, the term of ``explosive percolation transition'' was solely based on
the numerical evidence of a jump (a sort of discontinuity from a
visualization perspective) in the order parameter at the critical transition
point. Therefore, all  numerical results in Refs.~\cite{Ziff2009,Cho2009,Radicchi2009,Friedman2009,Radicchi2010,DSouza2010,Cho2010b,Ziff2010,Hooyberghs2011,Manna2011} initially focused on
searching for the existence of such a discrete jump.

\subsubsection{Finite-size scaling theory}

While the existence of a discontinuous jump in EP was confirmed since the earliest studies on the subject,
most other properties of a first-order transition (like cooperation, phase coexistence, and nucleation
\cite{Grassberger2011}) had to be proved.
From the physical point of view, the competitive choice
of a new link suppresses the formation of large clusters, which yields the
threshold value significantly delayed in comparison to
random percolation. Therefore, the competitive percolation transition can be
extremely abrupt with a large (seemingly discontinuous) jump in the connectivity
of the graph. Although systems of rather large sizes (i.e., tens of billions of
nodes) are used in the original AP model, the heuristic arguments of
discontinuity at the percolation transition disappears in the thermodynamic
limit. Therefore, the term ``explosive'' refers here to the unusual feature
which distinguishes the critical behavior of the AP from both ordinary and truly
discontinuous models \cite{Squires2013}.

Finite-size scaling theory is useful for
characterizing phase transitions \cite{Bruce1992,PrivmanBook}. Traditionally, the percolation transition has
been conceived as a typical continuous phase transition displaying a power law size distribution, and
a set of standard scaling properties. Therefore, a percolation transition is continuous if
one finds the signature of power law distribution of component sizes. In contrast,
it is discontinuous if a power law distribution does not exist.

The first systematic analysis of cluster size distributions for AP
 by finite-size scaling theory has been reported in Ref.~\cite{Radicchi2010},
including lattices, random, and scale free networks.
 For continuous phase transitions, every variable
$X$ of interest near the percolation threshold $p_c$ (critical average
connectivity of the network in the limit of infinite size) is scale independent,
due to the infinite correlation length of the system at $p_c$, so it has a  power-law form
\begin{equation}
X \sim |p - p_c|^\beta,
\end{equation}
where $\beta$ is a critical exponent. On a system of a finite size $N$, the
scaling of the variable $X$ is normalized as follows:
\begin{equation} \label{eq:scaleR}
X = N^{-\beta/\nu} F[ (p - p_c) N^{1/\nu}  ],
\end{equation}
where $\nu$ is another critical exponent, and $F(\cdot)$ is a universal function.
Both critical exponents $\beta$ and $\nu$ are associated with the magnetization
and the correlation length, respectively. For $p=p_c$, the variable shows a
simple scaling $X\sim N^{-\beta/\nu}$, which can be used to deduce $\beta / \nu$
by examining several systems of different sizes. Furthermore, if $p_c, \beta, \nu$
are known, the plot of $X N^{\beta/\nu}$ as a function of $(p-p_c) N^{1/\nu}$
yields the universal function $F$, which does not depend on $N$.

In the case of percolation, one chooses $X$ to be the average value $\left < S \right >$ of the
order parameter $S$, which measures the relative size of the largest
connected component with respect to the total system size $N$ (i.e., the number
of nodes belonging to the largest connected component divided by $N$), and which is
also called ``percolation strength'' \cite{Radicchi2010,daCosta2010,Grassberger2011}.

The identification of the percolation threshold $p_c$ is performed in several
independent ways \cite{Radicchi2010}. One way is to plot the order parameter
$\left < S \right >$ as a function of the system size $N$, for a given value of
$p$. A correct value of the percolation threshold can be determined by finding
the value of $p$ which yields the best power-law fit. The second method consists
in using the scaling of the pseudo-critical points $p_c(N)$, defined as
\begin{equation}
p_c = p_c(N) + bN^{-1/\nu},
\end{equation}
where $b$ is a constant to be determined from the fit, together with
the other critical parameters $\nu$ and $p_c$. The pseudo-critical value $p_c(N)$
can be simply obtained by the monitoring positions of the peaks of $S$ at different system
sizes $N$.

An alternative quantity that can be adopted as variable $X$ in Eq.~\eqref{eq:scaleR}
is the susceptibility, which again can have two slightly different definitions. The first is
to consider the mean cluster size $\chi_1 = \sum_s n_s s^2 / \sum_s n_s s$,
where $n_s$ stands for the number of clusters of size $s$ per node. The sums run
over all possible values of $s$, except for the one of the largest cluster
\cite{Radicchi2010,Cho2010b}. The second definition is obtained by considering the
amplitude of the fluctuations of the order parameter $S$, i.e. $ \chi_2 = N
\sqrt{\left< S ^ 2\right> - \left< S \right>^2}$ \cite{Radicchi2009,Cho2010b}.
When these susceptibility measures are used, the critical exponent is indicated
by $\gamma$, and the scaling ansatz is $N^{\gamma/\nu} F'[ (p - p_c) N^{1/\nu}
]$, where the universal function $F'$ is different from $F$. % Depending on the
% particular variables $X$ and specific percolation models on different topologies
% (i.e., lattices embedded in different dimensions, ER random networks, scale free
% networks etc.), one often obtains a set of critical exponents and scaling
% functions.

\begin{figure}
	\centering
	\includegraphics[width=0.4\textwidth]{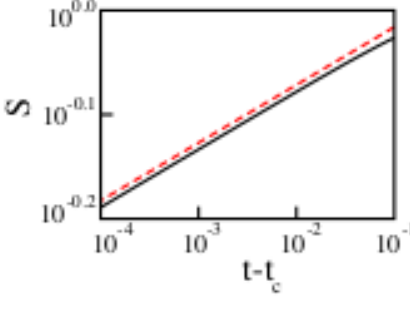}	
\caption{(Color online). Log-log plot of the relative size $S$ of a percolation cluster vs.
$t-t_c$ (black solid curve), showing a power law regime with exponent $\beta = 0.05555$ (red dashed line). Adapted with permission from Ref.~\cite{daCosta2010}. Courtesy of R. A. da Costa. \label{fig:daCostaDelta}}
\end{figure}

For ER networks, the order parameter $S$ starts to be defined at the percolation threshold
$p_c = 1/2$, and grows with time as $S \sim | p - p_c |^\beta$,
with $\beta = 1$. For randomly diluted $d$-dimensional lattices, $\beta$ depends
on $d$. At $p = p_c$, the cluster size distribution $n(s)$ (the fraction of finite
connected components of $s$ nodes) scales as $n(s) \sim s^{-\tau}$, with $\tau = 5/2$.
When the transition is  discontinuous, the finite size scaling theory does not apply anymore, and the scaling
relation of Eq.~\eqref{eq:scaleR} trivially holds with $\beta = 0$. The curves
of $S$ versus $p$  do not scale as well, and  $p$ approaches $p_c$ at increasing $N$
faster than a power law.

Simulations of  EP performed for AP-type processes
reveal power law distributions of cluster sizes as well, resembling the
results of continuous percolation transitions. It turns out that this
self-contradicting combination of discontinuity and scaling has made EP one of
the most challenging and urgent issues. The first contradicting result has been
reported in Ref.~\cite{daCosta2010},  which concludes that EP is actually
continuous for a modified version of the AP (the dCDGM model), and analytically derives
the critical scaling relations based on numerical observations of power-law
critical distribution of cluster sizes, with a unique small critical
exponent $\beta \approx 0.0555$. The numerical
simulation for a very large system ($2 \times 10^9$ nodes) shows a pronounced power
law regime, which indicates a continuous transition (see Fig.~\ref{fig:daCostaDelta}).
The small $\beta$ exponent  makes the
transition so sharp that it is almost indistinguishable  from a discontinuous one.

Further systematic analysis of finite size scaling was performed for four Achlioptas-type processes  \cite{Grassberger2011},
showing unusual  behaviors with non-analytic scaling functions which demonstrate that AP-like percolation
transitions are indeed continuous. More specifically, Ref.~\cite{Grassberger2011} characterized
the phase transitions in terms of the distribution $P_{p,N}(S)$ of the order parameter
$S$, where $p$ is the control parameter and $N$ is the system's
size. The distribution $P_{p=p_c, N}(S)$ at criticality scales, for continuous
transitions, as
\begin{equation}
P_{p=p_c, N} (S) \sim N^{\beta/\nu} f(S N^{1/\nu}),
\end{equation}
where $f$ is a universal function. This equation is directly related to the
finite size scaling relation of Eq.~\eqref{eq:scaleR}.
For a continuous phase transition, the universal function $f$ might be
double peaked, but then, as $N \to \infty$, the dip between two peaks usually
does not deepen and the horizontal distance between them shrinks to zero, so that
$P_{p=p_c,N}(S)$ becomes single humped. In contrast, in typical first-order
transitions, $P_{p=p_c, N}(S)$ is double peaked with a deepening valley between
the two peaks. Furthermore, the distance between the peaks tends to be equal to the value of the jump
observed in the order parameter $S$.

\begin{figure}
	\centering
	\includegraphics[width=0.7\columnwidth]{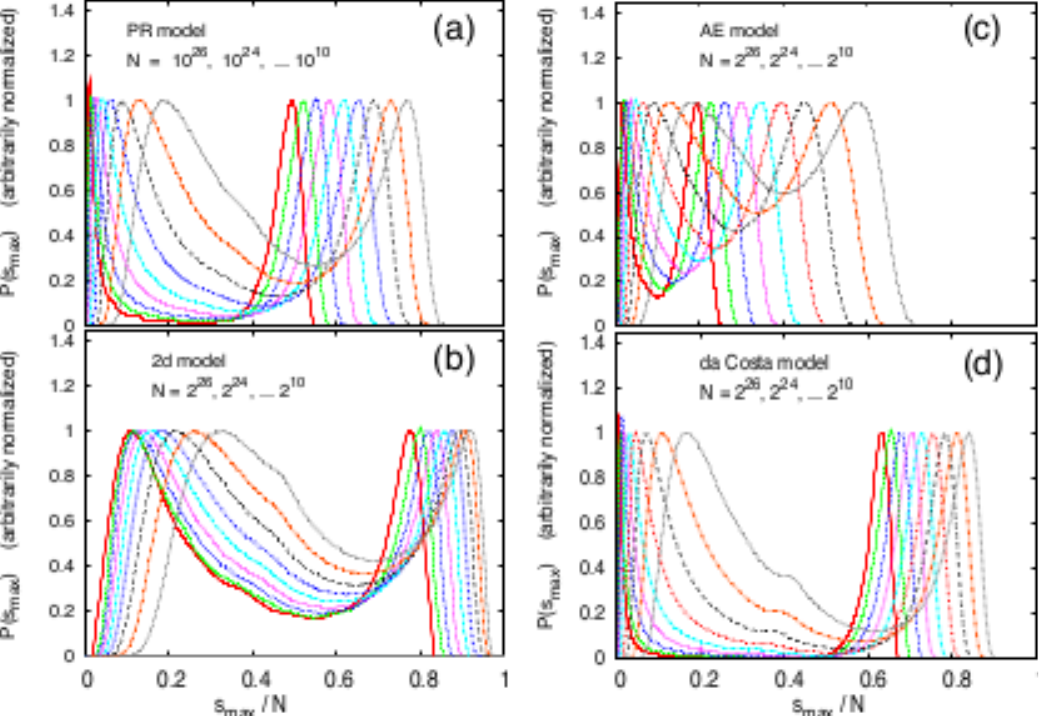}
\caption{(Color online). Distributions of the order parameter $S = s_{max}/N$ for four EP
model: (a) the original product rule from Ref.~\cite{Achlioptas2009}; (b) the product rule
on a 2D square lattice; (c) the adjacent edge rule; and (d) the dCDGM rule
\cite{daCosta2010}. Adapted with permission from Ref.~\cite{Grassberger2011}. \label{fig:grassberger}}
\end{figure}

The results are shown in Fig.~\ref{fig:grassberger}. Notice that, while the two peaks seem to indicate an indeed explosive character of the percolation transition,
they do not yet prove that EP is discontinuous. For the latter, one must also show that the distance between the peaks
does not vanish  for $N \to \infty$. It turns out that the valley
between the peaks deepens with increasing $N$ in all cases, but at the same time both peaks
shit to the left (see Fig.~\ref{fig:grassberger}). Ref.~\cite{Grassberger2011} also confirms
that the critical exponent $\beta$ is rather small. Furthermore, the distance between the peaks
decreases for large $N$. As a matter of fact, one may extract information on the  continuous transition
from the depth and position of the valley \cite{Tian2012}.

In the original AP model with the PR on a complete graph, a careful
finite size scaling analysis has shown that EP transition is indeed continuous
in the thermodynamic limit \cite{Lee2011}. For percolation
processes, the main question is whether the gap ($S |_{p \to p_c}$) vanishes or not.
To this end, the strategy consists in the following steps: (i) the
information above the transition point $(p > p_c)$ is used to set up lower and upper
pseudo-transition points, $p_l(N)$ and $p_u(N)$. This
provides an easy way to find $p_c$, since one would expect that both
pseudo-transition points converge to $p_c$ as $N \to \infty$.
(ii) The upper bound for the size increase of the largest cluster  between $p_l(N)$ and
$p_u(N)$ is found, namely $\Delta G = G_u(N) - G_l(N)$. This is equivalent to consider
the gap of the order parameter  between the two pseudo-transition points $\Delta S =
\Delta G / N$. (iii) Such an upper bound turns out to be sub-linear in $N$ for a continuous
percolation transition, and therefore the investigation of the  cluster size distribution leads to
conclude that the explosive percolation transition is indeed continuous.

Other developments regarding the application of finite size
scaling analysis to EP can be found in Refs.~\cite{Cho2010b,Li2012a}.
For a wide set of representative AP models, an exhaustive
investigation has been conducted in Ref.~\cite{daCosta2014b}, where the full set of
scaling relations between critical exponents was found, together with the scaling
functions and the upper critical dimension.

\begin{figure}
	\centering
	\includegraphics[width=0.5\columnwidth]{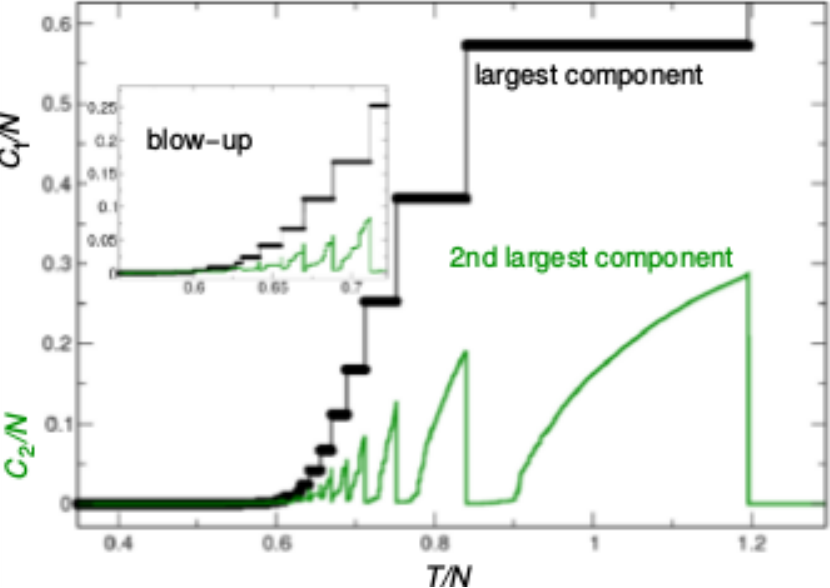}
\caption{(Color online). Staircase shape of the multiple discontinuous stochastic process for $N
= 655,360$. Figure taken from Ref.~\cite{Nagler2012}. Courtesy of Harold W. Gutch. \label{fig:SR_RPX2012}}
\end{figure}

\subsubsection{Discontinuity of the transition}

Finally, the continuity of the EP transition has been rigorously proven in Ref.~\cite{Riordan2011},
which indeed mathematically demonstrated that   continuity is an essential feature of all mean-field AP models,
and of a wide range of their variants. For instance, any competitive rule based on picking a fixed number of edges would lead
to a continuous transition. Furthermore, it is shown that the order parameter of
the AP percolation models has random fluctuations that do not disappear in the
thermodynamic limit \cite{Riordan2012a}. On the other hand, a discontinuous phase
transition may appear if the number of nodes sampled grows with the network
size \cite{Waagen2014}. Discontinuous transitions are observed in some models (such as, for instance, the BFW model
\cite{Chen2011}), and in a hierarchy of SW random bonds model
\cite{Boettcher2012}. Notice that the results of Ref.~\cite{Riordan2011} do not imply
that the percolation transition at $p_c$ is characterized by a power law
divergence of the order parameter, nor they mean that a globally continuous transition
cannot be entangled by multiple discontinuous transitions. For instance, for
certain $l$-vertex rules, the continuous percolation transition can have an
incomplete devil's staircase with multiple discontinuous steps in the vicinity
of $p_c$ \cite{Nagler2012}, as reported in Fig.~\ref{fig:SR_RPX2012}.

\subsubsection{Irreversibility of the transition}

The last point to be discussed is that EP processes are irreversible, in
contrast to traditional continuous percolation. In the latter case indeed, one can
reach any state either adding or removing connections. For EP, instead, only adding links
makes sense, while the inverse process is impossible \cite{daCosta2010}. This
argument suggests that any efforts in searching hysteresis behavior in EP would fail.
Typically, hysteresis is considered to be a
necessary condition of a first-order phase transition.  For AP processes on
2D square lattices of finite size, hysteresis behavior between the
forward (adding edges) and the reverse (removing edges) processes has been
numerically observed in Ref.~\cite{Bastas2011}. However, the hysteretic area tends
to 0 in the thermodynamic limit, which indirectly shows that the product
rule of EP processes generates a continuous transition.
In a different model embedded in a Euclidean space (that chooses the
spanning cluster instead of the giant cluster as the target pattern) a
hysteretic loop has been observed in discontinuous percolation transitions \cite{Cho2013}.

% %%%%%%%%%%%%%%%%%%%%%%%%%%%%%%%%%%%%%%%%%%%%%%%%%%%%%%%%%
\section{Synchronization as a continuous phase transition} % Section 3

\subsection{Introduction and historical overview}
Synchronization
is possibly one of the most ubiquitous natural phenomena. There, inanimate or living systems \cite{Berge1984} tend to adjust
their own rhythms due to an existing coupling interaction, or a driving force.
It was known for few centuries that two pendulum clocks can synchronize when hanging
from the same piece of wood. Since then, synchronization has captivated the
interest of scientists and, nowadays, it is investigated in fields as diverse as
natural and social sciences \cite{Pikovsky2001}, or engineering
\cite{Blekhman1988}.
Popular examples of systems exhibiting synchrony are: birds flocking together, fishes swimming
 in schools, male fireflies flashing together, people clapping in
unison after a performance, pacemaker cells in the
heart, $\beta$-cells in a pancreatic islet, and neurons in the brain. For comprehensive reviews
 providing the detailed description of synchronization phenomena we address the reader to Refs. \cite{Pikovsky2001,Boccaletti2002, Strogatz2003,Manrubia2004,Boccaletti2008}.

When the coupled systems are oscillators, the fate of synchronization always
depends on two ingredients: i) the frequency detuning (the more alike the systems are, the better is
synchronization), and ii) the coupling pattern (i.e. the strength, the network topology, the output function) through which the network's units communicate. It is not sufficient, indeed, that two systems are alike, it is also necessary that the
information exchanged between the systems is of good quality to allow them to synchronize \cite{Letellier2010,Bianco-Martinez2015,Sendina2016}.

\begin{figure}
  \centering
\includegraphics[width=0.8\textwidth]{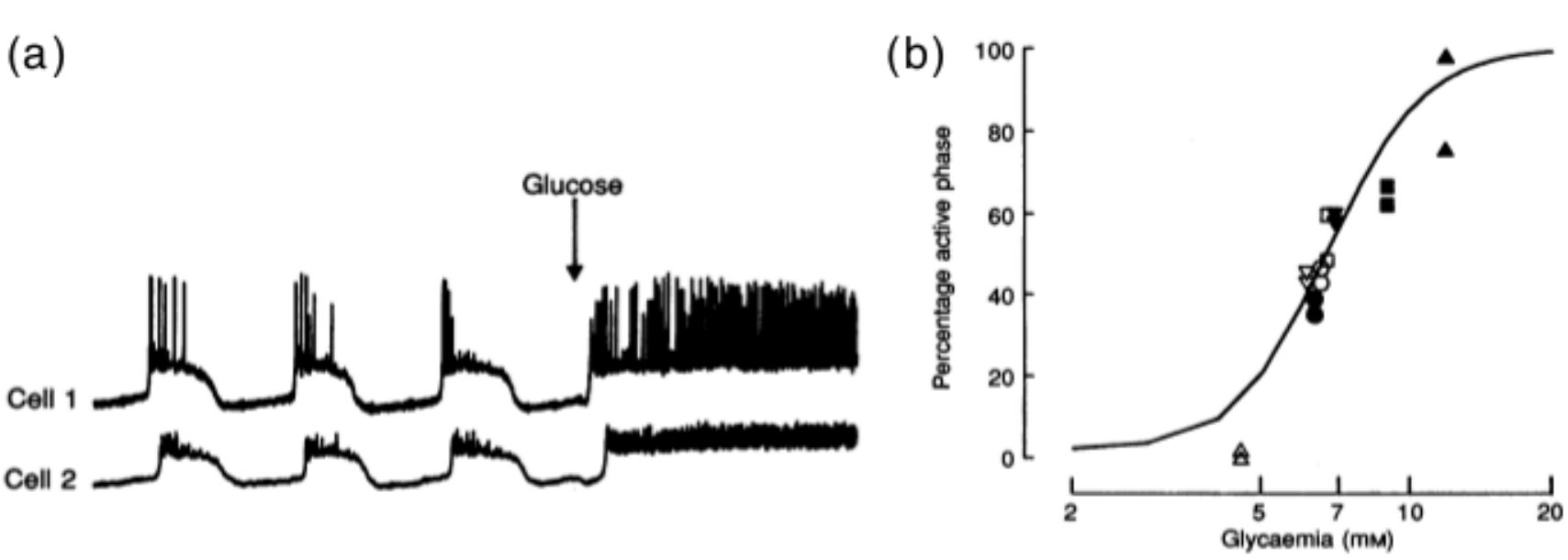}
\caption{(a) Simultaneous recordings of the electrical activity of two
  $\beta$-cells within a given islet of Langerhans of an Albino mice's
  pancreas show synchronous oscillations. (b) Relationship
  between blood glucose concentration and the percentage of time that
  pairs of islets spend in the active phase. Reprinted from Ref.~\cite{Valdeolmillos1996}, published
  under CC(Creative Commons)-BY license. }
  \label{fig31-1}
\end{figure}

As a coupling strength increases, a transition from an incoherent to a coherent phase takes place generically at a
critical value, where the elements follow the same dynamical behavior. The nature of such a
macroscopic rhythm can be described by an order parameter (in analogy with phase transitions in statistical mechanics
\cite{Daido1990}). As first pointed out by
Winfree \cite{Winfree1967}, in most of the cases the transition is second-order like,  with the order parameter exhibiting
power-law divergences at the onset of the synchronization. Figure
~\ref{fig31-1} shows an example of such a transition in the case of $\beta$-cells within a
pancreatic islet, which burst all in synchrony, and therefore produce a coordinated
insulin release in response to glucose \cite{Valdeolmillos1996}.

Abrupt (discontinuous) transitions to synchrony have been recently reported \cite{Kuramoto1984,Pazo2005,Wood2007} where an infinitesimal
variation of the coupling strength gives rise to a macroscopic synchronous phase exhibiting hysteresis, a feature
observed in theoretical models of coupled Josephson junctions \cite{Filatrella2007}, as well as in more complex coupled oscillators models
\cite{Tanaka1997,Bonilla2000,Choi2000,Wood2007,Giannuzzi2007,Rosenblum2007}.
These abrupt transitions will be the object of the next Chapter, whereas the present Chapter has the aim of reviewing
the classical works on
synchronization, the synchronization types and the main
indicators and models used to describe the spontaneous emergence of ordered states.

\subsection{Synchronization of two coupled dynamical systems}
\label{section_two_coupled}

The case of two coupled dynamical systems constitutes the building block for the later study of networked systems.
The problem is properly framed by considering each systems as
described by a $m$-dimensional ($m\ge 2$)
vector ${\bf x}_i(t)$ ($i=1,2$). In
general, the equation of motion is given by:
\begin{equation}
  \label{eq:onesys}
\frac{d{\bf x}_i}{dt}={\bf F}_i({\{\bf x\}}),
\end{equation}

\noindent where $\{\bf x\}\equiv ({\bf x}_1,{\bf x}_2)$. Each unit evolves
according to the velocity field  ${\bf F}_i$, and can reach either a stable fixed point, or a limit cycle, or a chaotic attractor \cite{Berge1984,Strogatz1994}.

Equation (\ref{eq:onesys}) accounts for the general framework in
which the two systems can be of different nature, have different
dimensions, and the interaction can be described by any generic function of their
dynamical states. However, in order to allow an analytical treatment of the problem, the autonomous dynamics is assumed
to be independent from the interaction between the units,
and a particular form of interaction has to be assumed. We here focus on the case in which each unit is coupled to a linear superposition of the outputs, and the interaction is symmetrical:

\begin{equation}
  \label{eq:dynsys}
\frac{d{\bf x}_i}{dt}={\bf f}_i({\bf x}_i) + \lambda\sum_{i=1}^N A_{ij} {\bf h}({{\bf x}_j}),
\end{equation}
\noindent where $\lambda$ represents the coupling strength, $
{\bf f}$ and ${\bf h}$ are vectorial functions, and $A_{ij}$ is a matrix satisfying the rules of a diffusive coupling ($\sum_j A_{ij}=0$). For two units, one has therefore
$\dot{\bf x}_{1,2}={\bf f}_{1,2}({\bf x}_i) + \lambda [{\bf h}({\bf x}_{2,1}) \mp {\bf h}({\bf x}_{1,2})]$.

\subsubsection{Complete synchronization}

If the systems are identical (${\bf f}_i({\bf x}_i)={\bf f}({\bf x}_i)$), the
synchronization is realized by the
equality of the state variables as they evolve in time ($\{{\bf x}_i(t) = {\bf s}(t), i=1, 2\}$), and is
termed as  {\em  complete} (or {\em identical} or {\em full}, or {\em conventional}) synchronization \cite{Afraimovich1986,Pecora1990,Boccaletti2002}.

Complete synchronization (CS), ${\bf x}_1 = {\bf x}_2$, is a threshold phenomenon and occurs beyond a critical value $\lambda_c$ of the coupling strength. CS can be observed by monitoring the mean synchronization error, defined as:

\begin{equation}
  \label{eq:syncerr}
  \langle e \rangle =\lim_{T\to \infty}\frac{1}{T}\int_0^T\left\lVert {\bf x}_1(t)-{\bf x}_2(t)\right\rVert dt
\end{equation}
i.e., the averaged distance to the synchronization manifold ${\bf x}_1(t)={\bf x}_2(t)$.
% \begin{equation}
%   \label{eq:dynsys2}
%   \frac{d{\bf x}_i}{dt}={\bf F}_i({\bf x_i})  + \sigma\sum_{j=1}^2  C_{ij} H_i({\bf x}_j)
% \end{equation}

One of the most popular approaches to assess the stability of CS is the {\em master stability function} (MSF) method, introduced for arrays of coupled systems in Ref.~\cite{Pecora1998}. The MSF is the largest Lyapunov exponent transverse to the synchronization manifold, and measures the exponential rate at which any infinitesimal perturbation decays or grows. Therefore, a necessary condition for CS to occur is that the MSF is negative.

\subsubsection{Phase synchronization}

Real systems, however, are not identical, and they are subjected in general to noise or parameter mismatch preventing CS. Weaker forms of synchronization exists, where correlation or equality of the two systems are limited to proper subsets, or functions, of the variables.

When two periodic oscillators are coupled, for instance, synchronization is observed as
the entrainment of their phases and frequencies. When the systems are
uncoupled, the two stable limit cycles are described by two phases
$\theta_{1,2}$ whose dynamics are governed by their respective natural
frequencies $\omega_1$ and $\omega_2$ as
$\dot\theta_{1,2}=\omega_{1,2}$,  which in general are incommensurate ($\Delta\omega=n\omega_1- m\omega_2\ne 0, \forall n,m\in \cal{N}$).
Phase locking means here that the phase difference ($|n\theta_1-m\theta_2|$) is bounded beyond a certain coupling strength, and the corresponding
instantaneous frequencies $\nu_{i}=\dot\theta_i$ are locked, i.e., $n\nu_1-m\nu_2=0$

The onset of synchronization is therefore a transition from a quasi-periodic motion
(with two incommensurate frequencies) to a motion with a single
frequency, and can be characterized by means of families of
Arnold tongues \cite{Arnold1961}, i.e. synchronization regions where the relationship
$|\frac{\Delta\omega}{\lambda}|\le 1$ holds. Inside those regions, the
oscillators mutually adjust their frequencies until attaining the relation
$n\nu_1=m\nu_2$, with the resulting synchronization
frequency depending on the initial frequency detuning $\Delta\omega$. Moving away from the
border of an Arnold tongue, the phase difference uniformly grows in time.

Phase synchronization has been later extended to chaotic oscillators  \cite{Rosenblum1996}.
When the chaotic oscillators have a
well defined center of rotation, as it is the case of phase coherent systems, the instantaneous phase can be reliably calculated by means of few mathematical techniques \cite{Boccaletti2002}. The mean square displacement of
the phase linearly diffuses in time (with a very small diffusion
constant), due to the pronounced variations of the amplitude modulating
the rotations. As two phase coherent chaotic
oscillators are coupled, a transition from a non-synchronous
regime (where the phase difference linearly grows in time) to a
synchronous regime (where the phase difference
$|\theta_1-\theta_2|$ is bounded and the mean frequencies
are equal) occurs.

When instead the two chaotic oscillators are not phase coherent (and
exhibit a broad distribution of time scales), phases do not fully synchronize, and synchronization episodes are
 interrupted by intermittent phase slips (jumps of $2\pi$ in the
phase difference). Such a latter phenomenon has been called {\em
  imperfect phase synchronization} \cite{Zaks1999} and
it has been observed experimentally in an electronic circuit representing a forced Lorenz
system \cite{Pujol2003}.

\subsubsection{Lag, anticipated, relay and generalized synchronization}

After the setting of a phase synchronization, intermediate values of the coupling strength drive the systems into a stronger degree of synchronization, where also amplitudes start to become correlated. The resulting regime corresponds to the dynamics of one of the oscillators which lags (in {\em lag synchronization} \cite{Rosenblum1997}) or leads (in {\em
  anticipated synchronization} \cite{Voss2000,Voss2001}) the dynamics
of the other.
 Lag synchronization was first shown in symmetrically coupled nonidentical oscillators \cite{Rosenblum1997}, and the time shifting $\tau$ between one and the other state's variable was detected as the minimum of a time dependent similarity function.

The existence of this {\em achronal synchronization} in mutually
delayed-coupled oscillators has been observed in laser systems
\cite{Heil2001}, and studied theoretically in
Ref.~\cite{White2002}. It is possible also to have two completely synchronized chaotic
units in the presence of a delayed coupling, phenomenon known as {\em zero-lag} or {\em isochronal} synchronization
\cite{Fischer2006,Wagemakers2008}. Such a regime can be achieved
via a self-feedback in both coupled units \cite{Klein2006,Peil2007}, or via an indirect
coupling through a relay \cite{Tiana-Alsina2012}, or through a third
driving  unit \cite{DaSilva2006,Zhou2007,Wagemakers2007}, or
when multiple coupling delay times are used \cite{Englert2010}.

In the so called {\em relay systems}, two units achieve CS as a result of an indirect coupling through a
third system acting as a relay. Experimental evidence of relay
synchronization has been obtained with lasers \cite{Fischer2006,Wu2011,Tiana-Alsina2012} and electronic
circuits \cite{DaSilva2006,Wagemakers2007}. Relay synchronization
has been further used as a method for chaos-based communications
\cite{Peil2007,Zhou2007,Vicente2007a,Wagemakers2008}, as well as a synchronizing mechanism among remote neuronal resources
\cite{Vicente2007b,Vicente2008,Esfahani2014}.

Anticipated synchronization, on the other hand, may occur in systems coupled in a drive-response configuration, and it relies on the use of time delay lines in the dynamics of the slave system. It has been studied theoretically and numerically in different contexts including
dissipative chaotic flows \cite{Voss2000,Voss2001,Voss2001b}, chaotic maps \cite{Masoller2001}, excitable spiking neurons
\cite{Ciszak2003,Toral2003,Ciszak2004,Matias2011}, inertial ratchets
\cite{Kostur2005}, and complex Ginzburg-Landau equation
\cite{Ciszak2015}. Experimentally, it has been observed in
electronic circuits \cite{Voss2002,Pethel2003,Ciszak2003,Zamora2014}
and in optical systems \cite{Sivaprakasam2001,Buldu2002}. Furthermore, the fact that behavior of one of the oscillator can be forecasted up to time scales larger than those characteristics of the system's dynamics has been used
recently  as a method to control and suppress unwanted spikes in
excitable systems \cite{Ciszak2009}, or to anticipate or suppress extreme
desynchronization events \cite{Zamora2014}.

A further form of synchronization, called {\em
  generalized synchronization} \cite{Rulkov1995,Kocarev1996}, involves
the existence of a time-independent functional relationship between the
states of the two coupled units (typically nonidentical), such that $\Psi ({\bf  x}_1, {\bf
  x}_2)=0$. Initially, this type of synchronization was found for unidirectional
coupling schemes and proved experimentally by means of the auxiliary system
method \cite{Pyragas1996,Abarbanel1996}. Recently, generalized synchronization has been
reported also for bidirectionally coupled oscillators
\cite{Landsman2007,Moskalenko2012,Gutierrez2013}. In particular, Guti\'errez et al.  provided evidence  that relay synchronization
corresponds in fact to the setting of generalized synchronization between the
relay  and the outer units \cite{Gutierrez2013}.

Finally, for very
large coupling or parameter mismatches a counterintuitive phenomenon may emerge
 that suppresses the oscillations, termed as {\em oscillation
  quenching}. Such a phenomenon can be manifested in two distinct ways, whose origin
is structurally different: amplitude death and
oscillation death. In the former case, the suppression of oscillations
results in the existence of a homogeneous steady state, in the latter a homogeneous unstable steady state splits, giving
rise to an inhomogeneous state with two coexisting steady states. Refs.~\cite{Saxena2012,Koseska2013} are two very recent reviews on the
subject.

\subsection{Synchronization in ensembles of coupled systems}\label{sec31-2}

The scenario described in Section~\ref{section_two_coupled} can be generalized to ensembles of spatially ordered oscillators,
 coupled (either locally or globally)  in chains and lattices, as well as to spatially extended oscillatory media.
Synchronization in these contexts, indeed, is very relevant for the study of cooperative behavior
in many physical \cite{Kuramoto1975}, biological \cite{Winfree1967,Winfree1980,Murray1989} and
chemical systems \cite{Kuramoto1984,Kapral1995,Kheowan2007,Leyva2011b}. For a review, the reader is addressed to  the
books in Refs.~\cite{Pikovsky2001} and \cite{Osipov2007}.

\subsubsection{Identical systems}

Assemblies of coupled identical systems can exhibit non trivial synchronization patterns. Precisely,
two types of synchronous behavior can be produced: {\em full} or {\em global}
synchronization where all elements evolve identically in time,
and {\em partial} or {\em cluster} synchronization in which groups of
mutually synchronized systems with different synchronization patterns emerge. The specific scenario that is produced (and its features) depends on many factors:
the number of units, the way they are mutually coupled, and the initial and boundary conditions.

Formally, the coupling conditions may be classified
as local (by setting  in Eq.~(\ref{eq:dynsys}) $A_{ij}=1$ for $j=i\pm 1$, and zero otherwise),
%$$\dot{x_i}=f(x_i)+\sigma h(x_{i+1}-2x_i+x_{i-1}),$$
global ($A_{ij}=1/N$ $\forall i\neq j$),
%$$\dot{x_i}=f(x_i)+\frac{\sigma}{N}\sum_{j=1}^N h(x_j-x_i}),$$
and non-local ($A_{ij}=g(|i-j|)/N$, where $g$ is a decreasing function of the distance between oscillators $i$ and $j$).

From now on, we will focus mostly on the case in which each unit of the ensemble is an oscillator.
In such a case, the amplitude of the mean field is taken as an indicator of the onset of
coherence. If all the frequencies in the ensemble are different,
then the instantaneous phases $\theta_i$  are uniformly distributed in the interval
$[0,2\pi)$, and the mean field vanishes. If instead some of the
oscillators are locked to the same frequency, the vectors associated to their phases sum up
coherently in the unit circle, and the mean field is non-zero. A weaker condition of synchronization is the coincidence of the
averaged frequencies $\Omega_i=\langle \dot\theta_i\rangle$.

Ensembles of locally {\em identical} coupled systems  have been studied
through chaotic maps \cite{Kaneko1985,Kaneko1993,Hasler1998} and flows
\cite{Heagy1994,Heagy1995,Kocarev1996,Belykh2000,Belykh2001}, and
limit-cycle oscillators. An example is the complex Ginzburg-Landau equation \cite{Cross1993,Aranson2002} which represents  the normal form
of an Hopf bifurcation, and describes a wide  variety of wave phenomena,
including phase turbulence (spatiotemporal chaos), plane waves, targets and spirals in which
oscillations are synchronized \cite{Cross1993}. As  for globally coupled systems, the main studies have
concerned chaotic maps \cite{Kaneko1991,Manrubia1999,Popovych2001,Pikovsky2001b} and flows
\cite{Heagy1994,Kocarev1996b,Zanette1998}, rotators with diffusive \cite{Han1995,Daido2006,Rosenblum2007,Yeldesbay2014} or pulsed \cite{Mirollo1990,VanVreeswijk1996,Mohanty2006} coupling.  Experiments instead have been conducted with chaotic electrochemical oscillators \cite{Wang2000}.

Regardless on the specific framework that is considered, a transition is generically found
from global (at large coupling) to cluster  (at intermediate coupling) synchronization.
Detailed analysis of the mechanisms determining the loss of stability of the global coherent state into a hierarchical cluster dynamics can
be found in Refs.~\cite{Popovych2001,Pikovsky2001b}, while Ref.~\cite{Pecora1998b}
studies the destabilization of the synchronous state for increasing coupling due to the so called short-wavelength
bifurcation. Ref. ~\cite{Manrubia1999} analyzed the cluster structure emerging in a system of randomly coupled identical maps, as a function of the average connectivity.

% {\bf Comentar sobre antiphase and in-phase-antiphase?
% Antiphase and in-phase-antiphase sync $[215,223,227,234,235]$}

\begin{figure}
  \centering
  \includegraphics[width=0.55\textwidth]{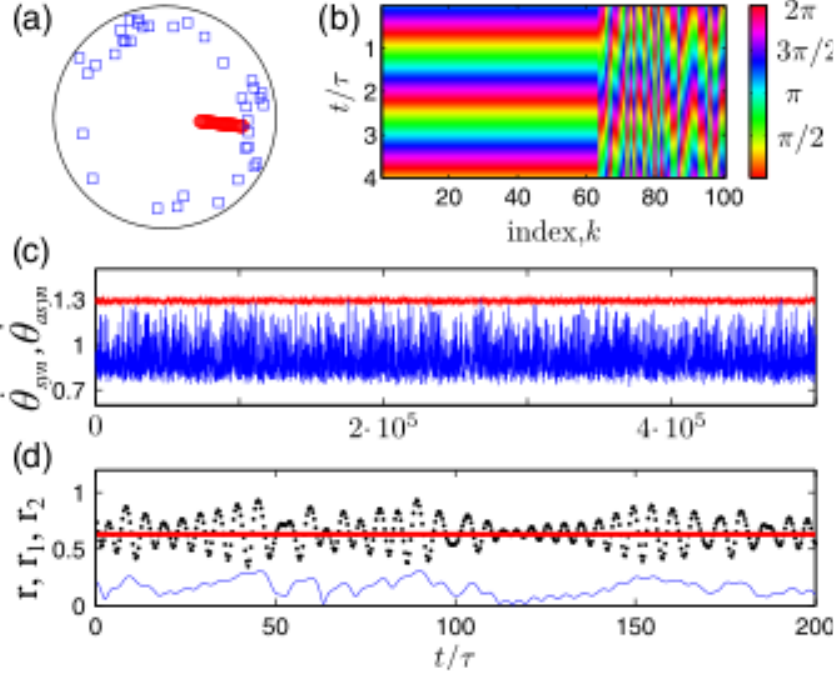}
  \caption{(Color online). Chimera states in a homogeneous ensemble of globally
    coupled oscillators with internal delayed feedback loop
    [$\dot\theta_k=\omega+\lambda \sin(\theta_{\tau,k}-\theta_k) + \epsilon
    Im(e^{\rm{i}\beta}Ze^{-\rm{i}\theta_k})$].  (a) Snapshot of the
    phases of the oscillators belonging to the cluster (red circles)
    and to the cloud (blue squares). The amplitude of the phasor in
    the circle is set proportional to the oscillator index $k$ to
    improve visibility. (b) Colormap of the instantaneous phases. (c)
    Instantaneous frequencies of an oscillator from the cluster
    (upper red curve) and of an oscillator from the cloud (lower blue
    curve). (d) Amplitude of the mean fields of the cluster ($r_1$, red
    thick line), of the cloud ($r_2$, blue thin line), and of the
    whole ensemble  ($r$, black dotted line). Adapted with permission from Ref.~\cite{Yeldesbay2014}. Courtesy of M. Rosenblum.   \label{fig:fig31-chimerasimul}}
\end{figure}

\subsubsection{Chimera states}

\begin{figure}
   \centering
  \includegraphics[width=0.55\textwidth]{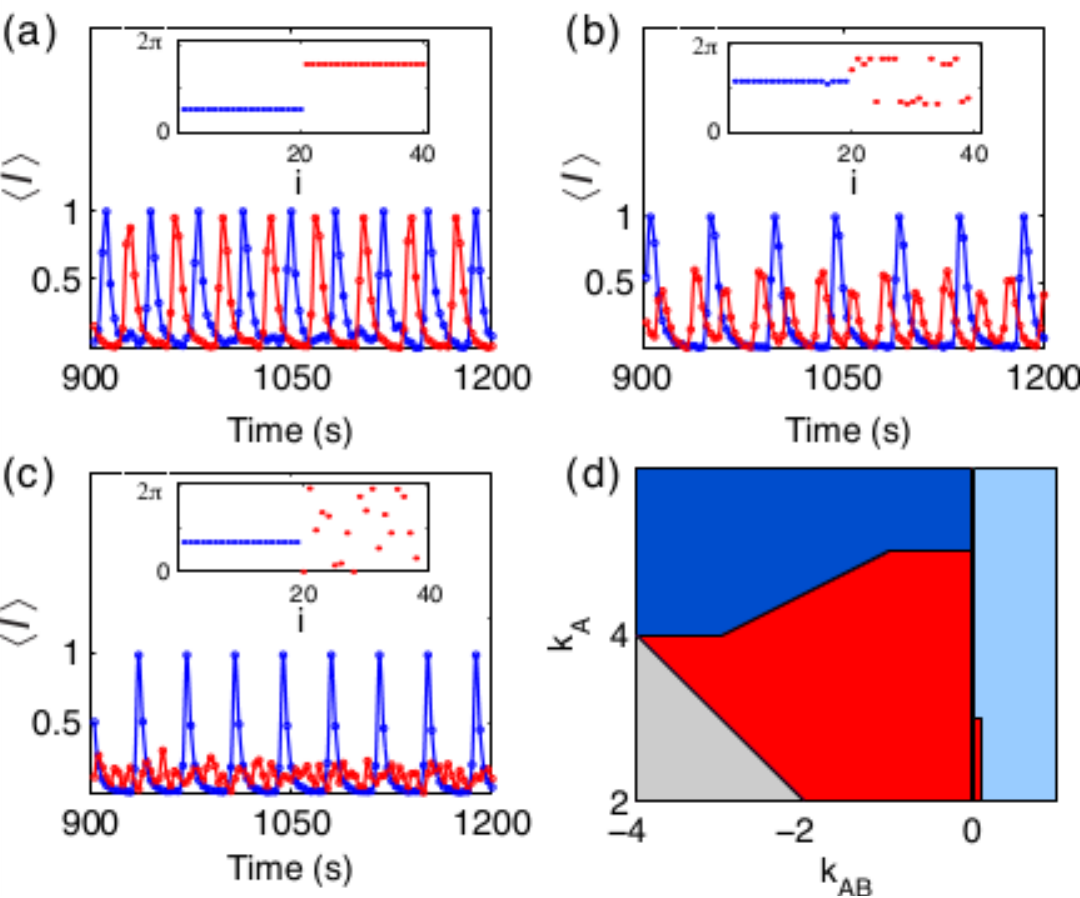}
\caption{(Color online). Chemical chimeras in the photosensitive Belousov-Zhabotinsky (BZ) reaction.
(a)-(c) Time evolution of the normalized mean intensity $\langle I\rangle$ measured in
the BZ particles of groups $A$ (blue) and group $B$ (red), and
snapshot of the phases of the oscillators in each group (insets). (a) Synchronized out-of-phase
state, $k_A=3.0$, $k_{AB}=-1.8$; (b) synchronized 1-2 phase cluster state, $k_A=4.0$,
$k_{AB}=-1.0$; (c) chimera state, $k_A=3.0$, $k_{AB}=-0.2$. (d) Regions of chimera (red),
in-phase (light blue), and out-of-phase (dark blue) synchronized states as a function of $k_A$ and
$k_{AB}$. Reprinted by permission from Macmillan Publishers Ltd: Ref.~\cite{Tinsley2012}, $\copyright$ 2012.
\label{fig:fig31-chimeraexper}}
 \end{figure}

Together with the emergence of clustered synchronous patterns, an even more surprising phenomenon
is the coexistence of clusters of coherent and incoherent phases  in ensembles of identical oscillators,
which was termed as {\em chimera state}
by Abrams et al. in 2004 ~\cite{Abrams2004}. As an example, Fig.~\ref{fig:fig31-chimerasimul}
shows the coexistence of a synchronized cluster of 64  oscillators and a cloud of 36 asynchronous ones, which emerge
in a homogeneous ensemble of  globally  coupled oscillators, due to an internal delayed feedback in individual
  units \cite{Yeldesbay2014}.  First noticed by Battogtokh and Kuramoto \cite{Battogtokh2000}, the
breakdown of spatial coherence in the presence of non-local interaction has been reported theoretically
\cite{Omelchenko2008,Abrams2004,Abrams2008,Bordyugov2010,Martens2010},
and numerically \cite{Omelchenko2010,Omelchenko2011}. Experimental evidences of chimera states
have been obtained in chemical \cite{Tinsley2012,Nkomo2013}, optical
\cite{Hagerstrom2012} and mechanical \cite{Martens2013}
systems. For instance, in Fig.~\ref{fig:fig31-chimeraexper}, a chimera
state is observed in a photosensitive Belousov-Zhabotinsky reaction, in which the catalyst is loaded
onto $N$ ion-exchange oscillatory particles. The $N$ oscillatory particles are
divided into two groups of equal size ($A$ and $B$), with each member $i$ of
group $\pi$ experiencing the feedback light intensity $P_i^\pi$:
$$P_i^\pi=P_0 + k_\pi [\bar I_\pi (t-\tau)-I_i(t)]+k_{\pi\pi'}[\bar I_\pi'(t-\tau)-I_i(t)],$$
being $\pi=A,B$ and $\pi'=B,A$. $\bar I_\pi$ is the mean intensity in group $\pi$,
$k_\pi$ and $k_{\pi\pi'}$ are the intra and inter-group coupling coefficients, $P_0$
is a  normalized background light intensity, and $I$ is the measured intensity scaled from $0$ to $1$.
The delay $\tau$ plays the role of the phase frustration term used in the model by
Abrams et al. \cite{Abrams2004}. With such a coupling scheme, group $A$ remains fully
synchronized, while $B$ can exhibit three basic states: fully synchronized
(Fig.~\ref{fig:fig31-chimeraexper}(a)), $n-$cluster ($n=2$ in Fig.~\ref{fig:fig31-chimeraexper}(b))
and the unsynchronized chimera (Fig.~\ref{fig:fig31-chimeraexper}(c)). Panel (d) shows
the behaviors found at different intra- and inter-group coupling strengths,
under the setting of $k_A=k_B$ and $k_{AB}=k_{BA}$.

Coexistence of coherence
and incoherence has also been observed in networks of coupled nonidentical phase
oscillators \cite{Laing2009,Laing2012}, and in globally coupled populations
of identical oscillators \cite{Kaneko1990} with internal delayed
feedback \cite{Omelchenko2008,Yeldesbay2014}. A recent review
on the subject \cite{Panaggio2015} gives a comprehensive history and overview on chimera states.

\subsubsection{Non-identical systems}

Real-world systems often feature noise and various degrees of heterogeneity. Therefore,
synchronization phenomena have been explored in large ensembles
of interacting non-identical oscillators \cite{Kuramoto1984}.

The same routes of two coupled
 oscillators can also be explored for ensembles of locally or globally
coupled (periodic or chaotic) systems.
As long as oscillators are phase coherent, there is no qualitative
difference between periodic and chaotic case \cite{Osipov1997,Osipov1998,Pikovsky1996},
with the exception that chaotic assemblies can display phase synchronization while their amplitudes
remain uncorrelated.

The transition to synchronization in chains of
periodic \cite{Strogatz1988,Osipov1998} and phase coherent chaotic
\cite{Osipov1997} oscillators with different frequencies
can occur in a soft or a hard way, depending on the
frequency mismatch of the oscillators. Namely, for a linear distribution  with a
small mismatch, the transition is smooth, while for large mismatches the
transition is characterized by an abrupt formation of clusters with phase slips
between them and whose size is increasing
with the coupling. In the case of randomly distributed frequencies,
only the hard transition is possible, and the onset of synchronization
appears for lower values of the coupling.

% {\bf Me falta global coupling En \cite{Matthews1990} ven que el comportamiento es mas rico cuando en vez de phase oscillators hay amplitud}
% Josephson \cite{Wiesenfeld1996}

\subsubsection{The Kuramoto model}
\label{kuramoto_model_section}

Despite the peculiarity of the system under study, a large population of weakly interacting oscillators
 can be described by considering their
phases, and assuming a homogeneous interaction among them. With such an assumption,
Kuramoto  formulated his celebrated equation \cite{Kuramoto1975,Kuramoto1984}  (based on the previous
Winfree's works with phase models \cite{Winfree1967}) which
transforms Eq.~(\ref{eq:dynsys}) into a mathematically tractable one.

The Kuramoto model \cite{Kuramoto1975,Kuramoto1984} considers a system of $N$ oscillators having
natural frequencies $\omega_i$ distributed according to a probability
density $g(\omega)$ which (for the time being) is uni-modal and
symmetric around its mean frequency ${\Omega}(t=0)=\sum \omega_i/N$.
The instantaneous phases  $\theta_i$ obey the following equations:

\begin{equation}
  \dot{\theta_i}=\omega_i+\frac{\lambda}{N}\sum_{j=1}^N
\sin(\theta_j-\theta_i)
\label{eq:kuramcomplete}
\end{equation}
\noindent where the normalization constant $1/N$ ensures the boundedness of the coupling term  in the
thermodynamic limit $N \rightarrow \infty$. A transition to synchronization occurs for values of the
coupling $\lambda$ above a critical threshold
$\lambda_c$, which depends on the distribution $g(\omega)$.
A motivating survey of the
Kuramoto model is offered in Ref. \cite{Strogatz2000}.
Tutorials and reviews of the main achievements on the Kuramoto model can be found in
Refs.~\cite{Acebron2005,Dorfler2014,Rodrigues2015}.

The monitoring of the synchronization onset is performed by introducing a complex mean field
of the population, known as the {\em order parameter}  \cite{Kuramoto1975}

\begin{equation}
  \label{eq:kuramparam}
  r(t)e^{{\rm{i}}\psi(t)}=\frac{1}{N}\sum_{j=1}^N e^{{\rm{i}}\theta_{j}(t)}.
\end{equation}

Equation ~(\ref{eq:kuramparam}) is the vector centroid
of all oscillators represented as points in the unit circle.
$\psi$ is the average phase, and $r$ its modulus, which implies $r\sin(\psi)=\sum_j\sin(\theta_j)$ and
$r\cos(\psi)=\sum_j\cos(\theta_j)$. $r$ measures the level of coherence in
the population: if $r\sim 1$ the population is synchronized, and if $r\sim 1/\sqrt{N}$
the points are uniformly distributed in the unit circle.

Using the parameters $r$ and $\psi$, Eq.~(\ref{eq:kuramcomplete}) can be rewritten as

\begin{equation}
  \label{eq:kurammeanfield}
  \dot{\theta_i}=\omega_i+\lambda r \sin(\psi-\theta_i).
\end{equation}

\noindent Now, the dynamics of each oscillator is  implicitly coupled to all the others through the mean field
in a positive feedback way: the more coherent is the population, the larger is $r$
and the more effective is the coupling strength $\lambda r$, which drives therefore more and more oscillators
into the synchronization cluster. The threshold value $\lambda_c$  can be derived by self consistency conditions applied
to Eq.~(\ref{eq:kurammeanfield}). If one assumes a priori a solution where $r(t)$ is constant and $\psi$ is
rotating at a constant frequency $\Omega$ ($\psi=\Omega t$),
then Eq.~(\ref{eq:kurammeanfield}) can be expressed as a function of the angle $\phi_i=\theta_i-\psi$
using a rotating reference frame at frequency $\Omega$:
\begin{equation}
  \label{eq:kurammeanfield2}
  \dot{\phi_i}=\omega_i - \Omega- \lambda r \sin \phi_i.
\end{equation}

The scalar vector field $f(\phi)=\omega-\Omega-\lambda r \sin\phi$ for $\phi \in [0,\pi]$ exhibits a saddle-node
bifurcation at $\lambda r = |\omega_i-\Omega|$ or $\phi_i=\pi/2$.  If $|\omega_i-\Omega|\le \lambda r$,
the solution $\phi_i=\sin^{-1}\frac{\omega_i-\Omega}{\lambda r}$ is stable, and oscillators are locked to the mean frequency $\Omega$.
The oscillators instead with $|\omega_i-\Omega|>\lambda r$ are uniformly spread around the unit circle.

A rigorous analysis of Eq.~(\ref{eq:kurammeanfield2}) has been performed in Refs.~\cite{Strogatz2000,Acebron2005}. Here we just
highlight the main steps in
such an analysis. By introducing the probability density function $\rho(\phi,t,\omega)$,  the fraction of oscillators lying within the arc $[\phi_a,\phi_b]$
with natural frequencies $\omega \in [\omega_a,\omega_b]$ is
$$\int_{\phi_a}^{\phi_b}\int_{\omega_a}^{\omega_b}\rho(\phi,t,\omega) g(\omega) d\omega d\phi,$$
and the order parameter can be expressed as:
\begin{equation}
  \label{eq:kuramparamintegral}
  r(t)e^{{\rm i} \psi(t)}=\int_0^{2\pi}\int_{-\infty}^{\infty} e^{{\rm i}\phi}\rho(\phi,t,\omega)g(\omega)d\omega d\phi.
\end{equation}
$\rho(\phi,t,\omega)$ must obey the continuity equation
\begin{equation}
  \label{eq:densityequation}
  \frac{\partial}{\partial t}\rho + \frac{\partial }{\partial \phi} (\rho v)=0,
\end{equation}
\noindent where $v=\omega-\Omega -\lambda r\sin(\phi)$
is the angular velocity of a given oscillator with phase $\phi$ (in the co-rotating frame) and natural frequency $\omega$.
One obtains therefore the following self-consistency condition
\begin{equation}
  \label{eq:kuramself-consistency}
  r=\lambda r \int_{-\pi/2}^{\pi/2}\cos^2\phi \cdot g(\Omega + \lambda r \sin \phi)d\phi.
\end{equation}

The saddle-node bifurcation manifests as a phase transition
from the incoherent state (at $r\sim0$ with density $\rho=1/2\pi$) to the partially synchronized state (with $0<r<1$)
characterized by a group of phase oscillators rotating in unison at $\Omega$ (those whose frequencies obey
$|\omega-\Omega|\le \lambda r$), while the remaining ones are rotating incoherently. These solutions satisfy
\begin{equation}
  \label{eq:kuramself-consistency-sol}
  1=\lambda  \int_{-\pi/2}^{\pi/2}\cos^2\phi \cdot g(\Omega +
  \lambda r \sin \phi)d\phi.
\end{equation}
By expanding  the integrand $g(\Omega+\lambda r\sin\phi)$ in a Taylor series, one finds
that the amplitude of the mean field $r$ (at the onset of synchronization) is given by
\begin{equation}
  \label{eq:parambifurc}
r^2\sim \frac{-16}{\pi\lambda_c^4 g''(\Omega)}(\lambda-\lambda_c),
\end{equation}
\noindent with

\begin{equation}
  \label{eq:coupcrit}
  \lambda_c=\frac{2}{\pi g(\Omega)}.
\end{equation}
Therefore, the transition is continuous and the order parameter behaves (in the limits $N,t \rightarrow \infty$) as
\begin{equation}
  \label{eq:rcrit}
 r =
  \begin{cases}
   0 &  \lambda < \lambda_c \\
   (\lambda-\lambda_c)^{\beta} & \lambda\ge \lambda_c
  \end{cases}
\end{equation}
which resembles a continuous second-order phase transition characterized by the critical exponent $\beta=1/2$.

Finally, exploiting the extensive symmetries of the continuum-limit model ~(\ref{eq:kuramparamintegral})-(\ref{eq:densityequation}),
Ott and Antonsen  derived
an exact reduction to a finite set of nonlinear differential equations for the evolution
of the order parameter  \cite{Ott2008,Ott2009}. This latter approach has been used in several recent studies \cite{Martens2009,Laing2009,Bordyugov2010,Pikovsky2011}.

\subsection{Synchronization in complex networks}
Units in the ensemble may, however, interact through specific coupling patterns (or connection topologies) which
can be conveniently encoded in an interaction network. % as the one given in Eq.~(\ref{eq:dynsys}).
The seminal paper by Watts and Strogatz \cite{Watts1998} proved that the connection topology of networks can be organized in
between a regular and a random architecture, and that such a topological arrangement (called SW property) endows the system
with enhanced synchronizability. Since then, a flurry of studies has been focused on synchronization in complex networks, and the reader is here addressed to a series of classic \cite{Boccaletti2006,Arenas2008,Barrat2008} and recent \cite{Dorfler2014,Boccaletti2014} reviews on the subject.

We here follow the same path of the previous two Sections, and
provide a schematic overview of synchronization on networks, for both cases of identical and non-identical oscillators.

\bigskip
\subsubsection{Identical systems}

%Generalized sync \cite{Shang2009}
Complete, cluster, and chimera states of synchronization have been
reported so far in networks of identical flows or maps. As far as complete synchronization is concerned, one of the main research lines
has been the investigation on the stability of this state for different network topologies and dynamical functions.

For a generic network of  $N$ diffusively coupled identical units there are two main approaches to assess the stability of the synchronous state. The first one is the extension of the MSF \cite{Pecora1998,Jalan2003,Huang2009} to complex networks
\cite{Barahona2002}, which related the local stability of the synchronization manifold ${\bf s}(t)={\bf x}_1(t)={\bf x}_2(t)=\dots={\bf x}_N(t)$ to the spectral properties of the Laplacian matrix associated to the underlying network structure. A comprehensive derivation of this formalism can be found in Ref.~\cite{Boccaletti2006}. Precisely, the analysis reduces to plot the maximum Lyapunov exponent $\Lambda_{max}$
transverse to the synchronization manifold as a function of a generic coupling parameter $\mu$ which encodes
both the local function ${\bf f}({\bf x})$ and the coupling function ${\bf h}({\bf x})$ in Eq.~(\ref{eq:dynsys}).
Different choices of local and coupling functions give rises to three possible classes of MSF. For instance,
for a bounded MSF (the so called case of class-III MSF), where $\Lambda_{max}(\mu)<0$ within a finite interval $(\mu_1,\mu_2)$,  the necessary condition for stability  can be attained only if $\frac{\lambda_N}{\lambda_2}<\frac{\mu_2}{\mu_1}$, being $\lambda_2$ and $\lambda_N$  the minimal and maximal non-zero eigenvalues of the Laplacian matrix. The interested reader is here referred to Ref.~\cite{Boccaletti2006} for the complete discussion on the three possible classes of MSF.

The second method is known as {\it the connection graph stability} method, and was derived in Ref.~\cite{Belykh2004}. It combines the Lyapunov function approach with graph  theoretical reasoning, and provides a bound for the minimum coupling strength  needed for synchronization
which explicitly depends on the average path length of the coupling graph.

MSF arguments led to the study of the propensity towards synchronization of different
topologies regardless on the nature of the coupled dynamical
systems. Ref.~\cite{Donetti2005} introduced a family of graphs, called
entangled networks, having an extremely homogeneous structure which
optimizes synchronizability: degree, node distance, betweenness, and
loop distributions are all very narrow. Therefore, synchronization is
shown to be affected by the heterogeneity of the degree distribution,
clustering, characteristic path length or betweenness.

Reference~\cite{Barahona2002} compared the synchronizability of a SW network against an interpolated ordered ring structure, and
a completely random ER graph, as implemented in
Ref.~\cite{Watts1998}. It was shown that SW graphs exhibit
better synchronizability than regular and completely random ones, when the system is far from
the complete graph limit. In contrast, Ref.~\cite{Belykh2005} showed that the onset and stability of
complete synchronization in networks of pulse coupled oscillators
(neurons) only depends on the number of inputs a neuron receives
regardless on the details of the network topology, as long as the degree
distribution is a $\delta$ function. The effect of link directionality in SW networks has been studied in
Ref.~\cite{Tonjes2010}.

Reference ~\cite{Nishikawa2003} further elaborated on the role of heterogeneity,
and realized that heterogeneous structures may display very large ratios
$\lambda_N/\lambda_2$ ($\lambda_N$ scales with the
hub's degree, i.e. the largest degree in the graph \cite{Goh2001}). The
fact that heterogeneous networks may be more difficult to synchronize
is known as the {\em paradox of heterogeneity} \cite{Nishikawa2003}. The effect is due
to the overloading of the hubs, which may receive independent inputs and may hamper effective
communication between oscillators. The paradox stimulated many scientists to
explore mechanisms to improve synchronizability, by the implementation of asymmetric weighting
procedures. In Refs.~\cite{Motter2005,Motter2005b}, enhanced synchronization is obtained
by normalizing the coupling to the power of the node degree $k_i$ (i.e. $\frac{\lambda}{k_i^{\beta}}\sum_jA_{ij}{\bf h}({\bf x}_j)$), while in Ref.~\cite{Chavez2005} the coupling
strength between two nodes is re-scaled to the load of the edge connecting them
through the edge's betweenness centrality $b_{ij}$ (i.e.
$\frac{\lambda}{\sum_jb_{ij}^{\alpha}}\sum_jb_{ij}^{\alpha}A_{ij}{\bf
  h}({\bf x}_j)$). Both methods implement an asymmetric arrangement of the
coupling, and promote a dominant interaction from the  hubs to
the non-hubs nodes \cite{Hwang2005}. This consideration can be straightforwardly generalized to
weighted networks, by showing that synchronizability decreases when
the node's strengths become more heterogeneous \cite{Zhou2006}.

The existence of degree
correlations may also affect the propensity for synchronization \cite{Chavez2006}. In
particular, synchronizability is enhanced in
disassortative SF networks when smaller nodes drive larger
degree ones, while the opposite holds for assortative networks if the
hubs are driving the small degree nodes \cite{Sorrentino2006,DiBernardo2007}.
Synchronizability is therefore enhanced if hubs are strongly
inter-connected and drive the nodes with smaller degrees, that is,
if the network displays a rich club of large degree nodes.
%Interestingly, Zhou and Kurths \cite{Zhou2006b} confirm this scenario by showing that,

\begin{figure}
  \centering
  \includegraphics[width=0.6\textwidth]{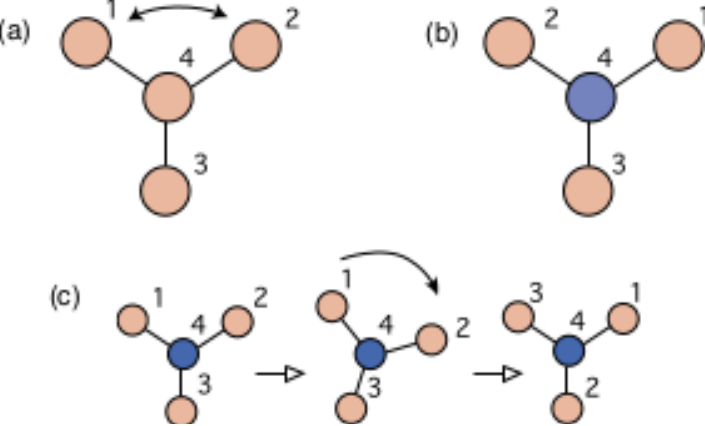}
  \caption{(Color online). Relationship between group symmetry and dynamics in
    networks of oscillators. (a) A star network  of four identical
    oscillators. (b) Nodes 1 and 2 are reflected. (c) The same network
    in (a) after a rotation operation. Adapted from Ref.~\cite{Sorrentino2016}, published under CC BY-NC license. \label{fig:fig31-symmetries}}
\end{figure}

At a more microscopic scale, Ref.~\cite{Lodato2007} showed that the
stability of the synchronous states supported by $3$-node and $4$-node
motifs  is correlated with their relative
abundance in undirected networks, while in directed graphs the
correlation only holds for some specific motifs.
Recently, furthermore, stability of synchronization has been investigated also in relationship with
topological criteria which apply to network's meso-scales \cite{annelido2012}.

Another extension is that covered by adaptive networks
\cite{Gross2008}, whose structure is shaped by the microscopic
organization patterns \cite{Zhou2006,Sorrentino2008,Yuan2013}. To
achieve global synchronization, one plausible assumption is to
consider that each node tries to synchronize to its neighbors by
increasing the coupling strength from them. Inspired by the anti-Hebbian learning rule in
neural systems, Ref.~\cite{Yuan2013} studies how the feedback from dynamical
synchronization shapes network structure by adding, at every time step, a link between two
randomly chosen nodes $i$ and $j$ with a probability proportional to
the rate of their dynamical differences with the mean activity of the
network. The evolution of the network spontaneously forms a SF
structure with negative degree-degree correlation, a typical feature
of technological and biological networks.

On the other
hand, Refs.~\cite{Zhou2006,Sorrentino2008} introduced a dynamical feedback principle
by means of a nonlinear differential equation for the
evolution of the weights which is coupled to the dynamics of the nodes while
the connectivity matrix is fixed a priori. In this latter case, when
complete synchronization is achieved, the coupling strength becomes weighted and correlated
with the topology due to a hierarchical transition to synchronization
in heterogeneous networks.

MSF approach has also inspired some pinning control strategy for synchronization \cite{Almendral2009,Sorrentino2007b} where nodes are sequentially perturbed to improve as much as possible the
eigenratio of $\lambda_N/\lambda_2$.

More recently, several studies revealed that the symmetry groups in the architecture of complex networks can determine constraints for the appearance (or stability) of a given
cluster solution \cite{Josic2006}. Cluster synchronization has
attracted considerable attention in SW networks of
coupled maps \cite{Jalan2003}, chaotic flows \cite{Belykh2008},
synaptically coupled networks of
bursting neurons \cite{Belykh2011} (for diffusively coupled neurons
only complete synchronization is possible \cite{Belykh2005}), or in
networks of Stuart-Landau oscillators \cite{Poel2015}.

\begin{figure}
  \centering
  \includegraphics[width=0.8\textwidth]{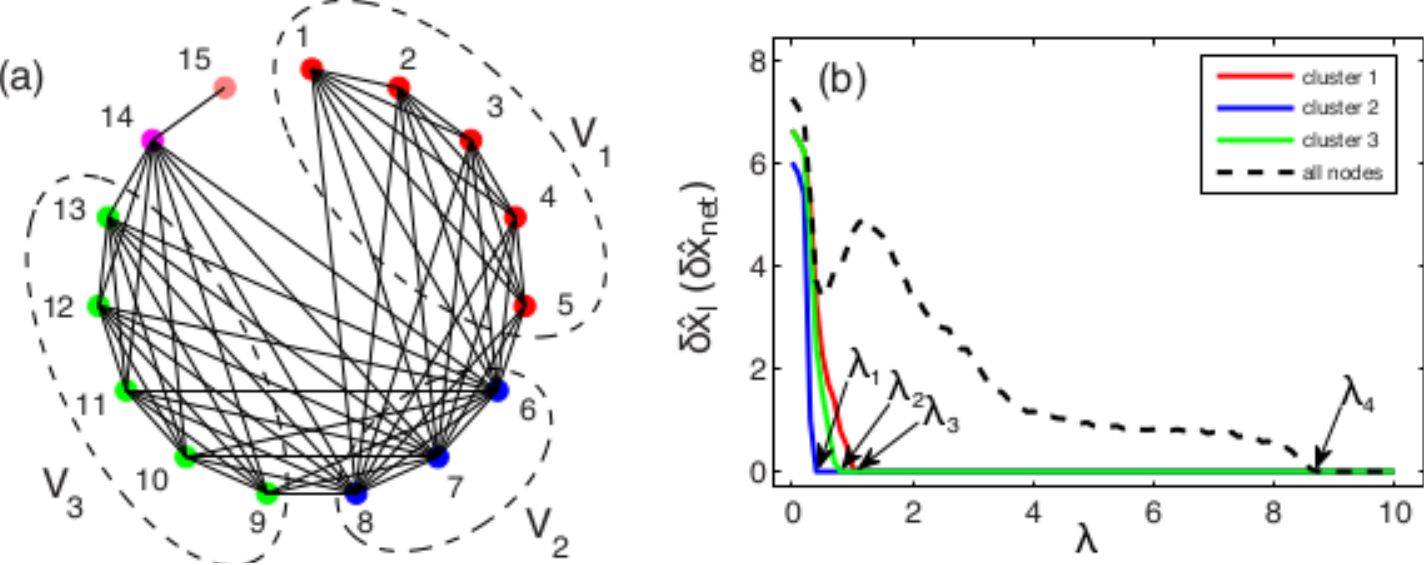}
  \caption{(Color online). Cluster synchronization in the
    Nepal power-grid network. (a) The structure of the Nepal
    power-grid consisting of 15 nodes (power stations) and 62 links
    (power lines). Computational group theory partitions the nodes
    into 5 clusters [$V_1 =\{1,2,3,4,5\}$ (red), $V_2 =\{6,7,8\}$ (blue),
    $V_3 =\{9,10,11,12,13\}$ (green) and two trivial clusters of just one
    node $V_4 =\{14\}$     (pink) and $V_5 =\{15\}$ (yellow)]. (b)  Using Lorenz chaotic
    oscillators as nodal dynamics and coupling function ${\bf
      h}=(0,x,0)$, the cluster
    synchronization error $\delta \hat x_l$  ($l=1,2,3$) and the network
    synchronization error $\delta \hat x_{net}$ (black dashed line) are reported as functions of the
    coupling strength $\lambda$. Clusters 2, 3, and 1 are
    synchronized at $\lambda_1\approx 0.4$, $\lambda_2 \approx 0.8$, and $\lambda_3 \approx1.1$, respectively. For
$\lambda> \lambda_4 \approx 8.9$, the network is globally
synchronized. $\delta \hat x_l=\sum_i\langle    |x_i-\bar
x_l|\rangle/n_l$, with $\bar x _l=\sum_ix_i/n_l$ being the
    averaged state of the oscillators in cluster $l$; $\delta \hat
    x_{net}=\sum_i\langle|x_i-\bar x|\rangle/N$, with $\bar x
    =\sum_ix_i/N$ being the network averaged state.
Adapted with permission from Ref.~\cite{Lin2016}. Courtesy of X. Wang.  \label{fig:fig31-Nepalpowergrid}}
\end{figure}

In particular, Refs.~\cite{Sorrentino2016,Pecora2014} have studied the emergence
and stability of cluster synchronization in networks of coupled oscillators for
arbitrary topologies and individual dynamics. Using the tools
of computational group theory, Ref.~\cite{Pecora2014} addressed the general case
of networks where the
intrinsic symmetries are neither {\em ad hoc} produced nor easily
observed. Figure ~\ref{fig:fig31-symmetries} schematically described a case where a strong
relationship exists between symmetry and dynamics: a symmetry
 (a permutation matrix like a reflection in
 Fig.~\ref{fig:fig31-symmetries}(b) or rotation in Fig.~\ref{fig:fig31-symmetries}(c))
applied to an adjacency matrix leaves it unchanged (i.e.
the equations of motion are the same when mapped into each other
by the symmetry transformation). In the example, nodes 1, 2, and 3 constitute one cluster
and node 4 another one. The main result is a technique, based on the
cluster variational equations, to evaluate the
stability of all the dynamically valid cluster synchronization
patterns. The conclusion is that the range of stability typically
becomes smaller for those synchronization cluster that are
characterized by higher symmetry. Moreover, the validation of such patterns is performed
by an optoelectronic experiment on a five-node graph. The same approach was independently developed
in Ref.~\cite{Fu2012}, and recently an application for the control of specific clusters
along the synchronization path has been reported in
Ref.~\cite{Lin2016} (see Fig.~\ref{fig:fig31-Nepalpowergrid}).

Reference~ \cite{Nicosia2013} also relates network's symmetries with
the emergence of cluster synchronization states in a Kuramoto model of
identical oscillators, when the interactions include a phase frustration preventing
full synchronization. In these conditions, the system organizes into a regime of
remote synchronization where pairs of nodes with the same network
symmetry are fully synchronized, despite their distance on the
graph. An application to brain networks suggests that anatomical
symmetry may play a role in neural synchronization by determining
correlated functional modules across distant locations.

In the framework of networks of communities \cite{Fortunato2010} and
networks of networks \cite{Boccaletti2014}, there are several works
seeking for the synchronization performance as a function of the
competition between individual communities and the whole network
\cite{Park2006,Zhao2011,Aguirre2014}. In particular, Ref.~\cite{Aguirre2014}
discusses how the degree of the
nodes through which two networks are connected influences the ability
of the whole system to synchronize. It is shown that connecting the
high-degree (low-degree) nodes of each network turns out to be the
most (least) effective strategy to achieve synchronization, and
the existence of the optimal connector link weights for the different
interconnection strategies is reported. On the other hand, Ref.~\cite{Zhao2011}
reaches similar conclusions, and observes also intermediate states of inter-community connectivity
for which there is a balance between segregation (clustered) and
integration (global synchronization) dynamics.

\begin{figure}
  \centering
  \includegraphics[width=0.6\textwidth]{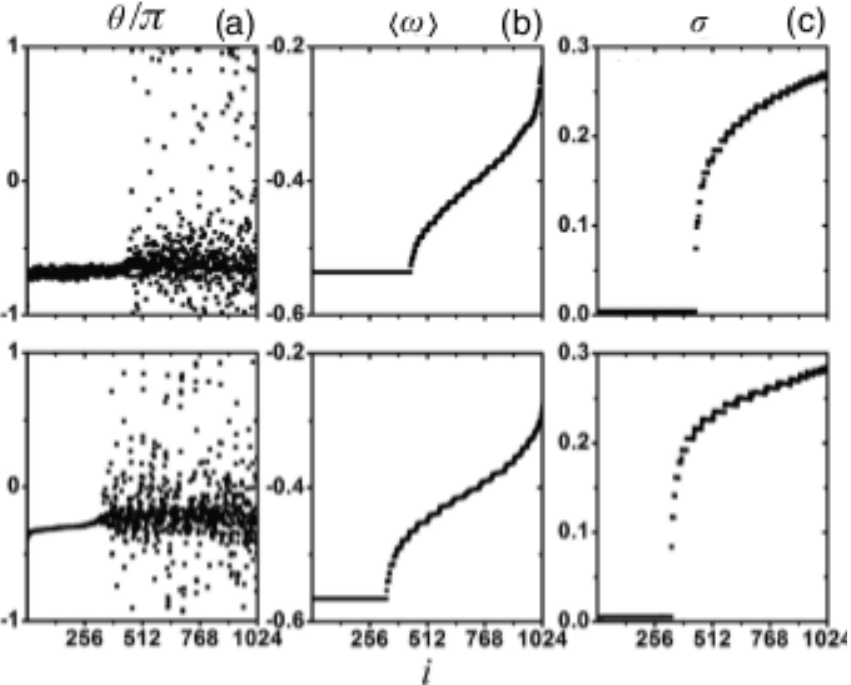}
  \caption{Chimera states in complex networks. (Column a)
      Snapshots of the oscillator phases $\theta_i$, (column b) Corresponding time
      average frequency, and (column c) frequency variance. Top (bottom) panels
      correspond to ER (SF) networks. The size is $N=1,024$, and the mean degree is $\langle k\rangle=4$. Other
      parameters are $\lambda=1$ and $\kappa=0.1$. Reprinted with permission from Ref.~\cite{Zhu2014}.  $\copyright$ 2014 by American Physical Society. \label{fig:fig31-chimeracomplexnet}}
\end{figure}

To conclude, chimera states have also been reported in complex networks (for an updated
compendium of references the reader is addressed to
Ref.~\cite{Panaggio2015}). In Ref.~\cite{Buscarino2015}, the Authors considered
two globally coupled populations of identical Kuramoto
oscillators with inter-population links randomly switched on or off,
resulting in the appearance of stable, breathing, and
alternating chimera states. Reference~\cite{Zhu2014}
investigated non-locally coupled identical phase oscillators in SF and
ER networks, where a coupling function dependence on the shortest length $d_{ij}$ between
oscillators  is introduced ($A_{ij}=\lambda e^{-\kappa d_{ij}}$, with $\lambda$ being the global
coupling strength and $\kappa$ the strength of the non-local
coupling). With these settings, chimera states spontaneously emerge
out of arbitrary initial conditions for both types of complex
networks and oscillators. These states tend to include nodes with
high degrees on SF networks, while there is no preference for degree for
oscillators on ER ones (see Fig.~\ref{fig:fig31-chimeracomplexnet}).

\subsubsection{Non-identical systems}

In real systems, populations of interacting oscillators are neither
homogeneous nor regular in frequencies, and we already saw that a widely studied paradigm is the Kuramoto model, described in Subsection
\ref{kuramoto_model_section}. The first attempt to investigate the role of non-regular topologies was performed in Ref.~\cite{Niebur1991}. It was shown there that introducing some variability in a lattice (by randomly rewiring
a few local interactions to long-range connections)
leads to a more rapid and robust phase locking than nearest-neighbor couplings, or locally
dense connection patterns. When the Kuramoto model is considered on top of
a complex network of interactions, Eq.~(\ref{eq:kuramcomplete}) translates to

\begin{equation}
  \dot{\theta_i}=\omega_i+\lambda\sum_{j=1}^N A_{ij}\sin(\theta_j-\theta_i),
\label{eq:kuramcomplex}
\end{equation}
\noindent
where $\lambda$ is again a global coupling parameter, and $A_{ij}$ is a matrix defining the oscillators' interactions (a complete graph is recovered by setting $A_{ij}=1/N$ for all pairs).

There are two important issues to be discussed about the extension of Kuramoto model to complex networks: the re-scaling factor in the coupling term of the dynamics, and the issue of properly defining an order parameter.
In Refs.~\cite{Gomez-Gardenes2007b,Arenas2008} the implications of different normalizations
used in the literature are investigated. One of the most used prescriptions consists in taking $A_{ij}=A_{ij}/N$, which however has the problem that (in the thermodynamic limit) the coupling term vanishes for all nodes except for those whose degree scales as $N$. Another prescription, used to solve the
{\em paradox of heterogeneity} (discussed in the previous Subsection), consists in taking $A_{ij}=A_{ij}/k_i$.
In particular, Ref.~\cite{Gomez-Gardenes2007b} considers $A_{ij}=A_{ij}/k_{max}$, being $k_{kmax}$ the largest degree in the network, this way decoupling the heterogeneity of the network and the interaction dynamics and, at the same time, having a non diverging limit for $N\to\infty$.
As for the  definition of the order parameter, it is worth noticing that the asymptotic magnitude $r$ defined in Eq.~(\ref{eq:kuramparam}) allows for a complete graph the measurement of the average performance of synchronization. Tailoring
$r$ for non-complete graphs is the object of many studies, as e.g. those realized by Ichinomiya \cite{Ichinomiya2004}, Restrepo et al. \cite{Restrepo2005}, or by  Sonnenschein and Schimansky-Geier \cite{Sonnenschein2012}  in the context of complex networks of noisy oscillators.

Early works investigating synchronization in Eq.~(\ref{eq:kuramcomplex}) are those by
Watts and Strogatz \cite{Watts1998} and Hong et al. \cite{Hong2002} for
SW networks, and by Ichinomiya \cite{Ichinomiya2004} and  Moreno and Pacheco \cite{Moreno2004}
for SF networks. In particular, Ref.~\cite{Hong2002} showed
(by means of extensive calculations) that
any finite probability $p$ of rewiring in the Watts-Strogatz model \cite{Watts1998} gives rise to a finite critical coupling  $\lambda_c$, and that the
scaling is indeed compatible with the mean-field of the globally connected
graph. Moreover, $\lambda_c$ diverges as $p\rightarrow 0$, and the time
needed to achieve the coherent state decreases with $p$. In the presence of large degree fluctuations, Ichinomiya proposed to rescale $r$ as follows
\begin{equation}
  \label{eq:Kuramparam2}
  r(t)=\bigg|\frac{1}{\sum_lk_l}\sum_jk_je^{{\rm i}\theta_j(t)}\bigg|,
\end{equation}
and found that the critical coupling for uncorrelated random networks with arbitrary degree distribution is
\begin{equation}
  \label{eq:coupcrit2}
  \lambda_c=\lambda_c^0\frac{\langle k\rangle}{\langle k^2\rangle},
\end{equation}
where $\lambda_c^0$ is the critical point for Eq.~(\ref{eq:coupcrit}).
When the degree fluctuations are bounded, both $\langle k^2\rangle$  and $\lambda_c$ are finite in the thermodynamic limit. However, for power-law degree distributions $k^{-\gamma}$, the critical coupling vanishes when $2<\gamma<3$. A refinement of the latter result has been provided by Restrepo
et al.~\cite{Restrepo2005}, who generalized the mean-field approach, and showed that the coupling strength at which the transition takes place is determined by the largest eigenvalue of the adjacency matrix ($\lambda_c=\lambda_c^0/\lambda_N$).
As for the role of the hubs
in the synchronization process on top of a SF network,
Ref.~\cite{Moreno2004} showed that  the relaxation time $\tau$
for the hubs is shorter than that of the less connected
nodes, scaling as $\tau \sim k^{-1}$. Hence, the more connected a node is, the more stable it is.

The mean-field approach allows also to compute the scaling of $r$ \cite{Lee2005}. For finite $\lambda_c$, one has
\begin{equation}
  \label{eq:rscaling}
  r\sim \Delta ^{\beta},
\end{equation}
where $\Delta=(\lambda-\lambda_c)/\lambda_c$. For SF networks with
$P(k)\sim k^{-\gamma}$, the exponent takes the value $\beta=1/2$ if
$\gamma >5$, and $\beta=1/(\gamma-3)$ for $3<\gamma<5$.

Finally, we briefly review clustering and modular synchronization
in networks of oscillators \cite{Restrepo2005,Oh2005,McGraw2005,Park2006,Sorrentino2007a,Li2008,Rad2012}. Since the route to complete synchronization is made up of groups of
synchronized oscillators that grow and coalesce, structural properties at the meso-scale
of the network can be detected by a fine tuning of the coupling
strength. In particular, Ref.~\cite{Gomez-Gardenes2007} compared the
synchronization patterns in ER and SF networks and showed that even in
the incoherent solution, the system self-organizes following different
paths: while in SF networks the giant component of the synchronized
pairs is gradually increasing with the coupling strength around the largest degree nodes, in ER
networks the coalescence of many small clusters leads to a giant
component whose size is of order of the system size once the incoherent state destabilizes.
Oh et al. \cite{Oh2005} studied
the Kuramoto model on two different types of modular  networks, finding that the synchronization transition crucially
depends on the type of inter-modular connections. McGraw and
Menzinger \cite{McGraw2005} found that the promotion
of dynamical clusters oscillating at different
frequencies hinders in general synchronization.

In Li et al. \cite{Li2008},  the interfaces between synchronized clusters emerging due to frustration are explored. An algorithm is there provided
able to detect the frustrated nodes located at the overlapping structures of real-world networks. The algorithm was then used to unveil protein functions in a protein-protein
interaction network in Ref.~\cite{Sendina-Nadal2011}. Further, an easily computable
measure was introduced in Ref.~\cite{Rad2012}  which locates the effective
crossover between segregation and integration in modular networks.

Relay (or remote) synchronization of pairs of nodes that are not
directly connected via a physical link (nor  via any sequence of synchronized
nodes) has been observed in star like motifs \cite{Bergner2012}, and in
arbitrary networks  through the action
of mismatched units \cite{Gambuzza2013,Gambuzza2016, Gambuzza2016b}.

New forms of synchronous solutions have also been found in multi-layer networks
\cite{deDomenico2013,Boccaletti2014}. In this latter context, several scenarios have been described: intra-layer synchronization, i.e. a state in which the nodes of a layer are synchronized between them without being necessarily synchronized with those of the other layer \cite{Gambuzza2015}; inter-layer synchronization, which, on the contrary, refers to a state where all nodes of a layer are synchronous with their replicas in the other layer regardless of whether or not nodes in each layer are synchronized with the other members of the layer
\cite{Gutierrez2012,Sevilla-Escoboza2015}; breathing synchronization (a regime resulting from the competition
between an instantaneous intra-layer and a delayed inter-layer coupling  \cite{Louzada2013}), where, depending on the couplings and delay, two frequency groups of oscillators emerge within the same layer, each
one synchronized with its mirror in a breathing mode.

Most of the works dedicated to the study of the interplay between network topology and coupled dynamics have been
performed considering dynamical models of pure phase oscillators and diffusive couplings. However, many biological
systems interact in a rather episodic, or pulse-like, way and the amplitude dynamics of the oscillators
are usually not constant, as it occurs in coupled neuronal networks. Lago-Fern\'andez et al. \cite{Lago-Fernandez2000}
compared the synchronization patterns exhibited by networks of nonidentical Hodgkin-Huxley neurons coupled through pulses and  different connectivity topologies. It is found that regular topologies produce coherent oscillations with long transients and slow response times, while completely random topologies give rise to a fast but not coherent response to external stimuli, and SW configurations sustain both coherent global oscillations with a fast response.

In a posterior work, it was shown that different types of neural excitability (different amplitude dynamics) endow neural assemblies with very different dynamical properties \cite{LagoFernandez2001}. Recently, Ref.~\cite{Gambuzza2016b} compared the behavior of networks with dynamical units having both amplitude and phase dynamics and that of networks of pure phase oscillators. The comparison highlights that synchronization is enhanced when the
dynamics of the units is close to a Hopf bifurcation, and that the emerging behavior crucially depends on the
topology and on the node frequency distribution \cite{LagoFernandez2001,Gambuzza2016b}.

Synchronization phenomena have been used as a useful (and easy to compute) topology probing tool.
Several works focused on how synchronization can unveil structural details of the underlying network where
the dynamics is developing \cite{Arenas2006,Boccaletti2007,Giron2016}. It turns
out that, in the transient of a synchronization process, topological
hierarchies are related to different time scales \cite{Arenas2006}. On
the other hand, based on cluster de-synchronization,
Ref.~\cite{Boccaletti2007} was able to fully detect the modular structure of a network
by dynamically changing the weights of the graph's links. Furthermore, when considering negative interactions (a fraction of which has been shown to be beneficial in Ref.~\cite{Leyva2006}) synchronization in signed networks can be used to unveil the organization of an
ecological system, and to analyze the role of each species \cite{Giron2016}.

Finally, several works dealt with synchronization in networks in which structural and dynamical
features coevolve \cite{Sendina-Nadal2008,Aoki2009,Aoki2011,Gutierrez2011,Assenza2011,Avalos-Gaytan2012,Makarov2016}. By introducing different mechanisms for governing the co-evolution (as, for
instance, the competition between the reinforcement of those interactions between pairwise synchronized units \cite{Sendina-Nadal2008} and the conservation of the total input strength received by each unit \cite{Gutierrez2011,Assenza2011}), the adaptive network evolves to a state characterized by local synchronization patterns together with an underlying network structure featuring SF and/or modular structures.

%%%%%%%%%%%%%%%%%%%%%%%%INMA section%%%%%%%%%%%%%%%%
\section{Explosive synchronization}\label{sec:expsync}

\subsection{Earlier studies on discontinuous transitions in the Kuramoto model}
%\subsection{Discontinuous transitions in the Kuramoto model}
\label{earlykuramoto} % Subsection 3.2

We start this Chapter with a journey on several early studies aimed at showing the possibility of discontinuous transitions in the
Kuramoto model of Eq.~(\ref{eq:kuramcomplete}). The main ideas and conclusions of these studies, indeed,
have stimulated the recent advancements on explosive synchronization in networked systems,
which will then be reviewed in Sections 4.2, 4.3 and 4.5.

\subsubsection{The importance of the frequency distribution}\label{sec:freq_dist}

In 1984, Kuramoto proved that there are always two branches of solutions in the thermodynamic limit of Eq.~(\ref{eq:kuramcomplete})~\cite{Kuramoto1984}. The first one is the trivial solution $r=0$ (corresponding to incoherence), and the second one bifurcates continuously at $\lambda = \lambda_c$, with $\lambda_c := \frac{2}{\pi \, g(\omega_0)}$, being $g(\omega)$ the distribution of natural frequencies, characterized by $g'(\omega_0)=0$. Kuramoto also found the first evidence of an abrupt synchronization~\cite{Kuramoto1984}: close to $\lambda_c$, there is a small-bifurcating solution~\cite{Acebron2005}
\[ r = \sqrt{- \frac{16 (\lambda - \lambda_c)}{\pi \, \lambda_c^4 \, g''(\omega_0)}}, \]
which is supercritical for $g''(\omega_0) < 0$ (i.e., when there is maximum at $\omega_0$) and subcritical if $g''(\omega_0) > 0$ (i.e., when $\omega_0$ is a local minimum).

\begin{figure}
\centering
\includegraphics[width=0.6\linewidth]{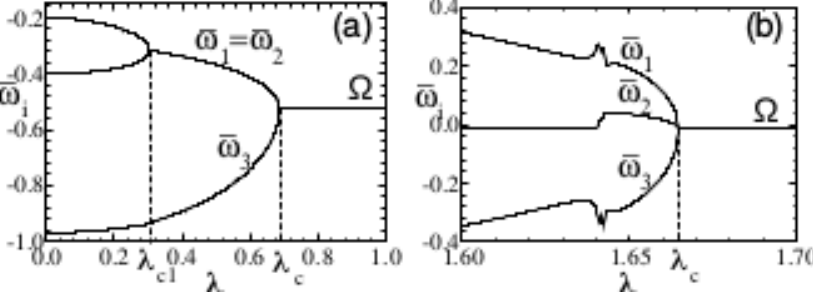}
\caption{Frequency-splitting bifurcation diagrams for the Kuramoto
  model with 3 oscillators. (a) When the natural frequencies are  $\{-0.97, -0.4, -0.2 \}$, and (b) when they are almost perfectly evenly
  distributed  ($\{ -1.0, 0.0, 0.953\}$). Adapted with permission from Ref.~\cite{Maistrenko2004}. Courtesy of Y. Maistrenko.\label{fig:maistrenko2004}}
\end{figure}

The behavior of Eq.~(\ref{eq:kuramcomplete}) with a discrete frequency distribution was then studied using a Lyapunov function formalism ~\cite{vanHemmen1993}. Precisely, the  Lyapunov function $\mathcal{H} := - \frac{\lambda}{2N} \sum_{i,j} \cos (\theta_i -\theta_j) - \sum_i (\omega_i - \Omega) \, \theta_i$ was proposed (with $\Omega$ being the average frequency) and interpreted as a Hamiltonian of a XY ferromagnet of strength $\lambda$ in a random field. The approach allowed to prove that the order parameter is always $r \ge 0.5$, and therefore  a continuous transition from the phase locked to the incoherent state is impossible.
Later on, Ref. ~\cite{Daido1996} proved that the appearance of a macroscopic cluster of mutually phase-locked oscillators takes place continuously when the bifurcation is supercritical, and discontinuously  (with hysteresis and bistability) otherwise.

Reference~\cite{Maistrenko2004} studied how the Kuramoto model de-synchronizes. A transition called frequency-splitting bifurcation is observed: when the coupling strength decreases below $\lambda_c$ only one of the phases splits off from the others, but if the natural frequencies are evenly distributed in an interval, the synchronized state splits into several clusters (see Fig.~\ref{fig:maistrenko2004}, where the effect is shown for three oscillators). The conclusion is that, although a discrete frequency distribution produces a discontinuous transition in the thermodynamic limit, the transition at any finite size depends on how the natural frequencies are chosen.

Following the same concepts and ideas, Paz\'o focused on understanding the finite-size behavior of the Kuramoto model~\cite{Pazo2005}. Ref. ~\cite{Pazo2005} proved (analytically and numerically) that several frequency distributions (with support on a finite interval) lead to a discontinuous phase transition. First, the Kuramoto model is analyzed in the limit $N\to \infty$ for a uniform frequency distribution $\mathcal{U}(-\gamma,\gamma)$, and the critical coupling $\lambda_c := 4 \gamma/\pi$ is deduced for which the order parameter undergoes an abrupt transition from zero to $r_c := \pi/4$ (associated with the emergence of meta-stable states~\cite{Pluchino2006}). Then, the finite size effects  are studied with natural frequencies that are evenly distributed in the interval $(-\gamma,\gamma)$. With such a choice of frequencies, Ref. ~\cite{Pazo2005} derives a self-consistent equation for a finite population of oscillators that accurately reproduces the numerical results.

\begin{figure}
\centering
\includegraphics[width=0.6\linewidth]{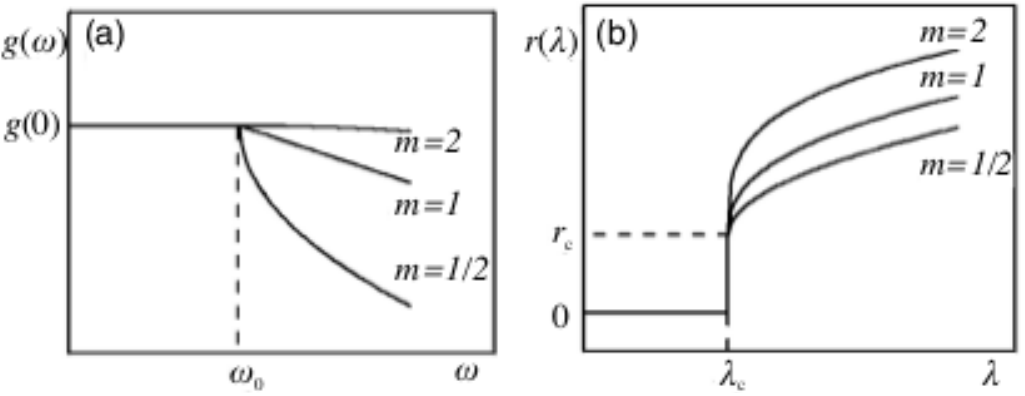}
\caption{(a) Frequency distributions considered
  in Ref.~\cite{Basnarkov2007}, for three values of the parameter $m$,
  and (b) the order parameter $r (\lambda)$ for each distribution in
  (a). Adapted with permission from Ref.~\cite{Basnarkov2007}. Courtesy of L. Basnarkov.\label{fig:basnarkov2007}}
\end{figure}

Reference ~\cite{Basnarkov2007} further studied the case of a unimodal frequency distribution $g(\omega)$ (depicted in Fig.~\ref{fig:basnarkov2007}(a)) with a plateau in the interval $(-\omega_0,\omega_0)$ and symmetric tails behaving as $g(\omega) := g(\omega_0) - C (\omega - \omega_0)^m \, H(\omega - \omega_0)$, where $C$ is a constant, $m>0$ is a parameter and $H$ is the unit step Heaviside function.  Figure~\ref{fig:basnarkov2007}(b) shows that a critical coupling value $\lambda_c$ exists where the system undergoes a discontinuous transition. The order parameter $r$ jumps from $0$ to $r_c := \frac{\pi}{2} \omega_0 \, g(0)$. The transition is discontinuous for any value of $m$, as long as the plateau is a finite flat region (i.e., $\omega_0 >0$), and only when $\omega_0 =0$ the transition becomes continuous.

The latter result was extended to asymmetric frequency distributions in 2008 ~\cite{Basnarkov2008}.
Going beyond earlier studies on the impact of asymmetries in the frequency distributions
~\cite{Ermentrout1985,Sakaguchi1986}, Ref.  ~\cite{Basnarkov2008} found that a necessary condition for a transition to be discontinuous is to have a flat interval in the region of the frequency distribution where the synchronization seed is generated.

\begin{figure}
\centering
\includegraphics[width=0.45\textwidth]{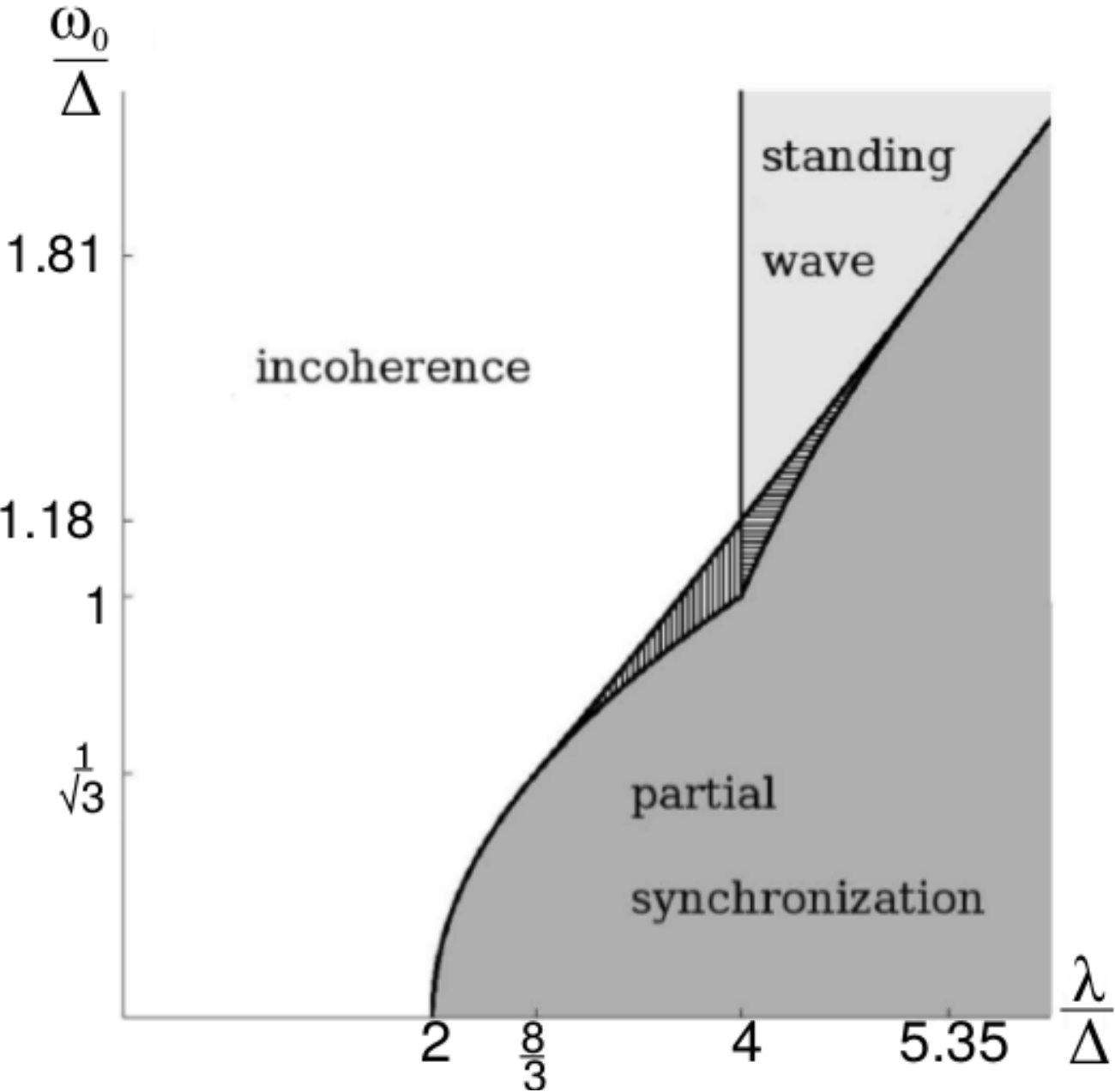}
\caption{Stability diagram of the Kuramoto model with a bimodal
  frequency distribution showing the long-term behavior in each region
  of parameter space. White, incoherence; dark gray, partial
  synchronization; light gray, standing wave; vertical lines,
  coexistence of incoherent and partially synchronized states;
  horizontal lines, coexistence of partial synchronization and
  standing waves. Adapted with permission from
  Ref.~\cite{Martens2009}. Courtesy of E.A. Martens.\label{fig:martens2009}}
\end{figure}

Originally, when Kuramoto considered the subcritical case, $g(\omega)$ was assumed to be a symmetric function around the local minimum at $\omega_0$, featuring two local maxima at $\omega_\pm$. The existence of the two local maxima implies that the incoherence solution, $r=0$, cannot persist up to $\lambda_c$ even as a meta-stable state. Therefore, the lower critical value for the onset of nucleation around $\omega_\pm$ is
$    \lambda'_c := \frac{2}{\pi g(\omega_\pm)} < \lambda_c$.
As $\lambda$ is increased, the two clusters grow independently of each other, like two giant oscillators, until for some large coupling value they eventually synchronize. An exact description of the thermodynamic limit of Eq.~(\ref{eq:kuramcomplete}) with a bimodal frequency distribution  was found in 2009 ~\cite{Martens2009} for $ g(\omega) := \frac{\Delta}{2 \pi} \left( \frac{1}{(\omega - \omega_0)^2 + \Delta^2} + \frac{1}{(\omega + \omega_0)^2 + \Delta^2} \right)$,
being $\pm \omega_0$ the center frequencies of two Lorentzians, and $\Delta$ the half width at half maximum,  which must verify $\omega_0/\Delta > 1/\sqrt{3}$.
Together with the well-known incoherent and partially synchronized states, a third state (consisting of two symmetric clusters of synchronized oscillators near the distribution's maxima) is found, which later on was termed as the standing wave state ~\cite{Crawford1994}. However, the bifurcation sequence suggested by Kuramoto (incoherence $\rightarrow$ standing wave $\rightarrow$ partial synchronization, as sketched in the upper part of Fig.~\ref{fig:martens2009})) is not the unique possibility. Actually, such a sequence occurs when the two maxima in the frequency distribution are sufficiently far apart. Furthermore, the conjectured formula for $\lambda'_c$ is incorrect in general, being only valid asymptotically in the limit of widely separated peaks. When the frequency distribution is barely bimodal, instead, the bifurcation sequence does not include the standing wave state, and there is hysteresis in the transition. In the intermediate regime, where the peaks of $g(\omega)$ are neither too far apart nor too close, the forward bifurcation occurs via the sequence predicted by Kuramoto, while the backward bifurcation skips the standing wave state and results directly from partial synchronization to incoherence.

These results were complemented by Paz\'o \& Montbri\'o~\cite{Pazo2009} with a study in which the minimum and the distance between maxima can be independently modified. In particular, a normalized distribution is considered of the form $ g(\omega) := \frac{1}{\pi (1- \xi)} \left( \frac{1}{\omega^2 +1} - \xi \, \frac{\gamma}{\omega^2 + \gamma^2} \right),$ and similar bifurcation sequences are obtained. The case $\xi = \gamma$ is of special interest (as there the value at the minimum is always $g(0)=0$, independently of how close the two maxima are), and results in a sequence incoherence $\rightarrow$ standing wave $\rightarrow$ partial synchronization without hysteresis (even for $\gamma$ extremely close to zero).

\subsubsection{Effects of noise \label{sec:noise}}

The study on noise affects on top of the Kuramoto model started with Ref.~\cite{Sakaguchi1988}, where an independent white noise stochastic source $\xi_i$ was added to Eq.~(\ref{eq:kuramcomplete}), with $\langle \xi_i(t) \rangle =0$ and $\langle \xi_i(t) \xi_j(t') \rangle = 2D \, \delta(t-t') \, \delta_{ij}$ ($D$ being the noise intensity and angular brackets denoting ensemble average over independent noise realizations). Assuming a Gaussian frequency distribution, Ref. ~\cite{Sakaguchi1988} found that noise makes the critical coupling strength larger but does not change the continuous nature of the transition.

Despite seeming an obvious fact, it is actually hard to prove that the solution $r=0$ is stable for $\lambda < \lambda_c$ and unstable for $\lambda > \lambda_c$. Reference~\cite{Strogatz1991} revealed that the incoherent solution is unstable for $\lambda > \lambda_c$, but neutrally stable for $\lambda < \lambda_c$. Surprisingly, when the problem is studied for small noise, and then the limit is considered of noise intensities going to zero, the results are completely different: the relaxation of $r(t)$ to its steady state follows an exponential law for $\lambda > \lambda_c$, whereas for $\lambda < \lambda_c$ the decay rate depends on the frequency distribution ~\cite{Strogatz1992}. If $g(\omega)$ has a compact support, $r(t)$ tends to zero slower than an exponential, but if $g(\omega)$ is supported on the reals, $r(t)$ is known only in some cases.

\begin{figure}
\centering
\includegraphics[width=0.7\textwidth]{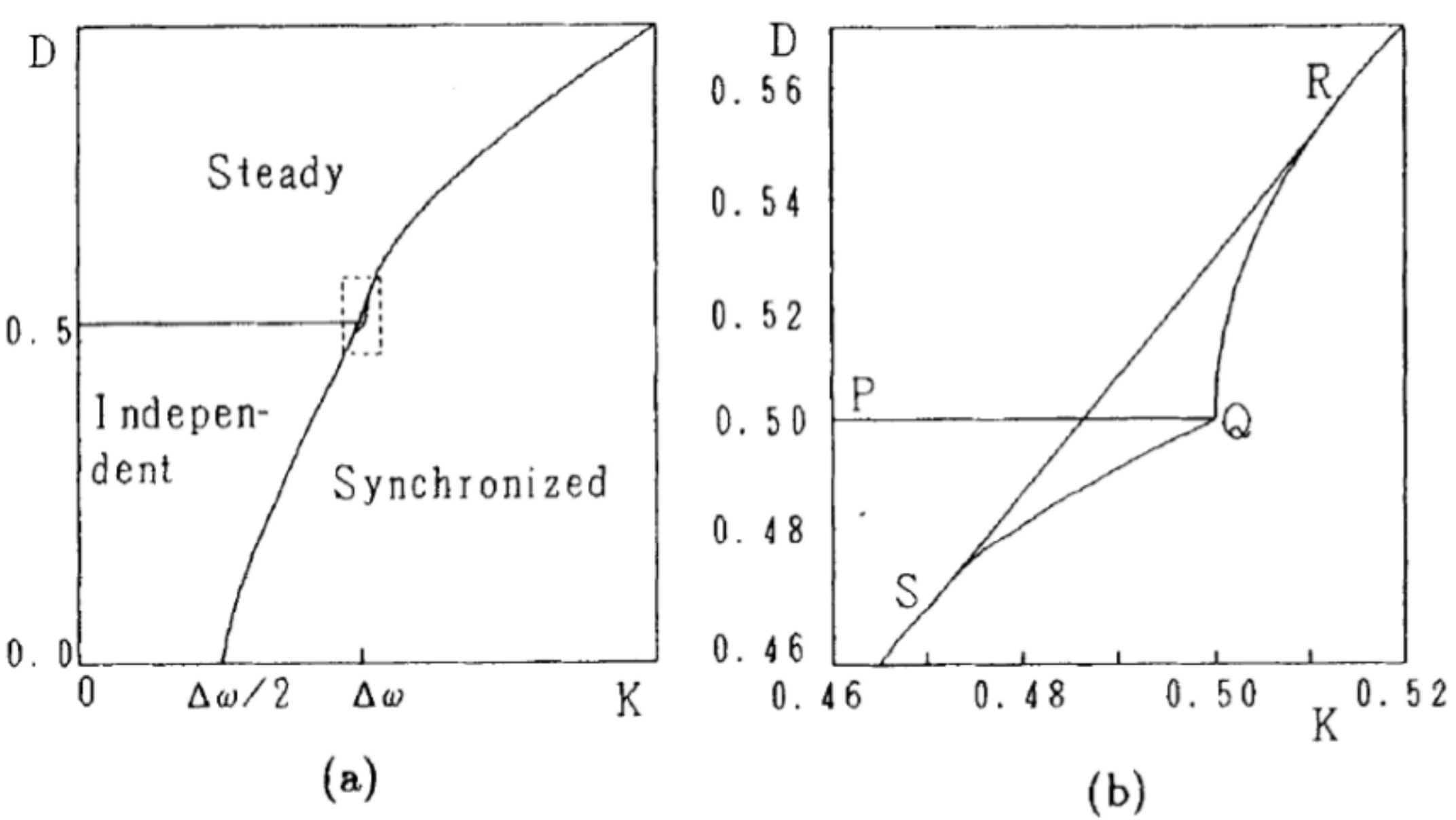}
\caption{(a) Bifurcation diagram of the Kuramoto model with Gaussian
  white noise of intensity $D$ and a discrete frequency distribution
  consisting of two equiprobable frequencies such that $\Delta \omega
  = \omega_2 - \omega_1 = 0.5$. (b) is a magnification of the square
  enclosed with broken lines in (a). The boundaries of steady (i.e.,
  asynchronous), synchronized and independent regimes are represented
  by the curves linking PQR, RS and PQS, respectively. Therefore, the
  system exhibits bistability in the region enclosed by QRS. Notice
  that here $K$ stands for the coupling strength $\lambda$. Reprinted
  from Ref.~\cite{Okuda1991} with permission of PTP.\label{fig:okuda1991}}
\end{figure}

Back in 1991, Okuda \& Kuramoto~\cite{Okuda1991} considered the case of Gaussian white noise and a discrete frequency distribution consisting of two equiprobable frequencies $\omega_2 > \omega_1$. A critical value $D_c$ of the noise intensity is found, below which the system synchronizes either partially (only oscillators with the same frequency synchronize) or globally. Consequently, the bifurcation diagram, shown in Fig.~\ref{fig:okuda1991}(a), has three regimes: asynchronous (referred to as steady), partially synchronized (referred to as independent), and globally synchronized, which overlap in the region QRS depicted in Fig.~\ref{fig:okuda1991}(b), indicating that the transition to synchrony occurs through a region with bistability.

\begin{figure}
\centering
\includegraphics[width=0.5\textwidth]{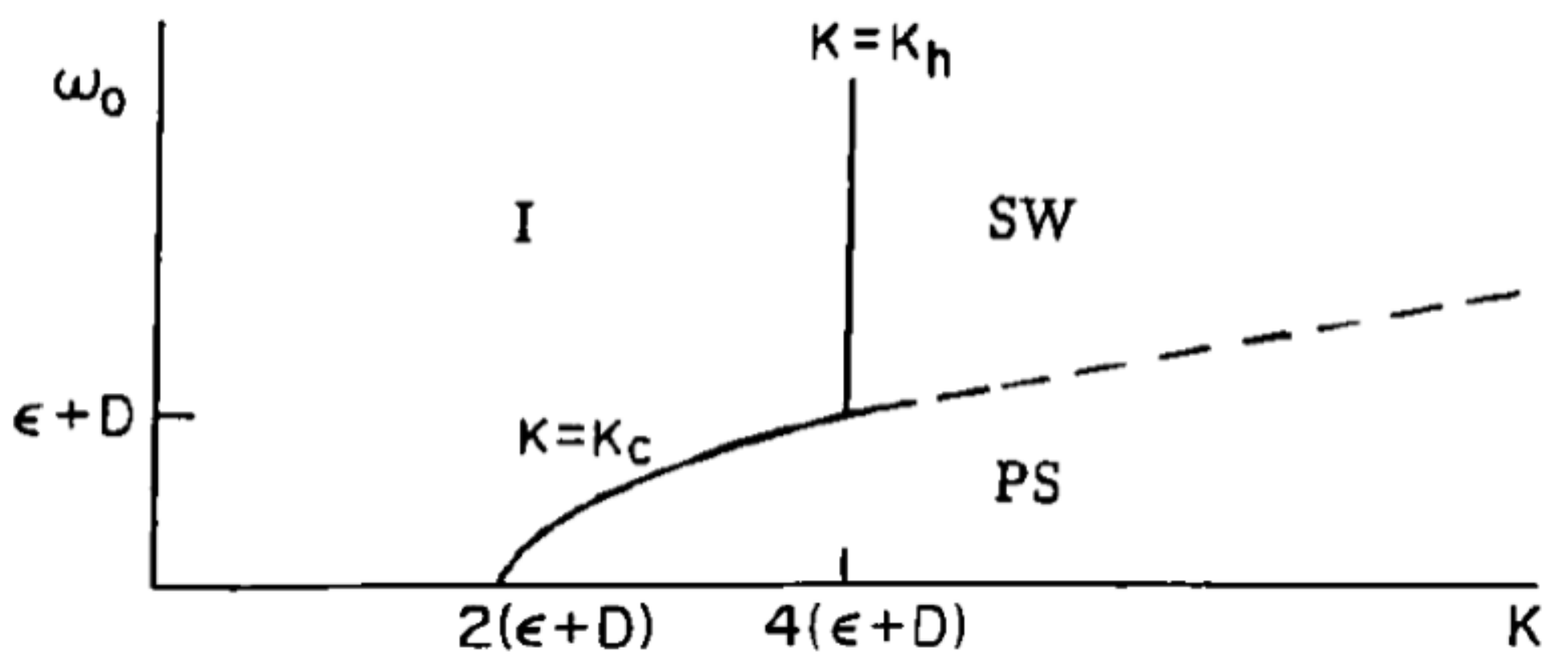}
\caption{States of the Kuramoto model with Gaussian white noise of intensity $D$ and a bimodal frequency distribution consisting of two Lorentzians centered at $\pm \omega_0$ with the same width parameter $\varepsilon$: the incoherent state (I), the partially synchronized state (PS), and the standing waves (SW). The boundary between the standing waves and the partially synchronized state is shown schematically as a dashed line because the precise nature and location of this boundary have not been determined. Notice
  that here $K$ stands for the coupling strength $\lambda$.  Reprinted from Ref.~\cite{Crawford1994} with permission of Springer.  \label{fig:crawford1994}}
\end{figure}

Therefore, with or without noise, the Kuramoto model presents a rather complex behavior as soon as the frequency distribution has two peaks. The case of two peaks equidistant to zero $\omega_2 - \omega_1 := 2 \, \omega_0$ was studied in Ref. ~\cite{Bonilla1992}. If the maxima are close, $\omega_0 < D/\sqrt{2}$, the transition to the synchronous state is continuous. When $\omega_0 \in (D/\sqrt{2}, D)$, the transition is discontinuous with hysteresis. Finally, when the peaks are separated enough, $\omega_0 > D$, the transition is also discontinuous with hysteresis, but now the stable synchronized state is characterized by having a time-periodic order-parameter.
The case of a more general frequency distribution $g_\varepsilon (\omega) := \frac{\varepsilon}{2 \pi} \left( \frac{1}{(\omega + \omega_0)^2 + \varepsilon^2} +  \frac{1}{(\omega - \omega_0)^2 + \varepsilon^2} \right)$ (which consists in two Lorentzians centered at $\pm \omega_0$ with the same width parameter $\varepsilon$) was studied in Ref. ~\cite{Crawford1994}.  A critical value for the separation between the maxima,
$ \omega_d := (\varepsilon + D) \sqrt{\frac{\varepsilon + 2D}{3\varepsilon + 4D}}$ is found, that determines the behavior of the system. For $\omega_0 < \omega_d$, the bifurcation is supercritical to a stable synchronized state. For $\omega_d < \omega_0 < \varepsilon + D$, there is a subcritical bifurcation to a synchronized state with no prior onset of synchronization in the peaks of the distribution. And for $\omega_0 > \varepsilon + D$, there is a transition to a state consisting of two symmetric clusters of synchronized oscillators, called standing waves. As the peaks move far apart (i.e., $\omega_0 \rightarrow \infty$), the oscillators in the peaks indeed synchronize, but as $\omega_0 \rightarrow \varepsilon + D$, they synchronize at frequencies that are substantially shifted away from the peaks of the native distribution. Actually, Fig.~\ref{fig:crawford1994}, where these results are schematically shown, is remarkably similar to Fig.~\ref{fig:martens2009}, but the effect of the noise added is to increase the width of each Lorentzian in $D$, having an effective width $\varepsilon +D$.

\begin{figure}
\centering
\includegraphics[width=0.6\linewidth]{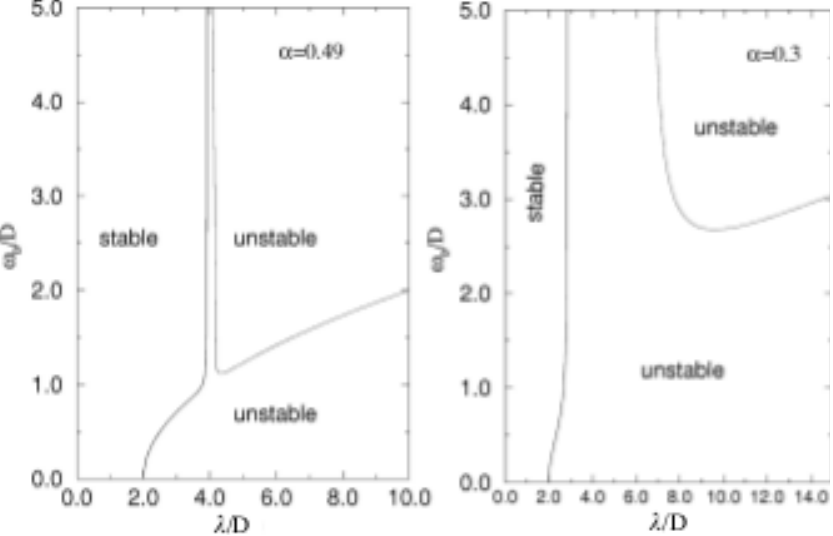}
\caption{Stability boundaries for the incoherent solution for the
  asymmetric bimodal frequency distribution when $\alpha =0.49$ and
  $\alpha =0.3$. Adapted with permission from
  Ref.~\cite{Acebron1998}. Courtesy of R. Spigler. \label{fig:acebron1998}}
\end{figure}

The asymmetric case was studied in Ref.~\cite{Acebron1998} which considered the bimodal distribution $g(\omega) := \alpha \, \delta (\omega - \omega_0) + (1- \alpha) \, \delta (\omega + \omega_0)$. Here, $\alpha \in (0,1)$ entails essentially different features with respect to the symmetric case ($\alpha = 1/2$). By comparing Fig.~\ref{fig:acebron1998} (which shows the stability boundaries for $\alpha =0.49$ and $\alpha =0.3$) with Fig.~\ref{fig:crawford1994}, one can appreciate how the asymmetry of the frequency distribution changes the stability boundaries: both the region of incoherence and standing waves becomes smaller, whereas the partially synchronized area grows.
Finally, Ref.~\cite{Balmforth2000} considered a discrete version of the Kuramoto model to explore the transition to synchronization in a population of weakly coupled oscillators. When the coupling strength exceeds some noise-dependent threshold, the population develops spike-like solutions with noisy boundary layers, and its behavior is similar to that of the continuum limit. However, a much more complicated behavior occurs for sub-threshold coupling strengths, where the finite population displays complicated temporal dynamics (ephemeral coherent structures, switching patterns).

\subsubsection{Generalizations of the coupling pattern}

Another streamline of research analyzed the effects of changing the
original (uniform, all-to-all) connectivity of
Eq.~(\ref{eq:kuramcomplete}). Reference ~\cite{Cohen1982}, for instance, generalized the Kuramoto model to account for intersegmental coordination of  neural networks, responsible for generating locomotion in the isolated spinal cord of lamprey. Reference ~\cite{Cohen1982} considered that each oscillator only has two nearest neighbors, and assumed that all upward and downward coupling strengths are equal. Not only most of the experimental observations were reproduced, but also stable phase locked motions (corresponding to traveling waves) as well as drifting motions were found.

A more general approach was followed in Ref. ~\cite{Sakaguchi1987}, which assumed a Gaussian $g(\omega)$ and an underlying topology of a $d$-dimensional hypercubic lattice. Namely,  $\lambda_{ij}= \lambda$ is taken only if $i$ and $j$ are nearest neighbors in the lattice. The $d$-lattices self-organize in clusters of entrained oscillators, for which the lattice dimension determines if some of them develop into a macroscopic size.

Reference ~\cite{Daido1988} revisited the problem with the aim to give more accurate predictions by means of a real-space renormalization-group analysis. Reference ~\cite{Daido1988} considered the case of a generic coupling pattern $h_{ij}(\theta , \lambda)$ satisfying $h_{ij}(-\theta , \lambda) =- h_{ij}(\theta , \lambda)$ (therefore the Kuramoto model is encompassed as the case $h_{ij}(\theta , \lambda) = \lambda \sin \theta$). Reference ~\cite{Daido1988} first showed that the global behavior of the lattice strongly depends on the frequency distribution. If the variance of $g(\omega)$ is not defined, its asymptotic behavior can be written as $ g(\omega) = \Theta (| \omega |^{-a-1})$,
with $a \in (0,2]$, while if a frequency distribution has finite variance, it can be studied by taking $a=2$. Assuming that the whole $d$-dimensional lattice is divided into a set of hypercubes with an equal linear scale $L$, Ref. ~\cite{Daido1988} demonstrated that synchronization only arises in a $d$-lattice when $d > a/(a-1)$. In particular, this implies that synchronization is impossible in any $d$-lattice if $a \in (0,1]$ (e.g., for a Lorentzian), while it is possible for any frequency distribution with finite variance (e.g., for a Gaussian) if the lattice has $d>2$.

Later on, however, Ref.~\cite{Strogatz1988b}  found that global synchronization is actually impossible for all $d$-lattice. Precisely,
given a Gaussian frequency distribution, and $N$ oscillators in a $d$-lattice coupled with strength $\lambda$, the probability $P(N, \lambda ,d)$ of phase-locking is found to verify $\lim_{N \rightarrow \infty} P(N, \lambda ,d) = 0$, for all finite $\lambda$ and $d$. Furthermore, dropping the assumption of a Gaussian distribution and assuming only that $g(\omega)$ has finite mean and finite non-zero variance, the former result remains valid. Therefore, there is no critical dimension for phase-locking,  because the probability tends to zero exponentially fast for any $d$-lattice.

Reference~\cite{Daido1987} proposed one of the first models in which the interactions among the oscillators are random. In order to have an analytically tractable model, it is considered that the interaction $\lambda_{ij}$ between oscillators $i$ and $j$ (which are placed on a lattice) is $\lambda_{ij} = \frac{\lambda}{z} s_i s_j$, where $\lambda$ is a control parameter, $z$ is the number of interacting neighbors of each oscillator, and $s_i$ are random parameters taken from a distribution $P(s)$. For the particular case $P_a(s) = a \, \delta (s-1) + (1-a) \, \delta (s+1)$ (being $a \in [0,1]$ a parameter and $\delta$ the Dirac function),  the order parameter $r_d$ comes out to be proportional to the original Kuramoto parameter $r$, $r_d = |2a-1| \, r$, thus randomness causes a decrease in the order parameter. The symmetric case (i.e., $a=1/2$) results in a disordered system even for $\lambda > \lambda_c$, which was called a spurious glass-like phase.
Reference ~\cite{Daido1992} proposed another model in which the interaction $\lambda_{ij}$ is assumed to be an independent random variable with normal probability distribution, $P(\lambda_{ij}) = \mathcal{N}(0,\lambda^2/N)$ (where $\lambda$ is the control parameter). Numerical evidence is obtained of a new type of ordered phase that is analogous to those observed in a variety of glassy systems.

Reference ~\cite{Lumer1991} studied the case in which the underlying network topology is a tree with branching ratio $b$, being the oscillators at the leaves, under the hypotheses that the coupling strengths vary with the distance and are distributed over a hierarchy of values. Specifically, it was assumed that intra-cluster interactions are weaker than inter-cluster ones. It is shown that distinct regimes are emerging depending on whether the branching ratio $b$ is smaller or larger than the critical value $b_c$.

\begin{figure}
\centering
\includegraphics[width=0.9\textwidth]{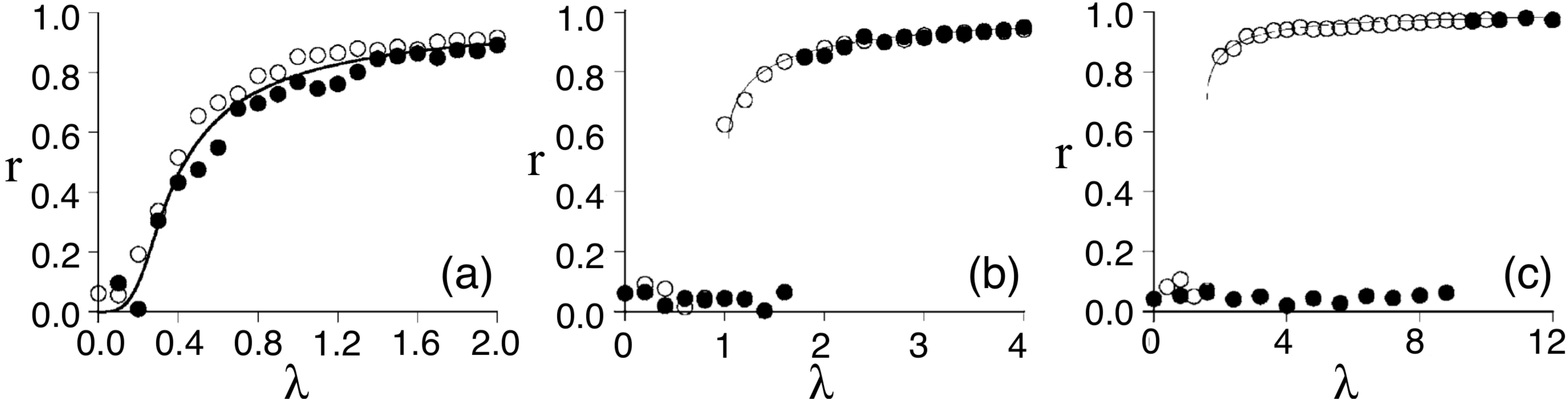}
\caption{Numerical and analytical results of the Kuramoto model
  proposed in Ref.~\cite{Filatrella2007} for various feedback
  strengths: (a) $z=0.7$, (b) $z=2$, (c) $z=3$. The solid lines
  represent the analytic prediction, the filled (open) circles show
  the forward (backward) transition numerically computed. The disorder
  is set to $\gamma =0.05$. Adapted with permission from
  Ref.~\cite{Filatrella2007}. Courtesy of G. Filatrella. \label{fig:filatrella2007}}
\end{figure}

Reference~\cite{Jadbabaie2004} provided a stability analysis of the Kuramoto model in terms of appropriate Lyapunov functions, recovering all the previous results for fully connected graphs, and obtaining new ones for generic topologies.
Finally, Filatrella et al.~\cite{Filatrella2007} proposed a
modification of the Kuramoto model to account for the effective change
in the coupling constant suggested by some experiments on Josephson
junctions~\cite{Wiesenfeld1996,Wiesenfeld1998}, laser arrays, and
mechanical systems. Namely, Ref.~\cite{Filatrella2007} proposed a
variation of the Kuramoto model that accounts for change of the
coupling with the number of active oscillators: the natural
frequencies are taken from a Lorentzian with zero average and $\gamma$
width, and the coupling  $\lambda_{ij}$ is taken to be $\lambda_{ij} =
\lambda (r) = \lambda \, r^{z-1}$, where the parameter $z$ is a
heuristic measure of the strength of the feedback mechanism. The
resulting model is analytically tractable, and predicts that both
first and second order phase transitions are possible, depending upon
the value of the parameter $z$ that tunes the coupling among the
oscillators. For $z \le 1$ the evolution from the incoherent value
$r=0$ to the partially coherent state is continuous (see
Fig.~\ref{fig:filatrella2007}(a)). When $z>1$, the critical value
where the transition to synchrony occurs changes for the forward and
backward transitions, yielding hysteresis, whose width is larger for
the larger $z$ (as it is evident by comparing the case $z=2$ in Fig.~\ref{fig:filatrella2007}(b) to the case $z=3$ in Fig.~\ref{fig:filatrella2007}(c)).

\subsection{Explosive synchronization in complex networks}

Unfortunately, the scientific community was initially somehow apathetic to the many results summarized in the previous Section, and such a situation (of almost indifference) persisted lamentably for longtime, up to when the scientific community started to be interested in the study of the transition to synchrony in complex networks.

\subsubsection{Explosive synchronization in networks of frequency-degree correlated oscillators}
\label{engineered_correlation}

In Ref.~\cite{Gomez-Gardenes2011}, an unweighted and undirected network of $N$ Kuramoto oscillators \cite{Kuramoto1975} is considered, whose equations of motion are
\be
\dot{\theta_i}=\omega_i+\lambda\sum_{j=1}^{N} A_{ij} \sin{(\theta_j-\theta_i)},
\label{eq:kuramoto}
\ee
where $\omega_i$ stands for the natural frequency of oscillator $i$, $\lambda$  is the coupling strength, and  $A_{ij}$ are the elements of the network's adjacency matrix. % where $A_{ij}$=1 when oscillators $i$ and $j$ are connected and 0 otherwise.

Reference~\cite{Gomez-Gardenes2011} actually introduced a specific
(ad-hoc) form of correlation between the  natural frequencies and the
degrees of the oscillators. Namely, the natural frequency $\omega_i$ of
node $i$ was imposed to be proportional to its degree $k_i$
($\omega_i=k_i$). Once such a correlation is artificially imposed,
different networks (generated by the algorithm of
Ref.~\cite{Gomez-Gardenes2006}) are studied, having the same average
connectivity but variable heterogeneity properties (i.e. ranging from
ER to Barab\'asi-Albert (BA) graphs, depending on the value of a single parameter $\alpha=[0,1]$).

\begin{figure}
\centering \includegraphics[width=0.6\textwidth]{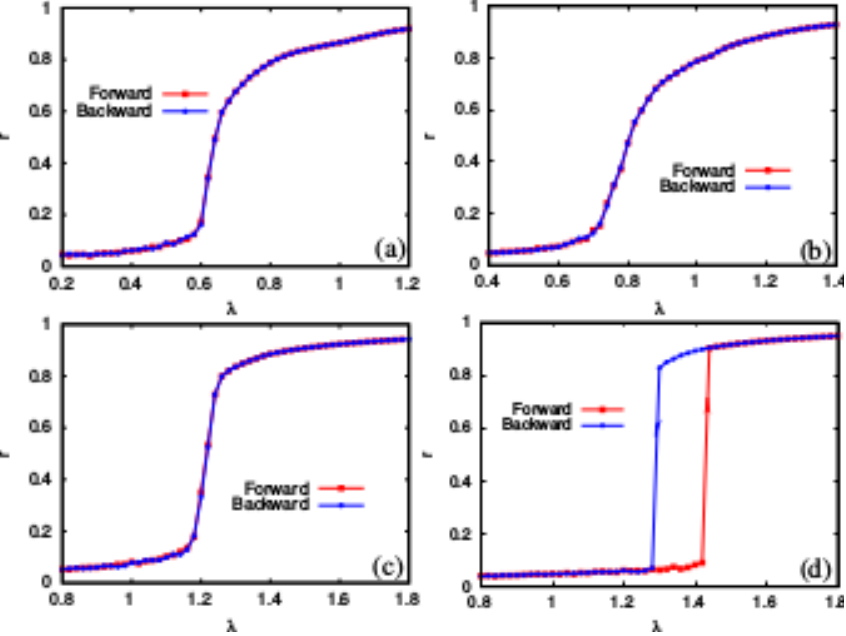}
\caption{(color online). $r(\lambda)$ for various networks constructed with the algorithm in Ref.~\cite{Gomez-Gardenes2006}. $N=10^3$ and $\langle k\rangle=6$. The $\alpha$ values are: $1$ (a, ER), $0.6$ (b), $0.2$ (c), and $0$ (d, BA). The panels report both the forward and backward transitions using steps $\delta\lambda=0.2$. Reprinted with permission from Ref.~\cite{Gomez-Gardenes2011}. $\copyright$  2011 by American Physical Society.}
\label{fig:Gomez-Gardenes2011-1}
\end{figure}

The results are summarized  in Fig.~\ref{fig:Gomez-Gardenes2011-1}. The synchronization diagrams for several values of $\alpha$ are plotted as a function of  the coupling $\lambda$, for both forward and backward transitions. The transition is soft and reversible for most values of $\alpha$, but when $\alpha=0$ (corresponding to a BA network) an abrupt behavior appears, with the parameter $r$ remaining close to zero  up to a point at which it suddenly jumps to  $r\sim 1$. As the critical coupling is different for the forward and the backward diagrams, the transition is irreversible and displays a hysteresis loop.

Reference~\cite{Gomez-Gardenes2011} offers also an analytical study of a specific star configuration, composed of a central node (the hub $h$) and $K$ peripheral nodes (the leaf nodes), each one
labeled by the index $i$ ($i=1,\dots,K$). By setting the frequencies of the central node ($\omega_h=K\omega$), and that of the leaves
($\omega_i=\omega$ $\forall i$), the equations of motion can be written as
\ba
\dot{\theta_i} &=&\omega+\lambda \sin(\theta_h-\theta_i),   \nonumber \\
\dot{\theta_h} &=&K\omega+\lambda \sum_{i=1}^{K} \sin(\theta_i-\theta_h) .
\label{eq:star}
\ea
Then, using  the average phase of the system ($\Psi(t)=\Psi(0)+\Omega t$, where $\Omega$ is the average frequency) as a rotating frame,  the equation for the hub evolution can be rewritten as
\be
\dot{\theta_h} = (K-\Omega)+\lambda (K+1) r \sin(\theta_h).
\ee
The condition for stability of the hub locking solution ($\dot{\theta_h}=0$) allows one to obtain the {\it backward} critical coupling value as $\lambda_c=(K-1)/(K+1)$, and the order parameter at the critical point is found to be $r_c=K/(K+1)$, that is $r_c>0$ at the critical point, confirming the existence of a discontinuity in the transition.

The same degree-frequency correlated star of Eqs.~(\ref{eq:star}) was studied  in Ref.~\cite{Zou2014}, where the origin of the hysteresis loop was rooted in the properties of the basin of attraction of the synchronization manifold. Ref.~\cite{Zou2014} rewrote Eq.~(\ref{eq:star}) as $\dot{\Phi}=\Omega_k+\lambda H(\Phi)$, where  $\Phi=(\theta_1...,\theta_{K+1})$,  $\Omega_k=(\omega,...\omega,K\omega)$ and $\mathbf{H}(\Phi)=(\sin(\theta_{k+1}-\theta_1),\sin(\theta_{k+1}-\theta_2),...,\sum_{j=1}^{k}\sin(\theta_j-\theta_{k+1}))$. Then, the locking manifold can be defined as $M_a=\lbrace \Phi  \in T^{K+1}: \theta_1=\theta_2=...=\theta_K ; \theta_{K+1}-\theta_1=a \rbrace$,
where $a$ is a constant. The condition $\dot{\Phi}=0$ yields the minimal coupling that enables the existence of $M_a$ : $\lambda^b_c=\frac{(K-1)\omega}{K+1}$, which corresponds to the \textit{backward} transition.

\begin{figure}
\centering 
\includegraphics[width=0.7\linewidth]{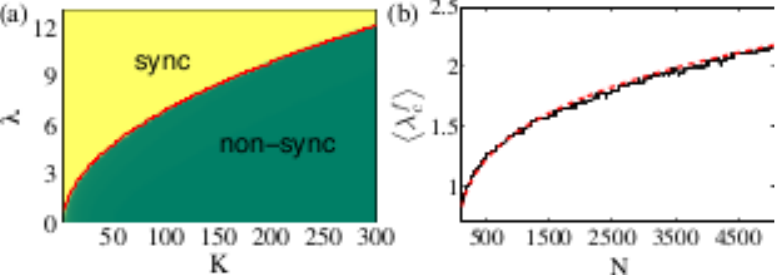}
\caption{(Color online). (a) Order parameter $r$  vs. the space of
  parameters $(K, \lambda)$ for the forward transition. (b) The red thick
  line indicates the solution of Eq.~(\ref{eq:lambda_f_Zou2014}), for
  $1/B=0.6989$. Adapted with permission from
  Ref.~\cite{Zou2014}. Courtesy of J. Kurths. \label{fig:Zou2014}}
\end{figure}

To obtain insight on the \textit{forward} transition, Ref.~\cite{Zou2014} provided numerical evidences that (for
$\lambda>\lambda_{c}^{f}$) the local attractiveness of $M_a$ becomes global, i.e. any state (even incoherent) is attracted to $M_a$ for coupling strengths above $\lambda_{c}^{f}$.
This fact (together with the theory developed in Ref.~\cite{Pereira2013}) can be used to reformulate the phase locking problem between the hub and leaves as a perturbation of an identical synchronization problem. In such a latter framework, the condition for coherence  gives
\be
\lambda^{f}_c \simeq \left(\frac{K-1}{\sqrt{K}} \frac{1}{B} \right) \omega,
\label{eq:lambda_f_Zou2014}
\ee
where $B$  is a constant which depends on the chosen initial conditions.  In Fig.~\ref{fig:Zou2014}(a) it is shown how numerical simulations are in excellent agreement with the predictions of Eq.~(\ref{eq:lambda_f_Zou2014}).

The results can be extended to the case of SF networks with small mean degrees, that can be considered, indeed, as a collection of stars of different sizes. For $K \gg 1$, $\lambda_{b}^{c} \rightarrow \omega$, but  $\lambda_{f}^{c} \rightarrow \sqrt{K}\omega$.
In a SF network with degree distribution $P(k)=k^{-\gamma}$ the expected degree of the largest hub scales as $N^{\frac{1}{\gamma-1}}$. Therefore, the expected $\lambda_{f}^{c}$ scales as
\be
\langle \lambda_{f}^{c} \rangle \approx N^{\frac{1}{2(\gamma-1)}}.
\ee
As it can be seen in Fig.~\ref{fig:Zou2014}(b) the approximation gives a very good fit for SF networks with $\gamma=3$.

The problem of calculating the forward critical point in a frequency-degree correlated star  has been very recently reconsidered by several groups, which applied dimensional reduction techniques \cite{Watanabe1993} and the Ott-Antonsen  method \cite{Ott2008}.
The first approach was applied independently by Vlasov et al. \cite{Vlasov2015}
and Xu et al. \cite{Xu2015a}. Taking $\phi_j=\theta_h-\theta_j$, the Eq.~(\ref{eq:star}) can be reduced to
\be
\dot{\phi_j}=\delta \omega -\lambda \sum_{i=1}^{K} \sin \phi_i -\lambda \sin \phi_j \hspace{1cm} j=1,...K,
\label{eq:reduced_star_1}
\ee
where $\delta \omega$ is the frequency difference between the hub and leaf node. By defining  $z=re^{i\Phi}=\frac{1}{K} \sum_{j=1}^{K} e^{i\phi_j}$, Eq.~(\ref{eq:reduced_star_1}) is rewritten as
\be
\dot{\phi_j}=fe^{i\phi_j} + g + \bar{f} e^{-i\phi_j},
\label{eq:reduced_star_2}
\ee
where $f=i\frac{\lambda}{2}$, and $g=\delta \omega -\lambda Kr \sin(\Phi)$. The Watanabe-Strogatz approach can be used in systems of identical oscillators driven by a common force (having the general form $\psi_j=f(t)+Im(F(t)e^{-i\psi_j}$), as it is the present case. The phase dynamics of the $K$ nodes can be constructed by $K$ constants as
\be
e^{i\phi_j}=\frac{\beta+e^{i(\psi+\epsilon_j)}}{1+\bar{\beta}e^{i(\psi+\epsilon_j)}},
\ee
where $\beta=\beta(t)$ and $\psi=\psi(t)$ are global variables, and
$\epsilon_j$ are constants depending on the initial condition of the systems. In the thermodynamic limit ($K\rightarrow \infty$) such a transformation allows to separate the evolution of the $\beta(t)$ and $\psi(t)$ variables, and gives $\beta(t)=z(t)$. Therefore, the equation of the order parameter can be written as \cite{Vlasov2015,Xu2015a}
\be
\dot{z}=-\frac{\lambda}{2}z^2 + i(\delta \omega - \lambda Kr \sin \Phi)z + \frac{\lambda}{2},
\label{eq:Ott_order}
\ee
which describes the collective dynamics of system (\ref{eq:star}) in terms of the ensemble order parameter, and can be used even for small values of $K$ in case of random initial distributions. In the phase space of the ensemble order parameter, the synchronous state corresponds to a fixed point (with $r=1$) and a fixed phase $\Phi$ . All the other solutions of Eq.~(\ref{eq:Ott_order}) represent various incoherent states. If Eq.~(\ref{eq:Ott_order}) is rewritten in its Cartesian coordinates $z=x+iy$, the null-clines $\dot{x}=\dot{y}=0$ of the system can be studied as a function of the coupling $\lambda$, allowing the calculation of the forward critical coupling \cite{Vlasov2015,Xu2015a}
\be
\lambda_{c}^{f}=\frac{\delta \omega}{\sqrt{K}}\frac{1}{\sqrt{2+K^{-1}}}.
\ee
It can be seen that the approximation (\ref{eq:lambda_f_Zou2014}) is a good estimation of the latter result for large $K$, where
$\lambda_{c}^{f}=\frac{\delta \omega}{\sqrt{2K}}$.

A similar dimensional reduction  has been provided in Ref.~\cite{Jiang2015}, in which an extension of the model is presented where random perturbations affect the frequencies of leaves. The results show that for small perturbations the synchronization phase transition is still discontinuous, while for large perturbations the transition becomes continuous.
\begin{figure}
\centering \includegraphics[width=0.4\linewidth]{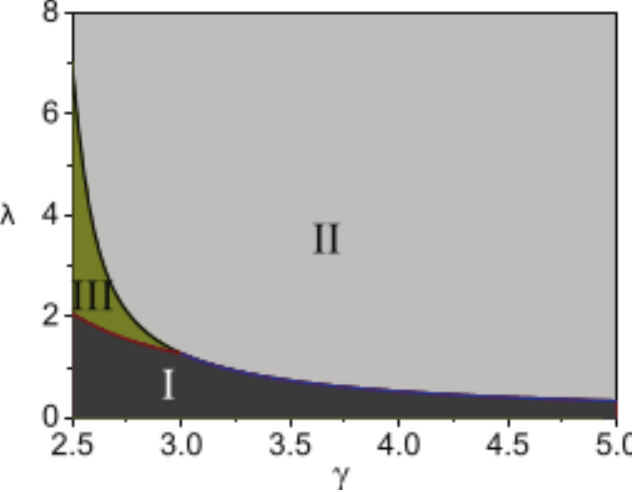}
\caption{ Phase diagram $\gamma-\lambda$ of the Kuramoto model for
  frequency-degree correlated SF networks. In region I there is no
  spontaneous synchronization, and the order parameter
  vanishes. Synchronization appears in region II, in which the order
  parameter is larger than zero. Region III is the region of
  hysteresis with a meta-stable state and one stable state.
Adapted with permission from Ref.~\cite{Coutinho2013}. Courtesy of S. N. Dorogovtsev.  \label{fig:Coutinho2013}}
\end{figure}

In Ref.~\cite{Coutinho2013} the case of a star graph with arbitrary natural frequency is studied. Then, the systems' equations become
\ba
\dot{\theta_i} &=&\omega_i+\lambda \sin(\theta_h-\theta_i) ,  \nonumber \\
\dot{\theta_h} &=&\omega_h+\lambda \sum_{i=1}^{K} \sin(\theta_i-\theta_h).
\label{eq:star2}
\ea
The transition to synchrony of this system is discontinuous when $\omega_h-\langle \omega_i \rangle > \omega_c$, where $\omega_c$ is a certain critical frequency difference.

Beyond a star graph, analytic results for general networks (under the assumption that $g(\omega)=P(k)$) were given by Peron et al. \cite{Peron2012}. In the continuous limit, the equation of the system is
\be
\dot{\theta}=k + \lambda k \int dk' \int d\theta \frac{k'P(k')}{\langle k \rangle} \rho(\theta'\vert k)\sin(\theta'-\theta).
\ee
Ref.~\cite{Peron2012} shows that (in such a limit) the forward critical coupling is $\lambda_{c}^{f}=\frac{2}{\pi \langle k \rangle P(\langle k \rangle)}$.

Notice that this critical coupling differs from the usual value $\lambda_{c}^{o}=\frac{2}{\pi g(0)} \frac{\langle k \rangle}{\langle k^2 \rangle}$ for symmetric $g(\omega)$.

Even though the calculations in Ref.~\cite{Peron2012} identify the frequency and degree distributions, they do not account for degree-frequency correlations, and therefore fail in predicting if the transition is explosive or continuous.  The issue was addressed in Ref. \cite{Coutinho2013}, where the annealed network approximation was used to solve Eq.~(\ref{eq:kuramoto}) under the assumption $\omega_i=k_i$. An implicit solution for the order parameter
\ba
\langle k \rangle -\omega &=& \int_{\vert k_i-\Omega\vert < \alpha r k} dk P(k) (k-\Omega) \sqrt{ 1-\left(\frac{\lambda r k}{k-\Omega} \right)^2}  , \\
r &=& \frac{1}{\langle k \rangle} \int_{\vert k_i-\Omega\vert > \alpha r k} dk P(k) k \sqrt{ 1-\left(\frac{k-\Omega}{\lambda r k} \right)^2},
\ea
is found. Following such a solution,  ER networks always undergo
second order transitions, whereas the nature of the transition in SF
networks depends on the heterogeneity parameter $\gamma$, as shown in
Fig.~\ref{fig:Coutinho2013}: a first-order-like transition is found
for $2<\gamma<3$, but a second-order  transition occurs for $\gamma >
3$. A  hybrid phase transition is found for $\gamma=3$, combining a
jump of the order parameter  and critical phenomena at the critical
point. This is actually a first indication that not all SF networks
are able to sustain explosive synchronization (ES), even when they are fully frequency-degree correlated, a subject that will be reconsidered afterwards \cite{Sendina-Nadal2015}.

In a number of works, various generalizations of the frequency-degree correlation have been analyzed. In Ref.~\cite{Skardal2013}, a more general joint distribution $P(\omega,k)=P(k)[\delta(\omega-\alpha k^{\beta})+ \delta(\omega+\alpha k^{\beta})]$ is considered. The phase-locking behavior of the system is examined, and it is found that the criterion for locking is $\alpha k^{\beta-1}<\lambda r$, being $r$ the order parameter. From this expression one can deduce that the nodes with low degree are the last ones to get phase-locked for sub-linear correlation ($\beta<1$), whereas  super-linear correlation ($\beta>1$) induce the last drifting nodes to be the hubs. These two qualitatively different behaviors are separated by the critical case of linear correlations  ($\beta=1$)  for which $k$ disappears, and the oscillators lock simultaneously.

\begin{figure}
\centering \includegraphics[width=0.8\linewidth]{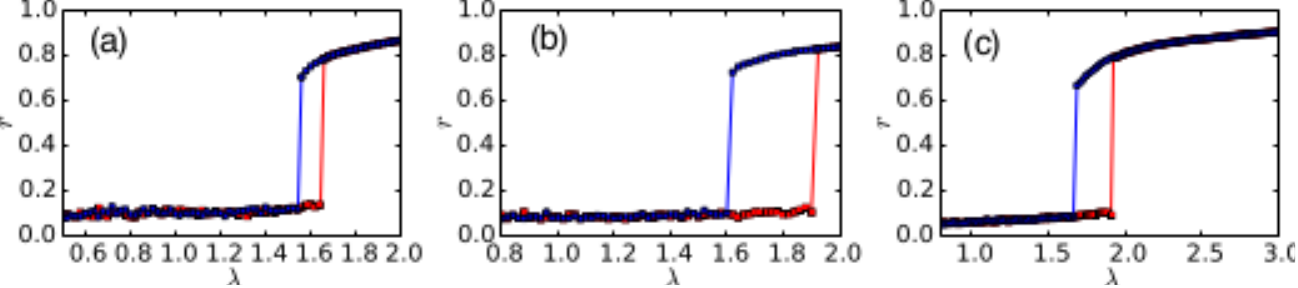}
\caption{(Color online). Synchronization diagrams $r(\lambda)$ for networks built with the mechanism proposed in Ref.~\cite{Gomez-Gardenes2006} with $\alpha$ = 0.2 (a), 0.1 (b) and 0 (c), where only 10$\%$ of the vertices with largest degree show degree-frequency correlations. Adapted with permission from Ref.~\cite{Pinto2015}. Courtesy of R. S. Pinto.\label{fig:Pinto2015}}
\end{figure}

In Ref.~\cite{DanYJun-Zhong2014}, instead,  the possibility of a partial frequency-degree correlation is considered, and it is shown that a small fraction of correlated nodes is enough to obtain ES  in a SF network. The issue of partial correlation has been  dealt with  also in Ref.~\cite{Pinto2015}, where only vertices with degree $k$ larger than a certain threshold $k_{∗}$ are supposed to exhibit the degree-frequency correlation, whereas the other vertices have random natural frequencies with distribution $g(\omega)$. This latter feature can be formally enunciated through a joint probability distribution (for a node of degree $k$ and natural frequency $\omega$) given by $G(\omega,k)=[\delta(\omega-k)P(k)-g(\omega)P(k)]H(k-k_*)+g(\omega)P(k)$, where $\delta(x)$ and $H(x)$ are, respectively, the Dirac delta
and the Heaviside step functions, and $P(k)$ is the degree distribution.  A mean-field calculation  shows that the correlation of the $10 \%$ highest degree nodes is enough  to promote ES in BA networks.   The results are shown in Fig.~\ref{fig:Pinto2015}:  by imposing a partial degree-frequency correlation not only it is possible to keep  ES in the cases where it already happens with full correlation (Fig.~\ref{fig:Pinto2015}(b)-(c)) but, unexpectedly, ES emerges also in cases in which full correlation would have prevented it (Fig.~\ref{fig:Pinto2015}(a)).

\begin{figure}
\centering \includegraphics[width=0.5\textwidth]{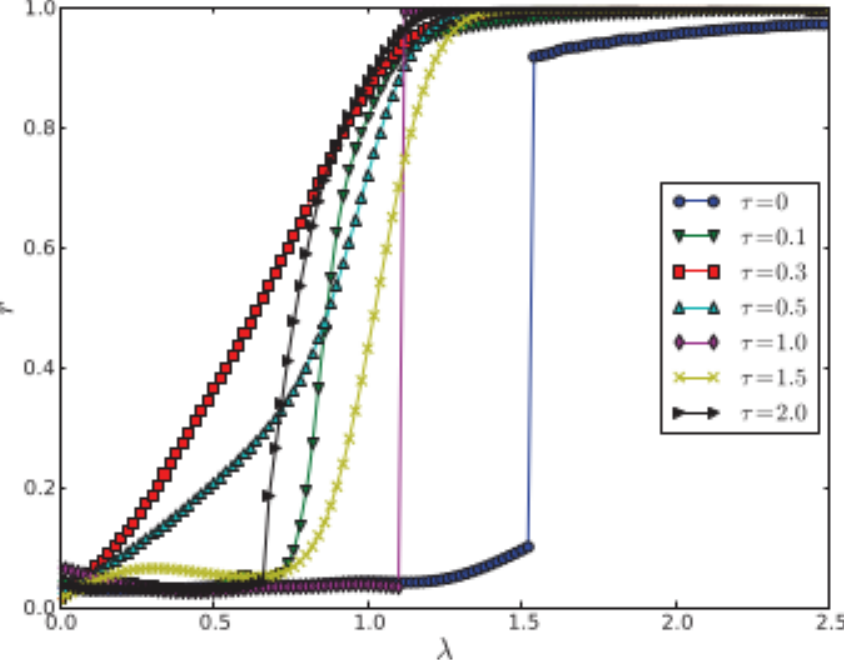}
\caption{(Color online). $r(\lambda)$ for a BA model with $N=1,000$
  and $\avk = 6$, for several values of the delay $\tau$ (see legend
  for color and symbol code). Adapted with permission from
  Ref.~\cite{Peron2012a}. Courtesy of F. A. Rodrigues. \label{fig:Peron2012a}}
\end{figure}

Other works have tested how ES is affected by modifications to the original Kuramoto model. In Ref.~\cite{Li2013}  it is reported that ES persists (and is even enhanced) in modular BA, $k-\omega$ correlated networks. In Refs. \cite{Peron2012a}, the model of Ref.~\cite{Gomez-Gardenes2011} is modified to include delay times phase shifts \cite{Xu2015a}, and it is found that an appropriated choice of parameters can enhance or suppress ES. An example of the effect of time delay is shown in Fig.~\ref{fig:Peron2012a}.

Reference~\cite{Liu2013} generalized the $k-\omega$ relationship as $\omega_i=k_i/\beta$, where $\beta=\sum_{i}^{N} k_i{^\alpha} / \sum_{i}^{N} k_i$ is a normalization to avoid homogenization of the frequencies when $|\alpha|$ is small, and $\alpha$ is in the range $[-1,1]$, thus including the possibility that $k_i$ and $\omega_i$ are anti-correlated. For positively k-$\omega$ correlated networks ($\alpha>0$) ES remains  unmodified for up to $\alpha=0.1$, but it transforms to a second-order transition for any $\alpha<0$ value. The effect of degree mixing in the model is also considered, and it is pointed out that assortativity has a strongly destructive effect on ES.  In  Ref.~\cite{Li2013a} a frequency-degree correlated BA network is rewired to introduce assortativity, and it is pointed out that a moderate disassortativity favors the ES, which instead is destroyed for assortative mixing.

A  detailed study of the role of the assortativity in ES is provided in Ref.~\cite{Sendina-Nadal2015}.  With the aim of inspecting whether or not ES depends on the chosen network model, Ref.~\cite{Sendina-Nadal2015} comparatively
considers ensembles of networks obtained with the BA algorithm and
networks displaying the same SF distributions,
but constructed by the so called configuration model (CM, \cite{Bender1978}). For both
networks, parameters were assumed to be $N=10^3$  and $\langle k \rangle= 6$, and the oscillators’
frequencies were distributed so as to determine a direct correlation with the
node degree ($\omega_i = k_i$).

\begin{figure}
\centering \includegraphics[width=0.5\textwidth]{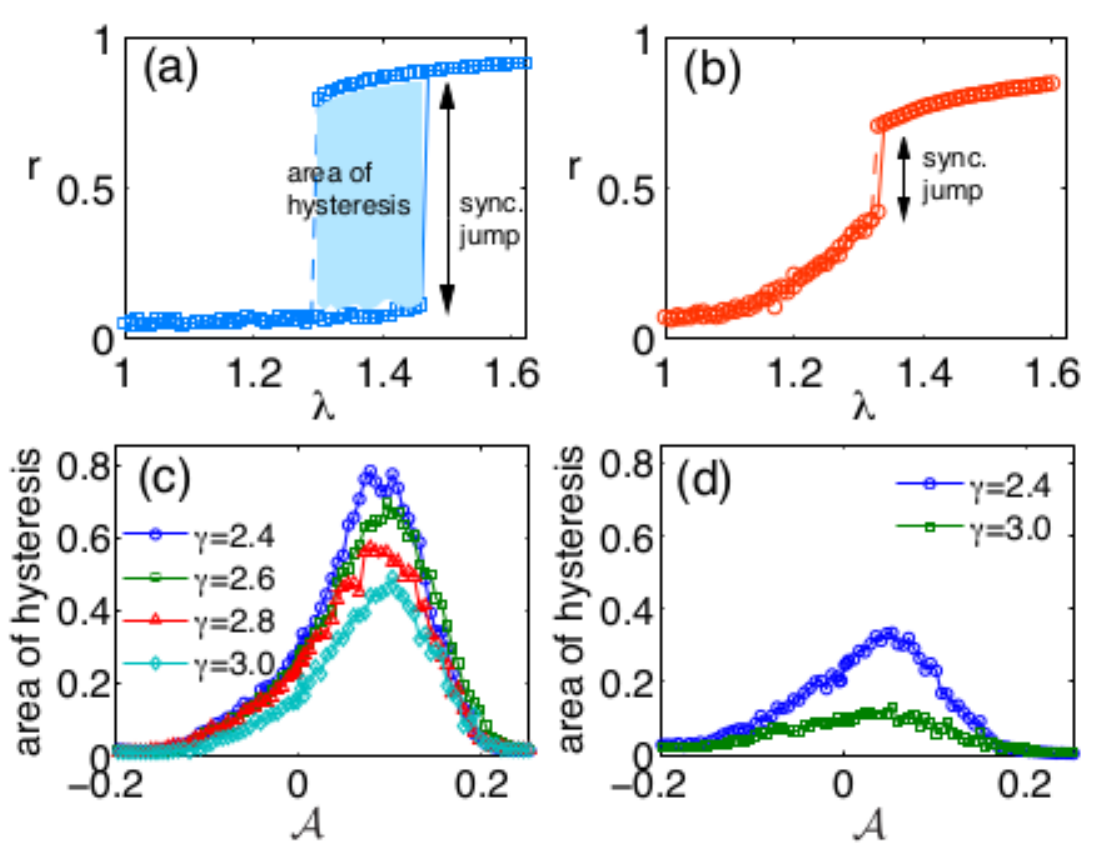}
\caption{(Color online). Comparison of ES between two SF
networks belonging to two different configuration ensembles:
(a),(c) preferential attachment (PA) and (b),(d) configuration (CM) models.
(a)-(b) Forward (solid lines) and backward (dashed lines)
synchronization curves for a PA (a) and CM (b) SF network, with exactly the same degree distribution.
The area of the hysteresis is depicted as a blue shaded area in (a). The synchronization jumps are marked in both
cases. (c)-(d) Area of the hysteretic region delimited by the forward and
    backwards synchronization curves  vs. the Pearson correlation
    coefficient $\cal A$
for PA (c) and CM (d) SF networks with different values
(reported in the legend) of the exponent $\gamma$ for the degree distribution $P(k)\sim k^{-\gamma}$.
Each point is an average over 10 different simulations.   Adapted with permission from  Ref.~\cite{Sendina-Nadal2015}. $\copyright$  2015 by the American Physical Society. \label{fig:Sendina2015_1}}
\end{figure}

The results reported in Fig.~\ref{fig:Sendina2015_1} show a dramatic
dependence of the ES behavior on the underlying SF network model used,
despite displaying the same $P(k)$. The forward and backward
continuations of the order parameter $r$ are totally different for BA
(Fig.~\ref{fig:Sendina2015_1}(a))
and CM (Fig.~\ref{fig:Sendina2015_1}(b)) networks, and indicate that a crucial condition to obtain irreversibility is having an underlying growing process through which the SF topology is shaped.

One customary way to quantify the amount of degree correlation with a
single parameter is using the Pearson correlation coefficient $\cal A$, which can be calculated as \cite{Newman2003}:
\be
{\cal A}=\frac{L^{-1}\sum_i j_i k_i  -[L^{-1}\sum_i\frac{1}{2}
  (j_i+k_i)]^2}{L^{-1}\sum_i \frac{1}{2}(j_i^2+k_i^2) -[L^{-1}\sum_i\frac{1}{2}
  (j_i+k_i)]^2},
\label{eq:assortativity_coef}
\ee
where $j_i$ and $k_i$ are the degrees of the nodes at the ends of
the $i$th link, with $i=1,\cdots,L$. Actually, one has that $-1\le {\cal A} \le 1$,
with positive (negative) values of ${\cal A}$ quantifying the level of assortative (disassortative) mixing. The BA  model does not
exhibit any form of mixing in the thermodynamic limit (${\cal A}\to 0$ as $(\log ^2N)/N$ as ${N \to \infty} $), while a random CM produces  highly disassortative networks.

Finally, Ref.~\cite{Sendina-Nadal2015} studies the impact of increasing/decreasing the assortativity
mixing on a network with a given degree sequence $\{k_i\}$ taken from a power law distribution $k^{-\gamma}$.
SF networks with given and tunable levels of degree mixing can be generated by
the rewiring  algorithm of Xulvi-Brunet and Sokolov \cite{Brunet2004}.
Two ensembles of BA and CM networks are then constructed, all of them having the same degree distribution, but different values of the assortativity coefficient ${\cal A}$.
Figures ~\ref{fig:Sendina2015_1}(c)-(d) illustrate the effect of degree
mixing on ES. Extensive numerical simulations were performed at
various values of ${\cal A}$, for slopes $\gamma$ ranging from $2.4$ to
$3.0$, and for the
same mean degree $\langle k \rangle= 6$. The most relevant result is that,
regardless of the specific SF network model, the hysteresis
of the phase transition is highly enhanced (weakened) for
positive (negative) values of ${\cal A}$,
and that there is an optimal (positive) value of $\cal A$ where the
irreversibility of the phase transition is maximum. The enhancement is far more pronounced in BA (Fig.~\ref{fig:Sendina2015_1}(c))
than in CM (Fig.~\ref{fig:Sendina2015_1}(d)) networks.

% \begin{figure}
% % \centering \includegraphics[width=0.3\linewidth]{fig33-Sendina2015-2a}
% % \centering \includegraphics[width=0.3\linewidth]{fig33-Sendina2015-2b}
% \centering \includegraphics[width=0.6\textwidth]{fig33-Sendina2015-2ab}
% \caption{(Color online). Adapted with permission from  Ref.~\cite{Sendina-Nadal2015}. \label{fig:Sendina2015_2}}
% \end{figure}

%%subsubsec332
\subsubsection{Explosive synchronization in frequency-disassortative networks.}

While earlier works concentrated on ad-hoc imposed correlations between the node degree and the corresponding oscillator’s natural frequency, posterior studies highlighted that a sharp and discontinuous phase transition is by no means restricted to such rather limited cases, but it constitutes, instead, a {\it generic} feature in synchronization of networked  oscillators. Precisely, Ref.~\cite{Leyva2013a} gave a condition for the transition from unsynchronized to synchronized states to be abrupt, and demonstrates how such a condition is easy to attain in many circumstances, and for a wide class of frequency distributions.

The basic idea, which evokes the Achlioptas process \cite{Achlioptas2009} described in Section \ref{gingillino}, is avoiding that
oscillators behave as cores of a clustering process, where neighboring units begin to aggregate smoothly and progressively. As the synchronization seeds are the aggregations of nodes with close frequencies, the procedure introduces then a {\it frequency disassortativity} in the networks. The practical realization of  such a condition is made by imposing explicit constrains in the frequency difference between each node $i$ and the whole set ${\cal N}(i)$ of oscillators belonging to its neighborhood \cite{Leyva2013a}:
\be
\vert {\omega_i}-{\omega_j}\vert > \gamma_c,  j  \in {\cal N}(i).
\label{eq:gap}
\ee

In practice, a threshold $\gamma_c$ for the frequency gap is set, and the network is then grown by means of a {\it conditional} CM approach: after having initially distributed the oscillator's frequency from a given frequency distribution, pairs of unconnected nodes are randomly selected, and a connection between the nodes is established only if they verify the condition (\ref{eq:gap}). The process is repeated until the network acquire a given, desired, mean degree $\langle k\rangle$, and the resulting adjacency matrix is used to simulate the Kuramoto model (\ref{eq:kuramoto}).

\begin{figure}
\centering \includegraphics[width=0.7\linewidth]{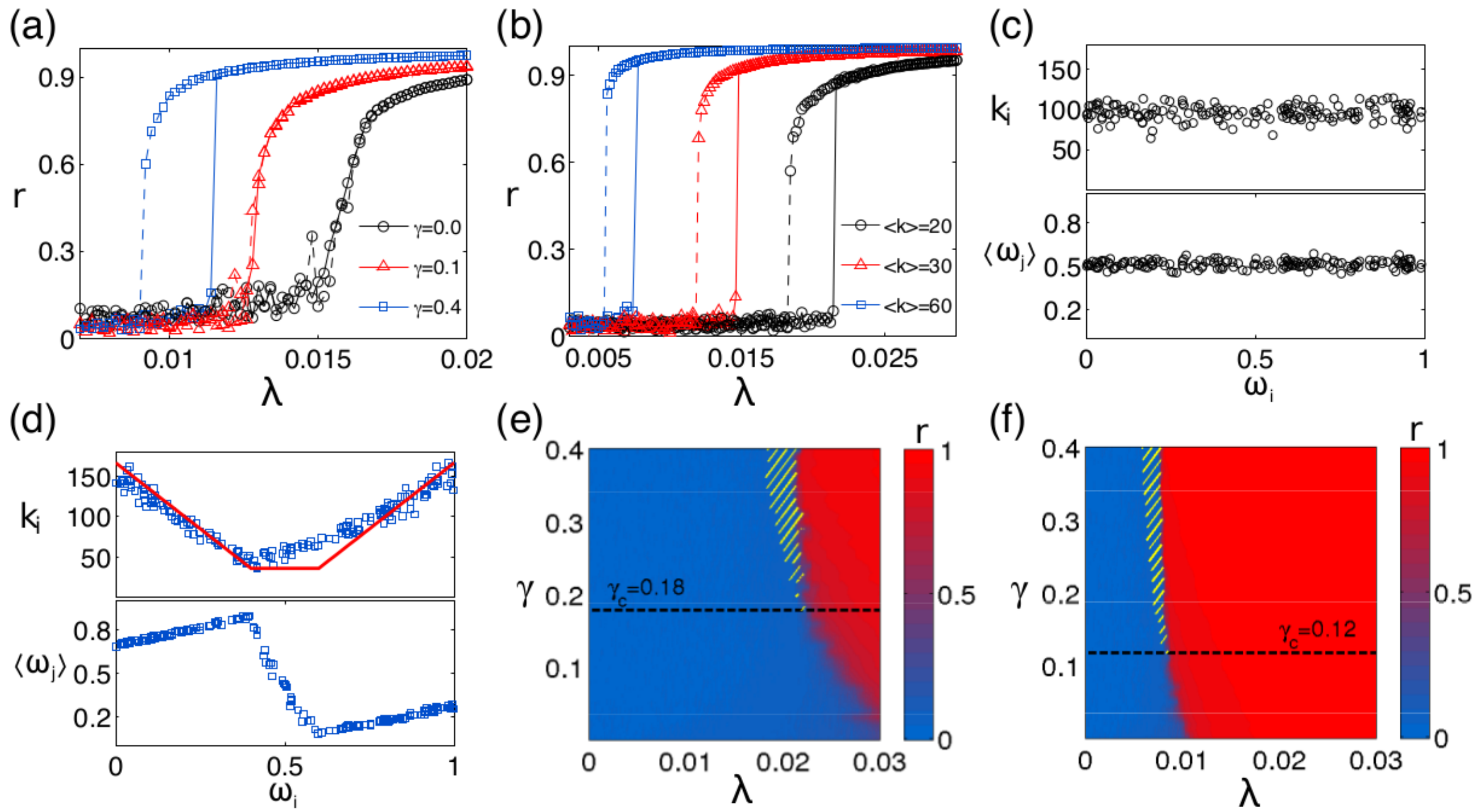}
 \caption{(Color online). (a) Phase synchronization level $r$  vs the
   coupling strength $\lambda$, for different values of the frequency
   gap $\gamma$ (see. Eq.~(\ref{eq:gap})) at $\langle k\rangle
   =40$. (b) Same as in (a), but for different values of the average degree  $\langle k\rangle$ at $\gamma=0.4$. In both panels, the legends report the color and symbol codes for the different plotted curves.
In (c) and (d), the degree $k_i$ that each node achieves after the network growth is completed (upper plots) and the average of the  natural frequencies $\langle \omega_j\rangle$ of the neighboring nodes ($j\in {\cal N}(i)$, bottom plots) are reported vs. the node's natural  frequency  $\omega_i$,   for $\langle k \rangle=100$ and   frequency gaps $\gamma=0.0$ (c), and  $\gamma=0.4$ (d). The red solid line in (d)  is a sketch of the theoretical prediction  $f(\omega)$. Panels (e) and (f) show  $r$ (color coded  according to the color bar) in the parameter
  space ($\lambda,\gamma$) for (e) $\langle k \rangle=20$ and (f) $\langle k \rangle =60$. The horizontal dashed lines mark the separation between the region of the parameter space where a second-order transition occurs (below the line) and that in which the transition is instead of the first order type (above the line). The yellow striped area delimits the hysteresis region. Reprinted from Ref.~\cite{Leyva2013a}, published under CC(Creative Commons)-DA license.  \label{fig:Leyva2013a_1}}
\end{figure}

Figure~\ref{fig:Leyva2013a_1} reports the results obtained by setting $g(\omega)$ as a
uniform frequency distribution in the interval $[0,1]$. Panels (a) and
(b) show the phase synchronization index as a function of the coupling
strength. In particular, Fig.~\ref{fig:Leyva2013a_1}(a)  (resp. (b)) illustrates the case of a fixed mean degree $\langle k\rangle =40$ (of a fixed frequency gap $\gamma=0.4$), and reports the results for the forward and backward simulations at different values of $\gamma$ ($\langle k\rangle$). A first important result is the evident first-order character acquired, in all cases, by the transitions for sufficiently high values of $\gamma$.

A second relevant result of Ref.~\cite{Leyva2013a} is the spontaneous emergence of degree-frequency correlation features associated to the passage from a second- to a first-order like phase transition. While such a correlation was imposed {\it ad hoc} in Refs. \cite{Gomez-Gardenes2011,Leyva2012}, here the condition (\ref{eq:gap}) creates for each oscillator $i$ a frequency barrier around $\omega_i$, where links are forbidden. The final degree $k_i$ will be then proportional to the total probability for that oscillator to receive connections from other oscillators in the network, and therefore
to $1-\int_{\omega_i-\gamma}^{\omega_i+\gamma} g(\omega') d \omega'$.
This is shown in (c) and (d) of Fig.~\ref{fig:Leyva2013a_1}, where the
degree $k_i$ that each node achieves after the network construction is
completed is reported as a function of its natural frequency
$\omega_i$, for $\langle k \rangle=100$. Precisely, the upper plot of Fig.~\ref{fig:Leyva2013a_1}(c) refers to the case $\gamma=0$ in which no degree-frequency correlation is present. In the upper plot of Fig.~\ref{fig:Leyva2013a_1}(d), instead, it is reported the case $\gamma=0.4$ (a value for which a first-order phase transition occurs) and the (conveniently normalized) function $f(\omega)=1-\int_{\omega-\gamma}^{\omega+\gamma} g(\omega') d \omega'$, with $g(\omega)=1$ for $\omega \in [0,1]$, and $g(\omega)=0$ elsewhere, which gives evidence of the emergence of a very pronounced V-shape relationship between the frequency and the degree of the network's
nodes. The latter fact marks a noticeable difference between the approach of Ref.~\cite{Leyva2013a} and those described in the previous Section. Here, indeed, a pattern of frequency-degree correlation emerges spontaneously, and is not artificially imposed. Moreover, the emerging pattern {\it i)} is clearly {non-linear}; {\it ii)}  it depends explicitly on the frequency distribution $g(\omega)$; {\it iii)}  it is in general V-shaped.

\begin{figure}
\centering
\centering \includegraphics[width=0.7\linewidth]{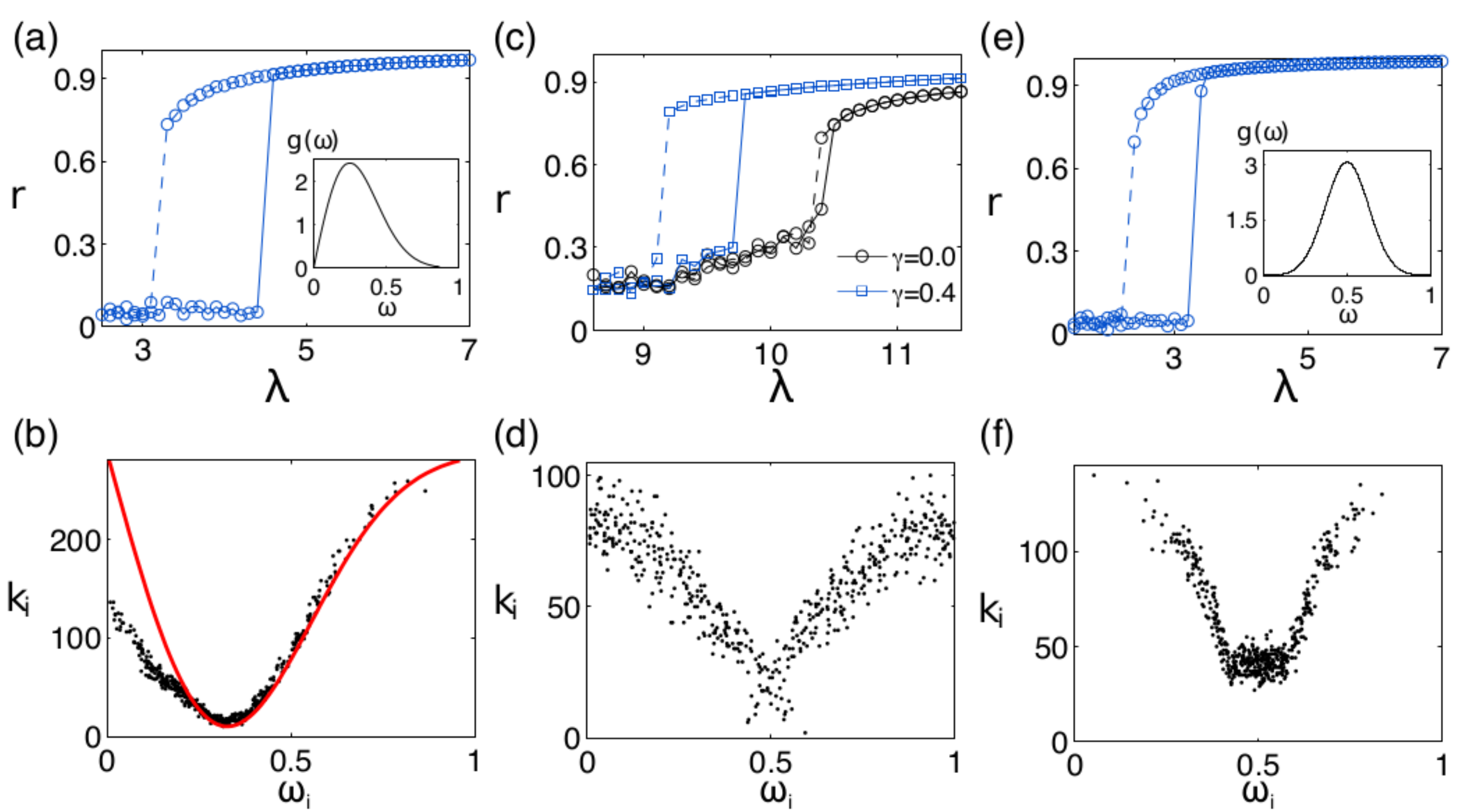}
\caption{(Color online). (Top row) $r$ vs. $\lambda$. Synchronization schemes for different frequency distributions or network construction rules. (Bottom row) The
corresponding distribution of the final node degree $k_i$  vs. the
corresponding oscillator's natural  frequency. ({a})-({b}) Rayleigh
distribution for $\gamma=0.3$. In (b), the red solid line depicts the theoretical prediction $f(\omega)$, obtained with the same method of the red solid line in panel (d) of Fig.~\ref{fig:Leyva2013a_1}; ({c})-({d}) uniform frequency distribution, but network constructed accordingly to a local mean field condition (see text) for $\gamma=0$, and $\gamma=0.4$ (see legend for color code). In panel (d)  $\gamma=0.4$; ({e})-({f}) Gaussian distribution with $Z=0.7$. The insets in panels {A} and {E} report the  corresponding distributions
$g(\omega)$. In all cases, $\langle k\rangle=60$. Adapted from Ref.~\cite{Leyva2013a}.
\label{fig:Leyva2013a_2}}
\end{figure}

Finally, Fig.~\ref{fig:Leyva2013a_1}(e) and (f) report $r$ in the
$\lambda-\gamma$ space, for $\langle k \rangle=20$ and $\langle k
\rangle =60$, and show that the rise of a first-order like phase
transition is, indeed, a generic feature in the parameter space. The
horizontal dashed lines in panels (e) and (f) mark the values of $\gamma_{c}$, separating the two regions where a second-order transition (below the line) and a first-order transition (above the line) occurs.  The fulfillment of Eq.~(\ref{eq:gap}) leads to an explosive transition for a very wide class of distributions of the oscillators' natural frequencies, as shown in Fig.~\ref{fig:Leyva2013a_2}(a)-(b) for a Rayleigh distribution.

Furthermore, Ref.~\cite{Leyva2013a}  discusses also several ways of even softening the condition of Eq.~(\ref{eq:gap}). For instance,  a frequency gap in the network growth can be introduced as $| {\omega_i}-\langle \omega_j\rangle | > \gamma_{c}$, where $\langle . \rangle$ indicates the average value over the ensemble ${\cal N}(i)$. The results of this (local mean field) gap in a homogeneous frequency distribution are shown in Fig.~\ref{fig:Leyva2013a_2}(c)-(d).
It is worth noticing that a strict application of the gap condition for non uniform frequency distributions implies that oscillators at different frequencies would in general have a different number of available neighbors in the network. That's why a natural extension is to consider a frequency-dependent gap $\gamma (\omega)$ defined by $\int_{\omega-\gamma}^{\omega+\gamma} g(\omega') d\omega' =Z$. The gap condition for the construction of the network is now
to fix the value of $Z$, and accept the pairing of nodes
when $|{\omega_i}-{\omega_j} | > \frac{1}{2}[\gamma(\omega_i)+\gamma(\omega_j)]$.
The generic case is again that of an explosive transition, with pronounced
frequency-degree correlation features, as long as $g(\omega)$ is
symmetrical. Panels (e) and (f) of Fig.~\ref{fig:Leyva2013a_2} report the case of a Gaussian distribution limited to the frequency range $[0,1]$, centered at $\omega=0.5$, and given by $g(\omega)=\frac{1}{\sigma\sqrt{2\pi}} e^{-\frac{(\omega-0.5)^2}{2\sigma^2}}$, with $\sigma=0.13$.

\begin{figure}
\centering
\centering \includegraphics[width=0.4\linewidth]{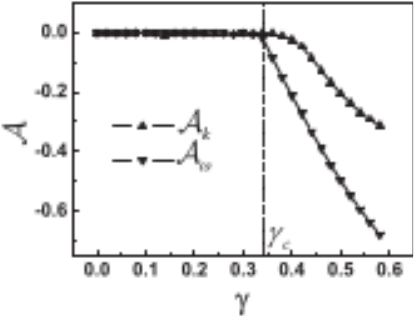}
\caption{ The degree- and frequency-assortativity coefficients $\cal
  A$ and ${\cal A}_\omega$  vs. the continuous frequency gap $\gamma$. Adapted with permission from Ref.~\cite{Zhu2013a}. Courtesy of D. Shi. \label{fig:Zhu2013a}}
\end{figure}

Reference~\cite{Zhu2013a} also focused in the idea that ES phenomenon is
rooted in preventing oscillators from behaving as the cores of
clustering processes. By the use of a constructive gap scheme similar
to that of Ref.\cite{Leyva2013a}, it is found that  frequency
disassortativity is accompanied  by  a strong increase of degree
disassortativity. The frequency-equivalent of the Pearson coefficient
${\cal A}$ in Eq.~(\ref{eq:assortativity_coef}) is
\be
{\cal A}_\omega=\frac{L^{-1}\sum_i p_i q_i  -[L^{-1}\sum_i\frac{1}{2}
  (p_i+q_i)]^2}{L^{-1}\sum_i \frac{1}{2}(p_i^2+q_i^2) -[L^{-1}\sum_i\frac{1}{2}
  (p_i+q_i)]^2},
\label{eq:freq_assortativity_coef}
\ee
where $q_i$, $p_i$ are now the frequencies of the nodes at the ends of
link $i$. From Fig.~\ref{fig:Zhu2013a}, it can be easily seen that, as
the gap parameter $\gamma$ increases, both degree- and
frequency-assortativity coefficients ${\cal A}$ and ${\cal A}_\omega$ stay close to zero
up to the critical value $\gamma_c$, and then become increasingly negative.

The same idea of a gap condition is used in Ref.~\cite{Zhang2014a} in a completely different context, i.e.  an ensemble of coupled maps:
\be
X^{n+1}_i=f(X^{n}_i)+\frac{d}{k_i} \sum_{j=1}^{N} a_{ij}\vert f(X^{n}_i)-f(X^{n}_j) \vert
\ee
where $f(X^{n}_i)=\mu_i X^{n}_i(1-X^{n}_i)$ is the logistic map, and the  parameter $\mu_i \in[3,4]$ is the dimension-less grow factor of element $i$. In   Ref.~\cite{Zhang2014a}, the connectivity on the ensemble is generated by imposing the gap restriction $\vert \mu_i - \mu_j \vert < \gamma$, and ES and hysteresis are observed as a consequence. This latter fact gives a clue on the vast generality of the gap approach, which provides, indeed, a rather good comprehension of the inner mechanisms of ES, and can be extended to a broad variety of systems.

\subsubsection{Explosive synchronization in frequency-weighted networks of oscillators}\label{freqwei}

Motivated by the results reviewed in the previous Subsection, Ref.~\cite{Leyva2013b} further extended the study to the case of a network with given frequency distribution and architecture, for which the only action that an external operator can perform is a weighting procedure on the already existing links.
Namely, Ref.~\cite{Leyva2013b} combined the information on the frequency mismatch of the two end oscillators of a link with that of
the link betweenness (in the more general case), and demonstrated that such a weighting mechanism has the effect of inducing (or enhancing) ES  for both homogeneous and heterogeneous graph topologies, as well as for any (symmetric or asymmetric) frequency distribution.

We follow the path of reasoning of Ref.~\cite{Leyva2013b}, and start by modifying Eq.~(\ref{eq:kuramoto}) as
\be
  \dot{\theta_i}=\omega_i + \frac{\lambda}{\left< k \right>}\sum_{i=1}^N \Omega_{ij}^{\alpha} \sin(\theta_j-\theta_i),
  \label{eq:weight1}
\ee
where
\be
\Omega_{ij}^{\alpha}=a_{ij}|\omega_i-\omega_j|^{\alpha},
\ee
is the weight factor to be applied to the (existing) link between nodes $i,j$,  being $a_{ij}$ the elements of the adjacency matrix that uniquely defines the network, and
$\alpha$ a constant parameter which possibly modulates the weights.  The strength of the $i^{th}$
node (the sum of all its links weights) is then
$s_i=\sum_{j} \Omega_{ij}^{\alpha}$.

\begin{figure}
\centering \includegraphics[width=0.8\textwidth]{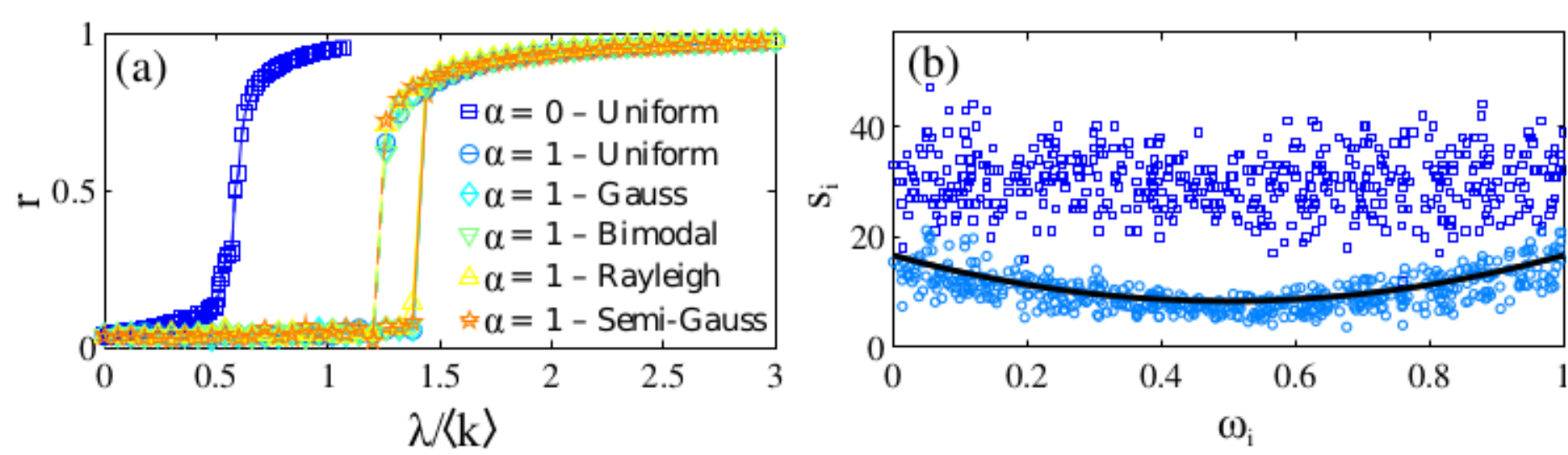}
  \caption{(Color online). (a) Synchronization transitions ($r$
    vs. $\lambda /\langle k\rangle$) for
    ER networks of size $N=500$, $\langle k\rangle $=30. The Figure reports the un-weighted case ($\alpha=0$)
    (blue squares), and linearly weighted cases ($\alpha=1$) for several frequency
distributions within the range $[0,1]$ (detailed in the legend). (b) Node strengths $s_i$ (see text for definition)  vs. natural frequencies $\omega_i$, for the un-weighted (dark blue squares) and weighted (light blue circles) networks
reported in (a). The solid line corresponds to the analytical prediction.
Adapted from  Ref.~\cite{Leyva2013b}. $\copyright$  2013 by the American Physical Society.
\label{fig:Leyva2013b_1}}
 \end{figure}

Figure~\ref{fig:Leyva2013b_1}(a) reports the results for a ER network of size $N=500$, and several
 frequency distributions $g(\omega)$ within the range $[0,1]$.
For the simplest case of a uniform frequency distribution, the un-weighted network ($\alpha=0$ in Eq.~(\ref{eq:weight1})) displays a
smooth, second-order like transition [dark blue curve in
Fig.~\ref{fig:Leyva2013b_1}(a)], whereas the effect of a linear weighting factor
($\alpha=1$)  is that of inducing ES in the system,
with an associated hysteresis in the forward
(solid line) and backward (dashed line) transitions.

More importantly, such a drastic change in the nature of the transition seems to be independent of the specific
frequency distribution $g(\omega)$, as long as they are defined in the
same frequency range $[0,1]$. The results are identical for symmetric
(homogeneous, Gaussian, bimodal derived from a Gaussian) and for asymmetric  (Rayleigh, a Gaussian centered at $0$ but with just having the positive half) frequency distributions, indicating the existence of a sort of {\it universal} behavior for homogeneous graph's topologies.

Figure \ref{fig:Leyva2013b_1}(b) shows the existence of a parabolic
relationship between the strengths and the natural
frequencies of the oscillators (associated with the passage from a
smooth to an explosive transition). This relationship has been
obtained analytically (see Eq.(\ref{fitstrength})), and {\it perfectly} fits the
numerical results. The emergent
correlation features shape a bipartite-like network, where low and high
frequency oscillators are the ones with maximal overall strength.

\begin{figure}
\centering \includegraphics[width=0.8\textwidth]{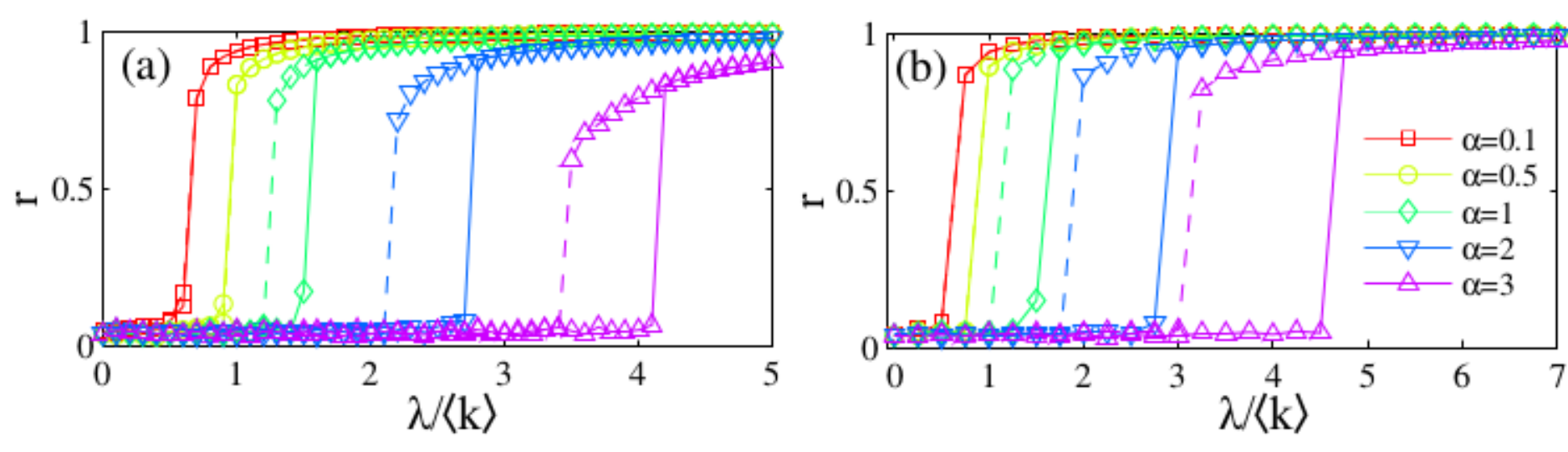}
    \caption{(Color online). Synchronization transitions for ER networks ($N=500$), with uniformly distributed frequencies in the
      $[0,1]$ range and nonlinear weighting functions
      $\Omega_{ij}^{\alpha}$. Reported curves refer to several $\alpha$
      values, from sub-linear to super-linear weighting (see legend in
      panel b). (a) ER networks, $\langle k\rangle =30$, (b) regular
      random networks, $k=30$. In all cases, forward and backward
      simulations correspond respectively to solid and dashed lines. Adapted from  Ref.~\cite{Leyva2013b}. $\copyright$  2013 by the American Physical Society.
\label{fig:Leyva2013b_2}}
 \end{figure}

Furthermore, the weighting procedure inducing ES appears to be really general, because it applies for a large family of detuning dependent functions. As an example, Fig.~\ref{fig:Leyva2013b_2} describes the case of nonlinear weighting procedures [$\alpha\neq 1$ in Eq.~(\ref{eq:weight1})]. There, $N=500$ and
$\langle k\rangle =30$ are set, and ER graphs (Fig.~\ref{fig:Leyva2013b_2}(a)), together with random networks (where each node has exactly the same number  of connections, $k_i=\langle
k\rangle =30$) are considered (Fig.~\ref{fig:Leyva2013b_2}(b)) (the latter case is
obtained by a simple {\it configuration model}, imposing a $\delta$-Dirac degree distribution).

%The results are shown in Fig.~\ref{fig:Leyva2013b_2}.
A generic non-linear
function of the frequency mismatch is able to induce ES in all topologies, and the effect of a super-linear ($\alpha>1$) weighting (a sub-linear ($\alpha<1$) weighting) is that of enhancing (reducing) the width of the hysteretic region.

\begin{figure}
  \centering
{\includegraphics[width=0.4\textwidth]{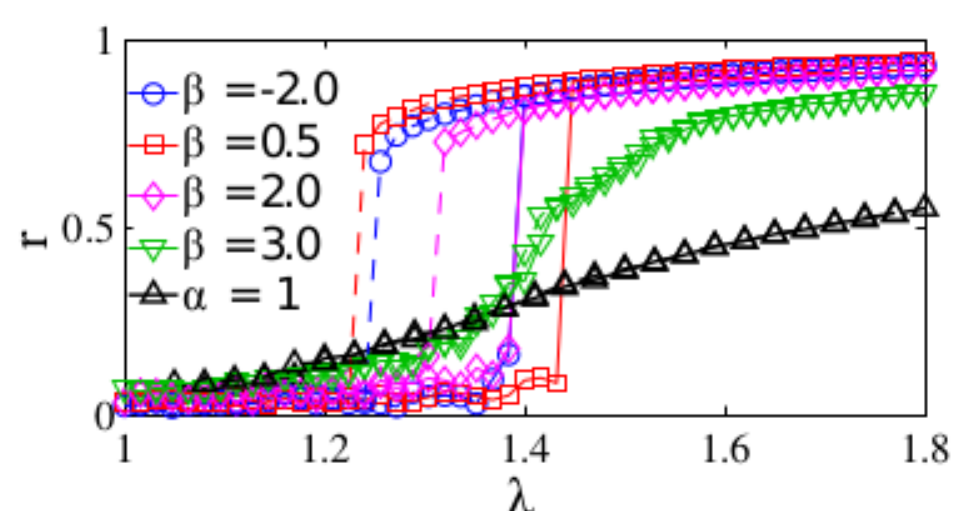}}
\caption{(Color online). Synchronization transition for SF networks using
  different values of the $\beta$ parameter [in
  Eq.~(\ref{eq:SF_weight})]. A clear passage from a second-order to a
  first-order like transition is seen. Adapted from  Ref.~\cite{Leyva2013b}. $\copyright$  2013 by the American Physical Society. \label{fig:betweeness}}
\end{figure}

When networks are heterogeneous, Ref.~\cite{Leyva2013b} suggests to use, as a new weighting function,
\begin{equation}
 \widetilde\Omega_{ij}=a_{ij}|\omega_i-\omega_j|\displaystyle\frac{\ell_{ij}^{\beta}}{\sum_{j\in
   {\cal{N}}_i}\ell_{ij}^{\beta}},
  \label{eq:SF_weight}
\end{equation}
\noindent
with $\beta$ being an extra tuning parameter, and $\ell_{ij}$ being the {\it edge betweenness} associated to the link $a_{ij}$, defined as the total number of shortest paths (between pairs of nodes in the network) that makes use of that edge. The results are reported in Fig.~\ref{fig:betweeness}. While the case $\beta=0$
corresponds to a smooth transition, moderate (positive or negative) values of $\beta$ establish abrupt
transitions to synchronization.

In order to rigorously predict the onset (and nature) of the explosive transition, Ref.~\cite{Leyva2013b} analytically examines the thermodynamic limit in which $N$ oscillators form a fully connected graph: $
\dot{\theta}_i = \omega_i + \frac{\lambda}{N} \sum_{j=1}^N |\omega_i-\omega_j| \sin(\theta_j - \theta_i)$.
By assuming the following definitions,
\ba
\frac{1}{N} \sum_{j=1}^N \Omega_{ij} \sin \theta_j &:=& A_i \sin \phi_i, \\
\frac{1}{N} \sum_{j=1}^N \Omega_{ij} \cos \theta_j &:=& A_i \cos \phi_i,
\ea
the evolution equations can be expressed in terms of trigonometric functions  as $\dot{\theta}_i = \omega_i + \lambda A_i \sin (\phi_i - \theta_i)$.  In the thermodynamic limit, the solution (in the co-rotating frame) is
\be
\omega = \lambda A_\omega \sin ( \theta_\omega - \phi_\omega ).
\label{eq:campomedio}
\ee
The definition of $A_\omega$ and $\phi_\omega$ implies that
\ba
F(\omega) := A_\omega \sin \phi_\omega = \int g(x) |\omega-x| \sin \theta (x) \, dx, \\
G(\omega) := A_\omega \cos \phi_\omega = \int g(x) |\omega-x| \cos \theta (x) \, dx, \nonumber
\ea
When all oscillators are close to synchronization, one can assume that $\cos \theta(x) \approx r$, and therefore
$G(\omega)\simeq r s(\omega)$, where $s(\omega)$ is the strength of a node with frequency $\omega$. As a consequence,
Eq.~(\ref{eq:campomedio}) takes the form
\be
\frac{2}{r \lambda} g(\omega) \omega = F''(\omega) s(\omega) - F(\omega) s''(\omega), \label{eq_ode}
\ee
which is a second order ordinary differential equation, whose integration yields $F(\omega)$. For instance, given a uniform distribution $g(\omega)$ in the interval $[-a/2,+a/2]$, the resulting strength is the second order polynomial
\begin{equation}
s(\omega) = a \left[ \left( \frac{\omega}{a} \right)^2 + \frac{1}{4}
\right],
\label{fitstrength}
\end{equation}
which fits the numerical results of Fig.~\ref{fig:Leyva2013b_1}(b). The analysis allows determining the dependence of $r$  on $\lambda$ as
$r = \int g(x) \cos \theta (x) \, dx = \int g(x) \sqrt{1 - \sin^2 \theta (x)} \, dx$,
where $
H(z) := \frac{4}{4 + \pi} \left[ \frac{z}{1 + z^2} + \arctan (z) \right]$.
In Ref.~\cite{Leyva2013b} it is shown that $r$ is indeed a multi-valued function (for values of $\lambda$ above a given threshold).   Similar results are also reported in Ref.~\cite{Zhu2013b}, which used a slightly modified weighting function
\be
\Omega_{ij}=\frac{a_{ij} N \langle k \rangle|\omega_i-\omega_j|^{\beta}}{\sum^{N}_{p=1} \sum^{N}_{q=1} a_{pq} |\omega_p-\omega_q|^{\beta}}.
\ee

In Ref.~\cite{Zhang2013}, Zhang et al. proposed a simpler alternative weighting method:
\be
  \dot{\theta_i}=\omega_i + \frac{\lambda|\omega_i|}{\left< k_i \right>}\sum_{i=1}^N a_{ij} \sin(\theta_j-\theta_i).
  \label{eq:weight2}
\ee
For symmetric frequency distributions $g(\omega)$ (such as a random, a
Lorentzian, or a Gaussian distribution), the model of Eq.~(\ref{eq:weight2})
induces ES.

\begin{figure}
\centering \includegraphics[width=0.5\textwidth]{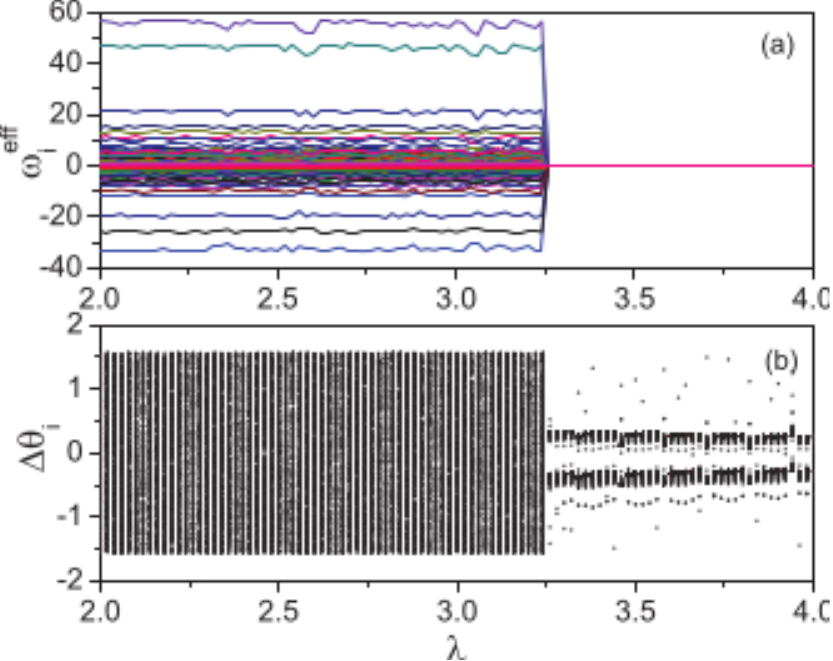}
    \caption{(Color online). The evolutionary dynamics along the forward continuation in the network model of Eq.~(\ref{eq:weight2}). (a) $\omega_{eff}$ vs. $\lambda$, (b) $\Delta \theta_i=\theta_i - \Psi$  vs. $\lambda$.
    Reprinted with permission from  Ref.~\cite{Zhang2013}. $\copyright$  2013 by the American Physical Society.
        \label{fig:Zhang2013-1}
    }
 \end{figure}

The model is, however, less effective in generating ES for asymmetric frequency distributions. In the mean field approximation,  Eq.~(\ref{eq:weight2}) yields $r \sin(\Psi - \theta_i)= \frac{1}{N} \sum_j \sin(\theta_j − \theta_i)$. Therefore, one can write
\be
\dot{\theta}_i = \omega_i + \lambda|\omega_i|r \sin(\Psi -\theta_i).
\label{eq:Zhang13_frame}
\ee
Then, setting as reference frame the one rotating with the average phase of
the system, we have $\Psi(t) = \Psi(0) + \langle\omega \rangle t$, where $\langle\omega\rangle$ is the average
frequency of the oscillators. For a symmetric $g(\omega)$, we have $\langle\omega\rangle = 0$. Letting $\Delta \theta_i = \theta_i − \Phi$, Eq.~(\ref{eq:Zhang13_frame})
becomes
\be
\Delta \dot{\theta_i} = \omega_i - \lambda|\omega_i|r \sin(\Delta\theta_i).
\ee
In the phase locking state, $\Delta \dot{\theta_i}=0$, and therefore one has
\be
\Delta \theta_i=  \left\lbrace
\begin{array}{ll}
\Delta \theta_{+}=\arcsin(\frac{1}{\lambda r}) & \omega_i>0 ,\\
\Delta \theta_{-}=\arcsin(-\frac{1}{\lambda r}) & \omega_i<0 ,
\label{eq:Delta_theta}
\end{array}
\right.
\ee
which implies that two equal clusters form: one including the clockwise, and the other the counterclockwise  oscillators. Both values of $\Delta \theta_{+}$ and $\Delta \theta_{-}$ gradually approach zero with increasing $\lambda$. This fully confirms the observation in Fig.~\ref{fig:Zhang2013-1}(b). Then, the order parameter takes the form $r=\frac{1}{2}(e^{i \Delta \theta_{+}} +  e^{i \Delta \theta_{-}})$.  Substituting Eq.~(\ref{eq:Delta_theta}) in such a latter expression, one obtains $r^2=\frac{\lambda + \sqrt{\lambda^2 -4}}{2\lambda}$,
which is independent on $g(\omega)$, provided it is symmetric. This result implies that ES exists only for $\lambda\geq \lambda_{c}^{b}=2$, where $r$  has a strong discontinuity from $r > \sqrt{5}\approx$ 0.707  to $r\approx$ 0.

In the case of asymmetric frequency distributions, a similar
calculation gives instead 
$$r(x)=\int \sqrt{1-\left(\frac{\omega-\langle\omega \rangle}{x \omega}\right)^2}g(\omega) H\left(1-\left\vert \frac{\omega-\langle\omega\rangle}{x\omega} \right\vert \right) d\omega,$$
where $x = r\lambda$, and $H(x)$ is the Heaviside step function. In this case, there is a strong dependence on $g(\omega)$, and Ref.~\cite{Zhang2013} shows that continuous or discontinuous transitions emerge for different frequency distributions.

%%%%%%%%%%%%%%%%%%%%%%%%%%%%%%%%%%%%%%%%%%%%%%%%%%%%%%%%%%%%%%%%%%%%%%%%%

The analytic study of  Eq.~(\ref{eq:weight2}) was tackled by Hu et al. in Ref.~\cite{Hu2014}, where the forward critical coupling
was calculated with the use of the method introduced in Ref.~\cite{Strogatz2000}. In the limit $N \rightarrow \infty$
a density function $\rho(\theta, \omega, t)$ can be defined, which
denotes the fraction of oscillators with frequency $\omega$ whose phases
have values between $\theta$ and $\theta + d\theta$ at time $t$. $\rho(\theta, \omega, t)$ satisfies
the normalization condition $\int_{0}^{2\pi} \rho(\theta, \omega, t) d\theta=1$
for all $\theta$ and $t$. The evolution of $\rho(\theta, \omega, t)$ is governed by the
continuity equation:
\be
\frac{\delta \rho}{\delta t}+\frac{\delta (\rho v)}{\delta \theta}=0.
\label{eq:continuity}
\ee
Using the expression obtained in Eq.~(\ref{eq:Zhang13_frame}), the velocity in the thermodynamical limit can be written as
\be
v=\omega + \lambda \vert \omega \vert r \sin(\phi-\theta).
\ee

When the coupling strength is relatively small, all oscillators are rotating in the unit circle  according (almost exclusively) to their natural frequencies. This corresponds to an incoherent state $\rho(\theta, \omega, t)= 1/(2\pi)$. Therefore, the point at which such a
 state loses its stability can be extracted, providing a prediction for the forward critical point. Specifically, let one consider a small perturbation from the incoherent state, i.e. $ \rho(\theta, \omega, t)= \frac{1}{2\pi} + \epsilon \eta(\theta, \omega, t)$.
Then, the linearized form of the continuity equation (\ref{eq:continuity}) is
\be
\frac{\delta \eta}{\delta t}=- \omega \frac{\delta \eta}{\delta \theta}+ \frac{\lambda r' \vert \omega \vert \sin(\phi-\theta)}{2\pi},
\label{eq:lin_continuity}
\ee
where
\be
r e^{i\phi}=\epsilon r' e^{i\phi} = \epsilon \int_{0}^{2\pi} \int_{-\infty}^{\infty} e^{i\theta} \eta(\theta, \omega, t) d\omega d\theta.
\ee
Now $\eta(\theta, \omega, t)$ can be expanded into a Fourier series as $\eta(\theta, \omega, t)=b(\omega) e^{\mu t} e^{i\theta} + b^{*}(\omega) e^{\mu t} e^{-i\theta} + O(\theta, \omega, t)$,
 where $O(\theta, \omega, t)$ stand for higher harmonics. The equation becomes
 \be
 \mu b(\omega)= -i \omega b(\omega)+ \frac{\lambda \vert \omega \vert}{2}  \int_{-\infty}^{\infty} b(v) g(v) dv,
 \label{eq:self-consistent}
 \ee
and can be solved in a self-consistent way. In particular, when $g(\omega)$ is an even function, Eq.~(\ref{eq:self-consistent}) becomes
\be
1=\frac{\lambda}{2}  \int_{-\infty}^{\infty} \frac{\mu\vert \omega \vert}{\mu^2 + \omega^2} g(\omega) d \omega,
\ee
\\
i.e., when $Re(\mu)$ changes from negative to positive, the incoherent state loses its stability. For any specific $g(\omega)$ one can then use this condition to determine the critical coupling strength $\lambda_f$ for the forward phase transition. Several examples of this calculation  are reported in Table~\ref{table:Hu2014} .

\begin{table}
\renewcommand{\arraystretch}{1.3}
\caption{Critical equations and critical points $\lambda^{c}_{f}$
  (calculated in Ref.~\cite{Hu2014}) for a triangular, a Lorentzian, a
  Gaussian and a bimodal Lorentzian frequency distribution
  $g(\omega)$. \label{table:Hu2014}}%\vspace*{0.25cm}
\begin{center}
\footnotesize 
\begin{tabular*}{\textwidth}{p{18em}@{\extracolsep{\fill}}p{25em}@{\extracolsep{\fill}}p{4em}}
%\begin{tabular}{ccc}
\hline
 Frequency distribution $g(\omega)$& Critical equation & $\lambda^{c}_{f}$ \\
 \hline
 $g(\omega)=(D-\vert \omega \vert)/D^2, \vert \omega \vert<D$, 0 otherwise & $1=\lambda \left[ \frac{z}{2} \ln\left(1+\frac{1}{z^2} \right) - z + z^2 \arctan \left( \frac{1}{z^2} \right) \right]$, $z=\mu/D$ & $\sim$ 2.65 \\
% \hline
 $g(\omega)=\frac{\Delta}{\pi (\omega^2 + \Delta)}$ & $\frac{\pi}{\lambda}=\frac{z \ln z}{z^2-1}$,$z=\mu/\Delta$ & 4 \\
% \hline
 $g(\omega)=\frac{1}{\pi \sigma} \exp \left( -\frac{\omega^2}{\pi \sigma^2} \right)$& $1=\frac{\lambda z}{2\sqrt{2\pi}} \exp(z^2/2)[-E_i(-z^2/2)]$,& $\sim$ 2.68 \\
% \hline
& $z=\mu/\sigma$, $E_i(z)=\int_{-\infty}^{z} \frac{e^u}{u} du$ & \\
 $g(\omega)=\frac{1}{2\pi} \left[ \frac{\Delta}{(\omega-\omega_o)^2+\Delta^2}+ \frac{\Delta}{(\omega+\omega_o)^2+\Delta^2} \right]$ & $\frac{2\pi}{\lambda}=\ln\left(\frac{\sqrt{1+\delta^2}}{\mu} \right) \left[ \frac{(1-\mu)\mu}{(1-\mu)^2 + \delta^2} +  \frac{(1+\mu)\mu}{(1+\mu)^2 + \delta^2} \right] $, $\delta=\omega_o/\Delta$ & $4/\sqrt{1+\delta^2}$ \\
 \hline
 \end{tabular*}
\end{center}
\end{table}

%%%%%%%%%%%%%%%%%%%%%%%%%%%%%%%%%%%%%%%%%%%%%%%%%%%%%%%%%%%%%%%%%%%%%%%%%

\begin{figure}
  \centering
{\includegraphics[width=0.5\textwidth]{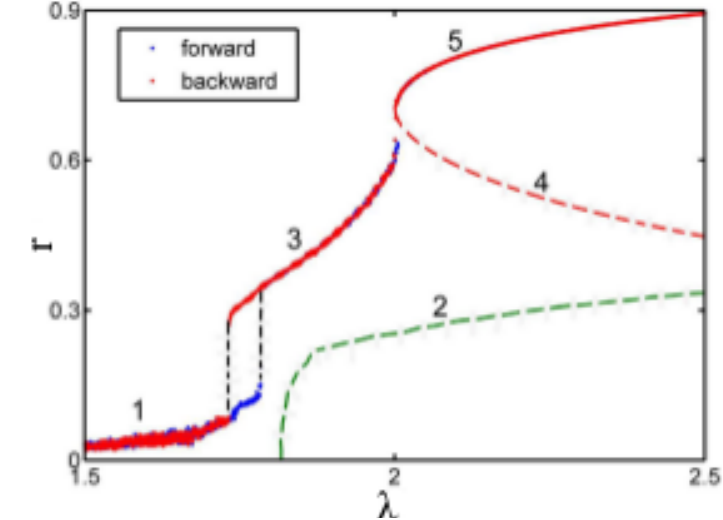}}
\caption{(Color online). Characterization of various coherent states in the phase diagram  $r$ vs. $\lambda$,
for a uniform frequency distribution $g(\omega)$ = 1/2 , $\omega \in$
(−1, 1), and for a fully connected $N$=50,000 Kuramoto ensemble. The
distinct branches correspond to: the incoherent state (1), the
(unstable) traveling wave (TW) state predicted by the mean-field theory (2), the standing wave state (3), the unstable (4) and the stable (5) two-cluster synchronous states. The blue and red lines denote the forward and  the backward transitions, respectively. Reprinted from Ref.~\cite{Xu2015b}, published under CC-NY-ND license.\label{fig:Xu2015b}}
\end{figure}

Xu et al. \cite{Xu2015b} completed the analysis initiated in Ref.~\cite{Hu2014}, and calculated all possible steady states in the model of Eq.~(\ref{eq:weight2}), which include the incoherent state and the two-cluster synchronous state already deduced in Ref.~\cite{Hu2014}, but also a traveling and a standing wave state. Reference~\cite{Xu2015b} demonstrated that the incoherent state is only neutrally stable below the synchronization threshold. The amplitude equations near the bifurcation point are derived according to the center-manifold reduction, and  non-stationary standing wave states are predicted to exist in the model. Figure~\ref{fig:Xu2015b} reports all possible solutions for a fully connected Kuramoto ensemble with uniform frequency distribution $g(\omega) = 1/2$, $\omega \in  (−1, 1)$.

%%%%%%%%%%%%%%%%%%%%%%%%%%%%%%%%%%%%%%%%%%%%%%%%%%%%%%%%%%%%%%%%%%%%%%%%%

\begin{figure}
  \centering
{\includegraphics[width=0.8\textwidth]{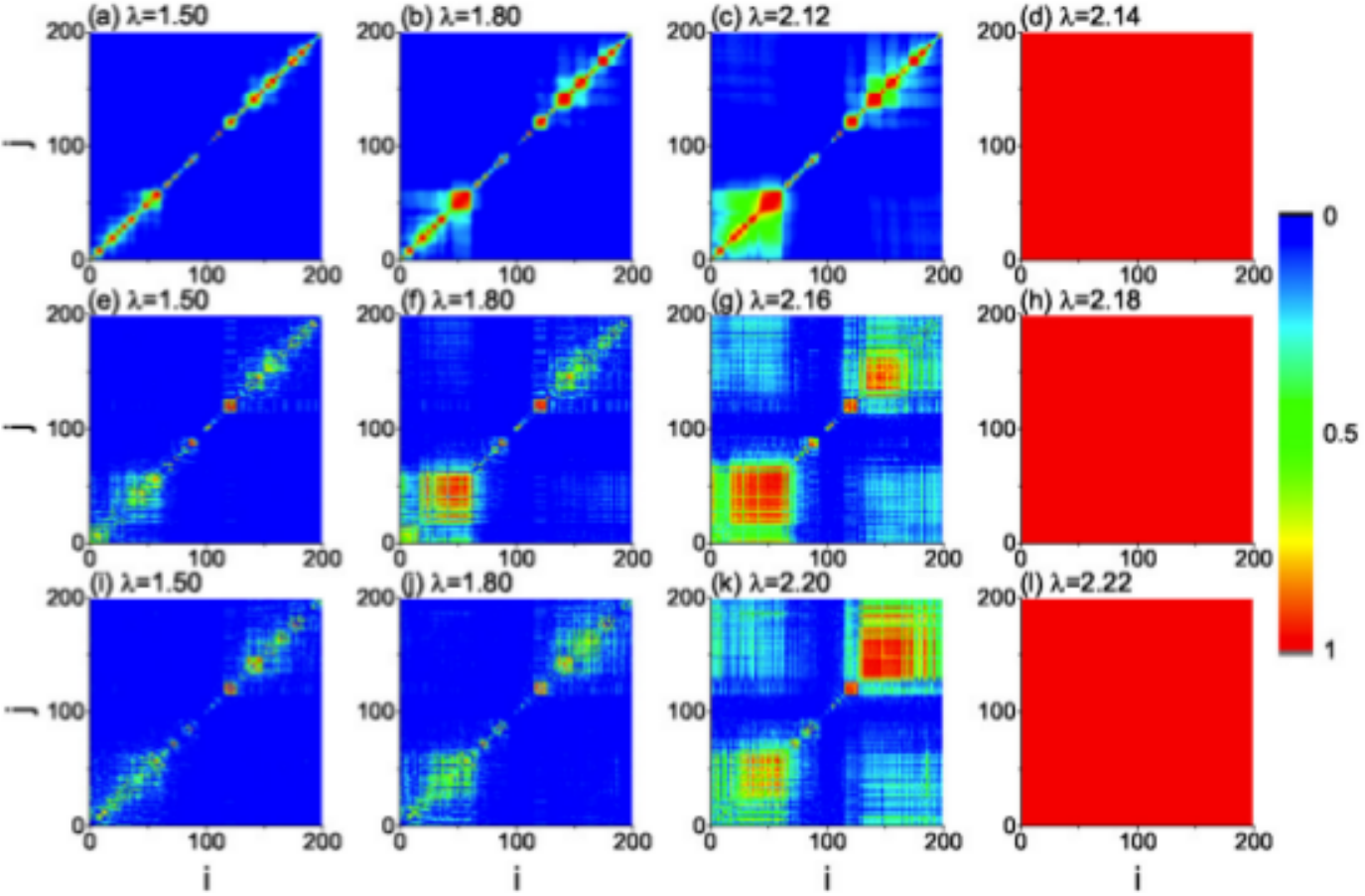}}
\caption{Plots of the matrix $r_{ij}$ (Eq.~(\ref{localpiro})) for fully
  connected (first line), ER (second line) and uncorrelated CM (third line) networks. Oscillator are labeled in the ascending order of the frequency $\omega_i$. The coupling strengths are $\lambda=1.5, 1.8, 2.12$, and $2.14$ in (a)-(d) ($\lambda_c$=2.13); $\lambda=1.5, 1.8, 2.16$, and $2.18$ in (e)–(h)
($\lambda_c=2.17$); and $\lambda=1.5, 1.8, 2.20$, and 2.22 in (i)–(l) ($\lambda_c=2.21$). Reprinted from Ref.~\cite{Zhang2014}, published under CC-NY-ND license.\label{fig:Zhang2014}}
\end{figure}

More detailed information on the microscopic mechanisms underlying ES in the model of Eq.~(\ref{eq:weight2}) can be found in Ref.~\cite{Zhang2014}, where a local order parameter, $r_{ij}$, is introduced
\be
r_{ij}=\left\vert \lim_{T\rightarrow \infty}  \frac{1}{T} \int_{t}^{t+T} e^{(\theta_i(t)-\theta_j(t))} dt  \right\vert ,
\label{localpiro}
\ee
where $T$ is a, sufficiently long, time window.  $r_{ij}$ is 1 for any two phase-locked
oscillators, zero for all pairs of fully uncorrelated oscillators, and will
take a value between 0 and 1 for any two partially correlated
oscillators. Figure \ref{fig:Zhang2014} reports the values of $r_{ij}$
for: (a)-(d) fully connected, (e)-(h) ER and (i)-(l)
uncorrelated CM networks, for four representative $\lambda$ values.  Only small synchronized clusters of oscillators exist for $\lambda<\lambda_c$, while a giant synchronized cluster shows up suddenly right after $\lambda_c$, indicating that the small synchronized clusters suddenly merge together right at $\lambda_c$.

The similarity with the Achlioptas process of EP is even clearer from looking at Fig.~\ref{fig:Zhang2014}, where it is seen that, below $\lambda_c$, links are always generated among those nodes with close frequencies. As a consequence, separated clusters  form for $\lambda$ up to  $\lambda_c$, where instead (and suddenly) all  clusters merge together to mold a giant one.
To quantify the analogy with EP, an equivalent suppressing rule is introduced. From Eq.~(\ref{eq:Zhang13_frame}), the evolution of the phase difference $\Delta \theta_{ij}=\theta_{i} - \theta_{j}$ is  given by
\be
\Delta \dot{\theta}_{ij}=\omega_{i}+|\lambda_{i}|r\sin(\Psi-\theta_i)-\omega_{j}+|\lambda_{j}|r\sin(\Psi-\theta_j).
\ee
When the two oscillators $i$ and $j$ are phase-locked, one has $\Delta \dot{\theta}_{ij}$=0 and therefore
\be
\omega_{i}-\omega_{j}=|\lambda_{i}|r\lbrace\sin(\Psi-\theta_i)-\sin(\Psi-\theta_j) \rbrace,
\ee
whose maximum value gives a necessary condition for the phase-locking between two oscillators:
\be
\frac{|\omega_{i}-\omega_{j}|}{|\omega_{i}|+|\omega_{j}|}\leq\lambda r.
\label{eq:Zhang2014_rule}
\ee
Equation~(\ref{eq:Zhang2014_rule}) can be considered as {\it a suppressive rule} for pair wise synchronization: as $r$ takes a rather small value when $\lambda<\lambda_c$, only those pairs of oscillators with smaller frequency differences can satisfy the
condition (\ref{eq:Zhang2014_rule}) and thus form synchronized clusters. When all the free oscillators have been attracted to the synchronized clusters, the further increase of $\lambda$ cannot make the synchronized clusters become larger, but makes the clusters to merge each other suddenly, this way producing the significant jump on $r$ observed in the explosive transition.

Reference~\cite{Zhang2014} shows also that the suppressive rule (\ref{eq:Zhang2014_rule}) can actually guide manipulative processes to enhance or quash ES. For instance, one can randomly pick a pair of nodes $i,j$ and then artificially re-adjust the network by exchanging their frequencies if the quantity
\be
S_{ij}= \sum_{l=1}^{\Gamma_i} \frac{|\omega_{i}-\omega_{l}|}{|\omega_{i}|+|\omega_{l}|} + \sum_{l=1}^{\Gamma_j} \frac{|\omega_{j}-\omega_{l}|}{|\omega_{j}|+|\omega_{l}|}\label{eq:sij}
\ee
 decreases (or increases), where $\Gamma_i,\Gamma_j$ stand for the neighborhoods of nodes $i,j$, resulting in a enhanced (suppressed) ES, as it occurs for the  Achlioptas algorithm in EP.

%%%%%%%%%%%%%%%%%%%%%%%%%%%%%%%%%%%%%%%%%%%%%%%%%%%%%%%%%%%%%%%%%%%%%%%%%

Reference~\cite{Zhou2015} considers the effect of using [in
Eq.~(\ref{eq:weight2})] frequency distributions $g(\omega)$ not
centered in 0, and specifically studies the case of uni-modal
frequency distributions (Lorentzian, Gaussian, triangle), displacing
their central frequency from $0$ to $\omega_0\neq 0$, which is then used
as a parameter.

\begin{figure}
  \centering {\includegraphics[width=0.5\textwidth]{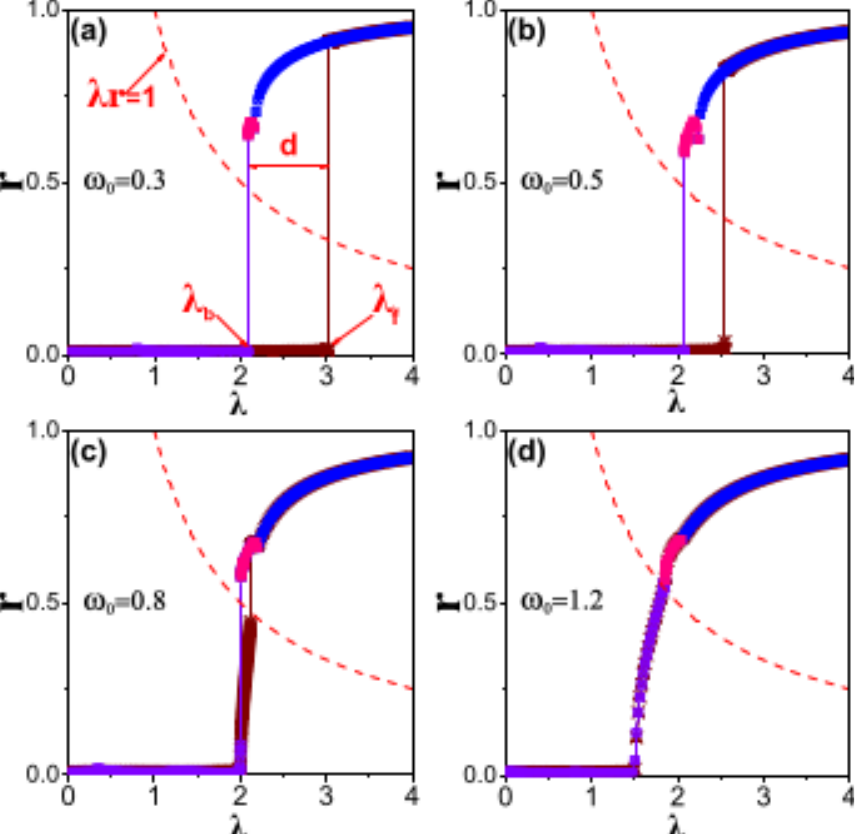}}
\caption{(Color online). $r$  vs. $\lambda$. With the increase of $\omega_0$ , the synchronization
transition converts from a first-order type into a second-order one.
From (a) to (d),  $\omega_0 = 0.3,0.5,0.8,1.2$. The dashed lines
correspond to $\lambda r= 1$. A Lorentzian frequency distribution is used, and $N = 10000$. Adapted from
Ref.~\cite{Zhou2015}. $\copyright$  2015 by the American Physical Society.\label{fig:Zhou2015-1}}
\end{figure}
Figure~\ref{fig:Zhou2015-1} plots the phase diagrams when increasing $\omega_0$, and shows that the hysteresis
area significantly shrinks, mainly due to the decrease of
the forward transition point $\lambda_{c}^{f}$, while the backward transition
point $\lambda_{c}^{b}$ remains unchanged.

\begin{figure}
  \centering
{\includegraphics[width=0.6 \textwidth]{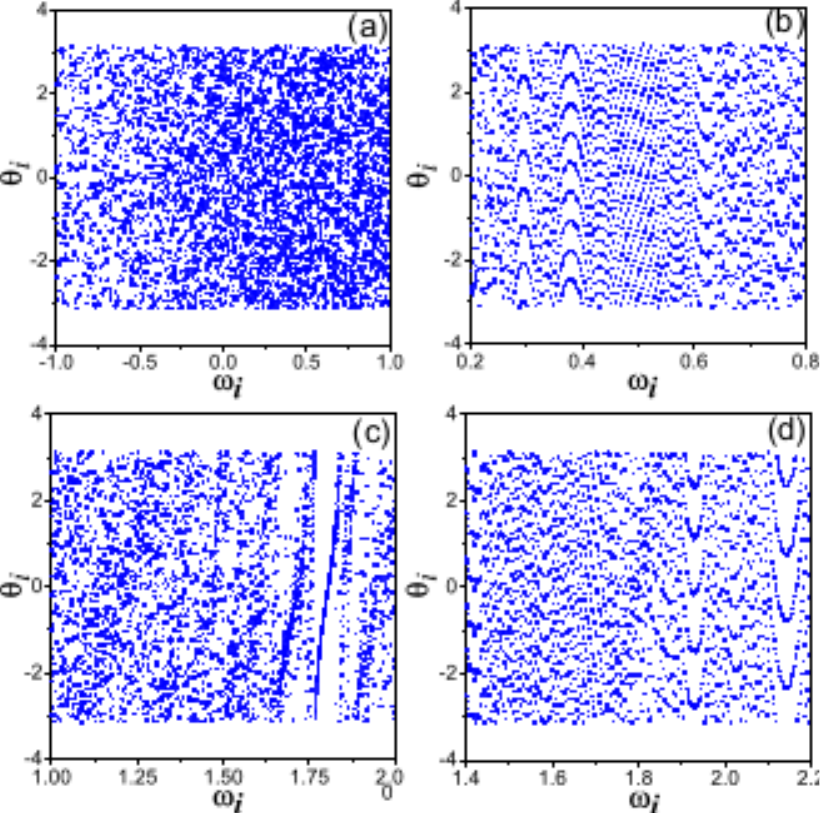}}
\caption{ Instantaneous phase distribution
  $\theta_i$ vs. natural frequency $\theta_i$ for all the nodes of a
  ER of $N=10000$ nodes, coupling strength $\lambda=1.5$, for Lorentzian
  frequency distribution centered in  $\omega_0=0.5$ ((a) and (b)  corresponding to a system able to achieve ES), and $\omega_0=1.2$
  ((c) and (d) corresponding to a system with second order
  transition).  Left (right) panels correspond to forward (backwards)
  continuations. Adapted from Ref.~\cite{Zhou2015}. $\copyright$  2015 by the American Physical Society. \label{fig:Zhou2015-2}}
\end{figure}

An interesting view on microscopic processes in both the first-order and second-order cases can be obtained from Fig.~\ref{fig:Zhou2015-2}. Here, the incoherent steady state is explored for two values of the central frequency: $\omega_0=0.5$ (Fig.~\ref{fig:Zhou2015-2}(a)-(b), which causes ES), and $\omega_0=1.2$ (Fig.~\ref{fig:Zhou2015-2}(c)-(d), for which the transition is reversible). In all the cases of Fig.~\ref{fig:Zhou2015-2} the coupling is $\lambda=1.5$, i.e. below the synchronization threshold, and therefore $r\simeq0$. However, essential microscopic differences exist  between these incoherent states. While the phases are randomly distributed just below the forward threshold in the ES case (Fig.~\ref{fig:Zhou2015-2}(a)), the incoherent state of the corresponding backward continuation (Fig.~\ref{fig:Zhou2015-2}(b)) is non-trivial, showing the remaining of small clusters with fine structures. In the second order case
with $\omega_0=1.2$, these microscopic structures appear in the incoherent state of both the forward and the backward (Fig.~\ref{fig:Zhou2015-2}(c)-(d)) continuations. Such  microscopic differences reveal that, as the frequencies distribution is not symmetric around $0$,   formation of  micro-clusters occurs, and prevents ES.

This is actually due to the  suppression rule of Eq.~(\ref{eq:Zhang2014_rule}). All the pairs of nodes $i,j$ whose frequencies $\omega_i,\omega_j,$ have different sign fulfill  $|\omega_{i}-\omega_{j}|/|(\omega_{i}|+|\omega_{j}|)=1$, and therefore, being $r\simeq$0 in the incoherent state, they do not synchronize. An asymmetrical frequency distribution
will reduce the number of possible pairs of nodes with frequencies of opposite sign, and therefore an increasing number of pairs will be able to form synchronization seeds, so that eventually ES is destroyed for large enough $\omega_0$.

 %%%%%%%%%%%%%%%%%%%%%%%%%%%%%%%%%%%%%%%%%%%%%%%%%%%%%%%%%%%%%%%%%%%%%%%%%

Following the same line of investigation of microscopic mechanisms underlying ES, Ref.~\cite{Navas2015} proposed the use of an effective topological network whose structure explicitly reflects the interplay between the topology and dynamics of the original system in an individual basis, and therefore, is able to easily  identify the nodes which actually  act as synchronization seeds. The  effective adjacency matrix is given by
\be
C_{ij}\equiv A_{ij}\left( 1-\frac{\Delta \omega_{ij}}{\Delta\omega_{max}} \right),
\label{eq:Navas2015_C}
\ee
where $\Delta\omega_{ij}=\vert\omega_i-\omega_j\vert$ is the frequency detuning, and $\Delta \omega_{max}$ the maximum possible detuning present in the system in order to guarantee $C_{ij}\geq0$.

In order to quantify the role of each node in the synchronization process, Ref.~\cite{Navas2015}  extracts the most important nodes in the network defined by ${\bf C}$, i.e. the standard eigenvector centrality measure of ${\bf C}$, obtaining the {\it effective centrality} vector $\bf\Lambda^C$, whose $i$-th component  quantifies the potential of node $i$ to behave as a seed of synchronization.

\begin{figure}
 \centering {\includegraphics[width=0.5\textwidth]{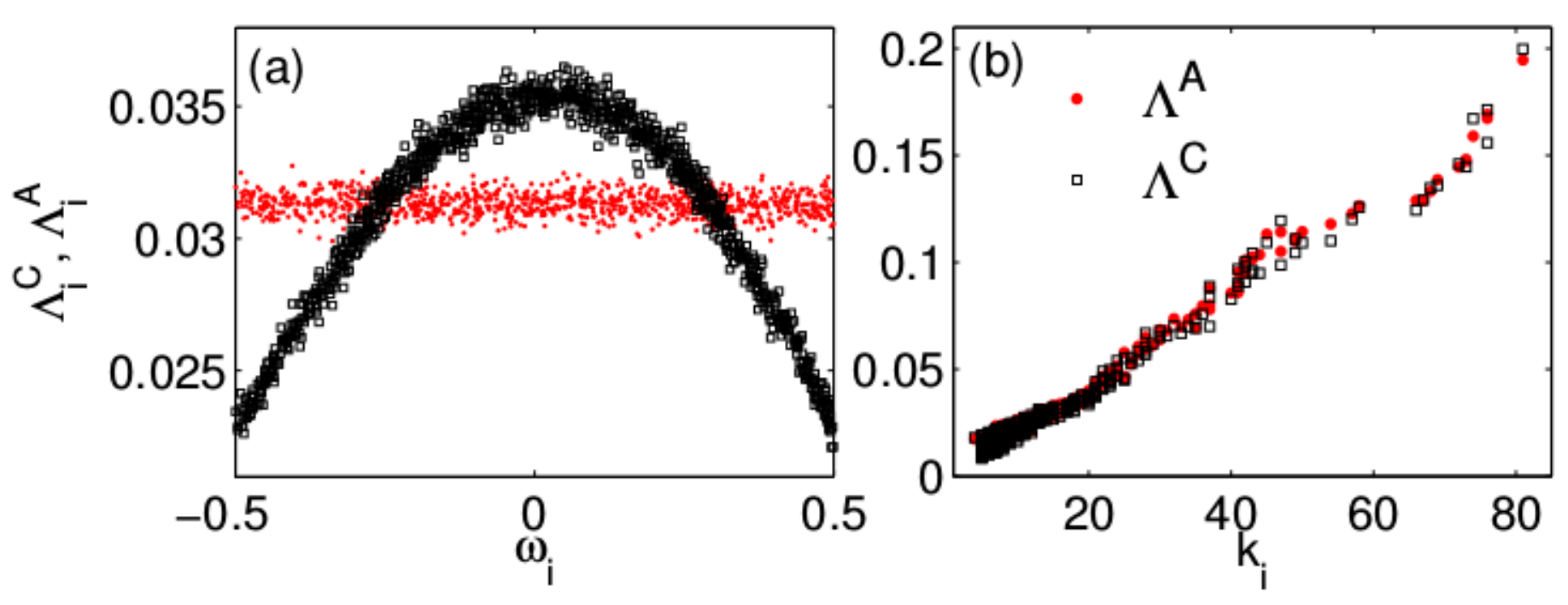}}
\caption{(Color online). Comparison between synchronization centrality
  $\Lambda^C_i$ (black squares) and topological centrality
  $\Lambda^A_i$ (red dots). (a) ER networks, $\langle k \rangle$=50,
  $\Lambda^C_i$ and $\Lambda^A_i$ are reported  vs. the nodes' natural
  frequencies $\omega_i$; (b) SF networks, $\langle k \rangle$=12,
  $\Lambda^C_i$ and $\Lambda^A_i$ are plotted  vs. the node degrees
  $k_i$. All data refer to ensemble averages over 100 different
  network realizations. Reprinted from  Ref.~\cite{Navas2015}. $\copyright$  2015 by the American Physical Society.
\label{fig:Navas2015-1}}
\end{figure}

Figure~\ref{fig:Navas2015-1} reports the comparison between $\bf\Lambda^C$ and its topological counterpart $\bf\Lambda^A$, the eigenvector centrality extracted from the original adjacency matrix ${\bf A}$. For ER networks (Fig.~\ref{fig:Navas2015-1}(a)), the distribution of the components of the vector $\bf\Lambda^C$ as a function of the corresponding node's natural frequencies  shows the existence of many seeds of synchronization with natural frequencies close to $\Omega_s=0$. This allows characterizing the connection between the micro-scale  and the macro-scale of the network in a much better way than $\bf\Lambda^A$, whose components are instead uniformly distributed. For SF networks (Fig.~\ref{fig:Navas2015-1}(b)), the synchronization seeds are the hubs, and therefore $\bf\Lambda^C$ and  $\bf\Lambda^A$ provide essentially the same information.

%%subsubsec336
\subsubsection{Noise induced explosive synchronization}
\label{ES-noise}

As discussed in Subsection~\ref{engineered_correlation}, a simple (yet artificially engineered) degree-frequency correlation gives explosive synchronization on sufficiently heterogeneous networks~\cite{Gomez-Gardenes2011,Leyva2012}.
An extension of such artificial process is proposed in Ref.~\cite{Skardal2014}, where the frequencies of oscillators are correlated with (but not fully determined by) the local network's properties, such as the degree. More precisely, the evolution of phase oscillator $\theta_i$ is taken to be
\begin{equation}
\dot{\theta_i}=\omega_i^{\prime}+ \lambda \sum_{i=1}^N A_{ij}\sin(\theta_j -\theta_i),
\label{eq-noise-1}
\end{equation}
where $\omega_i^{\prime}=k_i+ \xi_i$, where $\xi_i$ is a noise term
taken from a random homogeneous distribution in the range
[$-\varepsilon$, $\varepsilon$]. $\varepsilon=0$ returns the case of
Ref.~\cite{Gomez-Gardenes2011}.  Reference~\cite{Skardal2014} considers
realistic {\em C. elegans} neural networks, stretched exponential networks and SF networks with mild heterogeneity (i.e. $\gamma>3$ in the degree distribution $P(k) \sim k^{-\gamma}$).
Given that the synchronized solution is composed of a cluster with uniform angular velocity $\Omega$, the rotating reference frame involving the change of variable $\phi_i=\theta_i-\Omega t$ is defined as follows
\begin{equation}
\dot{\phi_i}=(\omega_i-\Omega)+ \lambda \sum_{i=1}^N A_{ij}\sin(\phi_j-\phi_i),
\label{eq-noise-2}
\end{equation}
and the local order parameter is defined as
\begin{equation}
r_ie^{i\psi_i}=\sum_{j=1}^N A_{ij}e^{i\phi_j},
\label{eq-noise-3}
\end{equation}
having a  magnitude $r_i\in [0, k_i]$. Substituting Eq.~(\ref{eq-noise-3}) into Eq.~(\ref{eq-noise-2}), one gets
\begin{equation}
\dot{\phi_i}=(\omega_i-\Omega)+ \lambda r_i \sin(\psi_i-\phi_i).
\label{eq-noise-4}
\end{equation}
From Eq.~(\ref{eq-noise-4}) it is clear that, if $|\omega_i-\Omega|\le
\lambda r_i $, $\phi_i$ can reach a steady point expressed
by $\sin(\phi_i-\psi_i)=(\omega_i-\Omega)/ \lambda r_i$ and becomes phase-locked; otherwise the oscillator $i$ drifts forever.

\begin{figure}
\begin{center}
\includegraphics[width=0.9\textwidth]{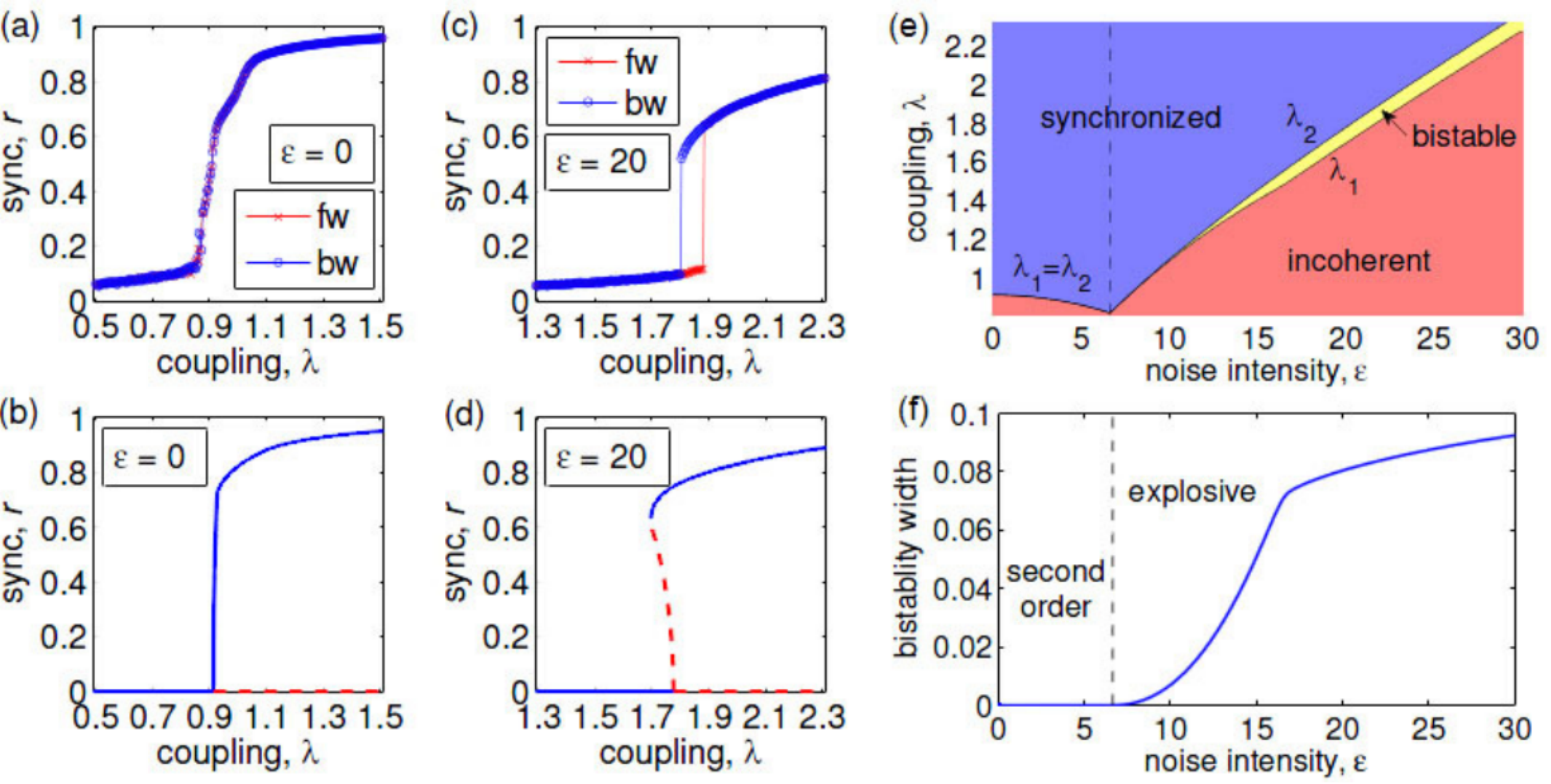}
\end{center}
\caption{(Color online). (a)-(d) $r$ vs. $\lambda$ for different values of $\varepsilon$ from simulations (top) and theoretical analysis (bottom). (e) Phase space with critical coupling strength $\lambda_{c}^{f}$ and $\lambda_{c}^{b}$ separating incoherent, synchronized and bistable regions. (f) Bistability width $\lambda_{c}^{f}-
\lambda_{c}^{b}$ as a function of noise $\lambda$. The used networks are SF networks with size $N=10^3$. Reprinted with permission from Ref.~\cite{Skardal2014}. $\copyright$  2014 by American Physical Society.  \label{ES-noise-fig}}
\end{figure}

The self-consistency condition for the steady-state order parameter $r$ is \cite{Skardal2014}
\begin{equation}
r=\langle k\rangle^{-1}\int \int_{|\omega-\Omega|\le \lambda rk}P(k,
\omega)k \sqrt{1-\Big(\frac{\omega-\Omega}{\lambda rk}\Big)^2}d\omega dk,
\label{eq-noise-5}
\end{equation}
which can be used to predict the evolution of synchronization, and where $P(k, \omega)$ is the so-called joint probability distribution of degree and frequency.

In the absence of noise,  the transition from incoherent to synchronized states seems to be  second-order like for both real and synthetic networks. The trend can be accurately predicted by the self-consistency condition (\ref{eq-noise-5}), and is shown in Fig.~\ref{ES-noise-fig}(b). For sufficient large noise, however, ES takes place with a clear bistable region on the less heterogeneous networks (see Fig.~\ref{ES-noise-fig}(d)). If $\lambda_{c}^{f}$ and $ \lambda_{c}^{b}$ denote the forward and backwards critical coupling strengths, the width of hysteresis loop is  $\lambda_{c}^{f}-\lambda_{c}^{b}$ From Figs.~\ref{ES-noise-fig}(e) and (f), one sees that the transition from incoherent to synchronized phases is second-order like until a critical value of the noise $\varepsilon_c$ (denoted by the dashed lines). Therefore, ES can be promoted by noise mechanisms.

Reference~\cite{Gupta2014} suggested to focus on the angular velocity $v_i$ of oscillator $i$. With Gaussian noise $\eta_i(t)$, the dynamics is ruled by
\begin{equation}
\begin{split}
\dot{\theta}_i&=v_i,\\
m\dot{v}_i&=-\gamma v_i+\lambda r \sin(\psi-\theta_i)+\gamma \omega_i+\sqrt{\gamma}\eta_i(t),
\end{split}
\label{eq-noise-6}
\end{equation}
where $m$ accounts for an inertia term, $\gamma$ denotes the friction
constant, and $r$ is the order parameter. In particular,
Ref.~\cite{Gupta2014} assumed 
$\langle \eta_i(t)\rangle=0,\langle
\eta_i(t)\eta_j(t^{\prime})\rangle=2T\delta_{ij}\delta(t-t^{\prime})$, where $T$ is a stochastic noise (like a temperature in units of the Boltzmann constant). Furthermore, a uni-modal $g(\omega)$ is considered, with width $\sigma$. At $\sigma=0$, the  dynamics resembles that of a Brownian mean-field  model~\cite{Chavanis2014}. In the absence of inertia, $T=0$  reproduces the Kuramoto model~\cite{Kuramoto1984,Strogatz2000,Acebron2005}, while $T \ne 0$ gives the model of Ref.~\cite{Sakaguchi1988}.

\begin{figure}
\begin{center}
\includegraphics[width=0.8\textwidth]{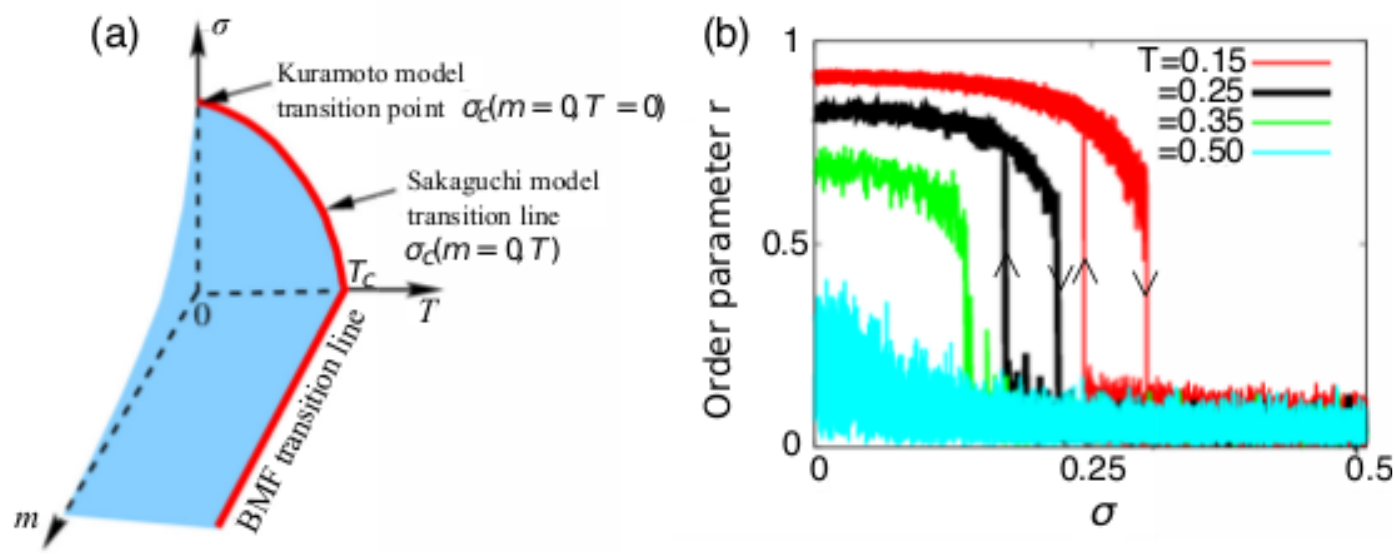}
\end{center}
\caption{(Color online). (a) Phase diagram of synchronization in the parameter space of inertia $m$, noise $T$ and width $\sigma$ of the frequency distribution. The shaded blue surface (red line) indicates the first-order (second-order) transition. (b) Synchronization level $r$  vs. $\sigma$ for different $T$ with fixed inertia $m=10$. Reprinted with permission from Ref.~\cite{Gupta2014}. $\copyright$  2014 by American Physical Society.  \label{ES-noise2-fig}}
\end{figure}

Figure~\ref{ES-noise2-fig}(a) reports the phase diagram  in the parameter space. The shaded blue surface is a first-order transition surface. Under (outside) this surface, the synchronization transition is abrupt and explosive (continuous). To furnish a better intuitive illustration, Fig.~\ref{ES-noise2-fig}(b) shows the behavior of $r$ as a function of noise $T$. For large $T$, the change of $r$ is continuous and overlapped between forward and backward transitions. With decreasing $T$, however, ES takes place and the width of hysteresis loop becomes larger.

The effect of noise on synchronization was actually observed in other types of synchronization models~\cite{Sonnenschein2013,He2003,Zhou2002,Toral2001,Neiman2002,Sanchez1997,Baier2000}. For instance, for a positive correlation and sufficient wide frequency distribution, an intermediate noise can lead to the maximal hysteresis loop in the synchronization diagram of FitzHugh-Nagumo oscillators operating on top of SF networks~\cite{Chen2013}.

\subsection{Explosive synchronization in generalized Kuramoto models}

In the original Kuramoto model (Eq.~(\ref{eq:kuramcomplete})), the sign of the coupling strength is assumed to be positive, implying that the interactions among oscillators are always attractive. However, in many practical circumstances,
the interaction among the constituents of an ensemble can be repulsive or suppressive, which would correspond to a negative coupling in the model \cite{Borgers2003,Qu2007,Daido1992,Galam2004,Lama2005}.

For this purpose, Ref.~\cite{Hong2011} generalized Eq.~(\ref{eq:kuramcomplete}), by allowing
the coupling strength $\lambda$ to assume both positive and negative values.
The model is described by
\begin{equation}\label{eq:model-CC}
 \dot{\theta}_i=\omega_i + \frac{\lambda_i }{N}\sum_{j=1}^{N}\sin(\theta_j-\theta_i),\quad i=1,...,N,
\end{equation}
with a Lorentzian frequency distribution $g(\omega) = \gamma/[\pi(\omega ^2 + \gamma ^2)]$.
Here, $\lambda_i$ (the coupling strength of the $i$th oscillator to the mean field)
is a binary variable, i.e., either $\lambda_i=\lambda_1<0$ or $\lambda_i=\lambda_2>0$.
Accordingly, oscillators in the ensemble are divided into two populations: those with positive $\lambda_i$ will behave like {\it conformists},
whereas those with negative $\lambda_i$ will tend to act as {\it contrarians}.
Initially, the system is set in the incoherent state where only contrarian oscillators exist.
Then contrarians are gradually flipped into conformists according to certain strategies.
When the proportion of conformists, denoted by $p$,  exceeds a certain threshold,
the system undergoes a transition to the coherent state.

\begin{figure}
\begin{center}
\includegraphics[width=0.5\textwidth]{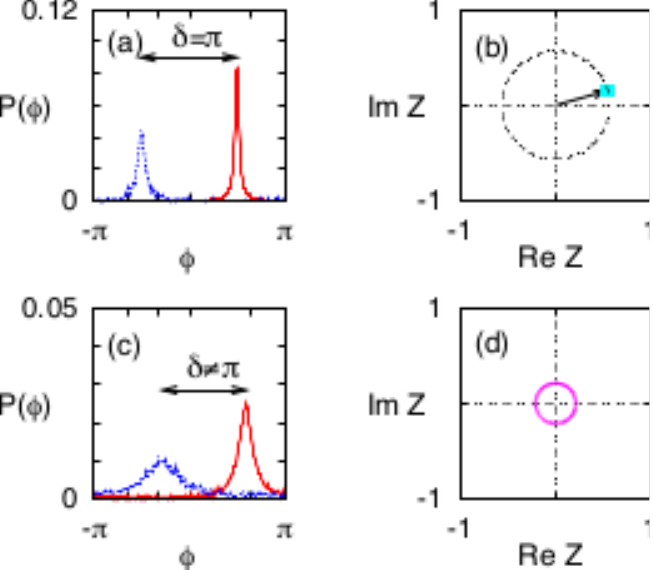}
\caption{(Color online). The $\pi$  and TW states in model
(\ref{eq:model-CC}).  $N=25,600$, $\gamma=0.05$, $\lambda_1=-0.5$, $\lambda_2=1.0$.
(a) Phase distribution for the $\pi$ state; the mean phase difference $\delta$ between the conformist (red) and contrarian (blue)
oscillators is given by $\delta=\pi$. (b) The order parameter for the $\pi$ state
corresponds to a fixed point. (c) Phase distribution for the
TW state; $\delta \ne \pi$. (d) The order parameter for the TW state traces a circle, implying a non-zero mean
phase velocity. Reprinted with permission from Ref.~\cite{Hong2011}. $\copyright$  2011 by American Physical Society.\label{fig:pi-state}}
\end{center}
\end{figure}

Specifically, there are three typical strategies for changing contrarians into conformists:
(1) a fraction $p$ of contrarians is randomly chosen and then flipped into conformists;
(2) contrarians are ranked according to the absolute value of their natural frequencies ($|\omega_i|$).
Then, they are flipped into conformists from the largest $|\omega_i|$ to a threshold $\omega_0$, i.e.
the coupling strength of the $i$th oscillators will be $\lambda_i=\lambda_2$ if $|\omega_i|>\omega_0$ and $\lambda_i=\lambda_1$ otherwise.
Therefore $1-p=\int_{-\omega_0}^{\omega_0}g(\omega)d\omega$;
(3) the opposite of the process described in (2), i.e. $\lambda_i=\lambda_2$ if $|\omega_i|<\omega_0$ and $\lambda_i=\lambda_1$ otherwise, which leads to $p=\int_{-\omega_0}^{\omega_0}g(\omega)d\omega$.
Mathematically, the three strategies correspond to
three different correlations between the coupling strength and the natural frequencies as follows:
\begin{equation}\label{eq:Case1}
Case \ 1:\ \ \Gamma_1(\lambda)=(1-p)\ \delta(\lambda-\lambda_1)+p\ \delta(\lambda-\lambda_2),
\end{equation}
\begin{equation}\label{eq:Case2}
Case \ 2:\ \ \Gamma_2(\omega,\lambda)
=H(\omega_0-|\omega|)\delta(\lambda-\lambda_1)+H(|\omega|-\omega_0)\delta(\lambda-\lambda_2),
\end{equation}
\begin{equation}\label{eq:Case3}
Case \ 3:\ \ \Gamma_3(\omega,\lambda)
=H(|\omega|-\omega_0)\delta(\lambda-\lambda_1)+H(\omega_0-|\omega|)\delta(\lambda-\lambda_2),
\end{equation}
where $H( )$ is the Heaviside function.

Reference~\cite{Hong2011} considers $Case$ 1, and shows that the system has two types of coherent states: the $\pi$ state and the
TW state (see Fig.~\ref{fig:pi-state}). In the $\pi$ state, two clusters of conformists and contrarians
keep a constant phase difference $\pi$. The $\pi$ state is stationary, and the order parameter is a fixed point in phase space.
For the TW state, the phase distribution spontaneously
travels at a constant speed along the phase
axis, always maintaining a constant separation $\delta \ne \pi$. As a result, the order parameter traces a
circle around the origin at a constant angular velocity.

\begin{figure}
\begin{center}
\includegraphics[width=0.4\textwidth]{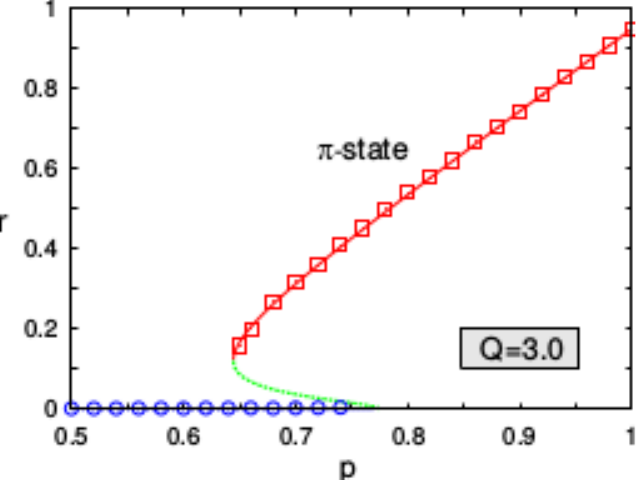}
\caption{(Color online). Order parameter $r$ vs. $p$ for $Q=3$,
computed by numerical integration (symbols) and compared to the
theoretical prediction (lines). The solid lines denote the stable states; the dotted line individuates the unstable $\pi$ state.
$N=25600$, $\gamma=0.05$, $\lambda_1=-3.0$, $\lambda_2=1.0$.
Reprinted with permission from Ref.~\cite{Hong2011}. $\copyright$  2011 by American Physical Society.\label{fig:transition}}
\end{center}
\end{figure}

As the control parameter $p$ varies, transitions are observed among the incoherent and the coherent states.
The results can be split into two scenarios, depending on whether the conformists or the contrarians are more strongly affected
by the mean field.
Figure ~\ref{fig:transition} reports the phase diagram
when the contrarians are more strongly affected by the mean field, i.e. $Q=-\lambda_1/\lambda_2>1$.
For small $p$, the
system is dominated by contrarians, and therefore is in the incoherent state. Once $p$ exceeds a certain threshold,
there are enough conformists for a consensus to
emerge. At that point, the system jumps discontinuously
to the $\pi$ state, where it is polarized into two groups (see Fig.~\ref{fig:pi-state}(a)).
Inversely, starting from the coherent state and adiabatically decreasing $p$ yields a discontinuous and hysteretic return
to the incoherent state. On the other hand, if $Q<1$, the transitions become continuous. Typically, as $p$ increases from $0$,
the system goes from the incoherent to the $\pi$  state, then  to the TW state, and finally settles on
the $\pi$ state (see Fig. 3 of Ref.~\cite{Hong2011}).

Based on the Ott-Antonsen ansatz \cite{Ott2008}, the dynamics of Eq.~(\ref{eq:model-CC}) can be effectively reduced into a low-dimensional system describing the complex order parameters for the contrarians and conformists.
By means of the analysis of the reduced system, the three steady
states and their bifurcation points can be analytically obtained
\cite{Hong2011}.

In Ref.~\cite{Yuan2014}, numerical simulations of $Case$ 2 and $Case$ 3 are presented, and  both the $\pi$ state and the TW state are found. In a certain parameter regime of $Case$ 3, furthermore,
a special explosive behavior is taking place: when the control parameter $\omega_0$ is decreased to a critical threshold, the system bifurcates from the incoherent state into the TW state via a first-order like transition, while starting from the TW state, if
$\omega_0$ increases, the order parameter exhibits
a continuous and  hysteretic return to the incoherent
state (see Fig. 7(b) in Ref.~\cite{Yuan2014}).

\begin{figure}
\begin{center}
\includegraphics[width=0.9\textwidth]{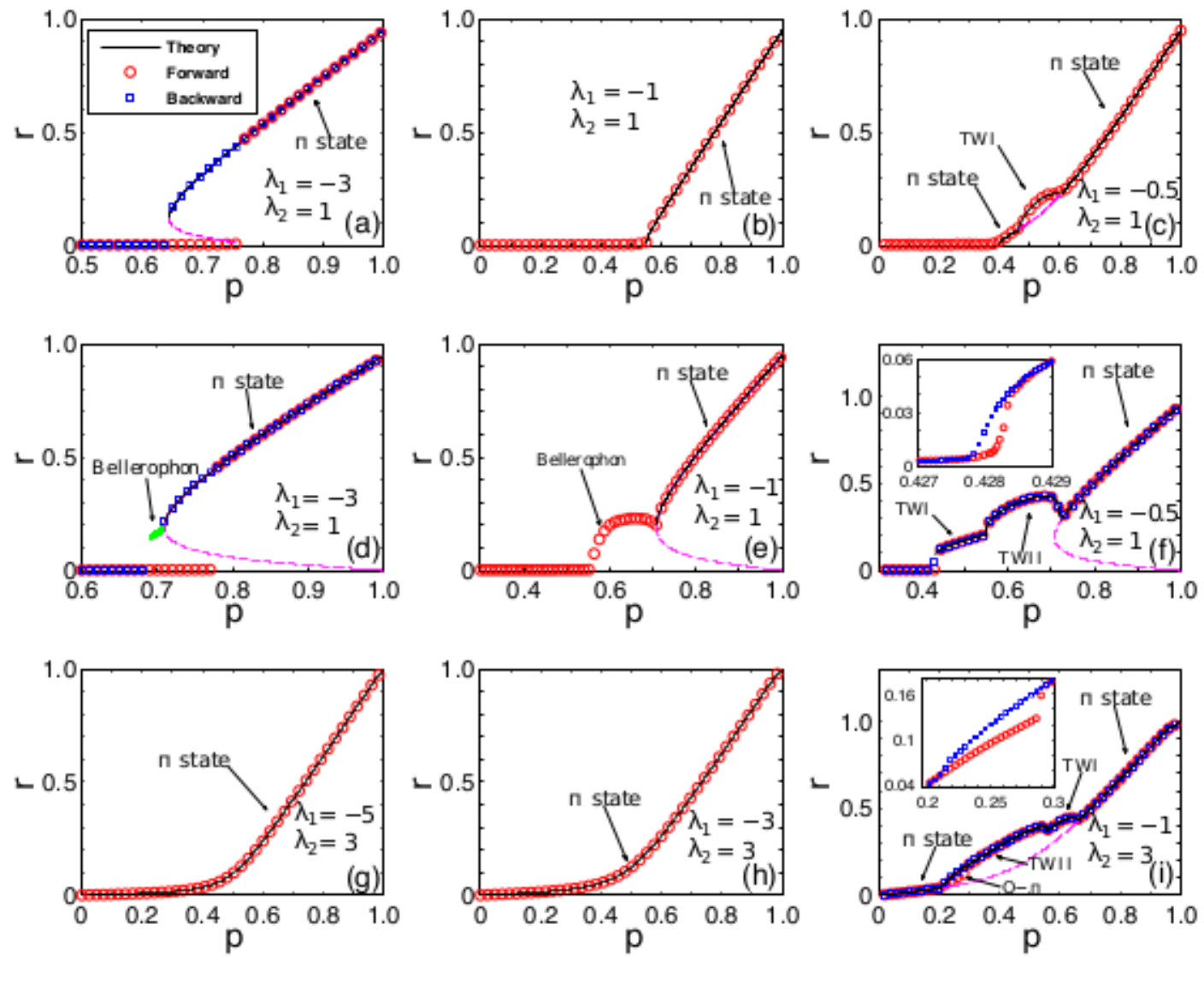}
\caption{(Color online). Scenarios of synchronization in Eq.~(\ref{eq:model-CC}) as the proportion of conformists increases. $N=50000$ and $\gamma=0.05$.
From top to bottom, three rows correspond to $Case$ 1, $Case$ 2 and $Case$ 3, respectively.
From left to right, the three columns correspond to the case of $|\lambda_1|>\lambda_2$, $|\lambda_1|=\lambda_2$,
and $|\lambda_1|<\lambda_2$, respectively. Both the forward (red circles) and the backward (blue squares) transitions are reported as obtained in an adiabatic way, while the black (pinkish red) curves correspond to the
theoretical predictions of the stable (unstable) stationary coherent states, including the $\pi$ state,
the TW-I state, and the TW-II states.
The O-$\pi$ state denotes the oscillating $\pi$ state.
The numerical results are perfectly consistent with the theoretical predictions. Adapted from Ref.~\cite{Qiu2016}\label{fig:diagram}}
\end{center}
\end{figure}

Recently, Qiu et al. provided a detailed and complete analytical treatment for model (\ref{eq:model-CC}), based on linear stability analysis and mean-field analysis \cite{Qiu2016}.
It has been proved that in this model the incoherent state is neutrally stable below the synchronization
threshold, and the critical points for synchronization are derived for all the three cases.
Moreover, all possible stationary coherent states in the model are predicted,
including the $\pi$ state and the two types of TW states.

In Fig.~\ref{fig:diagram}, a detailed phase diagram characterizing the synchronization transitions in model (\ref{eq:model-CC}) is reported. Depending on the correlations between the coupling
strength and the natural frequencies (as well as on whether the conformists or contrarians are more strongly
affected by the mean field), the system exhibits either first-order or second-order like
transitions among different steady states. Besides the two examples of ES studied in Refs.~\cite{Hong2011} and \cite{Yuan2014} (which corresponds to Figs.~\ref{fig:diagram}(a) and \ref{fig:diagram}(i), respectively)  it is further revealed that a first-order transition can also occur in $Case$ 2.
As illustrated in Fig.~\ref{fig:diagram}(d),
when $p$ increases, the system experiences a first-order transition from the incoherent state into the coherent state, i.e., the $\pi$ state. Inversely, when the system starts from
the coherent $\pi$ state, as $p$ decreases, the system does not directly go back to the incoherent state.
The $\pi$ state first bifurcates into a new coherent state named the Bellerophon state (see next Section for the full characterization of such a novel state) via a second-order transition. Then, as $p$ further decreases, the Bellerophon
state loses its stability and the system jumps into the incoherent state via a first-order transition.

\subsection{The Bellerophon state}\label{sec:bellerophon}

In the previous Sections, a variety of coherent states emerging in the classical Kuramoto model have been discussed and characterized,
such as the partially coherent state \cite{Strogatz2000}, the standing-wave  \cite{Martens2009,Pazo2009} and TW  \cite{Martens2009,Iatsenko2013} states, and the chimera state \cite{Kuramoto2002,Shima2004,Abrams2004,Omelchenko2008}. In such coherent states, oscillators inside each one of them display typically frequency-locked patterns, and therefore behave like a giant oscillator as a whole.
Here, instead, we describe the main properties of a recently discovered novel state: the Bellerophon state.

\subsubsection{The Bellerophon state in frequency-weighted Kuramoto models}

\begin{figure}
\begin{center}
\includegraphics[width=0.8\textwidth]{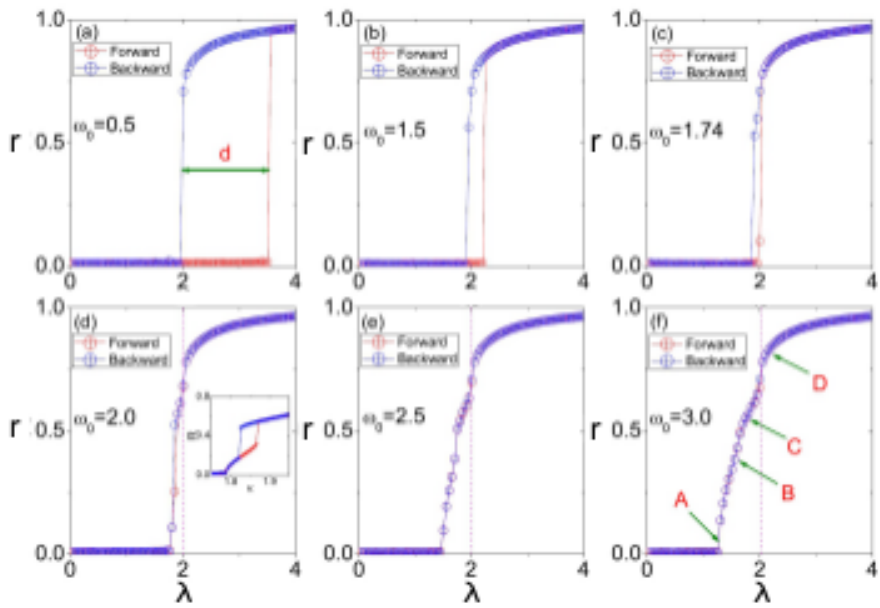}
\caption{(Color online). From explosive to continuous transition. $r$  vs. $\lambda$ for model (\ref{eq:model-weighted}). $N=10000$ and
$\Delta =1$.  $\omega_0=0.5$ (a), $1.5$ (b), $1.74$ (c), $2.0$ (d), $2.5$ (e), and $3.0$ (f), respectively.
Adapted from Ref.~\cite{Bi2016a}. \label{fig:bimodal}}
\end{center}
\end{figure}

In Ref.~\cite{Bi2016a}, ES was studied in a frequency-weighted Kuramoto model
\begin{equation}\label{eq:model-weighted}
\dot{\theta}_i=\omega_i + \frac{\lambda | \omega_i |}{N}\sum_{j=1}^{N}\sin(\theta_j-\theta_i),\quad i=1,...,N,
\end{equation}
with a bimodal Lorentzian distribution
\begin{equation}\label{eq:double-Lorentzian}
g(\omega)=\frac{\Delta}{2\pi} [\frac{1}{(\omega-\omega_0) ^2 + \Delta ^2} +\frac{1}{(\omega+\omega_0) ^2 + \Delta ^2  }].
\end{equation}
Here, $\theta_i$ and $\omega_i$  are the instantaneous phase and the natural frequency of the $i$th oscillator, respectively, and $\lambda$ is the coupling strength. $\Delta$ is the width parameter (half width at half maximum) of each Lorentzian distribution, and $\pm \omega_0$ are their central frequencies.
As shown in Fig.~(\ref{fig:bimodal}), system (\ref{eq:model-weighted}) sustains both a first- and a second-order transition to synchronization. According to the theoretical analysis \cite{Hu2014,Bi2016a},
the critical point for the backward transition $\lambda_{c}^{b}$ is always 2, while the critical point for the forward transition is
\begin{equation}\label{eq:kf}
\lambda_{c}^{f}=\frac{4}{\sqrt{1+(\omega_0 / \Delta)^2}}.
\end{equation}
As the parameter $\omega_0 / \Delta$ increases, the
characteristic hysteresis width $\lambda_{c}^{f}-\lambda_{c}^{b}$
shrinks, which induces eventually the conversion of a continuous transition to synchronization  into an ES.

A new coherent state is unveiled in  system (\ref{eq:model-weighted}), which is actually non-stationary
and  essentially different from all the coherent states studied before.
In such a state, oscillators form quantized, time-dependent clusters, where neither their phases nor their instantaneous frequencies are locked. Their instantaneous speeds are different within the clusters, but they
form a characteristic cusped pattern and, more importantly, they behave periodically in time so that their average
values are the same. Given its intrinsic specular nature with respect to chimera states, the new state has been named the Bellerophon state in Ref.~\cite{Bi2016a}.

\begin{figure}
\centering\includegraphics[width= 0.8\textwidth]{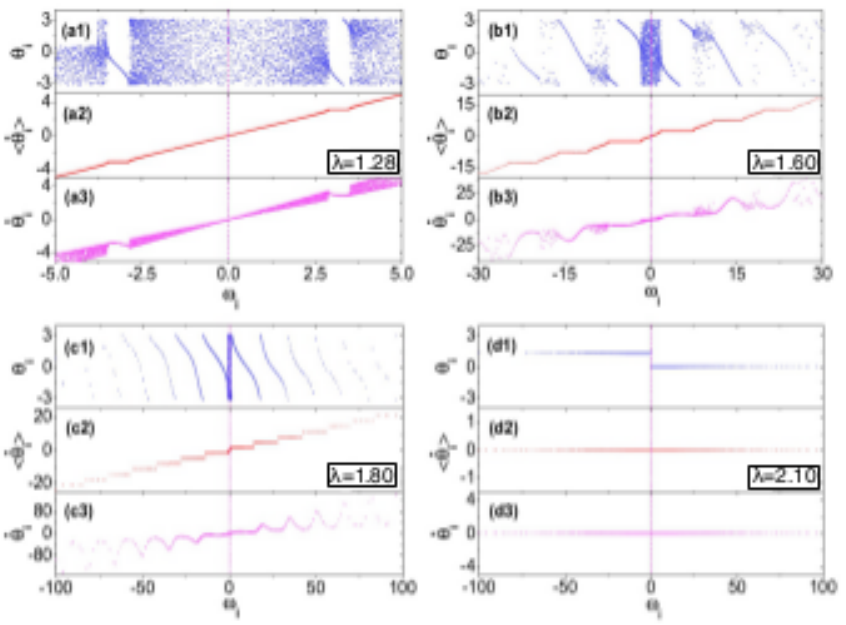}
\caption{(Color online). The Bellerophon states observed in the model (\ref{eq:model-weighted}) with bimodal frequency distribution. Snapshots of the instantaneous phase $\theta_i$ (upper plots), the average speed $\langle \dot{\theta}_i \rangle$ (middle plots), and  the instantaneous speed $\dot{\theta}_i$ (lower plots) vs. natural frequencies $\{\omega_i\}$ of the oscillators. $\lambda=1.28$ (a), $1.60$ (b), $1.80$ (c), and $2.10$ (d, the fully synchronized state).
Panels (a)-(c) refer to {\it Bellerophon} states. Adapted from Ref.~\cite{Bi2016a}.\label{fig:B-bimodal} }
\end{figure}

In Fig.~\ref{fig:B-bimodal} four typical steady states are illustrated, corresponding to the $\lambda$ values denoted by letters A, B, C, and D in Fig.~\ref{fig:bimodal}(f).
Among them, Figs.~\ref{fig:B-bimodal}(a)-\ref{fig:B-bimodal}(c) are Bellerophon states while Fig.~\ref{fig:B-bimodal}(d) is the fully coherent state. In  Bellerophon states,
oscillators form quantized clusters in terms of the average frequencies.
The crucial point here is that although the average speeds of oscillators inside each cluster are equal to each other, their instantaneous speeds are generally different and quite heterogeneous.
Furthermore, the instantaneous speeds of oscillators in each cluster are correlated and form the characteristic cusped pattern [Figs.~\ref{fig:B-bimodal}(a3), ~\ref{fig:B-bimodal}(b3), and \ref{fig:B-bimodal}(c3)] analogous to that featured by the average frequencies of the oscillators within the chimera state.
Therefore, inside each cluster, oscillators are neither phase- nor frequency-locked in a common sense.
Rather, they seem to correlate in a higher order collective way: in terms of average speed oscillators inside each cluster are locked, but in terms of instantaneous speed
each oscillator evolves uniquely.
In fact, very interesting collective motion of oscillators can be observed in the unit circle,
as illustrated in Ref.~\cite{Bi2016a} (and, more specifically, in the animated movies which form part of the Supplementary
Material of that Reference).

\begin{figure}
\begin{center}
\includegraphics[width=0.6\textwidth]{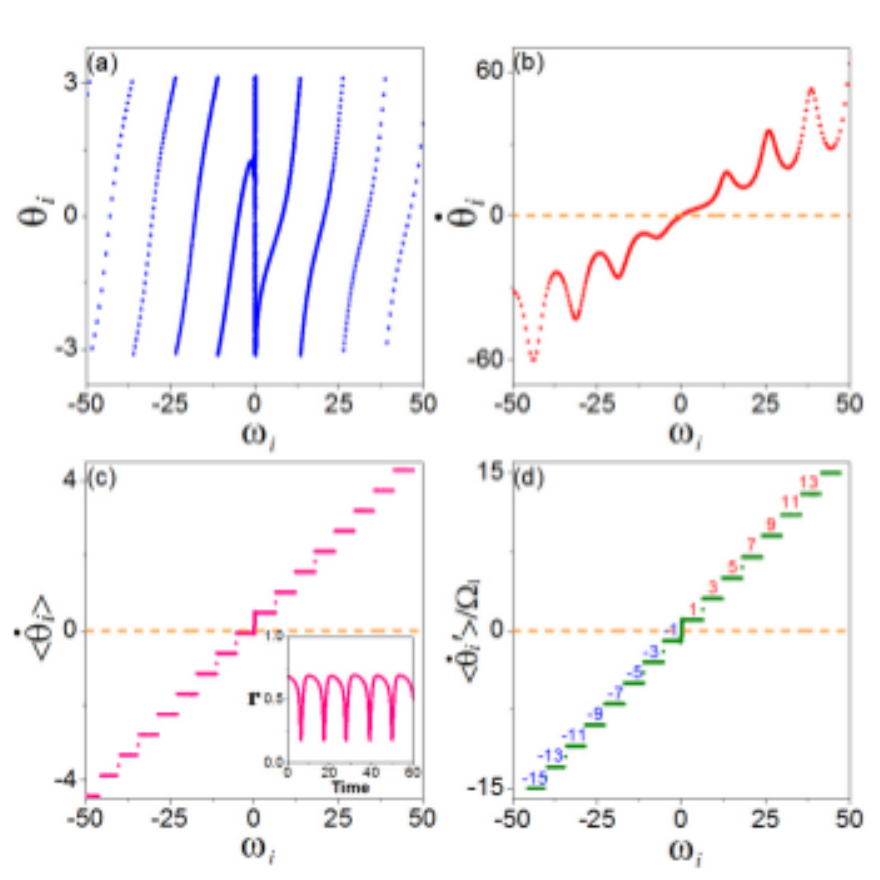}
\caption{(Color online).
The  Bellerophon states observed in the model (\ref{eq:model-weighted}) with asymmetric frequency distributions and $N=10000$,
$\omega_0=0.5$, and $\lambda=2.08$.
(a)-(c) The phases (a), the instantaneous frequencies (speeds) (b), and the average frequencies (average speeds) (c) vs.
the natural frequencies of oscillators.
(d) The counterpart of (c) in a suitable rotating frame. Notice that the average frequencies are normalized by the fundamental frequency $\Omega_1$. Reprinted from Ref.~\cite{Zhou2016}. \label{fig:B-asymmetric}
}
\end{center}
\end{figure}

Bellerophon states can occur in system (\ref{eq:model-weighted}) also for different frequency distributions. For instance, in Ref.~\cite{Zhou2016}, model (\ref{eq:model-weighted})
was investigated with an asymmetric Lorentzian distribution:
\begin{equation}\label{eq:Asymmetric}
g(\omega) = \frac{\Delta}{\pi [(\omega-\omega_0) ^2 + \Delta ^2]},
\end{equation}
where  $\Delta=1$ is kept as a constant, $\omega_0$ changes
in order to shift the frequency distribution along the positive axis.
It is found that in the intermediate regime between incoherence and full synchronization \cite{Zhou2016},
the Bellerophon states emerge, as illustrated in Fig.~\ref{fig:B-asymmetric}.
Furthermore, the evolution of oscillators exhibit periodic intermittency
following a synchronous pattern of bursting in short periods and resting in long periods \cite{Zhou2016}.

In Ref.~\cite{Bi2016b}, model (\ref{eq:model-weighted}) was investigated with a uniform frequency distribution:
\begin{equation}\label{eq:uniform}
g(\omega)=
\begin{cases}
\frac{1}{2\gamma} & \textrm{for} ~ |\omega|\leq \gamma,\\
0 & \textrm{for} ~ |\omega|>\gamma.
\end{cases}
\end{equation}
It is revealed that, as the coupling strength increases,
the system undergoes two transitions.
First, a first-order phase transition occurs from the incoherent state to a Bellerophon state, and then the Bellerophon state bifurcates into  the two-cluster synchronous state via a continuous transition, when the coupling strength further increases.

\begin{figure}
\begin{center}
\includegraphics[width=0.8\textwidth]{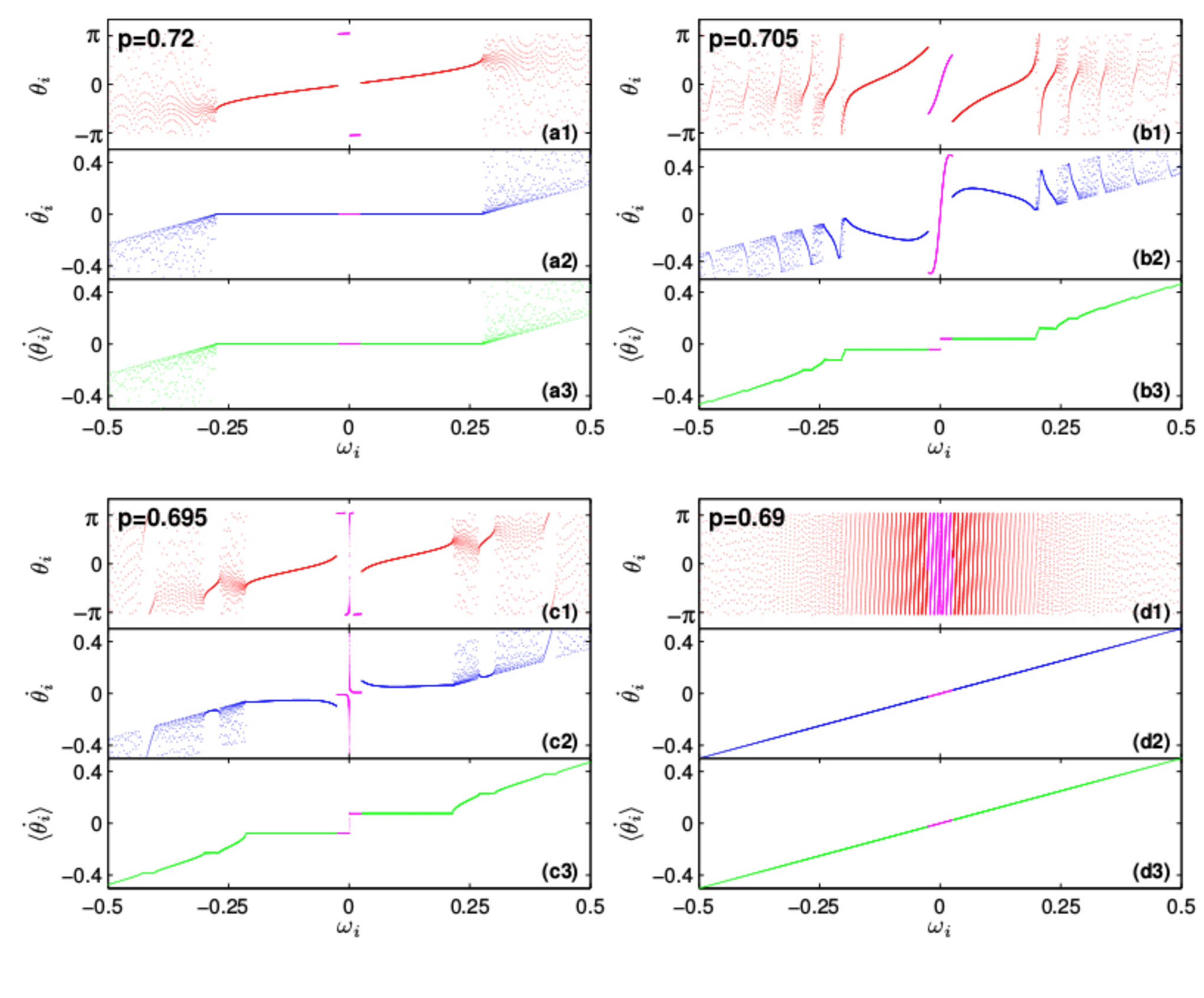}
\caption{(Color online). Typical coherent states and the incoherent
  state in $Case$ 2,
corresponding to the backward transition in Fig.\ref{fig:diagram}(d). $N=50000$ and $\gamma=0.05$.
Snapshots of the instantaneous phase $\theta_i$ (upper plots), the
instantaneous frequency (speed) $\dot{\theta_i}$ (middle plots), and
the average frequency (average speed) $\langle\dot{\theta_i}\rangle$
(lower plots) vs. natural frequencies $\{\omega_i\}$ of the
oscillators.  (a) The $\pi$ state with $p=0.72$. (b)-(c) The
Bellerophon states with $p=0.705$ and  $p=0.695$, respectively. (d)
The incoherent state with $p=0.69$. Reprinted from Ref.~\cite{Qiu2016}.  \label{fig:B-CC}}
\end{center}
\end{figure}

\subsubsection{The Bellerophon state in coupled conformist and contrarian oscillators}

%As shown in Fig.~\ref{fig:diagram}, phase transitions may occur among various steady states.
The Bellerophon state is also observed
in the model of Eq.~(\ref{eq:model-CC}), which is essentially different with respect to the frequency-weighted model (\ref{eq:model-weighted}).
Let us focus on a typical example, namely that of Fig.~\ref{fig:diagram}(d).
As the control parameter $p$ increases, the system experiences a first-order like transition from the
incoherent state to the coherent state, i.e. the $\pi$ state. Inversely, when the system starts from
the coherent $\pi$ state, as $p$ decreases, the system does not directly go back to the incoherent state. As shown in Fig.~\ref{fig:diagram}(d), the $\pi$ state first bifurcates into the Bellerophon state via a second-order transition. Then, as $p$ further decreases, the Bellerophon state looses its stability and the system finally jumps into the incoherent state via a first-order like transition.
In Fig.~\ref{fig:B-CC}, the typical Bellerophon states corresponding to Fig.~\ref{fig:diagram}(d) are characterized.
Except for the fact that there are clusters of both conformists and contrarians, the observed Bellerophon states
share all the other features of the frequency-weighted model described above.

\subsection{Explosive synchronization in multi-layer networks}
\label{ES-multi}

So far, our journey on explosive synchronization has been limited to the case of single-layer, or mono-layer, networks. Real complex systems, however, have far more complicated forms of interactions,  which  points to the fact that the hypothesis of a single-layer network
may actually result an overestimation (or an underestimation) of the problem under study. A more accurate
analysis resorts on multi-layer networks, and dynamical processes on top of them are attracting more and more attention
(see, for instance, Ref.~\cite{Boccaletti2014} for a recent and rather comprehensive review).
Therefore, examining ES on multi-layer networks became a topic of the utmost significance.
%If there exists explosive synchronization,  is the correlation feature between oscillators and coupling strength necessary?

\begin{figure}
\begin{center}
\includegraphics[width=0.4\textwidth]{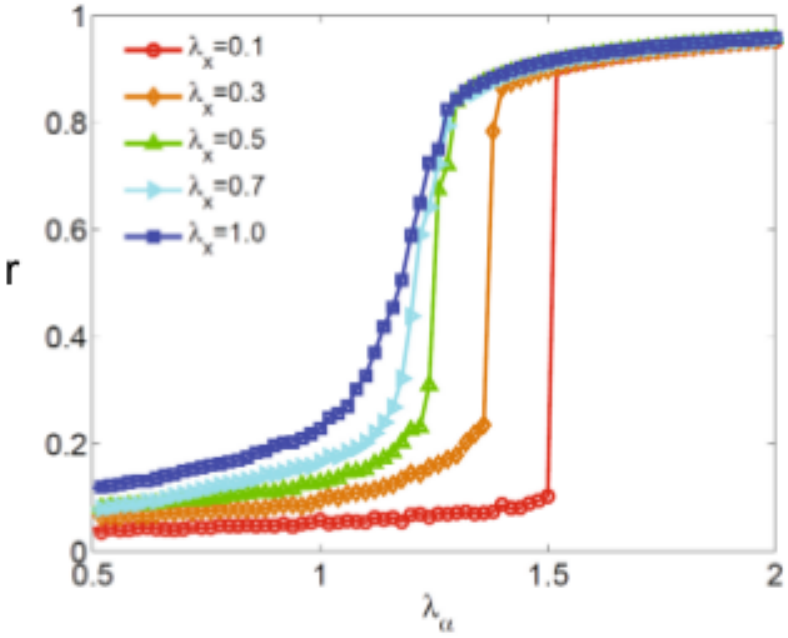}
\end{center}
\caption{(Color online). Synchronization level $r$ of the base layer vs. intra-layer coupling strength $\lambda_{\alpha}$ for different values of inter-layer
coupling strength $\lambda_{x}$. $N=10^3$, and $\lambda_{\beta}=1$.
Reprinted  from Ref.~\cite{Jiang2015b} with permission of Springer Ed.\label{ES-mul-1}}
\end{figure}

\subsubsection{Multi-layer networks with the correlation property $\omega_i=k_i$}
\label{ES-correlated}
 Jiang et al. recently proposed a Kuramoto model on multiplex  networks~\cite{Jiang2015b}, which are composed of one SF network $\alpha$ (as the base layer)
and one community structured network  $\beta$ (as the external layer), and where each node constructs  local links
with probability $1-\mu$ and long-range connections with probability
$\mu$ (which is then the mixing parameter). Reference~\cite{Jiang2015b} mainly focused on how the coupling between layers affects the phase transition of oscillators in the based layer $\alpha$, for which the evolution equation reads as
\begin{equation}
\dot{\theta}_i^{\alpha}=\omega_i+ \lambda_{\alpha}\sum_{i=1}^N A_{ij}^{\alpha}\sin(\theta_j^{\alpha}-\theta_i^{\alpha})+\lambda_{\alpha\beta}
 \sin(\theta_i^{\beta}-\theta_i^{\alpha}),
\label{eq-ku-mu-1}
\end{equation}
where $\omega_i^{\alpha}=k_i^{\alpha}$. The first and second terms are the same of those
referring to a single-layer network, and only involve intra-layer interactions in the base layer $\alpha$. The
third term accounts for the additional inter-layer interactions controlled by
 the coupling strength parameter $\lambda_{\alpha\beta}$. For simplicity,
the hypothesis of symmetric coupling between both layers ($\lambda_{\alpha\beta}=\lambda_{\beta\alpha}$) is assumed, and
$\lambda_{\alpha\beta}$ is denoted by $\lambda_{x}$.

The inter-layer coupling $\lambda_{x}$ plays a significant role in determining the phase transition of the base layer.
As shown in Fig.~\ref{ES-mul-1}, at weak coupling strengths ($\lambda_{x}=0.1$) ES emerges.
With increasing $\lambda_{x}$, however, the transition becomes more and more continuous:
there is an interim period at $\lambda_{x}=0.5$, and  a second-order-like transition occurs at $\lambda_{x}=1.0$, which indicates that strong inter-layer couplings hinder the emergence of ES.

Moreover, it was unveiled that other externalities
(mainly the parameters $\mu$ and $\lambda_{\beta}$) have great effect  on ES. In details,
increasing the mixing part $\mu$ puts off ES of the base layer, and requires larger critical coupling strengths $\lambda_{\alpha}^{c}$, irrespectively of other parameters.
This points  is contrary to the findings in traditional modular networks~\cite{Li2013,Park2006,Oh2005}, where long-range, inter-modular connections speed up the
coherent state. As for coupling strength $\lambda_{\beta}$, its increase can accelerate ES of the base layer, though this effect seems to be very restricted.

\subsubsection{Multi-layer
networks with partial and weak correlation}

A generative model of inter-dependent networks (including subnetworks $G_1$ and $G_2$) was proposed in Ref.~\cite{Su2013a}.
At each time step, a node $i$ enters into both subnetworks, and constructs $m$ different links on each subnetwork. That is to say,
nodes on both subnetworks have one-to-one interdependency with the same index. Then, like the preferential attachment rule in SF
networks \cite{Barabasi1999}, the probability of producing one new link reads as
$P_i^1 \sim \alpha_1k_i^1 + (1-\alpha_1)k_i^2$ and $P_i^2 \sim \alpha_2k_i^1 + (1-\alpha_2)k_i^2$, where $\alpha_1$ and $\alpha_2$
control the mutual correlation between networks. It is clear that the connectivity of each layer
depends on itself and its counterpart, and both subnetworks have the same power-law degree distribution.

The Kuramoto model is then introduced. Instead of a strong correlation, the natural frequency of oscillator $i$ in layer $G_1$ ($G_2$)
is taken to be $\omega_i^1=k_i^2$ ($\omega_i^2=k_i^1$), which means that the dynamics of oscillator is controlled by the local
topology and connectivity of the other layer.
Such a weak hypothesis of correlation can induce ES when $\alpha_1$ and $\alpha_2$ are sufficiently large. As $\alpha_1$ and $\alpha_2$ decrease, the abrupt transition is replaced by a continuous transition.
%These findings thus extend the condition of ES from strong correlation to weak correlation between $w_i$ and $k_i$.

\begin{figure}
\begin{center}
\includegraphics[width=0.6\textwidth]{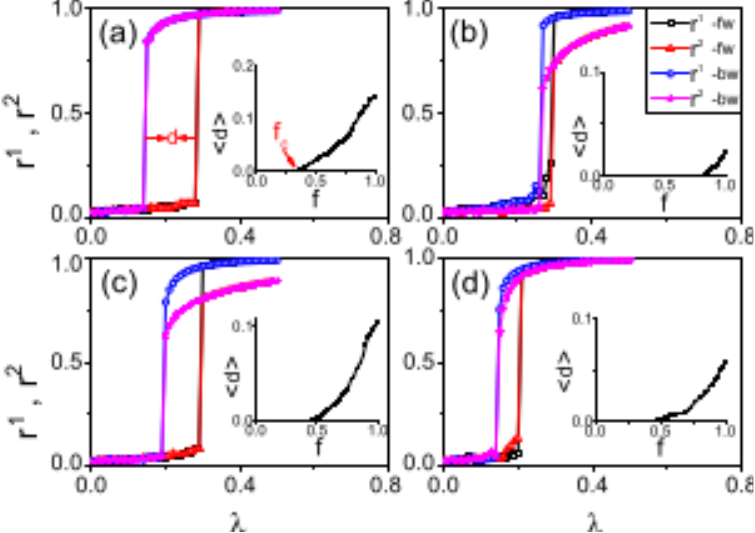}
\end{center}
\caption{(Color online). $r^{1}$ and $r^{2}$  vs. $\lambda$ for two
  inter-dependent networks with $N=10^3$ and $f=1$. Squares and
  circles (triangles and stars) denote the forward (backward)
  transitions, and the insets report the average width $\langle
  d\rangle$ of the hysteresis loop as a function of $f$. Here layer 1
  is a ER network with $\langle k\rangle=12$. From (a) to (b), layer 2
  is ER network with $\langle k\rangle=12$ and  $\langle k\rangle=6$,
  and $g(w_i^2)$ is a random homogeneous distribution in the range
$[-1, 1]$. From (c) to (d), layer 2 is ER network and SF network with
  $\langle k\rangle=12$, and $g(w_i^2)$ is Lorentzian distribution and random homogeneous fashion.
  Reprinted from Ref.~\cite{Zhang2015}. $\copyright$  2015 by American Physical Society.
\label{ES-mul-2}}
\end{figure}

\subsubsection{Multi-layer networks in the absence of correlation}

Zhang et al.~\cite{Zhang2015} proposed to consider the following system
\begin{equation}
\dot{\theta_i}=\omega_i+ \lambda \alpha_{i}\sum_{i=1}^N A_{ij}\sin(\theta_j-\theta_i), \label{eq-ku-mu-2}
\end{equation}
where the new parameter $\alpha_i$ accounts for an adaptive control of nodes.

Initially, a fraction $f$ of the nodes is randomly chosen, and
$\alpha_i=r_i$ is
assumed for each of the selected nodes, where $r_i$
is the instantaneous local order parameter of the $i$-{th} oscillator,
defined as
$r_i(t)e^{i\phi}=(1/k_i)\sum_{j=1}^{k_i}e^{i\theta_j}$, and $\phi$
denotes the phase averaged over the ensemble of neighbors. For the remaining fraction $1-f$ of the nodes, $\alpha_i=1$. Obviously, $f=0$ returns the traditional Kuramoto model~\cite{Kuramoto1984,Strogatz2000,Acebron2005,Moreno2004}, while $f>0$ indicates that a fraction of nodes are adaptively controlled by the local order parameter.
Besides, another significant point is the choice of the natural frequencies  $\omega_i$ of oscillators, which are taken from a random homogeneous distribution $g(\omega_i)$ in the range $[-1, 1]$. Based on these assumptions, it was reported that an abrupt (with a hysteresis loop) transition takes place if $f$ is over a critical value $f_c$. Namely, the absence of correlation features can also promote the occurrence
of ES in single-layer networks. %, similar to the observations in Refs.~\cite{Chen2015,Zhang2013}.
%Moreover, once there is explosive synchronization, the average width $<d>$ hysteresis loop monotonously increases with $f$.

In order to explore the case  of inter-dependent networks, Ref.~\cite{Zhang2015} extended the analysis to the situation where each node has a one-to-one partner with the same index $i$. The equations become
\begin{equation}
\begin{split}
\dot{\theta}_i^1&=\omega_i^1+ \lambda \alpha_i^1 \sum_{j=1}^{k_i^1} A_{ij}\sin(\theta_j^1-\theta_i^1),\\
\dot{\theta}_i^2&=\omega_i^2+ \lambda \alpha_i^2 \sum_{j=1}^{k_i^2} A_{ij}\sin(\theta_j^2-\theta_i^2),
\label{ eq-ku-mu-3}
\end{split}
\end{equation}
where the superscripts $1$ and $2$ denote the network layers 1 and 2, $\alpha_i^1$ and $\alpha_i^2$ involve the coupling of two layers via dependency links. In details, if nodes $i$ fall into the fraction $f$, $\alpha_i^1$ and $\alpha_i^2$ refer to the local order parameters $r_i^2$ and $r_i^1$, i.e. $\alpha_i^1=r_i^2$ and $\alpha_i^2=r_i^1$ ($\alpha_i^1=\alpha_i^2=1$ otherwise), where $r_i^1(t)e^{i\phi^1}=(1/k_i^1)\sum_{j=1}^{k_i^1}e^{i\theta_j^1}$ and $r_i^2(t)e^{i\phi^2}=(1/k_i^2)\sum_{j=1}^{k_i^2}e^{i\theta_j^2}$.
The frequency function $g(\omega_i^1)$ is still random homogeneous and in the range $[-1, 1]$, while $g(\omega_i^2)$ can change from a random to a
Lorentzian distribution.

Figure~\ref{ES-mul-2} shows the dependency of synchronization levels $r^{1}$ and $r^2$ on $\lambda$ for different choices of the layer
2 and $g(\omega_i^2)$. ES occurs and seems very robust against the difference in topology and frequency distribution,
which indicates that the correlation between oscillator's frequency and node's degree is not the essential
condition for the emergence of ES. If $f$ is below the critical value $f_c$, synchronization returns to a second-order transition. More importantly, these observations can be quantitatively verified via the mean-field theory.
In the traditional second-order transition, oscillators with close frequency firstly collapse into small synchronization clusters, which gradually converge towards a giant cluster at the critical coupling strength. However, under ES, small clusters are prevented and all free oscillators are abruptly attracted to the giant cluster.
Recently, furthermore, this kind of ES has been shown to coexist with the standard phase of the Kuramoto model
in the thermodynamic limit \cite{Danziger2016}.
% Here, the behaviors of the oscillators are divided into two groups: oscillators in the $f$ fraction (named as the controlled group) evolve similarly to the later case~\cite{Gomez-Gardenes2011}, while the remaining oscillators (named as the free group) are like the former case yet influenced by the controlled group. Only when $f$ is sufficient large, the controlled group can impede the emerging process of clustering in free group. In this sense, it is clear to provide an unified understanding for the intrinsic reason of explosive synchronization: suppressing the formation of giant synchronization cluster, regardless of interaction topology.

\section{Applications}

\subsection{Explosive percolation in real physical systems}
\label{gigimelo}

Most of the studies on Explosive Percolation reviewed in Section~\ref{sec:percol} are explicitly conducted in the thermodynamic limit, i.e.
under the hypothesis of an infinite
system size (or a size of the order of the Avogadro's number, $N\sim 10^{23}$). However,  real world systems
are always finite-sized. For instance, the Internet, world-wide airline network and online social
networks (just to mention three examples of very large sized real networks) have all values of $N$ considerably smaller than $10^{18}$ \cite{DSouza2015}.
In this sense, it is interesting
to reveal how the AP process may induce significant discrete jumps in the realm of real-world networks,
where EP is generally considered to be a phenomenon of
cascading failures \cite{Buldyrev2010,Parshani2011}.

One of the major events of electrical blackout was affecting much
of Italy on 28 September 2003: the shutdown of power stations directly led to the failure of nodes
in the Internet communication network, which in turn caused further breakdown of power stations.
This iterative process of
cascading failures can also occur in a single power-grid network such as the blackout in Northwestern America in
August 1996, and the blackout in Northeastern America and Canada in August 2003, where one overloaded node caused
other nodes to also become overloaded and thus disabled the entire network.
Other examples include the
financial crises (in the network of global financial market) and the spreading of information and rumors through
online social networks, such as Facebook or Twitter, where a failing node can cause its neighbors to fail as well.
A common feature of these cascading failures is that they represent an inverse process of EP, which
 has a devastating
effect on the network stability, and may cause huge economic losses.

% figure 1
\begin{figure}
\centering \includegraphics[width=0.45\linewidth]{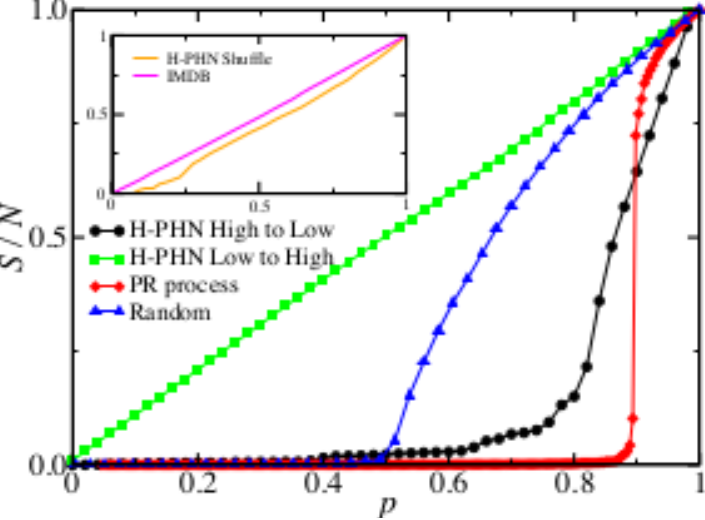}
\caption{(Color online). Fraction of nodes in the largest cluster of the H-PHN  vs. the percentage $p$ added links
according to different rules: (a) links are added in decreasing weight order (black circles), (b) links are added in
increasing weight order (green squares), (c) links are chosen according to the PR model (red diamonds), and (d) links are added randomly (blue triangles). Inset: same quantity as in the main figure but (a) the structure of the H-PHN remains the same and link weights are randomly
redistributed (orange line), and (b) a weighted IMDB co-acting network is considered (violet line).
 Reprinted from Ref.~\cite{Rozenfeld2010} with permission of Springer Ed. \label{Fig:Rozenfeld2010}}
\end{figure}

Although a general answer to the question of how to suppress or control a cascading failure in real systems
is still missing, some progress in this direction has been achieved
\cite{Rozenfeld2010,Pan2011,Cho2011a,Cho2012,Radicchi2015,Oliveira2014,Kim2010},
illustrating how the AP rule can actually be used in real systems.

For example, Rozenfeld et al. considered EP in the human protein homology network
(H-PHN) and showed that the emergence of a spanning cluster exhibits similar features to a AP process
\cite{Rozenfeld2010}. Their results indicate that the evolutionary-based processes shaping the topology of
the H-PHN through duplication-divergence events may occur in sudden steps, similarly to what is seen in first-order
like phase transitions.

Cho et al. considered the diffusion-limited cluster aggregation (DLCA) model of the sol-gel
transition, and found that a discontinuous percolation transitions can be observed when particles are Brownian, in
which cluster velocity depends on cluster size as $v_s \sim s^{\eta}$ with $\eta=-0.5$ \cite{Cho2011a,Cho2012}.

Pan et al. considered social networks community structures and assortativity (such as a mobile phone call network and a
large arXiv coauthorship network), and found that the percolation transition depends on the structural properties
of the network \cite{Pan2011}.

Oliveira et al. discovered that the electric breakdown due to pollution with
metallic powder can become explosive if the inhibition of adsorption due to a local electric field becomes too
strong \cite{Oliveira2014}.

Later on, Radicchi considered the case of coupled networks and found that the
dependency links have an important role in the phase transition \cite{Radicchi2015}. For a high density of
dependency links, the network disintegrates in a form of a first-order phase transition, whereas for a low density
of dependency links, the network disintegrates in a second-order transition. In the following, we
briefly linger in each one of these cases.

\subsubsection{The human protein homology network}

First, we discuss the case of H-PHN. Based on real data\footnote{The
  H-PHN was obtained from the Similarity Matrix of Proteins (SIMAP)
  project, http://boinc.bio.wzw.tum.de/boincsimap}, Rozenfeld et al. found
that H-PHN is composed of highly connected clusters of homologous nodes, while links between nodes of low homology
generate inter-cluster connections, which is similar to the presence of strong links within communities and weak
links between communities \cite{Rozenfeld2010}. Further, they found that H-PHN is a weighted modular network, with the
weights denoting the degree of similarity (homology) between two proteins. Ref.~\cite{Rozenfeld2010}
claimed that such a weighted modular network can be explained on the basis of
a product rule (PR) model. The idea is the following: suppose that H-PHN is grown up from a spanning skeleton
of the network which connects all the different network areas. A new protein has a much larger weight
to generate links within a module
rather than with proteins that are further away. These dense modules are then connected with each other through
weaker links. In the terminology of the PR model, this corresponds to an increased probability of connections between
small clusters, if compared to the growth of already large clusters.

A quantity $0\le \zeta\le 1$ is introduced to detect the degree
of similarity between two optimally aligned proteins \cite{Rozenfeld2010}. $\zeta = 1$ indicates a perfect alignment between the two proteins, or in other words, a short genetic
distance between the proteins. The values of $\zeta$ for a protein pair in the network is taken to be the weight of
the corresponding link. Then, the concrete process to construct the H-PHN can be implemented by considering
an empty network of all proteins, and adding one link at a time in decreasing order of the weight, quantified by $\zeta$.
The process initially leads to well-connected families of highly
homologous proteins, that are inter-connected at later stages by links with smaller values of $\zeta$, resembling the
requirements for EP in Ref.~\cite{Achlioptas2009}.

The results are reported (black line with ``circles'') in Fig.~\ref{Fig:Rozenfeld2010}. $S/N$ represents the fraction of
nodes in the largest cluster of the H-PHN, and $p$ denotes the percentage of links. One clearly sees that the largest
cluster in the network remains small until $p$ reaches $0.8$, and then quickly increases for  $p>0.8$. The behavior
is very similar to an AP, i.e. links are added locally and without a significantly large spanning cluster
before $p_c$, while the subsequent addition of links after $p_c$ causes the small clusters to merge into a spanning
entity. For comparison, the red line with ``diamonds'' in the same figure shows the results of a pure
AP, for the same number of nodes. One sees that the two curves have a high-degree similarity, and the PR model
has a sharper transition than the H-PHN.

% figure 2
\begin{figure}
\centering \includegraphics[width=0.85\linewidth]{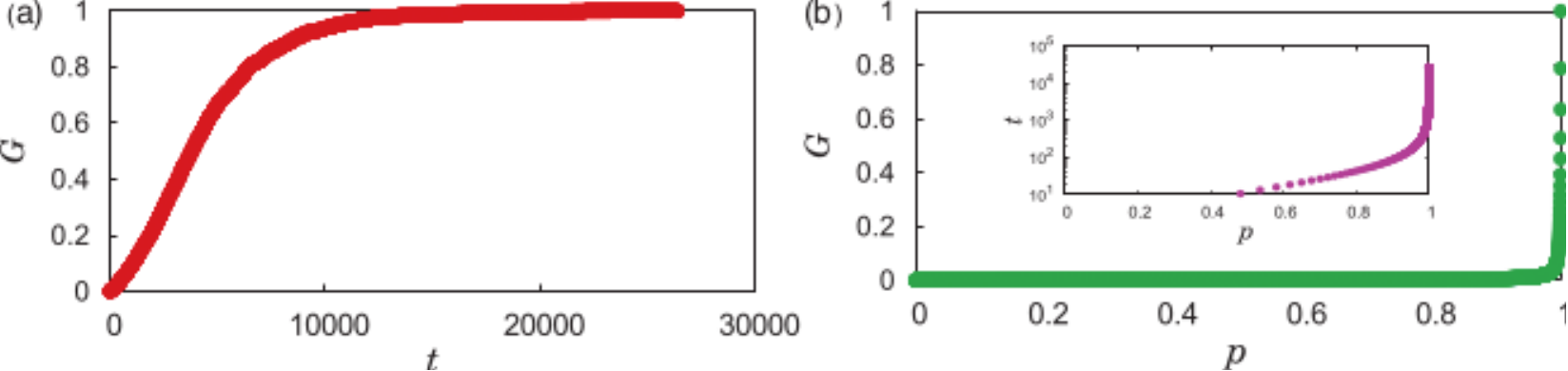}
\caption{(Color online). (a) The giant cluster size $G$ vs. $t$ for Brownian particles. Notice that here the giant cluster grows continuously from $t =0$. (b) $G$ vs. $p$. Here $G$ grows drastically near $p_f =1-1/N$. Inset: plot of the relationship between
$t$ and $p$. $t$ increases drastically, but in a power-law manner as $p$ approaches $p_f$. Simulations are carried out
with $N =8,000$ mono-particles at $t =0$ on a $400\times 400$ square lattice.
Reprinted with permission from Ref.~\cite{Cho2011a}. $\copyright$  2011 by American Physical Society.  \label{Fig:Cho2011}}
\end{figure}

For a deep understanding of the nature of the transition, Rozenfeld et al. reported the results for the cases of adding
links in increasing order of $\zeta$ (green curve with ``squares'' in Fig.~\ref{Fig:Rozenfeld2010}) and adding links randomly (the blue curve with ``triangles'' in Fig.~\ref{Fig:Rozenfeld2010}).
The former strategy induces a percolation transition at a very early stage, and the
largest cluster $S$ increases linearly with $p$, i.e. almost every new link attaches to the largest cluster.
In the latter strategy, the largest component $S$ remains small before its critical point $p_c$,
and then increases approximately linearly after $p_c$, i.e. following a second-order transition.
By comparison of all the four curves in Fig.~\ref{Fig:Rozenfeld2010} one concludes that the case of decreasing weight in the H-PHN is much steeper than the
case of random addition, and is much closer to the case of PR processes, indicating that the evolution of a modular
network can be explained through the idea of EP.

The sharp transition of the H-PHN is not a universal property ascribed to weighted
networks. The inset of Fig.~\ref{Fig:Rozenfeld2010} shows that, when the link weights of the H-PHN are shuffled
without modifying the structure of the network, a smooth percolation transition is observed at an early stage.
Similar growth patterns can be observed in other modular networks, such as the network of
movie actors from IMDB in which two actors are connected if they co-acted in a movie (see the inset of
Fig.~\ref{Fig:Rozenfeld2010}).

\subsubsection{The DLCA model}

As a second case, we briefly discuss the case of the DLCA model. Initially, $N$ single particles
are placed randomly in a $L\times L$ square lattice. Assume that the particles are Brownian and the cluster velocity
depends on the cluster size as $v_s \sim s^{\eta}$. For convenience of numerical simulations, one first lets the velocity of
a cluster be inversely proportional to the square root of its size. The simulations are performed as follows: at each
time step, (i) a $s$-sized cluster is selected with the probability $q \equiv s^{0.5}/(\sum_s N_ss^{0.5})$, and is
moved to the nearest neighbor, where $N_s$ is the number of $s$-sized clusters. When two distinct clusters are placed
at the nearest-neighbor positions, these clusters are regarded as being merged, forming a larger one. (ii) The time
is advanced by $\delta t =1/(\sum_s N_ss^{0.5})$. Steps (i) and (ii) are repeated until the network percolates.

Let $p$ be the number of cluster aggregation events per total particle number. Whenever two clusters merge, $p$ is
increased by $\delta p =1/N$. Since $N-1$ aggregation events occur during all aggregation processes, the aggregation
event stops at $p_f =1-\delta p$ \cite{Cho2011a}. Figure \ref{Fig:Cho2011}(b) shows the dependence of the giant
cluster size $G$ on $p$. One sees that it increases drastically, exhibiting a discontinuous phase transition. For
comparison, the dependence of $G$ on $t$ is shown for the original DLCA model in Fig.~\ref{Fig:Cho2011}(a), where sees
that $G$ increases monotonically with $t$. The different behavior between Figs.~\ref{Fig:Cho2011}(a) and (b) is
caused by the nonlinear relationship between $t$ and $p$, as shown in the inset of Fig.~\ref{Fig:Cho2011}(b). When $p$
is small, $t$ increases almost linearly with $p$. At variance, as $p$ approaches $p_f$, $t$ increases drastically
in a power-law manner, resulting a discontinuous phase transition.

% figure 3
\begin{figure}
\centering \includegraphics[width=0.5\linewidth]{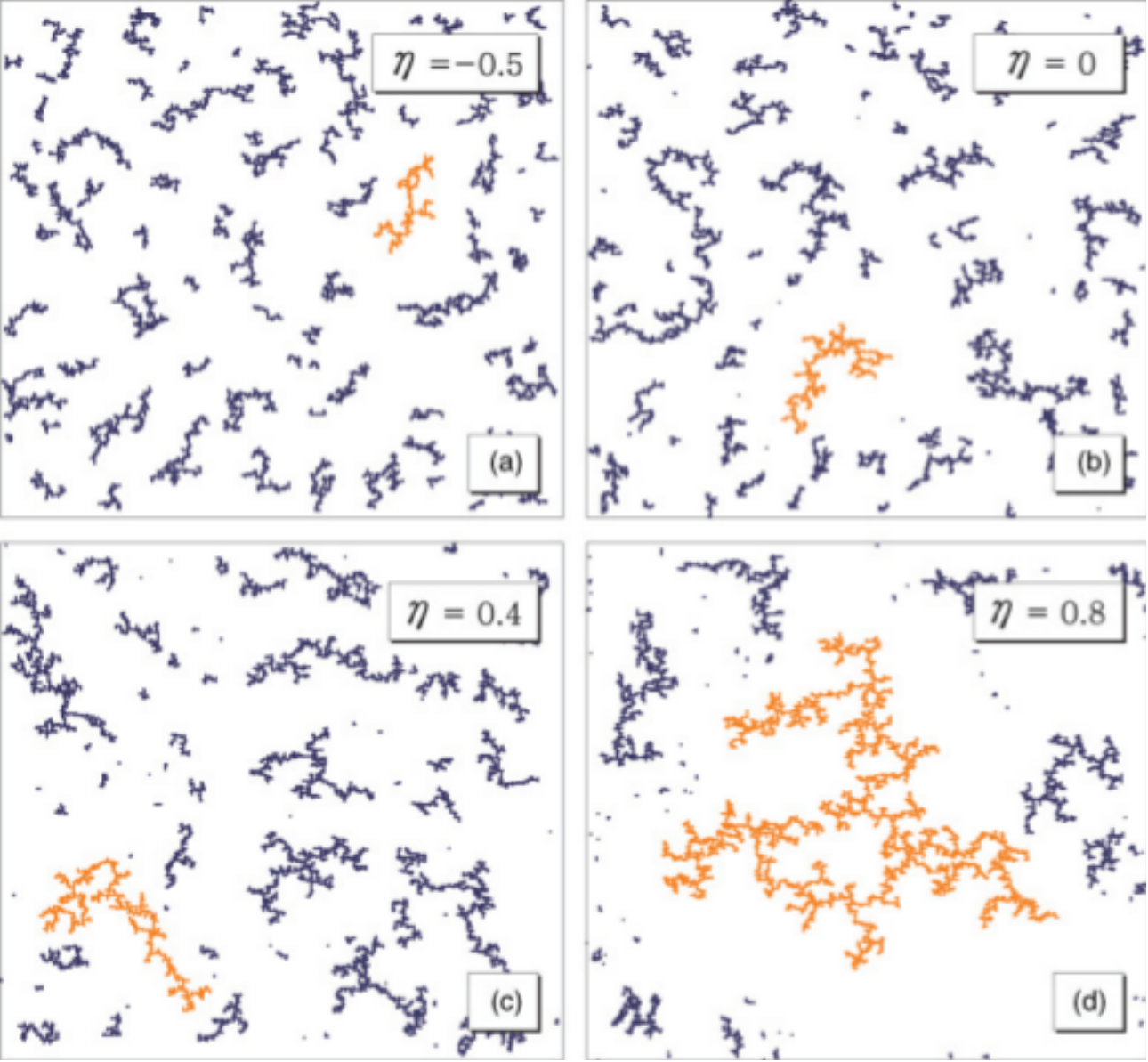}
\caption{(Color online). Snapshots of the system for various values of $\eta$ at $p =0.99$.
The velocity of each cluster is given as $v_s \sim s^{\eta}$, where $s$ is the cluster size.
Numerical simulations are carried out for $N =8,000$ particles on
a $L\times L =400 \times 400$ square lattice. Since $p$ is fixed, the number of clusters for each case is equal.
The giant cluster is represented in a different color (gray/orange). The cluster-size distribution becomes more
heterogeneous as $\eta$ increases. Reprinted with permission from Ref.~\cite{Cho2011a}. $\copyright$  2011 by American Physical Society. \label{Fig:Cho2011eta}}
\end{figure}

Let us now move to the general case of $v_s \sim s^{\eta}$. Here, a cluster is picked up with a probability proportional
to $s^{\eta}$ while the other rules in the numerical simulations are held. When clusters merge, $p$ is
increased by $\delta p =1/N$. The time is given by $\delta t =1/(\sum_s N_ss^{\eta})$. Intuitively, when $\eta$ is small or
negative, fewer large-sized clusters are selected, so that their growth is suppressed. Medium-sized clusters are abundant even
close to $p_f$, and they merge suddenly, resulting in a discontinuous transition. In contrast, when $\eta$ is positively large,
larger-sized clusters are selected and they have larger chances of colliding with other ones,
merging into a bigger cluster. The result is that these latter clusters can grow faster than smaller ones, so the giant cluster grows continuously and the transition becomes continuous. The presence of a tri-critical point $\eta_c$ is expected across which the transition type is changed. Figure~\ref{Fig:Cho2011eta} shows snapshots of the system for different values of $\eta$ at $p =0.99$, supporting the
above arguments.

The DLCA model was later extended to a reaction-limited cluster aggregation  model in which clusters diffuse
following Brownian motion, and when two clusters come into contact with each other, they merge with a certain probability
$r$ and remain separated with the remaining probability $1-r$ \cite{Cho2012}. It is found that a discontinuous transition
can be observed in two and three dimensions, while the transition remains continuous in four
dimensions. %In numerical simulations, they use $r = 10^{-3}$, in which the dynamics of cluster aggregations of the RLCA
%observed is different from that of the DLCA.

% figure 4
\begin{figure}
\centering \includegraphics[width=0.9\linewidth]{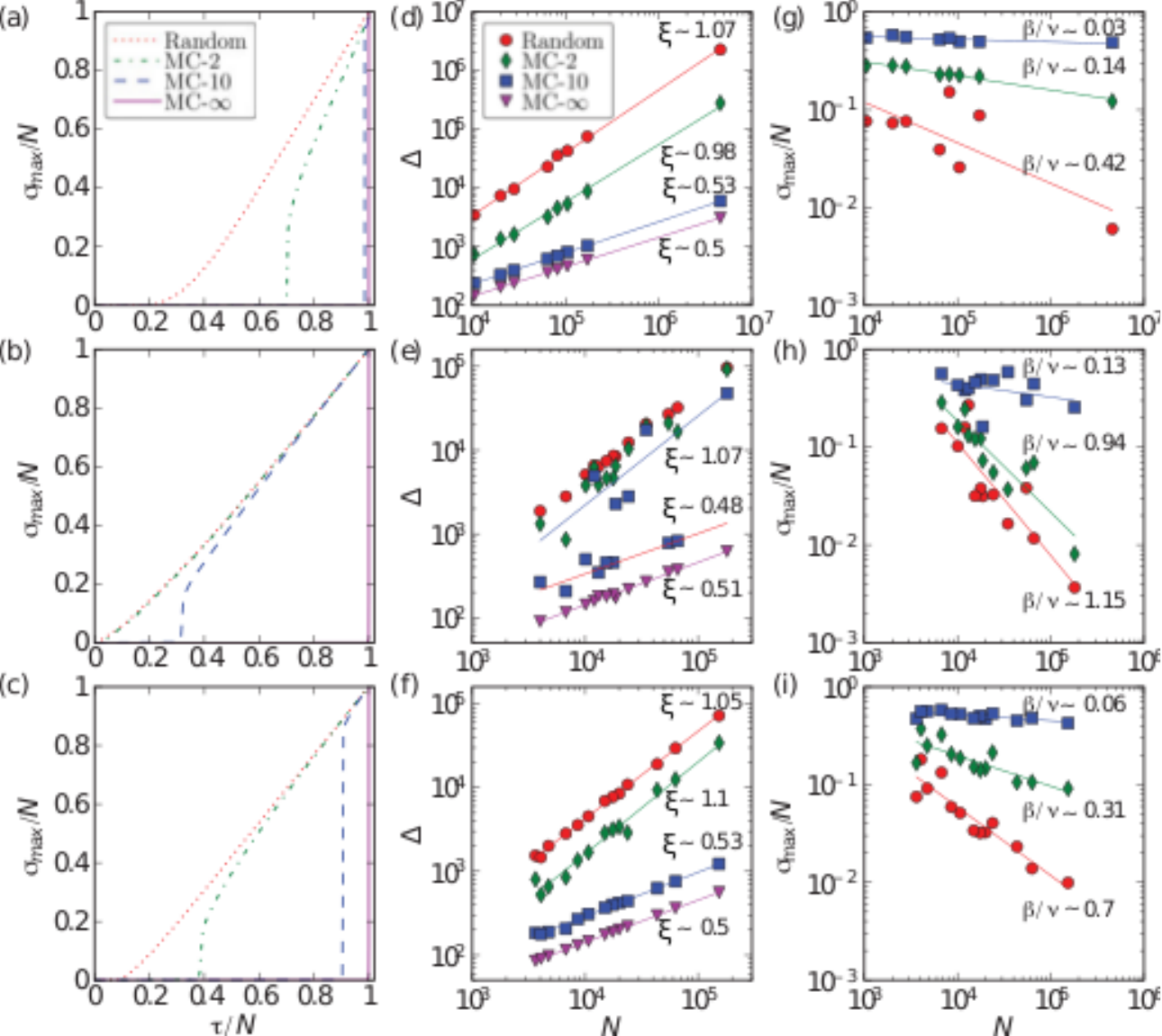}
\caption{(Color online).  Variation of the relative size of the giant component
($s_{max}/N$)  vs. the scaled number of inter-cluster edges $\tau/N$
for the (a) MPC, (b) CA, and (c) small CA networks. The corresponding variations in the gap,
$\Delta \equiv \tau(N/2)-\tau (\sqrt{N})$, as a function of system sizes are shown in (d), (e), and (f), for the Random,
MC-2, MC-10, and MC-$\infty$ rules (see legend for color codes). Solid
lines indicate fitted scaling exponents $\xi$. The variation of the order parameter ($s_{max}/N$) as a function of the system size $N$ is shown for (g) MPC, (h) CA, and (i) SCA networks. For each system the order parameter is calculated at the critical point. The solid line indicates the best fit obtained with a power-law scale with  exponent
$\beta/v$. All curves are averaged over $10^3$ runs.
Adapted with permission from Ref.~\cite{Pan2011}. Courtesy of M. Kivel\"a.  \label{Fig:Pan2011}}
\end{figure}

\subsubsection{Social networks}

Next, we linger on the case of social networks \cite{Pan2011}, such as the empirical mobile phone call (MPC)
network, and a large arXiv coauthorship (CA) network, where the nodes represent people and links denote their
interactions. These social networks share common features such as community structures and assortativity.

The MPC data consist of $325\times 106$ voice calls over a period of $120$ days.
An aggregated undirected weighted network can be constructed, where edges represent bidirectional calls between users and
weights represent the total number of calls. The largest connected component (LCC) is then extracted,
with $4.6\times 10^6$ nodes and $9.1\times 10^6$ edges.

The collaboration data is from the arXiv \footnote{http://arxiv.org} and
contains all e-prints in ``physics'' until March 2010. There are $4.8 \times 10^5$ article headers, from which one extracts
the authors. In the CA network two authors are connected if they signed together articles whose number determines the link
weight. The LCC is extracted, with $1.8\times 10^5$ nodes and $9.1\times 10^6$ edges. In addition, a filtered version of
the CA can be constructed, where articles with more than $10$ authors ($\sim 2\%$ of the total number) are ignored.
This is to remove the very large cliques from papers with $\sim 10^3$ authors in fields such as hep-ex or astro-ph,
where the principles behind
collaboration network formation appear different. The LCC of the resulting smaller CA network has
$1.5\times 10^5$ nodes and $9.1\times 10^5$ edges. Note that, although the number of nodes is not much smaller than for that of
CA, the number of edges is an order of magnitude smaller.

For describing the percolation process, Pan et al. used the min-cluster (MC-$m$) SR,
with different values of $m$ \cite{Pan2011}.
Initially, all the edges of the empirical network are considered unoccupied.
Then, at each time step, $m$ unoccupied edges are drawn at random.
Out of these edges, the one minimizing the size of the component formed if it were occupied is chosen.
Intra-component edges are always favored against inter-component edges, as they do not increase the
size of any cluster. When comparing two inter-component edges,
the one for which the sum of cluster sizes that
it connects is minimized is selected. As intra-cluster edges do not affect cluster growth, one considers
only the number of inter-cluster edges $\tau$. In particular, three variants of the MC rule (MC-2, MC-10, and MC-$\infty$) as well as random link percolation are compared  \cite{Pan2011}. Figures~\ref{Fig:Pan2011}(a), (b) and (c) show the variation of the fraction $s_{max}(\tau)/N$
against the scaled number of inter-component edges, $\tau/N$. For all the three networks, the transition of the order parameter is smooth for the random case, while for the extreme case, MC-$\infty$,
the transition appears abrupt. For MC-2
and MC-10, the situation is more complicated, and in the following will be described in more detail.

One can study the dependence of the width of the transition window
$\Delta \equiv \tau(N/2)-\tau (\sqrt{N})$ on the system size, where $\tau(N/2)$ and $\tau (\sqrt{N})$ are the lowest values
of $\tau$ for which $s_{max} > N/2$ and $s_{max} > \sqrt{N}$, respectively \cite{Achlioptas2009}. In general, the width
scales as a power law with the system size ($\Delta \sim N^{\xi}$). For classical percolation, $\xi =1$. It is argued that
$\xi <1$ for EP,  and the rescaled width of the transition region ($\Delta/N \sim N^{\xi-1}$) vanishes in
the limit of large $N$. In order to apply finite-size scaling to empirical networks, samples of different sizes are needed.
As call networks are geographically embedded, one extracts sub-networks of users in chosen cities, based on postal codes of their
subscriptions. For the CA networks, one extracts sub-networks of authors with articles in the same subject class. It turns out that,  for all networks $\Delta\tau \sim N^{\xi}$, with $\xi \sim 1$ for random and $\xi \sim 0.5$ for the MC-$\infty$ case (see
Figs.~\ref{Fig:Pan2011}(d), (e), and (f)). Thus, the exponent $\xi$ clearly differentiates the explosive transition from
random-link percolation. Furthermore, for all three networks, $\xi \sim 1$ for the MC-2, resembling an ordinary percolation
transition. However, for MC-10, the scaling exponent behaves
differently for the three networks. For the MPC and smaller CA
networks, $\xi \sim 0.5$, indicating EP. For the CA collected data points seem to indicate that there is no scaling. However, a closer inspection shows that they cluster around two straight lines with $\xi \sim 1$ and
$\xi \sim 0.5$.

According to Ref.~\cite{Radicchi2010}, the order parameter $s_{max}/N$ has a finite-size scaling
\begin{equation}
\label{order-scaling}
 \frac{s_{max}}{N} = N^{-\beta/v}F[(\tau-\tau_c)N^{1/v}],
\end{equation}
where $F$ is some universal function, $\tau$ is the control parameter, $\tau_c$ is the critical point of the transition, $\beta$ is
the critical exponent of the order parameter, and $v$ is another critical exponent. One chooses the critical value $\tau_c$
of the control parameter as the value of $\tau$ where the susceptibility (that is, the average cluster size) has its maximum.
For the MPC network [Fig.~\ref{Fig:Pan2011}(g)], the scaling at $\tau_c$ of $s_{max}/N$
yields a very small exponent $\beta/v\sim 0.03$ for the MC-10 case, while for MC-2 and random percolation, the exponents
are larger ($\beta/v\sim 0.14$ and $\beta/v\sim 0.42$, respectively). The exponents for the SCA network behave similarly
[Fig.~\ref{Fig:Pan2011}(h)] with a low value $\beta/v\sim 0.06$ for the MC-10 case and relatively high values
$\beta/v\sim 0.31$ and $\beta/v\sim 0.70$ for MC-2 and random percolation, respectively. In contrast, for the CA network,
the exponents have higher values for all cases [Fig.~\ref{Fig:Pan2011}(i)] ($\beta/v\sim 0.13$, $\beta/v\sim 0.94$, and
$\beta/v\sim 1.15$) for MC-10, MC-2, and random percolation, respectively.

\subsubsection{Inter-dependent networks}

Recently, percolation in real systems was extended to the case of inter-dependent networks \cite{Radicchi2015}. The
functioning of a real network depends not only on the reliability of its own components, but also on the
simultaneous operation of other real networks coupled with it. Percolation transitions in
inter-dependent networks can be understood by decomposing these systems into uncoupled graphs: the intersection
among the layers, and the remainders of the layers. When the intersection dominates the remainders, an inter-connected
network undergoes a smooth percolation transition. Conversely, if the intersection is dominated by the contribution
of the remainders, the transition becomes abrupt even in small networks.

As an example of real systems, Radicchi
showed the transition for two inter-connected systems of interest in
the biology: the {\em H. sapiens}
protein-protein interaction network and the {\em C. elegans} connectome, finding that the inter-connected systems undergo
smooth percolation transitions \cite{Radicchi2015}. The reason is that these organisms have developed inter-connected
networks sharing a core of ``high quality'' edges to prevent catastrophic failures. The same
properties seem to characterize also the multi-layer air transportation network within the USA major airports, where
the set of ``high'' quality' edges that avoid truly catastrophic changes in the connectedness of the entire inter-dependent system are constituted.
%This finding may be significant in the aspects of suppressig the occurrence of cascading failures in real systems such as the controlling of power-grid networks or epileptic seizures.

\subsection{Explosive synchronization in real physical systems}\label{sec:expsyncapp}

\subsubsection{Power grids}
\label{sec:powergrid}
%\subsubsection{ Modelling of power grids }

Among the many real-world systems where ES can be observed, power grids are actually the ones in which cascading
of failures has been experienced for numerous times.

One way to produce a cascade of failures
is by the overload of lines: an outage of a line may lead to overload on other lines, thereby eventually leading to their outage. For instance, in August 2003,
an undetected initial power line failure in Ohio caused a massive power blackout rolled across Northeastern USA
and Canada, spanning eight states and two provinces, and affecting some 55 million people \cite{Amin2007}.

Another way  is overloading the power stations: an outage of a node may lead to overload on other nodes, also leading to their outage. For example, a
small initial power shutdown in El Paso (Texas) caused the August 1996 blackout in Northwestern America
in which the power outage spread through six states between  Oregon and California, leaving $7.5$ million customers
without electricity.

Around the world, the number of outages affecting large populations has steadily risen in the past decades, causing enormous economic losses.

The cascade of failures in power grids can be also explained in terms of their
underneath inter-dependent network structure. For example, in September 2003, Italy was affected by a country-wide
blackout: the shutdown of power stations directly led to the failure of nodes in the Internet
communication network, which in turn caused further breakdown of power stations and left
57 million Italians in the dark \cite{Buldyrev2010}.

\begin{figure}
\centering \includegraphics[width=0.5\linewidth]{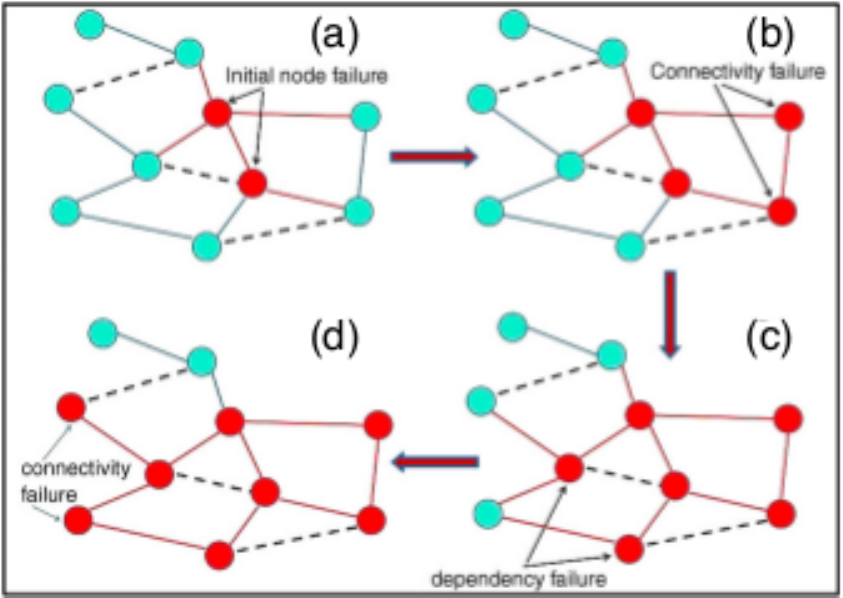}
\caption{(Color
online). Synergy between the percolation process and
the failures caused by dependency links. The network contains two types of
links: connectivity links (solid lines) and dependency links (dashed lines).
(a) The process starts with the initial failure of two nodes (marked in
red). The connectivity links attached to them also fail (marked in red).
(b) A percolation process takes place where all the nodes and the connectivity
links attached to the failed elements and that are not connected to giant cluster
(largest cluster) by connectivity links, also fail (marked in red). (c) The nodes connected by dependency
links to the failed nodes also fail  (marked in red). (d) Another step of connectivity
failure in which two more nodes fail because they are not connected to
the largest cluster (currently containing only two nodes). Reprinted with permission
from Ref.~\cite{Parshani2011}. Courtesy of R. Pashani.\label{Fig:Parshania2011}}
\end{figure}

Large-scale and/or
long term failures have devastating effects on almost every aspect in modern life, and especially on inter-dependent systems (such as telecommunication, gas and water supply, and transportation systems). The iterative process of
cascading failures can be described by the scheme of Fig.~\ref{Fig:Parshania2011} where the outbreak in panel (d)
is initiated by the node failure in panel (a) through a few intermediate steps \cite{Parshani2011}.
% figure 1

\begin{figure}
\centering \includegraphics[width=0.6\linewidth]{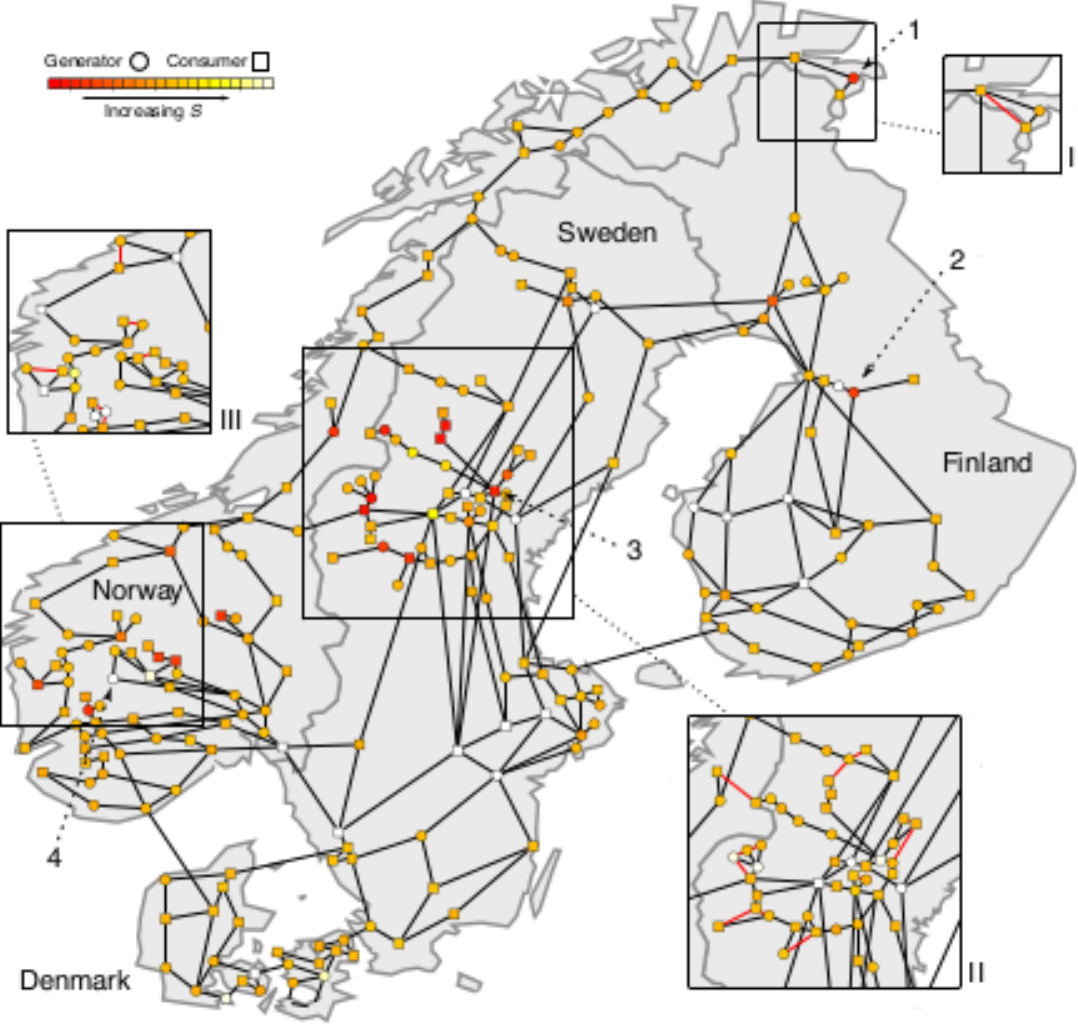}
\caption{(Color online) Schematic representation of the Northern European power grid. The grid has $N=236$ nodes and $L=320$ transmission lines.
For simplicity, the load scenario was chosen randomly, with squares (circles) depicting $N/2$ net consumers (net generators). Reprinted with permission from Macmillan Publishers Ltd: Ref.~\cite{Menck2014}, $\copyright$ 2014. \label{Fig:European}}
\end{figure}

A power grid is a network where links are power lines and nodes may be either generators (power plants) or consumers.
The generators produce the electric energy which needs to travel long distances in high-voltage transmission lines (e.g.
400 kV) in order to arrive the consumers or loads \cite{Nardelli2014,Chiang1995,Filatrella2008,Chiang1990}.
The nodes may therefore  consume, produce or distribute power, whereas the links transport power
and may include passive elements with resistance, capacitance and inductance.
A power grid is embedded in a 2D real space. Figure \ref{Fig:European} is a representation of the Northern
European power grid with $N=236$ nodes and $L=320$ connections \cite{Menck2014}.

A fundamental requisite for a power grid to operate in a stable way is
the maintenance of a phase-locked state within the entire system,
thus synchronization is understood as the ability of a power grid to keep a global state where
generators and loads (with intrinsic different frequencies) run with the same effective frequency
(e.g. 50/60 Hz). If loads are too strong (or unevenly distributed) or if some major
fault (or a lightening) occurs, a node may lose synchronization. In that situation, the synchronization
landscape may change drastically and a blackout may occur.

Therefore, power grids are generally
planned and operated to withstand the occurrence of certain disturbances. For this purpose, the
following conditions of dynamic security are being guaranteed \cite{Chiang1995}: 1) when any of
a specified set of disturbances occurs, the system survives the ensuing transient and moves into
a steady-state condition; 2) no bus voltage magnitudes during transients are outside their
permissible ranges; 3) in this new steady-state condition, no control devices, equipment or
transmission lines are overloaded and no bus voltage magnitudes are outside their permissible ranges
(say $5\%$ of nominal).

Now, it happens that a power system is continuously experiencing disturbances, such as outages,
short-circuits, sudden large load changes, or a combination of such events. Because of economic and
environmental pressures, power grids tend to run at their full utilization,
which increases the effects of these disturbances on their security.
For example, when several voltages collapse, a power system must operate even closer to the limits
of stability and then the risk of a major breakdown is increased
\cite{Chiang1990}.

Several models have been proposed to analyze the stability of power grids
\cite{Rohden2012a,Rohden2014,Dorfler2014,Filatrella2008,Lozano2012,Nardelli2014,Motter2013}.
In these models, each node-$i$ has the same equation of motion, and is characterized by a parameter $P_i$
quantifying the generated ($P_i>0$) or consumed ($P_i<0$) power. The state of each node is determined by its phase angle $\phi_i(t)$ and velocity $\dot{\phi}_i(t)$. During regular operations, generators and consumers run
with the same frequency $\Omega=2\pi\times 50$ Hz or $\Omega=2\pi\times 60$ Hz. The phase of each
node-$i$ is then written as
\begin{equation}
\label{grid-phase}
\phi_i(t)=\Omega t+\theta_i(t),
\end{equation}
where $\theta_i$ is a phase fluctuation.

For simplicity, one may assume that the nodes can be modeled as having the same dissipative coefficient
$\alpha$, the same moment of inertia, and that power is not lost during transmission.
By energy conservation, each generated or consumed power $P_i$
must be equal to the sum of the power taken from the grid ($P_i^{trans}$) plus the accumulated ($P_i^{acc}$)
and dissipated  ($P_i^{diss}$) power, i.e.
\begin{equation}
\label{power-grid1}
P_i = P_i^{trans} + P_i^{acc} + P_i^{diss}.
\end{equation}
Now, during its rotation a turbine dissipates energy at a rate proportional to the square of its angular
velocity \cite{Filatrella2008}
\begin{equation}
\label{power-grid2}
P_i^{diss} = k(\dot{\phi}_i)^2,
\end{equation}
and it accumulates kinetic energy $1/2 I(d\phi_i/dt)^2$ at a rate
\begin{equation}
\label{power-grid3}
P_i^{acc} = \frac{1}{2}I\frac{d}{dt}(\dot{\phi}_i)^2,
\end{equation}
where $I$ is the moment of inertia. The condition for power transmission is that the two devices do
not operate in phase, i.e. there is a phase difference between the two nodes connected by a line
$\Delta \theta_{ij}=\theta_j-\theta_i$. The transmitted power is proportional to the sinus of the phase
difference between the voltages of the nodes at the two end-points of the line \cite{Lozano2012},
which gives
\begin{equation}
\label{power-grid4}
P_i^{trans} = \sum_j-P_{ij}^{max}\sin(\phi_j-\phi_i),
\end{equation}
where $P_{ij}^{max}$ is an upper bound for the transmission capacity of a line. Substituting Eqs.~(\ref{power-grid2}), (\ref{power-grid3}) and ~(\ref{power-grid4}) into Eq.~(\ref{power-grid1}) one has
\begin{equation}
\label{power-grid5}
I\Omega \ddot{\theta}_i=P_i-k\Omega^2-2k\Omega\dot{\theta}_i+\sum_jP_{ij}^{max}\sin(\theta_j-\theta_i),
\end{equation}
which can be written into normalized units as
\begin{equation}
\label{power-grid6}
\ddot{\theta}_i=P'_i-\alpha\dot{\theta}_i +\sum_{j=1}^n W_{ij} \sin(\theta_j- \theta_i),
\end{equation}
where $P'_i, \alpha$ and $W_{ij}=\lambda A_{ij}$ are the normalized parameters, and $A_{ij}$ are the
elements of the adjacency matrix which accounts for the heterogeneity of the network. $P'_i$
is related to both generated  and dissipated power at the node, while $\lambda$ (the strength of the coupling) is related to the maximum transmitted power. Eq.~(\ref{power-grid6}) is
the basic equation considered in power grids dynamics. Notice that, when the inertia $I$ is negligible, one may ignore
the term $\ddot{\theta}_i$ in Eq.~(\ref{power-grid6}), which then returns to the classical 1970 Kuramoto model.

% figure 3
\begin{figure}
\centering \includegraphics[width=0.6\linewidth]{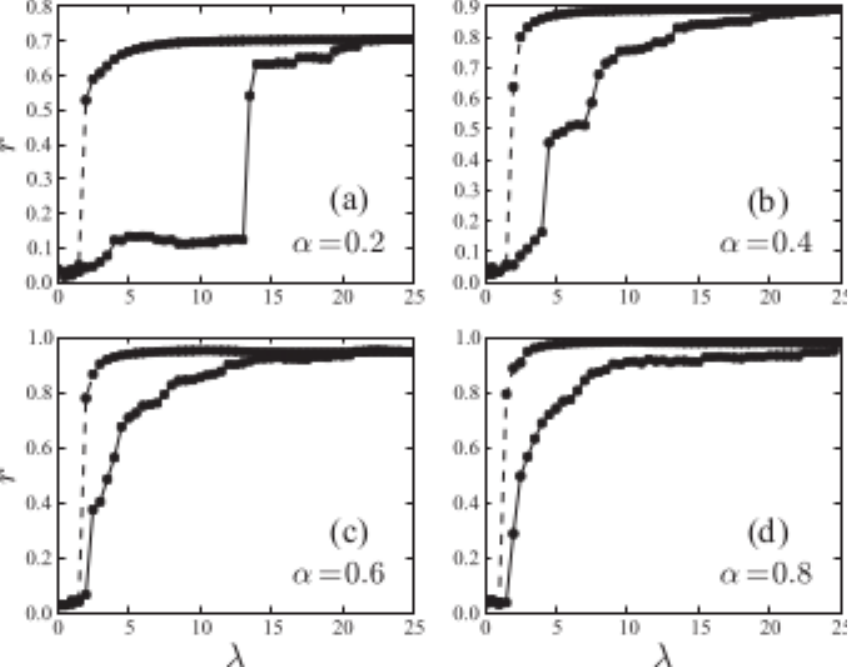}
\caption{Synchronization diagram $r(\lambda)$ for (a) $\alpha=0.2$,
(b) $\alpha=0.4$, (c) $\alpha=0.6$, and (d) $\alpha=0.8$. The network has  assortativity coefficient $A =-0.3$.
With increasing $\alpha$, the onset of synchronization and the hysteresis decrease.
The natural frequency of each oscillator is $P'_i=k_i-\langle k\rangle$ and the
networks have $N = 10^3$ and $\langle k\rangle= 6$. The degree distribution follows
a power law $P(k) \sim k^{-\gamma}$, where $\gamma=3$. Curves in which points are
connected by solid (dashed) line correspond to the forward
(backward) continuations of the coupling strength $\lambda$. Reprinted with permission
from Ref.~\cite{Peron2015}. $\copyright$  2015 by American Physical Society.\label{Fig:power1}}
\end{figure}

%\subsubsection{ Explosive synchronization in power grids }
In addition to load balance, synchronization between all elements of the grid is a crucial aspect to the system.
Recently, ES in Eq.~(\ref{power-grid6}) has been extensively studied \cite{Peron2015,Olmi2014,Ji2013,Ji2014}.
The stable operation of a power grid requires that all machines run at the same frequency $\Omega$. The
phases of the machines will generally be different, but the phase differences are constant in time. To study
the influence of dynamics and structure on global synchronization, $P'_i$ is generally assumed to satisfy a
distribution with zero average. For example, $P'_i$ can be explained as the natural frequency of node $i$, and
assumed to be of the form \cite{Peron2015,Ji2013,Ji2014}
\begin{equation}
\label{ji1}
P'_i=D(k_i-\langle k\rangle),
\end{equation}
where $k_i$ is the degree of node $i$, $\langle k\rangle$ the network average degree, and $D$ a proportionality
constant. That is, the natural frequency $P'_i$ of a node $i$ is assumed to be proportional to its degree.

The choice of Eq.~(\ref{ji1}) assumes that, in SF topologies, many nodes play
the role of consumers (nodes with $k_i<\langle k\rangle$), while nodes with high degrees play the role of
power producers (nodes with $k_i>\langle k\rangle$). In this framework, the relation $\sum_i P'_i=0$
is satisfied, which means that the total consumed power is equivalent to the total generated power.
In this latter case, all oscillators try to rotate independently at their own natural frequencies, while
the coupling $\lambda$ tends to synchronize them to a common phase. Besides the degree distribution $P(k)$, it is know that real
world network can display either assortativity or disassortativity degree mixing behavior.
When degree-degree correlations are accounted for, it is found that the synchronization diagrams have a strong
dependence on the network assortativity, indicating that one is able to control the hysteretic behavior
of the second-order Kuramoto model by tuning the network properties \cite{Peron2015}.

% figure 4
\begin{figure}
\centering \includegraphics[width=0.6\linewidth]{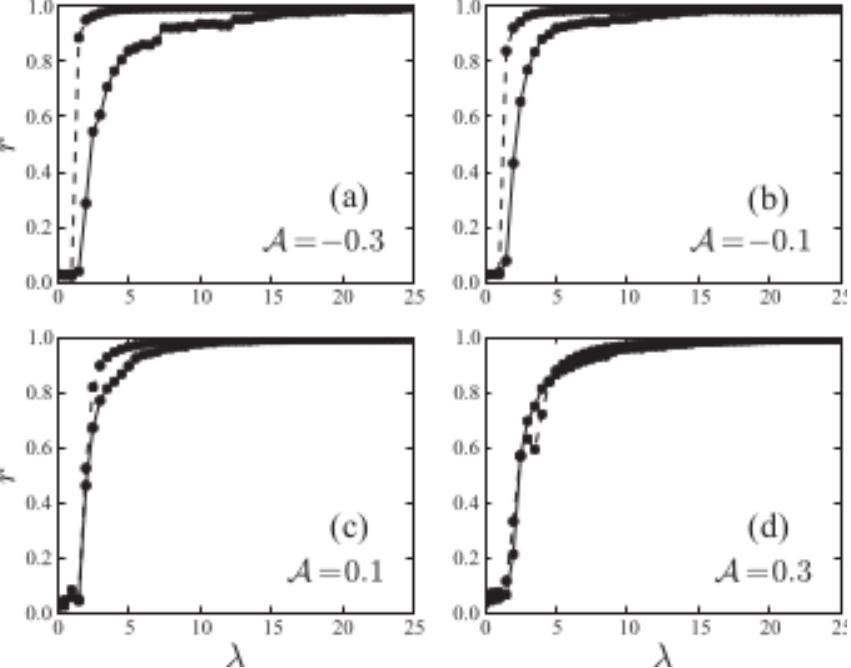}
\caption{Synchronization diagram $r(\lambda)$ for (a) ${\cal A}=
  -0.3$, (b) ${\cal A} =-0.1$, (c) ${\cal A} = 0.1$, and (d) ${\cal A} = 0.3$.
The dissipation coefficient is fixed at $\alpha= 1$. All other network parameters are in the caption of
Fig.~\ref{Fig:power1}. Reprinted with permission from Ref.~\cite{Peron2015}. $\copyright$  2015 by American Physical Society.
\label{Fig:power2}}
\end{figure}

In numerical simulations, all the initial networks are constructed through the BA model with $\langle k\rangle= 6$
and $N = 10^3$. Equation~(\ref{power-grid6}) is integrated and the
Kuramoto
order parameter $r(t)$ (Eq.~(\ref{eq:kuramparam}))
% \begin{equation}
% \label{order}
% r(t)e^{i\psi(t)}=\frac{1}{N}\sum_{j=1}^Ne^{i\theta_j(t)}
% \end{equation}
is calculated after a transient process. %To observe its ES, we calculate the dependence of $r$ on the coupling $\lambda$
%by increasing the value of $\lambda$ adiabatically.
Figure \ref{Fig:power1} shows the forward and
backward synchronization diagrams $r(\lambda)$ for different values of $\alpha$ within the
interval $[0.2,1]$. Panels (a)-(d) represent the cases of $\alpha=0.2, 0.4, 0.6$ and $0.8$, respectively.
It is easy to see that there is a hysteresis loop in all the four
cases, indicating that ES is taking place in the network.
From Fig.~\ref{Fig:power1} we also see that the area of hysteresis and
the critical coupling for the onset of synchronization in the forward branch tends to decrease as $\alpha$ is
increased, which also contributes to increase the maximal value of the order parameter.

The  network's degree-degree correlations can be quantified by the
assortativity coefficient $\cal A$ defined in Eq.~(\ref{eq:assortativity_coef}).
In order to inspect the influence of network's degree-degree correlation on ES in Eq.~(\ref{power-grid6}),
one needs to change the assortativity of the network, which can be implemented as follows \cite{Peron2015}:
at each step, two edges are selected at random and the four
nodes associated to the selected pair of edges are ordered from the lowest to the highest degree.
Given a target assortative mixing (${\cal A} > 0$), with probability $p$ a new edge is formed between the first and second nodes
and another one between the third and fourth nodes. If one of the two new
edges already exists, the step is discarded and a new pair of edges is chosen.
A similar heuristic mechanism can also generate disassortative
networks (${\cal A} < 0$). Figure \ref{Fig:power2}
shows the synchronization diagram $r(\lambda)$ for networks with different values of assortativity, where panels
(a)-(d) represent the cases of ${\cal A}=-0.3, -0.1, 0.1$ and $0.3$,
respectively. As ${\cal A}$ increases,
the hysteresis becomes smaller and smaller, and the critical coupling of the increasing branch is weakly affected.

The natural frequencies can also be chosen from different distributions. For instance, Olmi et al. considered
the case of a Gaussian distribution $g(P') = \frac{1}{\sqrt{2\pi\sigma^2}}e^{-P'^2/2\sigma^2}$ with zero
average and unitary standard deviation $\sigma$ \cite{Olmi2014}. It is observed that, in the hysteretic
region, clusters of locked oscillators of various sizes and different levels of synchronization coexist.

\begin{figure}
\centering \includegraphics[width=0.6\textwidth]{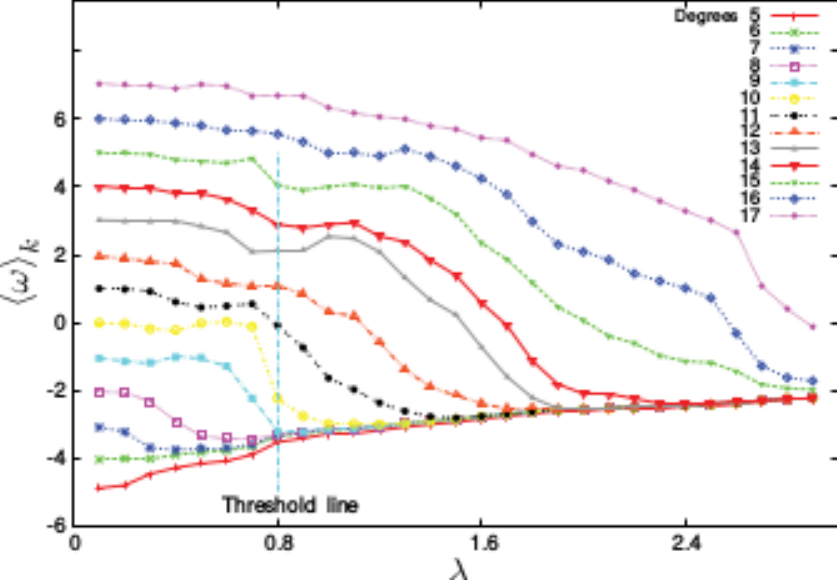}
\caption{(Color online). Cluster explosive synchronization in uncorrelated SF networks. Vertical axis reports the average frequency
of nodes of various degrees vs. the coupling strength $\lambda$ (see the legend for the color codes of the different curves). Reprinted with permission
from Ref.~\cite{Ji2013}. $\copyright$  2013 by American Physical Society. \label{Fig:cluster}}
\end{figure}

It is important to remark that the process of ES in Eq.~(\ref{power-grid6}) is quite different from that of the Kuramoto
model. It is revealed that in the case of uncorrelated SF networks, nodes join the synchronous
component progressively, grouped into cluster of nodes with the same degree, starting from small degrees.
The phenomenon was therefore called {\sl cluster explosive synchronization} \cite{Ji2013,Ji2014}, and the difference
with the case of the Kuramoto model is that in the latter the average
frequency $\langle \omega\rangle_k$ jumps to $\langle k\rangle$ for all $k$ at the same time.
Figure~\ref{Fig:cluster} reports the synchronization process and highlights the progressive presence of clustering.
% figure 5

\subsubsection{Smart grids}

An imperative condition for the reliable functioning of power-grids is that the generators
remain synchronized. This condition is relatively easy to be attained in traditional power
grids, where the power plants are mainly based on coal, gas, oil or nuclear power. However, in
recent years, more and more distributed energy sources and self-sufficient micro-grids are included in the network,
such as wind turbines, photovoltaic arrays, bio-gas power generators, and other
renewable energy sources. Such modern grids often consist of thousands of power substations and
generators linked across thousands of kilometers, and are therefore much larger (and more complicated) than
the traditional ones.

They are called {\it smart grids}, because they make use of information and communication technologies (ICTs)
to collect information (such as the behaviors of suppliers, consumers, and prosumers) and to take actions in an automated fashion
to improve the efficiency, reliability, economics, and sustainability of the production and
distribution of electricity \cite{Rohden2012,Nardelli2014,Rohden2014,Motter2013,Lozano2012}.

A key problem of smart grids is their stability. For instance, any person who has a solar panel at home,
may become an energy trader, i.e. buy energy when the price is low, store it and then resell when
the price is high, resulting in a new market within smart grids. As markets are known to be unstable,
their influence to power grids is serious. Another element of instability is the fluctuation of renewable
energy sources, as for instance wind turbines which depend significantly on the environment and season.
All these factors introduce disturbances and fluctuations in production and
demand, which may trigger desynchronization of power generators. Once it happens, the connected
generators cannot be in pace and thus induce a cascading failure, which may be the reason behind the increasing number of power
blackouts reported in recent years. %Understanding this mechanism will be the key for the question of how to ensure stable operation of the entire grid.

A characteristic feature of smart grids is their large fluctuations in energy.
Notice that in the case of traditional power grids, we have a balance between the generators and the consumers, i.e. from
Eq.~(\ref{ji1}) we have $\sum P'_i=0$. However, in the case of smart grids, this balance may not hold,  and thus one has
\begin{equation}
\label{power-grid8}
\sum P'_i>0 \quad or \quad \sum P'_i<0.
\end{equation}
In the following, without loss of generality, we describe the case of $\sum P'_i>0$.
%To address
%its effect, we first consider the case of two coupled nodes, i.e. one generator
%coupled with one consumer \cite{Rohden2014}.

\begin{figure}
\centering \includegraphics[width=0.6\textwidth]{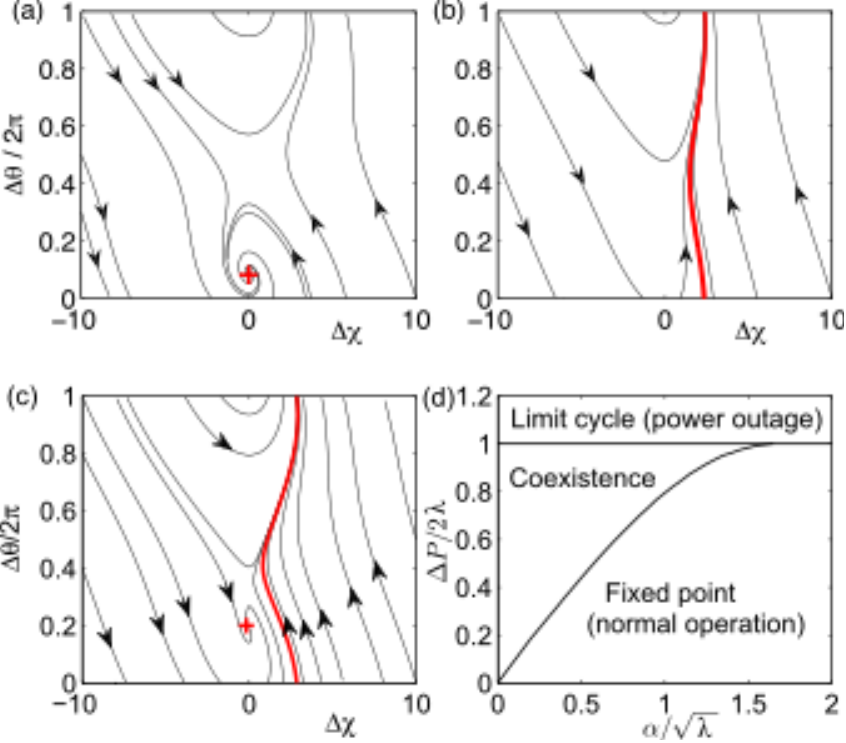}
\caption{(Color online). Scenarios of the dynamics for an elementary network with one generator and one
consumer for $\alpha=1$. (a) Globally stable phase locking for $\Delta P=\lambda=2$.
(b) Globally unstable phase locking (limit cycle) for
$\Delta P=2$ and $\lambda=0.5$. (c) Coexistence of phase locking (normal operation)
and limit cycle (power outage) for $\Delta P=2$ and $\lambda=1.1$. (d)
Stability phase diagram in the parameter space. Both stationary
solutions (fixed point and limit cycle) are reported in red. Reprinted from Ref.~\cite{Rohden2014}, with the permission of AIP Publishing. \label{Fig:two-nodes}}
\end{figure}

The simplest non trivial grid is a two-element system consisting of one generator and one consumer
\cite{Rohden2014}. In the phase-locked state, both derivatives
$\dot{\theta}_i$ and $\ddot{\theta}_i$ are zero, and from Eq.~(\ref{power-grid6}) one has
\begin{equation}
\label{power-grid90}
0=P'_i +\sum_{j=1}^2 W_{ij} \sin(\theta_j- \theta_i).
\end{equation}
For the sum over all the equations gives,
\begin{equation}
\label{power-grid9}
0=\sum_i P'_i + W_{ij} \sin(\theta_j- \theta_i)+ W_{ij} \sin(\theta_i- \theta_j)=\sum_i P'_i.
\end{equation}
Therefore, the system can only reach equilibrium when Eq.~(\ref{power-grid9}) is satisfied, i.e. when $P'_1+P'_2=0$.
When instead $P'_1+P'_2>0$, one defines $\Delta P=P'_1+P'_2$ and $W_{ij}=\lambda a_{i,j}$, with $\lambda$ being the
coupling strength. Equation ~(\ref{power-grid6}) can be then rewritten as
\begin{eqnarray}
\label{power-grid10}
\Delta \dot{\chi} &=&\Delta P-\alpha \Delta \chi -2\lambda \sin(\Delta \theta),\nonumber \\
\Delta \dot{\theta} &=& \Delta \chi,
\end{eqnarray}
where $\Delta \theta=\theta_2-\theta_1$ and $\Delta \chi=\Delta \dot{\theta}$.
Figure~\ref{Fig:two-nodes}(a)
shows the case of $2\lambda>\Delta P$, where there are two fixed points. One fixed point is stable and the other
is unstable, therefore all trajectories converge to the stable fixed point. Figure~\ref{Fig:two-nodes}(b) shows the
case of $2\lambda<\Delta P$, where there are no fixed points, and all trajectories converge to a limit cycle.
Figure~\ref{Fig:two-nodes}(c) shows the case of $2\lambda\approx \Delta P$, where the fixed point and the limit
cycle coexist such that the dynamics depends crucially on the initial conditions. Figure~\ref{Fig:two-nodes}(d)
shows the parameter space of the systems. % In the upper area (for $2K<\Delta P$), one only has the limit
% cycle; below for $2K>\Delta P$ is the coexistence regime on the left hand side for small $\alpha/\sqrt{K}$,
% on the right hand side the fixed point.
The majority of power grids are operating close to the edge of stability,
i.e. in the region of coexistence, at least during periods of high loads \cite{Rohden2014}. Therefore,
the dynamics depends crucially on the initial conditions, and static power grid models are yet insufficient to fully capture the complexity of the system.

% figure 6

\subsubsection{Brain dynamics}

Synchronization plays an important role in sustaining basic brain functions such as emotions, complex thoughts,
memory, language comprehension, consciousness, etc. \cite{Buzsaki2006}. For instance, the importance of
synchronization of oscillatory phases between different
brain regions in memory processes has been demonstrated \cite{Fell2011}. On the other side, clinical evidence points out
that abnormal synchronization of a small group of neurons may result in ruining some of the brain functions,
leading to pathological behaviors such as Parkinson disease, tremor, and epilepsy.

Let us take epilepsy as an example. The disease manifests itself with recurrent unprovoked seizures resulting from
a wide variety of causes, and is also the world's most prominent serious brain disorder. Unless other
neurological problems exist, patients with epilepsy typically have normal neurological function
in between the seizures. Understanding the transition between the normal function and the seizure is useful
for both improving the treatment of epilepsy and providing early warnings to patients.

Epilepsy has four phases: the interictal period, the onset of the seizure, and the propagation and termination phases. It is
found that the coupling between brain areas during seizures changes with time, increasing or decreasing at the seizure onset in
different cases \cite{Kramer2012}. At variance, when seizure termination is approached, the coupling of brain activity always increases.
Both the onset and termination processes are very fast, indicating a jumping
phase transition at critical coupling. The sudden emergence and termination of seizures suggested
therefore that ES may play a role \cite{Yaffe2015}.

A seizure originates in a focal region, and then spreads to other brain regions. The individuation of the focal region from
which an epileptic seizure is originated was performed in Ref.~\cite{BenJacob2007}
in terms of phase synchronization indicators. During seizures,
there is an overall increase of synchrony between the thalamus and temporal lobe structures. The
phenomenon of seizure can be studied by means of a network approach, where a node can be either an individual neuron, or a
population of neurons within a given structure, or an entire brain structure or region \cite{Yaffe2015}.

Guye et al. selected thirteen patients undergoing pre-surgical evaluation of drug-resistant
temporal lobe epilepsy, and checked their recordings of intracerebral electrodes \cite{Guye2006}. % The
% upper part of Fig.~\ref{Fig:Guye:2006} shows the time series from one electrode, and the lower part
% shows its corresponding time-frequency representation (see the figure's caption for its definition).
Figure 2 of Ref.~\cite{Guye2006} reveals that, during the seizure, synchronization values are significantly
higher than those of the background period, indicating a jumping transition. The abruptness of the transition
in the seizure has been also confirmed by many other studies \cite{Kramer2012,Yaffe2015,Perucca2014,Schindler2007}.

\begin{figure}
\begin{center}
\includegraphics[width=0.7\textwidth]{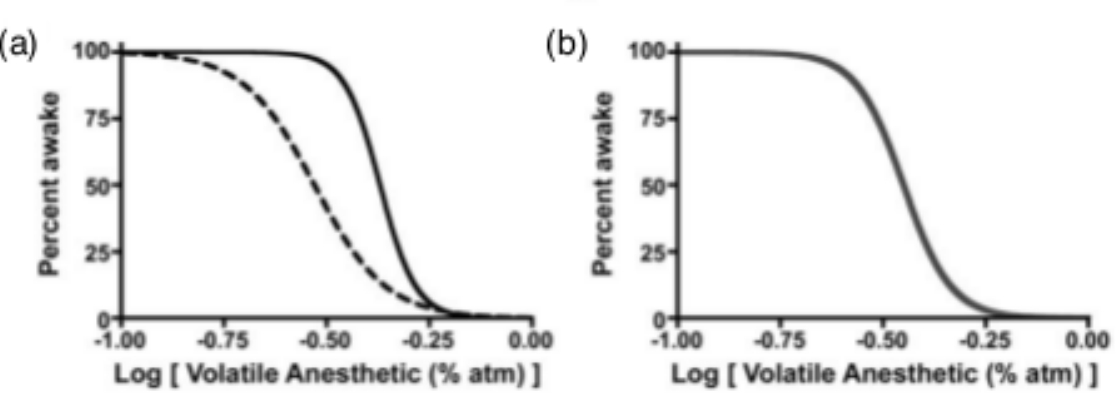}
\end{center}
\caption{Path-dependent and path-independent state transitions. (a) The solid black curve represents
a population of individuals entering the state of unconsciousness as a function of the anesthetic dose.
The dashed black curve depicts the reverse transition. The situation is here clearly characterized by the presence of
a hysteretic loop. (b) In the absence of hysteresis, the forward and reverse paths are superimposed (thick gray curve).
Reprinted with permission from Ref.~\cite{Friedman2010}, published under CC BY license.
\label{Fig:Friedman:2010}}
\end{figure}

Another example is the anesthetic-induced unresponsiveness, which may result from specific interactions
of anesthetics with the neural circuits regulating sleep and wakefulness. One common belief is that
emergence from anesthesia is the inverse process of induction. Friedman et al.
made experiments on transitions between conscious and unconscious states \cite{Friedman2010}. By
generating anesthetic dose-response data in both insects and mammals, Ref.~\cite{Friedman2010} demonstrated that the
forward and backward paths through which anesthetic-induced unconsciousness arises and dissipates are
not identical. Instead, a hysteresis is exhibited that isn't fully explained by pharmacokinetics, as
previously thought. Figure~\ref{Fig:Friedman:2010} shows the results
for both the cases of path-dependent (a) and path-independent (b) state transitions. In particular, Fig.~\ref{Fig:Friedman:2010}(a) displays
a hysteresis loop, in contrast to the common belief.

% figure 2

\begin{figure}
\begin{center}
\includegraphics[width=0.7\textwidth]{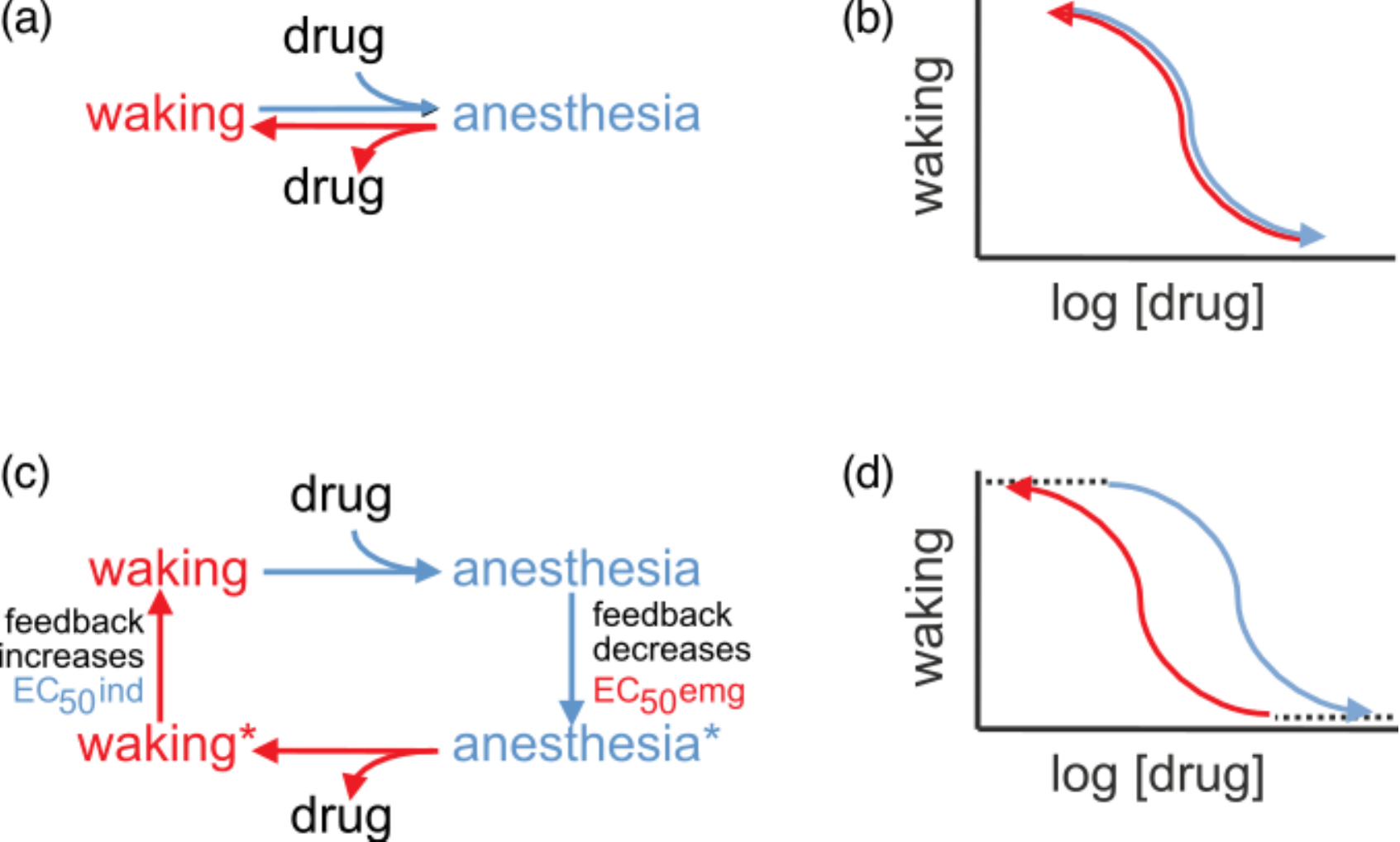}
\end{center}
\caption{(Color online). (a) Simple kinetic model describing drug-dependent behavioral state changes in the absence of
bistability. (b) In the absence of feedback and bistability, dose-response curves for anesthesia induction and
emergence are independent of the history of the prior behavioral state, and thus coincide. (c) Addition of feedback upon
binding or unbinding of drug leads to additional, more stable, anesthesia and waking states. (d) Feedback and
bistability lead to hysteresis in dose-response curves for anesthesia
induction and emergence. Reprinted with permission from Ref.~\cite{Joiner2013}, published under CC BY license.
\label{Fig:Joiner:2013}}
\end{figure}

Joiner et al. provided further experimental evidence of the existence of hysteresis in the
anesthetic-induced unresponsiveness \cite{Joiner2013}. The study proposes that processes selectively contributing
to neural inertia may be impaired in pathophysiological conditions. Certain features exist in a minimal neural
circuit that underly neural inertia, and can be used to model the dependence of anesthesia on feedback bistability. Figure \ref{Fig:Joiner:2013} represents a sketch of the models from Ref.~\cite{Joiner2013}.
The sketch (a) is a simple kinetic model
describing the transitions between two states, one unbound and the other bound to drug; (b) is the resulting
dose-response curves for the forward and reverse reactions; (c) is the bistable situation where distinct
feedback mechanisms are activated to shift drug sensitivity toward stabilization of the state; and (d) is
the hysteresis in dose-response curves for anesthesia induction and emergence.

% figure 3

Kim et al. recently hypothesized that the conditions for ES in human brain
networks would be present in the anesthetized brain just over the threshold of unconsciousness
\cite{Kim2016}. To test this hypothesis, Ref.~\cite{Kim2016} constructed functional brain networks from multi-channel
electroencephalogram
 recordings in seven healthy subjects across conscious, unconscious, and
recovery states, and demonstrated for the first time that the network conditions for ES (especially the suppressive rule of Ref.~\cite{Zhang2014}) are present in empirically-derived
functional brain networks.

Together with the experimental studies for the hysteresis loop in brain functioning, some theoretical
works have also tried to understand the underlying mechanism of ES in
neural systems \cite{Chen2013}. For instance, Chen et al.  investigated the property of phase
synchronization transition of coupled FitzHugh-Nagumo  oscillators in BA SF networks,
and reported the existence of ES. The model reads as

\begin{eqnarray}
\label{FitzHugh-Nagumo}
\epsilon \dot{x}_i &=&x_i-x_i^3-y_i+\lambda\sum_{j=1}^NA_{ij}(x_j-x_i)+\xi_i^{(x)}(t),\nonumber \\
\dot{y}_i &=& x_i+a_i+\xi_i^{(y)}(t),
\end{eqnarray}
where $i=1,2,\cdots,N$, $\epsilon= 0.01$, and $x$ and $y$ are the fast and the slow variables, respectively. The
parameter $a_i$ describes the excitability of the $i$th unit. If $|a_{i}| > 1$, the system is excitable,
while $|a_{i}| < 1$ implies that the system is oscillatory. $A_{ij}$ are the elements of the adjacency matrix of the
network, $\lambda$ is the coupling constant, and $\xi_i^{(\alpha)}(t)$ is a Gaussian noise that is independent
for different units and satisfies $\langle \xi_i^{(\alpha)}(t)\rangle =0$ and
$\langle \xi_i^{(\alpha)}(t)\xi_j^{(\alpha)}(t')\rangle =2D_{\alpha}\delta_{ij}\delta(t-t')$ with
noise intensity $D_{\alpha}$, and $\alpha\in \{x,y\}$. The natural frequency $\omega_i$ is assumed to
be an increasing function of the degree $k_i$, by taking $a_i$ as follows
\begin{equation}
\label{excitability}
a_i=0.99-\delta\frac{k_i-k_{min}}{k_{mas}-k_{min}},
\end{equation}
where $k_{max}$ and $k_{min}$ are the maximum and minimum degrees
in the network, respectively. The factor $\delta$ determines the
slope of the linear expression. The larger is $\delta$, the wider the distribution
of $a_i$. % has, or equivalently, a wider frequency distribution.
% figure 4
\begin{figure}
\begin{center}
\includegraphics[width=0.4\textwidth]{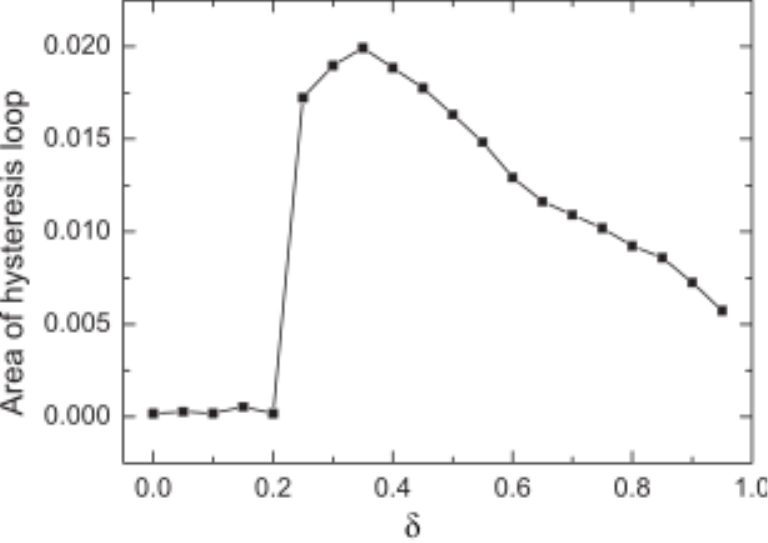}
\end{center}
\caption{Area of hysteresis loop (as a function of $\delta$) in the synchronization diagram
of Eqs.~(\ref{excitability}) for BA SF networks. Parameters are $N=200, \epsilon=0.01, D_x=0, D_y= 0.005$,
and the average degree $\langle k\rangle=6$. Reprinted with permission from Ref.~\cite{Chen2013}.
\label{Fig:Chen:2013}}
\end{figure}

The synchronization diagram can be obtained by performing both the forward and the backward simulations. Chen
et al. showed that, for BA networks, the nature of synchronization transition drastically changes with
$\delta$ \cite{Chen2013}. When $\delta$ is relatively small, the synchronization transition is continuous.
Increasing $\delta$ to a middle value, one can see that as $\lambda$ increases the order
parameter $r$
abruptly jumps from $r\approx 0$ to $r\approx 1$ at $\lambda_{c}^{f}=0.044$, and therefore a sharp transition takes place
at the onset of synchronization. On the other hand, the curve corresponding to the backward simulations also
shows a sharp transition from the synchronized state to the incoherent one at $\lambda_{c}^{b}=0.021$. The two sharp
transitions occur at different values of $\lambda$, determining a strong hysteresis loop. To measure the influence of $\delta$ on
ES, Chen et al. calculated the area of hysteresis loop, and found that,
increasing further $\delta$ to $\delta=0.9$, the first-order nature of the phase transition is still
present, but the area of the hysteresis loop becomes smaller \cite{Chen2013}. Figure ~\ref{Fig:Chen:2013}
shows how the area changes with $\delta$:  one can see that the
area vanishes when $\delta=0.2$, implying that the synchronization transition is there of
second-order type. When $\delta$ is increased to $\delta=0.25$, this area drastically changes to a non-zero value,
indicating that the transition changes from a second-order type to a first-order one at around
$\delta\approx 0.25$. With further increasing $\delta$, this area shows a non-monotonic dependence
on $\delta$, and a maximum area occurs at $\delta=0.35$.

\subsubsection{Biological systems}

Together with the example of the FitzHugh-Nagumo model of Eq.~(\ref{FitzHugh-Nagumo}), other models of relevance for biological systems have been shown to display ES. Recently Chen  et al. \cite{Chen2015}
reported that ES (as well as frequency-degree correlation properties)
can take place in a very wide range of oscillatory networks, without
any particular structure constraint. In Ref.~\cite{Chen2015},
diffusively coupled SF networks of complex Ginzburg-Landau oscillators (CGLE) \cite{Nakao2009} are used,
\be
\dot{z}_j=z_j-(1+i\omega_o) \vert z_j \vert ^2 z_j + \lambda (1+i\beta) \sum_{l=1}^{N} a_{lj} (z_l-z_j),
\label{eq:GL}
\ee
where $\omega_0$ is the oscillator's natural frequency, $\beta$ is a dispersion parameter. Both parameters are
taken to be identical for all nodes.  CGLE can generate rather complicated dynamics, including chaos, and different ways exist for the definition of frequency. In Ref.~\cite{Chen2015} the {\it principal frequency} of node $j$ ($\omega_j$) is defined as the value where the Fourier spectrum of a long time series of the dynamics of node $j$ has the highest peak.

\begin{figure}
\centering \includegraphics[width=0.95\textwidth]{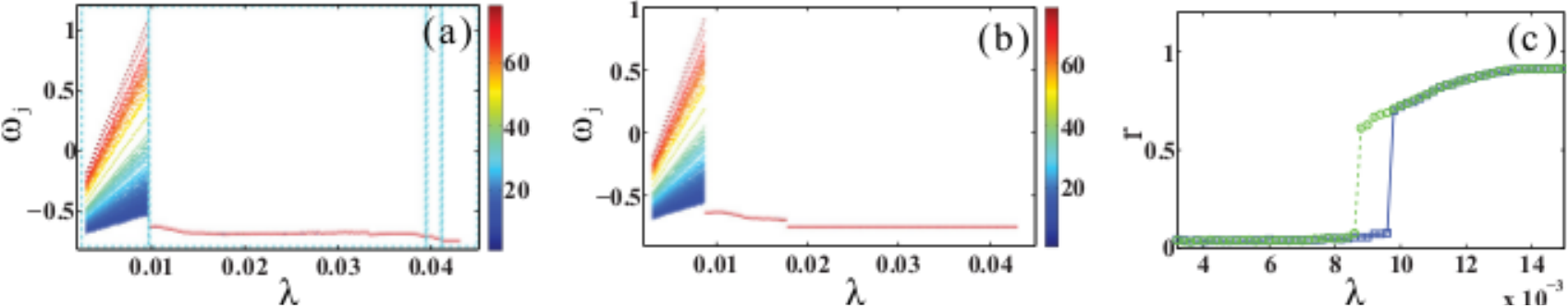}
\caption{(Color online). Principle frequencies
  $\omega_j$ vs. the coupling $\lambda$ for all nodes corresponding to
  (a) the forward transition, and (b) the backward transition for SF networks. The colors
  represent the degree of the nodes. Panel (c) reports the
  corresponding phase order parameter of panels (a) and (b) $r$ for
  the forward (blue squares) and backward (green circles)
  transitions. Parameters used in Eq.~(\ref{eq:GL}): $\omega_0=0.75$,
  $\beta=-1.7$, $N=200$, and $\langle k\rangle=20$.  Adapted with permission from Ref.~\cite{Chen2015}. Courtesy of S. Wang.\label{fig:Chen2015}}
\end{figure}

Figures~\ref{fig:Chen2015}(a),(b) report the main results of the integration of Eq.~(\ref{eq:GL}). The principal frequencies $\omega_j$ of all nodes are there reported vs. the coupling strength $\lambda$. Panel (a) corresponds to the forward transition, panel (b) to the backward  continuation. One observes a ES transition in the frequency domain. Before the transition, increasing of the coupling strength spontaneously induces a strongly inhomogeneous and well-ordered frequency distribution, with a positive correlation with the node degree. Defining the phase of node $j$ as $ \theta_j=\arctan[Im(z_j)/Re(z_i)]$, the authors compute also the phase order parameter $r$.  The result is shown in Fig.~\ref{fig:Chen2015}(c), where one notices the presence of a well pronounced hysteresis.

%%%%%%%%%%%%%%%%%%%%%%%%%%%%%%%%%%%%%%%%%%%%%%%%%%%%%%%%%%%%%%%%%%%%%%%%%%%%%%%%%%%%%%

 In Ref.~\cite{Bi2014}  Bi et al. investigated explosive oscillation death in  another well known model of amplitude oscillators.
 The oscillation death  refers to the complete  suppression of oscillations in coupled systems. This phenomenon has been extensively studied, both theoretically and experimentally \cite{Matthews1991,Atay2003}.  Ref.~\cite{Bi2014} considered a model of globally coupled Stuart-Landau oscillators with the frequency-weighted coupling term inspired by Ref.~\cite{Zhang2013}:
\be
\dot{z}_j=(a+i\omega_j-\vert z_j \vert ^2) z_j + \lambda\frac{\vert \omega_j \vert}{N}  \sum_{l=1}^{N} (z_l-z_j),
\label{eq:SL}
\ee
where $z(t)_j=x(t)_j+iy(t)_j$ is a complex variable, and $a$ acts as the control parameter for the individual Stuart-Landau oscillator:
the dynamics settles on a limit cycle if $a$ > 0, and on a
fixed point if $a$ < 0. At variance with Ref.~\cite{Chen2015},  oscillators here are not identical, and their natural frequencies are randomly taken from a frequency distribution $g(\omega)$.

\begin{figure}
\centering \includegraphics[width=0.3\textwidth]{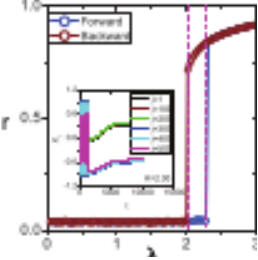}
\caption{(Color online). Characterization of explosive oscillation
  death in a Stuart-Landau model (Eq.~(\ref{eq:SL})). Order parameter $r$ vs.
coupling strength $\lambda$ for a triangular $g(\omega)$. $N$ = 500. Inset: evolution of the variable $x_j(t)$ for all the nodes at the transition point. Adapted from Ref.~\cite{Bi2014} with permission of IOP.}
\label{fig:Bi2014-1}
\end{figure}

Explosive oscillation dead was reported to occur in the solutions of Eq.~(\ref{eq:SL}). Fig.~\ref{fig:Bi2014-1} shows a clear discontinuous jump in the forward and backward transitions for the order parameter $r e^{i\psi}=\sum_{j=1}^{N} z_j(t)/N$. The inset reports the evolution of the variable $x_j(t)$ for all the nodes at the transition point. It can be observed that not all the nodes collapse to the same fixed point.

 %%subsubsec334
\subsubsection{Controlled laboratory experiments}

Some few experimental evidences of ES have been given in controlled laboratory systems, which actually are of importance as they ultimately demonstrate the robustness and ubiquitousness of the phenomenon.

The first of these experiments was published by Leyva et al. in Ref.~\cite{Leyva2012}, and involved the search for explosive phase synchronization in frequency-degree correlated heterogeneous networks of electronic circuits. Initially, Ref.~\cite{Leyva2012} performed numerical simulations of an ensemble of $N=1,000$ piece-wise linear R\"ossler units, interacting in a SF network via a bidirectional diffusive-like coupling \cite{Pisarchik2006,Pisarchik2009}:
\begin{eqnarray}
\dot{x_i} &=& -\alpha_i \left[ \Gamma \left( x_i-\lambda\sum_{j=1}^N a_{ij}(x_j-x_i) \right) + \beta y_i + \rho z_i \right] \nonumber \,, \\
\dot{y_i} &=& -\alpha_i(- x_i + \nu y_i) \,, \\
\dot{z_i} &=& -\alpha_i(-g(x_i)+z_i) \nonumber \,,
\label{eq:Ross_model}
\end{eqnarray}
where the piece-wise linear part is given by
\begin{equation}
g(x_i)=\left\lbrace \begin{array}{cc}
0 & \mbox{if $x_i\leq 3$} \\
\mu(x_i-3) & \mbox{if $x_i > 3$}
\end{array} \right. \ .
\end{equation}

Every node (indexed by $i=1,\ldots,N$) is here represented by an associated three-dimensional vector ${\bf x}_i (t) \equiv (x_i(t), y_i(t), z_i(t))$. The parameters are: $\Gamma=0.05$, $\beta =0.5$, $\rho=1$, $\mu = 15$,  and $\nu = 0.02-\frac{10}{R}$, where $R$ is a tunable quantity that regulates the dynamical state of the system. In particular, $R$ induces a chaotic dynamics  in the range $R = [55, 110]$ \cite{Pisarchik2006}.

Finally, the natural oscillation frequency of node $i$ depends linearly on the parameter $\alpha_i$. To impose a degree-frequency positive correlation in this system, the $\alpha_i$ values (and therefore, the oscillators' frequencies) are distributed following the relation
\begin{equation}
\alpha_i=\alpha \left(1+\Delta \alpha \frac{k_i-1}{N}\right)\,,
\label{eq:exp_corre}
\end{equation}
where $\alpha=10^4$, and $\Delta \alpha$ is a factor that determines the slope of the linear distribution. The range of frequencies in the ensemble becomes wider and wider as $\Delta \alpha$ and $k_{\max}$ (the maximum degree in the network) are increased, i.e. as more heterogeneous degree distributions (and/or steeper slopes) are considered.

\begin{figure}
\centering\includegraphics[width=0.7\textwidth]{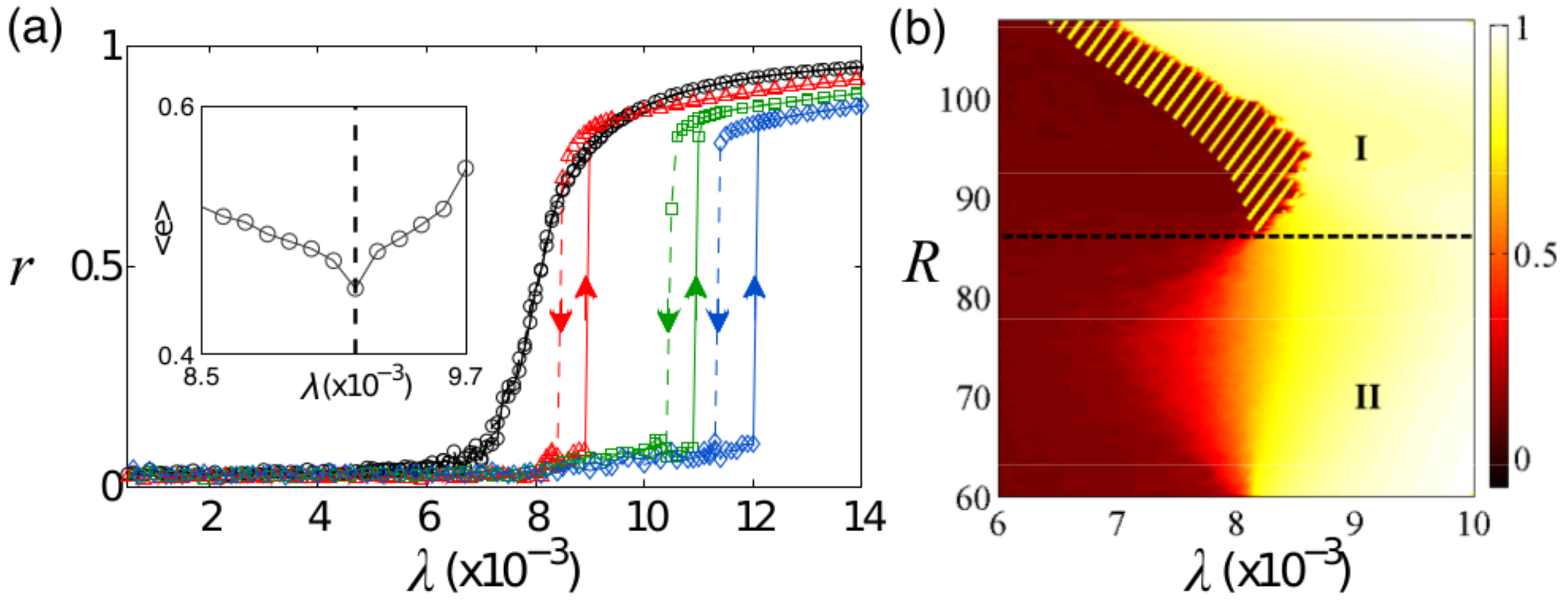}
\caption{ (Color online). (a) Phase  synchronization degree $r$ as a function of the coupling strength $\lambda$ for different SF networks of size
$N=1,000$, and average degree $\langle k\rangle = 6$, $\gamma=2.2$ (red
triangles), $\gamma=2.5$ (green squares), $\gamma=3.0$ (blue diamonds
and black circles). The correlation between node degree and natural
frequency is set via Eq.~(\ref{eq:exp_corre}). $\Delta\alpha=6.0$ and $R=70$ for
the black circles case, while $\Delta\alpha=10.0$ and $R=100$ for the other
networks. Continuous (dashed) lines mark the forward (backward)
simulations, as $d$ is increased (decreased) in steps of $\Delta
\lambda=3 \times 10^{-4}$. The inset reports the average synchronization error $\langle e \rangle$ vs.~$\lambda$ in the proximity of the first-order transition occurring for $\gamma=3.0$, $R=100$ and $\Delta\alpha=8.0$. (b) Mean synchronization degree $r$ in the parameter space $\lambda-R$. A second-order transition occurs for values of $R$ below the horizontal dashed
line. Above the same dashed line, the transition is instead of the first-order type,
exhibiting the typical hysteresis (striped area) in the forward and
backward simulations with $\Delta \lambda=5 \times 10^{-5}$. The entire phase diagram refers to the case of a scale free network with $N=1,000$, $\langle
  k\rangle = 6$, $\gamma=3$ and $\Delta\alpha=6.0$. Reprinted from Ref.~\cite{Leyva2012}. $\copyright$  2012 by the American Physical Society.
  }
\label{fig:Leyva2012-1}
\end{figure}

Reference~\cite{Leyva2012} shows that the system (\ref{eq:Ross_model}) gives rise to a ES transition, for an appropriate selection of the parameters.
Precisely, the instantaneous phase for each oscillator $i$ is defined
as $\theta_i(t) = {\arctan} (y_i(t)/x_i(t))$, and the usual order
parameter $S$ is monitored as a function of $\lambda$. The results are
shown in Fig.~\ref{fig:Leyva2012-1}(a), for a $N=1,000$,
$\langle k\rangle=6$ SF network obtained by the BA growing algorithm
for different values of the power-law exponent $\gamma$ of the degree distribution. The inset reports the amplitude average synchronization error $\langle e \rangle$ in the proximity of the transition for $\gamma=3.0$, $R=100$ and $\Delta\alpha=8.0$, to confirm that the system does not reach full synchronization when ES occurs.

The specific chaotic state of the nodes is a relevant variable, as it can be seen from the comparison (see Fig.~\ref{fig:Leyva2012-1}(a)) of the case $\Delta\alpha=6.0, R=70$ (black circles, second order phase transition) with the case $\Delta\alpha=10.0, R=100$ (blue diamonds, first order phase transition) for $\gamma=3.0$. A more exhaustive description of ES emerges from the exploration of the full parameter space $\lambda-R$. The results are shown in Fig.~\ref{fig:Leyva2012-1}(b), where the values of $r$, for both forward and backward simulations are reported. As it can be seen, the plane $\lambda-R$ can be clearly divided in two areas (denoted as $I$ and $II$ in the Figure) where the transition is of the first and second order, respectively. The striped portion of the area where the transition is of the first order marks the region where the hysteresis phenomenon is observed.

\begin{figure}
\centering\includegraphics[width=0.9\textwidth]{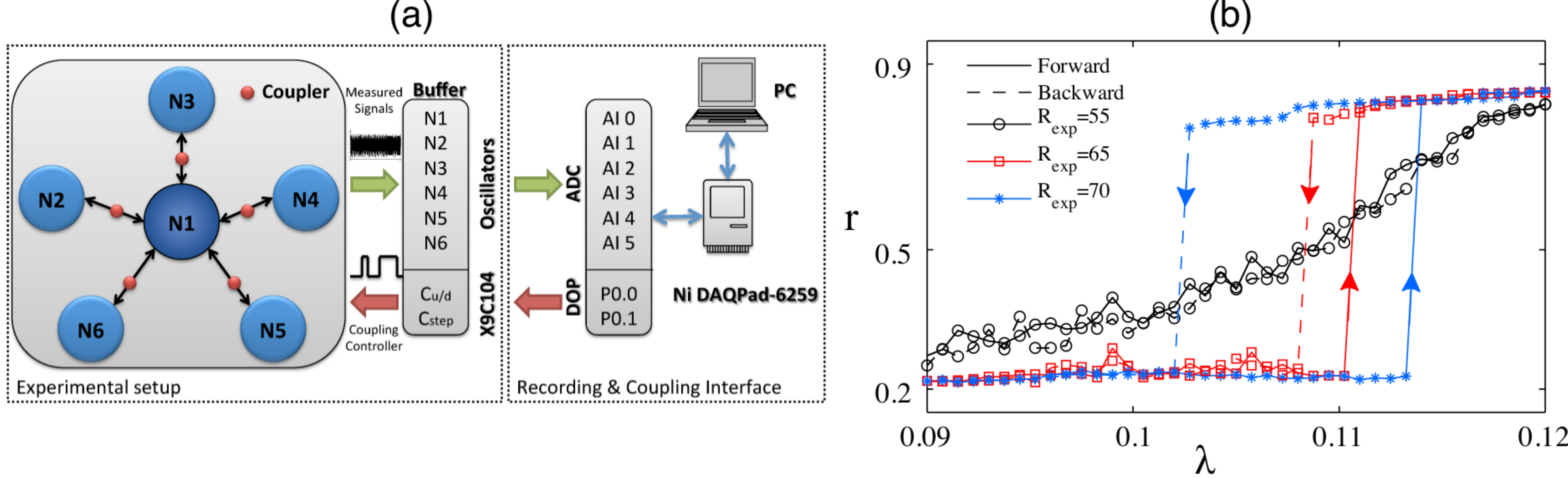}
\caption{ (Color online). (a) Sketch of the experimental setup, where six R\"ossler circuits (blue nodes) are bidirectionally connected in a star
configuration. (b)  $r$ vs.~$\lambda$ for the configuration shown in panel (a). The values of the internal resistance  are: $R_{\mbox{\scriptsize exp}}=55$ (black circles,  second-order phase transition), $R_{\mbox{\scriptsize exp}}=65$ (red squares, first order phase transition with narrow hysteresis), and $R_{\mbox{\scriptsize exp}}=70$ (blue stars, first order phase transition with wide hysteresis).
Reprinted from Ref.~\cite{Leyva2012}. $\copyright$  2012 by the American Physical Society.}
\label{fig:Leyva2012-2}
\end{figure}

Motivated by the numerical study of Eqs. (\ref{eq:Ross_model}), Ref.~\cite{Leyva2012} provided the first experimental evidence of ES, with the electronic network schematically shown in Fig.~\ref{fig:Leyva2012-2}(a). The experiment consists of six piecewise R\"ossler circuits operating in the chaotic regime, which are labeled as $N1$, $N2$, \dots, $N6$. The details of the circuit construction, as well as the qualitative equivalence with the model of Eqs.~(\ref{eq:Ross_model}) are available in Refs.~\cite{Pisarchik2006,Pisarchik2009}.

The circuits are arranged in a star-like configuration, which, on its turn, represents {\it the maximally heterogeneous structure} available for small ensembles. All the chaotic oscillators have the same internal $R_{\mbox{\scriptsize exp}}$, to ensure that they work in an almost identical dynamical regime. By a fine tuning of the values of the capacitors, circuits are configured such that the central node $N1$ oscillates with a mean frequency of $3333$~Hz, and the leaves nodes $N2$, \dots, $N6$ are set with frequencies in the range of $2240$~$\pm$~$200$~Hz. Notice that, due to the experimental variability, the frequencies of the oscillators suffer also from unavoidable dispersion.

Once the data are stored, processing of them is made to obtain the equivalent instantaneous phases. These latter quantities are calculated by defining the instantaneous phase $\theta_i(t)$ of each oscillator $i$ as $\theta_i(t)=\ 2 \pi l_i + 2 \pi \frac{t-t_{l_i}}{t_{l_i+1}-t_{l_i}}$ in each interval $t_{l_i} \leq t < t_{l_i+1}$, where $t_{l_i}$ is the instant at which the $l_i^{th}$ crossing of the $i$-th oscillator with its Poincar\'e section occurs.
Fig.~\ref{fig:Leyva2012-2}(b) reports the experimental synchronization diagram for three values of the parameter $R_{\mbox{\scriptsize exp}}$, both for forward and backward variations of the coupling strength $\lambda$. As in the numerical case, the dynamical regime tuned by $R_{\mbox{\scriptsize exp}}$ determines whether the transition is of the first or second order.

Experimental evidence of ES has been recently given also in coupled
chemo-mechanical systems, and more specifically in mercury
beating-heart (MBH) oscillators
\cite{Kumar2015}. Ref. \cite{Kumar2015} considered a star network configuration, where a central MBH oscillator is connected  with other three MBH leaves oscillators. Furthermore, the experimental parameters are tuned properly, in a way that the natural frequencies of each oscillator is proportional to the number of its links.
Not only Ref.~\cite{Kumar2015} gives evidence that a gradual increase
of the coupling strength results in an abrupt and irreversible
(first-order like) transition, but it also proved how to engineer magnetic-like states of synchronization, by the use of an external signal.

\subsubsection{Control of explosive synchronization}

A characteristic feature of continuous phase transitions is that after the transition point, the order parameter
increases gradually with the coupling strength, which makes the dynamics of the system rather predictable. In contrast,
the order parameter of a discontinuous transition has a jump at a critical point, which renders predictability
of the dynamics more difficult to achieve. Control of ES means to change the
transition from first-order to second-order like, avoiding big variations at the transition point.

In the first studies of ES, it was shown that the phase
transition is  first-order-like if the degree distribution exponent $\gamma$ of the SF network satisfies
$2<\gamma<3$, is of the second order if $\gamma>3$, while for $\gamma=3$ a hybrid phase transition is
found \cite{Gomez-Gardenes2011,Coutinho2013}. Later on, it was revealed that the conversion can be also induced
by other factors such as the degree-degree correlation, degree-frequency correlation, disassortativity, etc.
\cite{Sonnenschein2013,Li2013,Hu2014,Sendina-Nadal2015,Zhou2015,Pinto2015}.
For instance, Su et al. extended the condition of ES from strong correlation
($\omega_i=k_i$) to weak correlation ($\omega_i=f(k_i)$), and revealed that there is a conversion from
continuous to discontinuous takes place depending on the choice of the correlation function $f$ \cite{Hu2014}.
Li et al. showed that local degree-degree correlations (and precisely the degree of disassortative mixing)
contribute primarily to ES \cite{Li2013}. Sendi\~na-Nadal  et al. discussed the effects of degree correlations, and showed that high levels of
positive and negative mixing consistently induce a second-order phase transition,
while moderate values of assortative mixing enhance the irreversible nature of ES in
SF networks \cite{Sendina-Nadal2015}. Zhou et al. discussed explosive synchronization with
asymmetric frequency distributions and found that the synchronization transition converts from the first order
to the second order as the central frequency shifts toward the positive direction \cite{Zhou2015}. Pinto and Saa
found a conversion when only a small part of the vertices of the network is subjected to a degree-frequency correlation \cite{Pinto2015}.

However, in real situations the network properties (such as the degree distribution and assortativity) are often
unchangeable, and therefore, other control techniques have to be implemented to attain the conversion.
A first work in this line was done by Zhang et al., where the control is
implemented by breaking the suppressive rule of Ref.~\cite{Zhang2014} already introduced in Subsection~\ref{freqwei} (Eq.~(\ref{eq:Zhang2014_rule})).

According to the suppressive rule, local synchronization can only occur for those neighboring pairs of
nodes satisfying Eq.~(\ref{eq:Zhang2014_rule}), while those nodes violating Eq.~(\ref{eq:Zhang2014_rule})
will actually suppress the formation of local synchronized clusters. Reference~\cite{Zhang2014} showed that an
effective way to break the suppressive rule is by randomly exchanging the frequencies of two nodes $i$ and $j$.

\begin{figure}
\begin{center}
\includegraphics[width=0.5\textwidth]{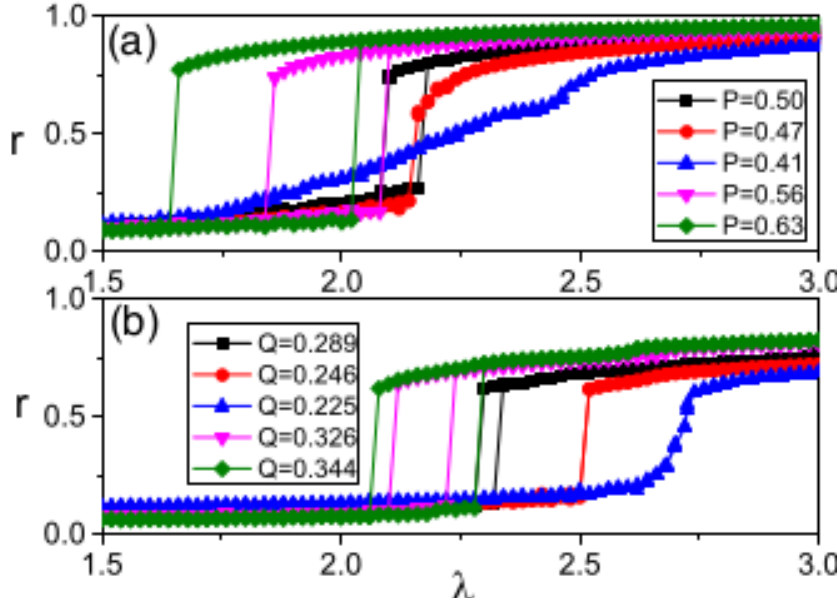}
\end{center}
\caption{(Color online). Synchronization diagrams obtained by exchanging the natural frequencies of
 connected nodes (see the text for the used exchanging strategy). (a) $g(\omega)$ is taken as a Gaussian distribution
 with both positive and negative $\omega$. ``Squares'' denote the
case where no exchanges of frequencies are performed, ``circles'' and ``up triangles''
represent the cases of decreasing $P$ to $P=0.47$, and $0.41$, respectively, and
``down triangles'' and ``diamonds'' represent the cases of increasing $P$ to $P=0.56$, and
$0.63$, respectively. (b) $g(\omega)$ is taken as a Lorentzian distribution with only $\omega>0$.
``Squares'' denote the case where no exchanges of frequencies are performed with $Q=0.289$, ``circles'' and
``up triangles'' represent the cases of decreasing $Q$ to $Q=0.246$, and $0.225$, respectively,
and ``down triangles'' and ``diamonds'' represent the cases of increasing $Q$ to $Q=0.326$,
and $0.344$, respectively. Reprinted from Ref.~\cite{Zhang2014}, published under CC-NY-ND license. \label{Fig:Zhang:2014}}
\end{figure}

The idea of Ref.~\cite{Zhang2014} is as follows:
if exchanging $\omega_i$ and $\omega_j$ would
result in a value of $S_{ij}$ [Eq.~(\ref{eq:sij})] smaller than the original one, the exchange procedure is accepted. Otherwise,
no operation is made. Ref. ~\cite{Zhang2014} further introduced the quantity
\begin{equation}
\label{measure-global}
P=\frac{1}{2}-\frac{1}{2N\langle k\rangle}\sum_{i,j}A_{ij} \sgn(\omega_i\omega_j),
\end{equation}
which represents the fraction of the pairs of connected oscillators
with opposite sign in $\omega$ and which measures the effect of the exchanging.
Figure~\ref{Fig:Zhang:2014}(a) reports the transitions to synchronization observed at different
values of  $P$. It is easy to see that decreasing $P$ initially  destroys the hysteretic
loop (see the curve with ``circles''), and eventually produces a transition with no jump (see the curve
with ``up triangles''), i.e. it converts ES into a second-order phase transition.
% figure 5

Together with $P$, Zhang et al. also introduced the quantity
\begin{equation}
\label{measure-Lorenz}
Q=\frac{1}{N\langle k\rangle}\sum_{i,j}A_{ij}Y_{ij},
\end{equation}
where $Y_{ij}\equiv\frac{|\omega_i-\omega_j|}{|\omega_i|+|\omega_j|}$. The quantity $Q$  measures the average connection
between nodes with large and small $\omega$, and repeating the
exchanging process driven by Eq.~(\ref{eq:sij}), it is possible to decrease or increase $Q$.
Figure~\ref{Fig:Zhang:2014}(b) shows the resulting synchronization transitions at different values of $Q$.
Similarly to Fig.~\ref{Fig:Zhang:2014}(a), the progressive decreasing of $Q$ has the effect of eventually
leading to a second-order transition.

In some circumstances, however, the natural frequencies of the oscillators are also not changeable.
To overcome this limitation, Zhang et al. recently presented another efficient method to control ES \cite{Zhang2016}.
The idea is to introduce a small
fraction of contrarians to suppress the growing of larger clusters, in contrast to the conformists.
There are two approaches to define a conformist or a contrarian. In the first method, a contrarian
oscillator will receive interactions from its neighbors via a negative coupling strength, while a conformist
oscillator will receive interactions from its neighbors via a positive coupling strength
\cite{Hong2011,Louzada2012}. The model is as follows
\begin{equation}
\label{Kuramoto1}
\dot{\theta_{i}}=\omega_i+\frac{\lambda_i|\omega_i|}{k_i}
\sum_{j=1}^NA_{ij}\sin(\theta_j-\theta_i), \quad
i=1,\ldots,N
\end{equation}
where the conformists have positive $\lambda_i$ while the contrarians have negative $\lambda_i$. For
simplicity, the coupling strength $\lambda_i$ is set at the same amplitude $\lambda>0$ with
$\lambda_i=\lambda$ for all conformists and $\lambda_i=-\lambda$ for all contrarians.

In the second method, a contrarian oscillator gives negative coupling to each of its neighbors, while
a conformist oscillator gives a positive coupling to each of its neighbors
\cite{Borgers2003,Restrepo2006,Zanette2005}. The model can be written as follows
\begin{equation}
\label{Kuramoto2}
\dot{\theta_{i}}=\omega_i+\frac{|\omega_i|}{k_i}
\sum_{j=1}^N\lambda_jA_{ij}\sin(\theta_j-\theta_i), \quad
i=1,\ldots,N
\end{equation}
where the conformists contribute positive $\lambda_j$, while the
contrarians contribute negative $\lambda_j$. Once again, all strengths
$\lambda_j$ are set at the same amplitude $\lambda>0$, with
$\lambda_j=\lambda$ for all conformists and $\lambda_j=-\lambda$
for all  contrarians.

% figure 6
\begin{figure}
\begin{center}
\includegraphics[width=0.5\textwidth]{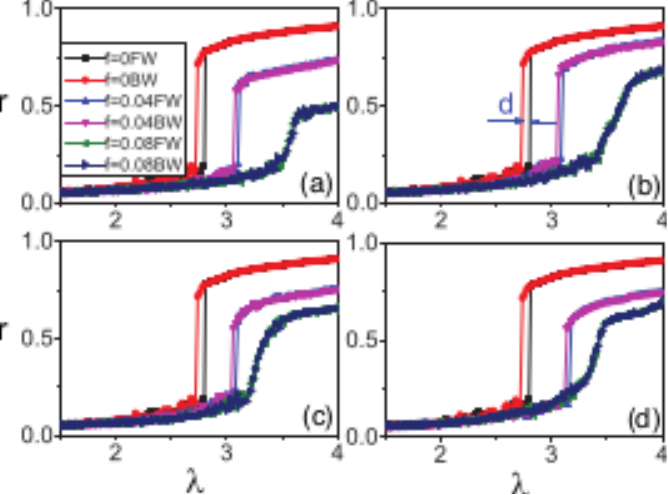}
\end{center}
\caption{(Color online). Forward and backward
synchronization transitions for a Erd\H{o}s-R\'{e}nyi network with $N=500$
and $\langle k\rangle=6$, where ``squares'' and ``circles'' represent the
forward and backward curves at $f=0$, the ``up triangles'' and
``down triangles'' represent the forward and backward curves at $f=0.04$,
and the ``left triangles'' and
``right triangles'' represent the forward and backward curves at $f=0.08$,
respectively. The panels report the following cases: the random- case1 (a), the random-case2 (b), the
hetero-case1 (c), and the hetero-case2 (d). See text for definition. Reprinted from Ref.~\cite{Zhang2016} with permission of IOP.
\label{Fig:Zhang:2016}}
\end{figure}

The first case considered is that of random uniformly distributed contrarians, with the definition of contrarian
in Eq.~(\ref{Kuramoto1}) (this setting is called {\sl the random-case1}). Let $f$ represents the fraction of
contrarians, i.e. the ratio between the number of contrarians and the total number of oscillators.
Figure~\ref{Fig:Zhang:2016}(a) shows how the order parameter $r$ changes with $\lambda$
for $f=0, 0.04$ and $0.08$, respectively. It is easy to see that there are hysteretic loops for the
cases of $f=0$
and $f=0.04$, but no loop is present for the case of $f=0.08$, indicating that the first-order transition to
synchronization has been changed into a second-order transition. A similar scenario, reported in Fig.~\ref{Fig:Zhang:2016}(b), characterizes
the case of random
uniformly distributed contrarians with
the definition of Eq.~(\ref{Kuramoto2}), called {\sl the random-case2}.
In order to inspect how the distribution of contrarians influences the controlling effect,
contrarians are heterogeneously distributed in the network. Figure~\ref{Fig:Zhang:2016}(c) (Eq.~\ref{Fig:Zhang:2016}(d))
shows the case of heterogeneously distributed
contrarians with the definition given in Eq.~(\ref{Kuramoto1}) (Eq.~(\ref{Kuramoto2}))
called {\sl the hetero-case1} ({\sl the hetero-case2}). Comparing the
four panels of Fig.~\ref{Fig:Zhang:2016}, it is clear that in all cases a
 second-order phase transition is obtained when $f=0.08$.

From Fig.~\ref{Fig:Zhang:2016}, it can be also noticed  that the transition point $\lambda_{c}$
increases with the increase of $f$. In this way, the induced second-order transition point $\lambda_{c2}$
 will be much larger than the original first-order transition point $\lambda_{c1}$.
This means that not only the first-order transition is suppressed but also the
transition point will be postponed, which is a relevant result in real systems with potential cascading
risk.

This finding may be also useful in explaining the normal functioning of brain. It is well known that there are
both excitatory and inhibitory neurons in cortical neural networks. The excitatory neurons comprise the
majority ($80\sim 90\%$) of the neuronal population, and are largely homogeneous. The inhibitory neurons
comprise only $10\%$ of the neuron population, but are extremely heterogeneous, and take the role
of controlling and coordinating the activity of large populations of local neurons
\cite{Freund1996,Klausberger2008}. One can then imagining, that the excitatory and inhibitory
neurons are in fact the conformists and contrarians, respectively, and that the existence of inhibitory neurons in brain network
is necessary not only for sustaining its normal behavior, but also for preventing its abnormal
functioning.% such as the epileptic seizures.

%%%%%%%%%%%%%%%%%%%%%%%%%%%%%%%%%%%%%%%%%%%%%%%%%%%%%%%%
\section{Conclusions and future perspectives}

After having revisited the main theory (and some applications) of explosive percolation and synchronization in complex networks,
we end this review with a few concluding remarks, that have the aim of spotting problems that still remain open to future progresses, as well as of offering a few thoughtful considerations about what are the questions of relevance that should (in our opinion, hopefully soon) attract  the attention of scientists in this area.

Explosive percolation was largely examined in Chapter~\ref{sec:percol}, and its main applications in physical systems discussed in Section~\ref{gigimelo}.
The process of percolation refers to the emergence of large-scale connectivity on an underlying network, or lattice.
The extensive studies of EP only started on 2009, following a seminal work by Achlioptas, where it was pointed out that introducing competitive mechanisms in the selection of bonds to be added to a graph,  the threshold of the percolation process can be significantly postponed, and moreover, the giant cluster emerges after a number of steps that is much smaller than the system size. As a
consequence, the order parameter exhibits an extremely abrupt jump at the percolation point, which makes this situation drastically different from classical, random, percolation.

As reported in full details in Chapter~\ref{sec:percol},  many other algorithms and models have been introduced and studied from 2009, and so far
EP has been intensively investigated on many networks topologies, including  2D lattices, 3D or high-D lattices, Bethe lattices, ER networks, SF networks, modular networks,  and real-world networks.
All these works have triggered a big debate in the community, aiming
at clarifying whether the observed transitions are really
discontinuous or not. While the existence of a seemingly discontinuous
jump in the order parameter of the original Achlioptas model is
universally recognized, careful numerical investigations of the
cooperation, phase coexistence, and nucleation (as well as strict theoretical proofs) have demonstrated that EP in that model is actually continuous in the thermodynamic limit, but it belongs to a universality class different from any other previously considered.

Furthermore (and notably),  consensus has been found on the fact that some other models of percolation (such as the k-core percolation model, the spanning cluster-avoiding model, the two-species cluster aggregation model, or the models of jamming on low-dimensional lattices) lead to truly discontinuous transitions. Therefore, it is now fully accepted that the term ``explosive''  has to refer to the unusual feature which distinguishes the critical behavior
of Achlioptas processes from both ordinary and truly discontinuous models.

There are issues that we believe still need efforts from the scientific community. One is that, undoubtedly, more theoretical and numerical studies on the scaling properties and critical exponents in EP are in order, to clarify which new features characterize the critical behavior associated to explosive percolation.
Another is the need of exploring EP in other realistic situations (real-world systems or experiments), such as, for instance the cascading failure processes in power grids. And finally, the hope is that the gathered knowledge will help devising methods to control EP, i.e. to enhance or suppress it, when necessary or when desired.

On the other side, ES has been largely discussed in Chapter~\ref{sec:expsync}, and its main applications described in Section~\ref{sec:expsyncapp}.
From a historical point of view, we saw that the discovery of ES went through a somehow tortuous
path: the community was initially apathetic to the early evidences of first order transitions to synchronization in the thermodynamic limit of a model of globally coupled phase oscillators.
Such a situation of guilty indifference persisted for long time, up to when it was clarified that a positive correlation between the natural frequencies of oscillators and the degrees of nodes could produce abrupt cooperative behaviors.
For the first time, it was realized that a complete understanding of the nature of phase transitions
intimately depended on an explicit interaction between the
network topology and the characteristic dynamics of oscillators, and therefore much attention has been
recently paid to ES, which begot today great progresses, including  theoretical understandings and experimental
confirmations.

Let us concentrate on a few issues that possibly will become hot topics on ES.

The first is extending the study of explosive transitions to the case of partial and cluster synchronization, a large area of research, indeed, having to do with cases (like brain functioning, or stock market dynamics, to quote just two examples) for which synchronization on the whole system scale is not possible nor desirable, but rather collective arrangements emerge at the system's micro- and meso-scales.
How to realize explosive transitions to these latter states from the incoherent behavior is still an open question.

The second is bridging ES with specific mixed states of coupled
oscillators.  Recently, indeed, Zhang et al. proposed a bridge
between ES and chimera state
\cite{Zhang2016b}, and found that they can both  coexist within the same coupled system.  Further studies on the relationship between ES and chimera state are certainly to be expected soon.
Furthermore, since it is now understood that ES can be considered as a
dynamical percolation process in phase space, an issue of the utmost
relevance is clarifying how one can use the results of explosive
percolation for a better control of ES. Therefore, further studies on
the relationship between ES and explosive percolation are certainly
necessary.

It is convenient to spend a few words on the novel, non-stationary, state that has been described in Section~\ref{sec:bellerophon}, the {\it Bellerophon} state.  In globally coupled nonidentical oscillators, this state can indeed emerge, made of quantized clusters where oscillators are neither phase- nor frequency-locked. Oscillators' instantaneous speeds are different within the clusters, but they are correlated and, more importantly, they behave periodically in time so that their average values are the same. So far, it has been  found that
the Bellerophon states occur in fact in a pretty wide context of globally coupled oscillators, and for various typical frequency distributions.
Physically, such a state can be regarded as a weaker form of coherence achieved by the coupled oscillators when the control parameter is at an intermediate value, or (in other words) as a transitional state between the incoherent and the phase-locked state: on the one hand the control parameter is not strong enough to completely entrain the system into the phase-locked state, on the other hand it is adequately large to maintain certain correlations among the instantaneous frequencies of oscillators.

Certainly, a lot of future work is in order to unveil Bellerophon states in other oscillatory models, as well as to check if (and under which conditions)
the hypothesis of a global coupling can be relaxed to that of a complex network architecture.
On the other side, we are totally confident that the discovery of Bellerophon states will stimulate the search for a better theoretical description and characterization, as well as a special care in seeking for such novel states in  experimental and natural systems.

Finally, another interesting issue is deepening and enhancing the knowledge of ES in physical systems, where the space embedding (and therefore the
lengths of physical links, such as synapses in real neural systems or power lines in power grids), or the multiple level character of interactions between the units (which would require a multi-layer network representation) may play a key role. The study of ES in physically embedded and multi-layer networks is still in its infancy, and we firmly believe that it should instead attract soon a lot of attention, due to its evident relevance for practical applications.

Although beyond the scope of this review, it is worth mentioning that other abrupt transitions
have been recently described in networked systems, having to do with spreading processes (like epidemics with co-infection \cite{Chen2013b,hebert2015} or re-infection \cite{Gomez-Gardenes2016} mechanisms), spontaneous recovery in dynamical networks \cite{majdandzic2014}, and social games using the inertial majority-vote model \cite{Chen2016}.

\section{Acknowledgements}

We would like to acknowledge gratefully all colleagues with whom we maintained (and are currently
maintaining) interactions and discussions on the topics covered in the report.

In particular, we would like to thank A. Amann, R. Amritkar,
F.T. Arecchi, A. Arenas, S. Assenza, V. Avalos-Gayt\'an, V. Beato, E. Ben Jacob,
G. Bianconi, B. Blasius, I. Bonamassa, J. Bragard, J.M. Buld\'u, J. Burguete, T. Carroll,
R. Criado, M. Courbage, M. Danziger, A. D\'iaz-Guilera,
A.L. Do, S.N. Dorogovtsev, E. Estrada, J. García-Ojalvo, C. del Genio,
S. G\'omez, J. G\'omez-Garde\~nes, C. Grebogi, P. Grigolini, R. Guti\'errez,
M. Hassler, S. Havlin, H.G.E. Hentschel, A. Hramov, D-U. Hwang,
R. Jaimes-Re\'ategui, S. Jalan, M. Jusup, A. Koronovskii, J. Kurths,
V. Latora, C. Letellier, D. Li, R. Livi,
H. Mancini, J.H. Martinez, D. Maza, J.F.F Mendes, R. Meucci, E. Mihaliuk, G.
Mindlin, Y. Moreno, O. Moskalenko, A. Navas, M. Newmann, Y. Ofran,
S. Olmi, D. Papo, D. Paz\'o, L.M. Pecora, M. Perc, T. Pereira, A. Pikovski,
A.N. Pisarchick, I. Procaccia, T. Qiu, A.A. Rad, M. Romance, M. Rosenblum, R. Roy, D. de
Santos-Sierra, J.R. Sevilla Escoboza, K. Showalter, M. Small,
S. Solomon, F. Sorrentino, R. Stoop, S. Strogatz, K. Syamal Dana,
A. Torcini, V. Tsioutsiou, J.A. Villacorta-Atienza, V. Vlasov,
L. Wang, D. Yu, M. Zanin, X. Zhang.

The huge number of inspiring discussions with them (and their sharing with us of
unpublished results on the subject) opened up our minds, and are largely responsible for having encouraged us to write this review.

Furthermore, we would like to acknowledge the financial support received by:
1)  the Spanish MINECO under projects FIS2012- 38949-C03-01 and FIS2013-41057-P;
2) the NNSF of China under Grant Nos. 11305062, 11135001, 11375066,
and 61374169; JA, IL, and ISN acknowledge support from GARECOM, Group of Research Excellence URJC-Banco de Santander.

\newpage
\section*{References}
%\bibliography{references}

\end{document}